\documentclass[11pt]{ncsudoctoralthesis}

\usepackage{amsfonts}
\usepackage{amssymb}
\usepackage{fancyvrb}
\usepackage{epsfig}
\usepackage{graphicx}
\usepackage{latexsym}
\usepackage{bm}
\usepackage{amsmath}
\usepackage{multirow}
\usepackage{rotating}
\usepackage{pdflscape}
\usepackage{mathrsfs}
\usepackage{appendix}
\usepackage{wrapfig}

\newcount\listnorom
\newcommand\listromanDE{\global\advance \listnorom by 1
{\lowercase\expandafter{(\romannumeral\listnorom)}\ }}
\newcommand\newlistroman{\listnorom=0}

\newcommand{\xx}[1]{\!\times\!10^{#1}}

\newcommand{\FP}{Analytic Precursor Approximation}
\newcommand{\FPshort}{APA}

\newcommand{\MFA}{magnetic field amplification}
\newcommand{\rel}{relativistic}
\newcommand{\NL}{nonlinear}
\newcommand{\CR}{cosmic ray}

\newcommand{\nonrel}{non\-rel\-a\-tiv\-is\-tic}

\newcommand{\mc}{Monte Carlo}
\newcommand{\MC}{Monte Carlo}
\newcommand{\syn}{synchrotron}

\newcommand{\kmps}{km~s$^{-1}$}
\newcommand{\pcc}{cm$^{-3}$}
\newcommand{\muG}{$\mu$G}

\newcommand{\SC}{self-consistent}
\newcommand{\SCly}{self-consistently}
\newcommand{\Pcr}{P_\mathrm{cr}}

\newcommand{\Pcrhat}{P_\mathrm{cr}}
\newcommand{\Pth}{P_\mathrm{th}}
\newcommand{\Pthzero}{P_\mathrm{th0}}
\newcommand{\Pthtwo}{P_\mathrm{th2}}
\newcommand{\Pwzero}{P_\mathrm{w0}}
\newcommand{\Pwone}{P_\mathrm{w1}}
\newcommand{\Pwtwo}{P_\mathrm{w2}}
\newcommand{\Fwzero}{F_\mathrm{w0}}

\newcommand{\Fwtwo}{F_\mathrm{w2}}
\newcommand{\Pp}{P_\mathrm{cr}}
\newcommand{\Ppzero}{P_\mathrm{p0}}

\newcommand{\Pptwo}{P_\mathrm{p2}}
\newcommand{\wpzero}{w_\mathrm{p0}}

\newcommand{\wptwo}{w_\mathrm{p2}}
\newcommand{\Rtot}{r_\mathrm{tot}}
\newcommand{\rtot}{r_\mathrm{tot}}
\newcommand{\Rsub}{r_\mathrm{sub}}
\newcommand{\rsub}{r_\mathrm{sub}}
\newcommand{\Mazero}{M_\mathrm{A0}}
\newcommand{\Mszero}{M_\mathrm{s0}}
\newcommand{\Msone}{M_\mathrm{s1}}

\newcommand{\Dfeb}{d_\mathrm{FEB}}
\newcommand{\xfeb}{x_\mathrm{FEB}}
\newcommand{\xFP}{x_\mathrm{APA}}
\newcommand{\xtr}{x_\mathrm{tr}}
\newcommand{\fcr}{f_\mathrm{cr}}

\newcommand{\pmax}{p_\mathrm{max}}

\newcommand{\rgzero}{r_\mathrm{g0}}
\newcommand{\rgone}{r_\mathrm{g1}}
\newcommand{\tacc}{\tau_\mathrm{acc}}

\newcommand{\fAlf}{f_\mathrm{Alf}}
\newcommand{\Beff}{B_\mathrm{eff}}
\newcommand{\Bseed}{\Delta B_\mathrm{seed}}
\newcommand{\Bism}{B_\mathrm{ism}}
\newcommand{\Befftwo}{B_\mathrm{eff2}}
\newcommand{\Btrend}{B_\mathrm{trend}}
\newcommand{\Qesc}{Q_\mathrm{esc}}
\newcommand{\qesc}{q_\mathrm{esc}}
\newcommand{\taucoll}{\tau_\mathrm{coll}}

\newcommand{\momentumflux}{\Phi_P}
\newcommand{\upstreammomentumflux}{\Phi_\mathrm{P0}}

\newcommand{\energyflux}{\Phi_E}
\newcommand{\upstreamenergyflux}{\Phi_\mathrm{E0}}

\newcommand{\gammabar}{\bar{\gamma}}
\newcommand{\deltabar}{\bar{\delta}}
\newcommand{\pvector}{{\bf p}}
\newcommand{\vvector}{{\bf v}}
\newcommand{\kdiss}{k_d}
\newcommand{\rg}{r_g}
\newcommand{\MS}{M_\mathrm{s}}

\newcommand{\hotism}{hot ISM ($T_0=10^6$~K)}
\newcommand{\coldismNoT}{cold ISM}
\newcommand{\hotismNoT}{hot ISM}

\newcommand{\heatpar}{\alpha_H}
\newcommand{\Alf}{Alfv\'{e}n}
\newcommand{\alf}{Alfv\'en} 
\newcommand{\Malf}{M_\mathrm{alf}} 
\newcommand{\Lbar}{L}

\renewcommand\Im{\operatorname{Im}}

\newcommand{\lcor}{l_\mathrm{cor}}

\newcommand{\lbohm}{\lambda_\mathrm{Bohm}}
\newcommand{\Dbohm}{D_\mathrm{Bohm}}
\newcommand{\lres}{\lambda_\mathrm{res}}
\newcommand{\kres}{k_\mathrm{res}}
\newcommand{\pres}{p_\mathrm{res}}
\newcommand{\lss}{\lambda_\mathrm{ss}}
\newcommand{\rss}{r_\mathrm{ss}}
\newcommand{\Bss}{B_\mathrm{ss}}
\newcommand{\kmin}{k_\mathrm{min}}
\newcommand{\kmax}{k_\mathrm{max}}

\newcommand{\gfunc}{\mathcal{M}}

\newcommand{\nrgrowthrate}{\Gamma_\mathrm{nr}}
\newcommand{\resgrowthrate}{\Gamma_\mathrm{res}}

\newcommand{\parcompamp}{\alpha_g}
\newcommand{\parcompwav}{\beta_g}
\newcommand{\paramplif}{\gamma_g}
\newcommand{\parcasc}{\delta_g}
\newcommand{\pardiss}{\varepsilon_g}

\newcommand{\kstar}{k_\mathrm{*}}
\newcommand{\Bls}{B_\mathrm{ls}}
\newcommand{\kwhirl}{k_\mathrm{v}}
\newcommand{\ptr}{p_\mathrm{tr}}

\newcommand{\pinj}{p_\mathrm{inj}}

\newcommand{\Pmax}{p_\mathrm{max}}
\newcommand{\fValf}{f_\mathrm{alf}}
\newcommand{\Pwtot}{P_w}

\newcommand{\TP}{test-particle}
\newcommand{\Diff}{D(x,p)}
\newcommand{\BFA}{$B$-field amplification}
\newcommand\Qlin{quasi-linear}
\newcommand{\Valf}{v_A}

\newcommand{\SNRJ}{SNR~RX~J1713.7-3946}
\newcommand{\Unmod}{unmodified}

\newcommand{\Epmax}{E^\mathrm{max}_p}
\newcommand{\EpmaxUM}{E^\mathrm{max}_p|_\mathrm{UM}}
\newcommand{\EpmaxNL}{E^\mathrm{max}_p|_\mathrm{NL}}

\newcommand{\pNL}{p_\mathrm{NL}^\mathrm{max}}
\newcommand{\pUM}{p_\mathrm{UM}^\mathrm{max}}
\newcommand{\ppmax}{p^\mathrm{max}_p}
\newcommand{\Lfeb}{L_\mathrm{FEB}}
\newcommand{\usk}{u_\mathrm{sk}}
\newcommand{\Bsk}{B_\mathrm{sk}}
\newcommand{\uBwt}{\left< u(x)B(x) \right >}

\newcommand{\gamray}{$\gamma$-ray}
\newcommand{\gamrays}{$\gamma$-rays}

\begin{document}

\newcommand{\puttitle}[0]{Modeling Magnetic Field Amplification in
      Nonlinear Diffusive Shock Acceleration}

\title{\puttitle}
\author{Andrey Vladimirov}

\degreeyear{2009}
\numberofmembers{4}

\degree{Doctor of Philosophy}
\chair{Dr. Donald Ellison}
\memberII{Dr. Stephen Reynolds}
\memberIII{Dr. James Selgrade}
\memberIV{Dr. Albert Young}

\newpage

\thispagestyle{empty}
\abstractname
\vskip .5cm
\ssp
\noindent 
VLADIMIROV, ANDREY. \puttitle. (Under the direction of Dr. Donald C. Ellison.)

\renewcommand{\baselinestretch}{2}

\vspace{0.2in}

\dsp

This research was motivated by the recent observations
indicating very strong magnetic fields at some supernova
remnant shocks, which suggests in-situ generation
of magnetic turbulence.
The dissertation presents a numerical model of collisionless 
shocks with strong amplification of stochastic magnetic fields,
self-consistently coupled to efficient shock acceleration of charged particles.
Based on a Monte Carlo simulation of particle transport and 
acceleration in nonlinear shocks, the model describes
magnetic field amplification using the state-of-the-art analytic models
of instabilities in magnetized plasmas in the presence of
non-thermal particle streaming.
The results help one understand the complex nonlinear connections
between the thermal plasma, the accelerated particles and the stochastic
magnetic fields in strong collisionless shocks. 
Also, predictions regarding the efficiency of particle acceleration 
and magnetic field amplification, the impact of magnetic field amplification on
the maximum energy of accelerated particles, and the compression
and heating of the thermal plasma by the shocks are presented. Particle
distribution functions and turbulence spectra derived 
with this model can be used to calculate
the emission of observable nonthermal radiation.

%
%

\field{Physics}

\campus{Raleigh, North Carolina}

\maketitle

\begin{frontmatter}

\begin{dedication}
\vskip .5cm

This dissertation dedicated to my family.
To my mother, whose hard work and care
have made my walk through the early life an easier one.
To my father, who, by personal example, has set the highest
standards for me in education
and achievement. And to my treasured wife, whose love, beauty and
support has sustained my inspiration and fostered our happiness.
Her patience and understanding in 
my graduate school years were truly heroic.

\end{dedication}

\biographyname
\vskip .5cm

I was born in 1982 in the vast and beautiful Eurasian country of Kazakhstan, 
which was one of the 15 Soviet Union republics at that time, 
and now it is an independent state. 
By nationality I am Russian, and my native language is Russian.

I earned my B.S. (2002) and M.S. (2004) in physics 
from St.~Petersburg State Polytechnical University in Russia, 
where my concentration was physics of space, 
and I did my research under Prof.~Andrei M. Bykov 
at the Department of Theoretical Astrophysics 
of Ioffe Physical-Technical Institute.

In 2004--2009 I was a graduate student at the Department of Physics
of North Carolina State University,
working on a theoretical research project in the field of
astrophysical plasmas with Prof.~Don Ellison.

\begin{acknowledgements}
\vskip .5cm

\vskip .5cm

I am deeply grateful to my adviser, Prof.~Don Ellison, who
not only commited to educating, supervising and directing
me in this work, but also was a great source of encouragement
and support throughout my graduate work at NC State University.

This project was carried out in a close collaboration with
Prof.~Andrei Bykov from the Ioffe Physical-Technical Institute
in Russia. I value very much the priviledge of working with
him and wish to thank him for his participation in this work.

I am also appreciative of the help of the members of the 
advisory committee, who agreed to contribute their
diverse expertise and time for evaluating this 
research.

I cannot praise enough many of the NCSU staff members,
especially in the Department of Physics and the Office of
International Services, who made my graduate school experience, 
even in the more complicated situations, stressless and memorable.

Finaly, my heartfelt thanks go to the American
and international friends whom I have met in the past five years in the United
States, and whose kindness and hospitality made me feel welcome 
in this country.

\end{acknowledgements}

\tableofcontents

\newpage
\listoftables
\newpage
\listoffigures

\end{frontmatter}

\chapter{Introduction. Interstellar Shocks, Cosmic Rays and Magnetic Fields}
\label{chap-one}

What happens after a massive star explodes at the end of its life cycle
as a supernova (SN)? Why are the rims of supernova remnants (SNRs) so
thin and luminous in the radio, X-ray and gamma ray spectral ranges? 
Where and how are cosmic rays (CRs) produced? What does it take
to explain the dynamics
of matter in the most energetic systems in space, including
the cosmological large scale structure of the Universe?
The current state of affairs in astrophysics makes it clear that, 
in order to answer these questions, the phenomenon of shocks
must be studied in detail. The low gas densities 
in many cosmic environments make the shocks collisionless
(see Section~\ref{sec_collisionless}), which gives
them properties different from those of
the collisional terrestrial shocks.

Understanding shocks is as important for astrophysicists as describing
electromagnetic waves is for radio engineers. Shocks are born whenever
gases or fluids are forced to move at a supersonic speed.
They compress and heat the interstellar matter (ISM), 
transfer energy and momentum,
produce cosmic rays that fill and affect the Universe, and, as recent
observations show, shocks may produce and strongly amplify turbulent
magnetic fields. Electromagnetic radiation from processes in shocks 
is a powerful diagnostic of the conditions in the shock-generating
systems.

\newpage

\section{Shocks in hydrodynamics}

I would like to illustrate shocks with a phenomenon that we encounter
on a daily basis -- a standing shell shock in a kitchen sink formed by
the quickly running water from the tap.

\begin{figure}[hbtp]
\centering
\vskip 0.5in
\includegraphics[width=5.0in]{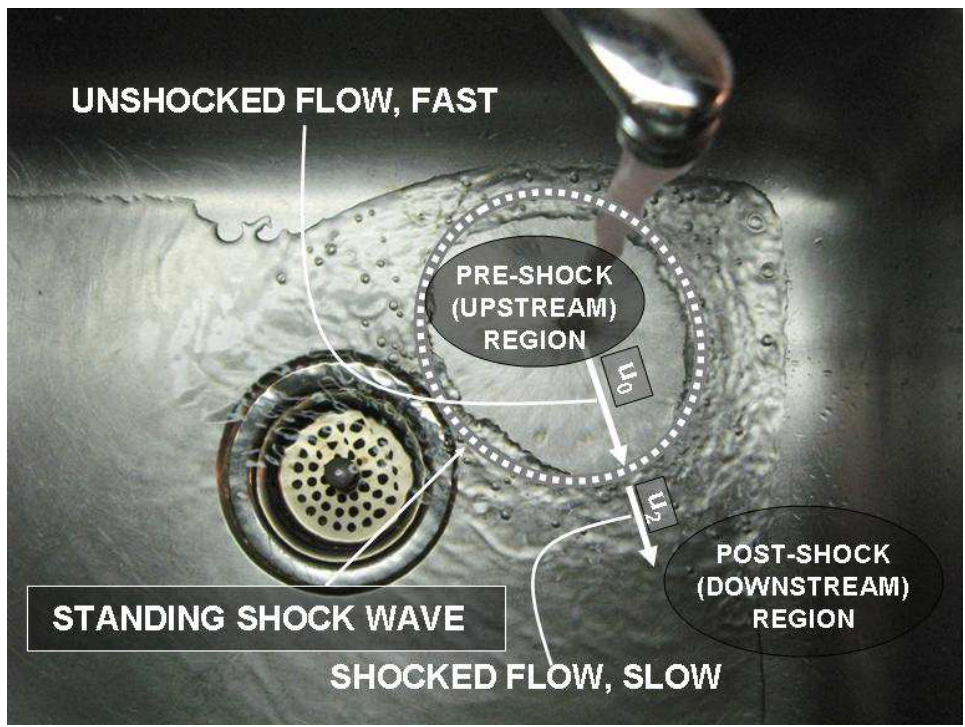}
\caption{$ $ Tap water in a sink forms a shock.}
\label{fig-shockinasink}
\end{figure}

As seen in Figure~\ref{fig-shockinasink}, 
the falling water hits the bottom of the sink and moves outward
at a speed that exceeds the speed of the surface waves in water
of the local depth. This makes a shock form, a relatively stationary enclosed
boundary, at which the speed of the water flow abruptly drops, and the depth
increases. The direction in of the shock's apparent motion depends on the choice
of the observer's reference frame, but let us adopt a convention that
unambiguously determines the {\it direction of shock propagation}. 
I will define the latter as the direction
in which the boundary between the unshocked and shocked media
moves with respect to the unshocked medium.
In the case of the shock in a sink, the unperturbed medium
is inside of the circular shell, and it moves outward. Therefore,
the shock is directed inward (i.e., any small arc of the shock boundary
is moving towards the center with respect to the
water inside the boundary). The arrows show the
velocity of the water with respect to the shock.

A similar inward-directed shock exists in the Solar System: 
the Solar wind, composed of fast charged particles emitted by the Sun, 
moves radially outward and collides with the cold 
interstellar material approximately 80-100 AU from the Sun
(an astronomical unit, $1$~AU$\approx 1.5\cdot 10^{11}$~cm,
is close to the distance between the Sun and the Earth). The so-called
termination shock forms there. At this thin boundary, the Solar
wind becomes compressed and heated, and its speed drops by a factor of 2-5.
Both Voyager spacecrafts recently passed through the termination shock
on their way out of the Solar System 
\cite{Voyager1_1, Voyager1_2, Voyager2_1, Voyager2_2}.


\section{Forward shock of SNRs}

After a star with an initial mass greater than approximately 8~$M_{\odot}$
($M_{\odot}\approx 2\cdot10^{33}$~g is the mass of the Sun) runs out of
its fusion fuel, or a white dwarf accreting mass from another star in 
a binary system reaches the critical mass and ignites, an explosion
will occur. This explosion, powered
either by gravity, or by thermonuclear fusion, 
is known as a supernova, and ejecting up to $10^{51}$~ergs in kinetic energy,
it can be bright enough to see with the naked eye thousands of light
years away. A remnant of a supernova in our Galaxy may remain visible to radio,
optical and X-ray telescopes for hundreds or
thousands of years after the explosion, as it expands into the interstellar
medium, cools and gradually fades. 

\begin{figure}[htb]
\centering
\vskip 0.5in
\includegraphics[width=4.0in]{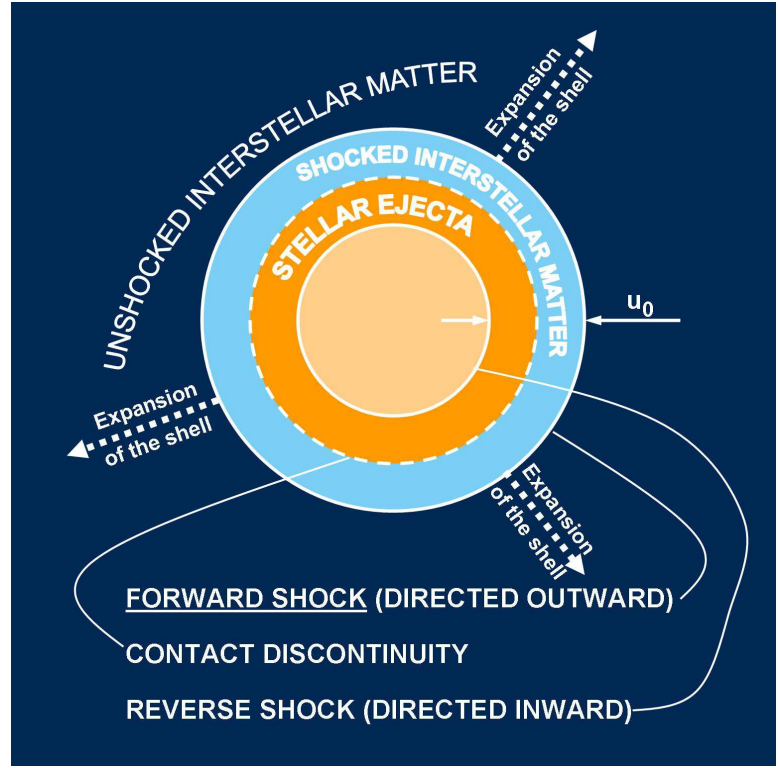}
\caption{$ $ Schematic structure of an SNR.}
\label{fig-snrstructure}
\end{figure}

Hydrodynamic simulations and observations show a common structure of flow that 
forms in SNRs, as shown in Figure~\ref{fig-snrstructure}.
The metal-rich material (ejecta) is thrown out from the star at speeds of several
thousand kilometers per second. It ploughs through the low-density ISM, and eventually
forms a strong forward shock in front of it, directed outward. 
A contact discontinuity separates the metal-rich ejecta material 
from the low-metallicity shocked ISM. Simulations
show that a reverse shock, may form in the ejecta. While the inverse
shock is directed inward (i.e., it shocks the material coming from
the interior of the reverse shock boundary), it may be physically
moving outward or inward at different stages of the SNR evolution
(e.g., \cite{EDB2005}).
Note that in Figure~\ref{fig-snrstructure}, the solid arrows show the
velocities of the unshocked medium with respect to the
forward and the reverse shocks. The dotted lines indicate the expansion of
the forward shock in time.

Of particular importance to us is the forward shock, because it can be
very strong, sonic Mach number reaching the values of several hundred.
There are two important differences between the shell shocks in 
Figure~\ref{fig-shockinasink} and Figure~\ref{fig-snrstructure}.
First, the shock in the sink is directed inward (it sweeps up water coming 
from the interior of the circle), while the SNR forward shock is
directed outward (sweeping up the interstellar matter outside of it).
The second difference is that the sink shock is stationary, i.~e., its
radius remains constant in time, but the SNR expands into a stationary
unshocked ISM, increasing the radius of the forward shock.

\begin{figure}[hbt]
\centering
\vskip 0.5in
\includegraphics[width=4.0in]{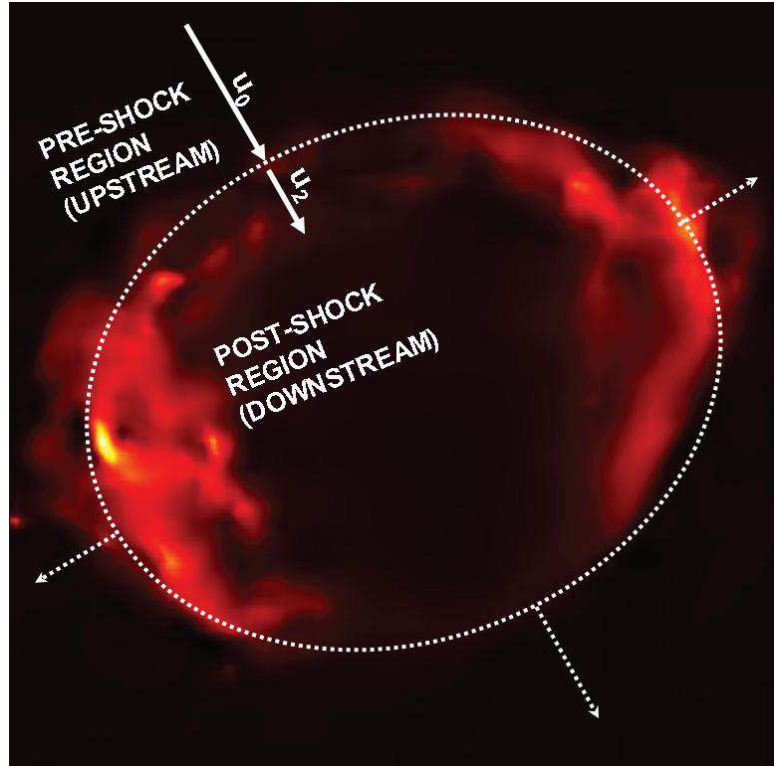}
\caption{$ $ The youngest known galactic SNR, G~1.9+0.3.}
\label{fig-g19}
\end{figure}

One of the hundreds of known and carefully observed SNRs, G~1.9+0.3, 
stands out as the youngest known supernova remnant in the Galaxy.
It was very recently identified as one by an international
team led by NCSU astronomers 
\cite{Reynolds2008_G19} and \cite{Green2008_G19}.
It provides an illustration of a typical spatially resolved
SNR imaged in X-rays. In Figure~\ref{fig-g19} (image credit: Prof.~S.~Reynolds,
NCSU, \cite{Reynolds2008_G19}), the dotted line maps the approximate
location of the forward shock\footnote{The determination of the location
of the shock is a complicated problem, and the contour 
in Figure~\ref{fig-g19} should be perceived as an artist's
impression.}, the dotted arrows indicate the direction
of the shock movement with respect to the interstellar
medium, and  the solid arrows indicate the directions
and the relative magnitudes of the velocities of the
unshocked (the long arrow) and the shocked (the short
arrow) plasma, with respect to the shock. Note that
in the following text we usually adopt the reference frame
in which the shock is at rest, and the 
plasma is flowing into the shock at a supersonic speed.
This approach corresponds to the flow directions
shown in Figures~\ref{fig-shockinasink} and \ref{fig-g19}.


\section{The concept of a collisionless shock}

\label{sec_collisionless}

Generally, gas flowing into a shock gets compressed and heated 
in a narrow region. But how narrow can this region be for a
shock in an astrophysical plasma?
In order to change the density, bulk speed and temperature, the gas particles
must experience a few strong collisions, and the thickness of the shock can
therefore be estimated as the mean free path of particles between collisions. Indeed,
for shocks in dense gases (for example, air) a particle mean free path
is comparable to the shock thickness. However, in an attempt
to apply the same reasoning to interstellar or interplanetary
shocks, one runs into a complication.

For a plasma consisting of fully ionized hydrogen,
the cross section of Coulomb collisions between protons
is formally infinite \cite{SpitzersBook}, 
but we can roughly estimate the cross
section of collisions that are strong enough to change
the energy of the particles significantly. If by 'significantly' one means
that the change energy due to collision must be comparable to
the thermal energy, then the protons must approach each other
within a distance $r_c$ such that
\begin{equation}
  \label{eq_sigcoll}
  \frac{e^2}{r_c} = k_B T.
\end{equation}
Here and in the rest of the equations in this dissertation, 
the CGS system of units is adopted. The quantity
$e$ is the elementary charge, and the left-hand side of
Equation~(\ref{eq_sigcoll}) is the electrostatic potential energy of two
protons separated by the distance $r_c$. The right-hand side
is the characteristic thermal energy of protons in a gas of
temperature $T$ ($k_B$ is the Boltzmann constant).
This gives a rough estimate of the collision cross section
\begin{equation}
  \sigma = \pi r_c^2=\frac{\pi e^4}{k_B^2 T^2}
\end{equation}
and of the mean free path
\begin{equation}
  \Lambda = \frac{1}{\sigma n}=\frac{k_B^2 T^2}{\pi e^4 n},
\end{equation}
where $T$ is the temperature and $n$ is the number density of the gas.
For conditions typical for the Solar Wind in the 
near Earth space, $n \sim 4$~cm$^{-3}$, $T\sim 10^6$~K, 
which gives $\Lambda \sim 3\cdot 10^{16}$~cm$=2\cdot 10^3$~AU.
This distance is much greater than the size of the Solar System,
which means that shocks just do not have room to form in the Solar wind 
near the Earth. However, spacecraft observations clearly indicate
numerous interplanetary shocks of various strengths traversing
the Solar System. Measurements reveal that the interplanetary
shocks are much thinner then the number above: observed 
thicknesses are around $\Lambda \sim 10^7-10^{10}$~cm 
\cite{Smith1983}.

These observational data are successfully explained by the theory
of collisionless shocks, which assumes that in the transition region
of the shock, particles collide not with each other, but
with inhomogeneities of magnetic fields. This shrinks the thickness
of the transition region down to the scales of (multiple) 
proton gyroradii (see Section~6.4 of \cite{Kulsrud2005}). 
A shock in which collisions between particles
play a negligible role compared to the dynamics of the particles
in stochastic magnetic fields is called \textit{a collisionless shock},
and the term \textit{collisionless plasma} is widely used
to define the systems in which similar conditions exist.

An interesting property of collisionless plasmas is that,
due to the absence of particle-particle collisions, the 
time scales of thermalization of non-equilibrium
energy distributions of particles are extremely large.
This allows for the existence and sustainability of a superthermal
component in the particle distribution (i.e., energetic particles).
Present research, along with other
models, shows that the superthermal particles may be not just
a minor admixture to the thermal particle pool, but, on the contrary, 
they may dominate the dynamics of a collisionless shock. This
assertion is explained in the following two sections.

\section{Cosmic rays}

Cosmic rays (CRs) are charged particles, first seen as radiation coming
from space in a balloon experiment performed by Victor Hess in 1912,
and identified as charged nuclei by Phyllis Freier and 
others in 1948 \cite{Freier1948}.
The spectrum of these particles spans many decades in particle energy
($10^{7}$ to $10^{20}$~eV (!) per nucleus) as well as in
flux (from $1$~cm$^{-2}$s$^{-1}$sr$^{-1}$ for energies
of $1$~GeV and above, down to 1 particle per square kilometer per
century for energies over $10^{20}$~eV \cite{Auger2008}).

From the multitude of observational data on CRs, it is known that
the lower energy CRs come from the Sun, and the higher
energy CRs (over $1$-$10$~GeV) are of Galactic origin.
CRs are therefore the second most important source
of information about deep space after electromagnetic radiation.
The problem of measuring and explaining the spectrum, 
composition, temporal variation and directional distribution
of CRs is extensive and longstanding. It requires answering
two major questions: how CRs are produced, and what happens
to them en route from the source to the detector on Earth.

\begin{figure}[p]
\centering
\vskip 0.5in
\includegraphics[height=6.8in]{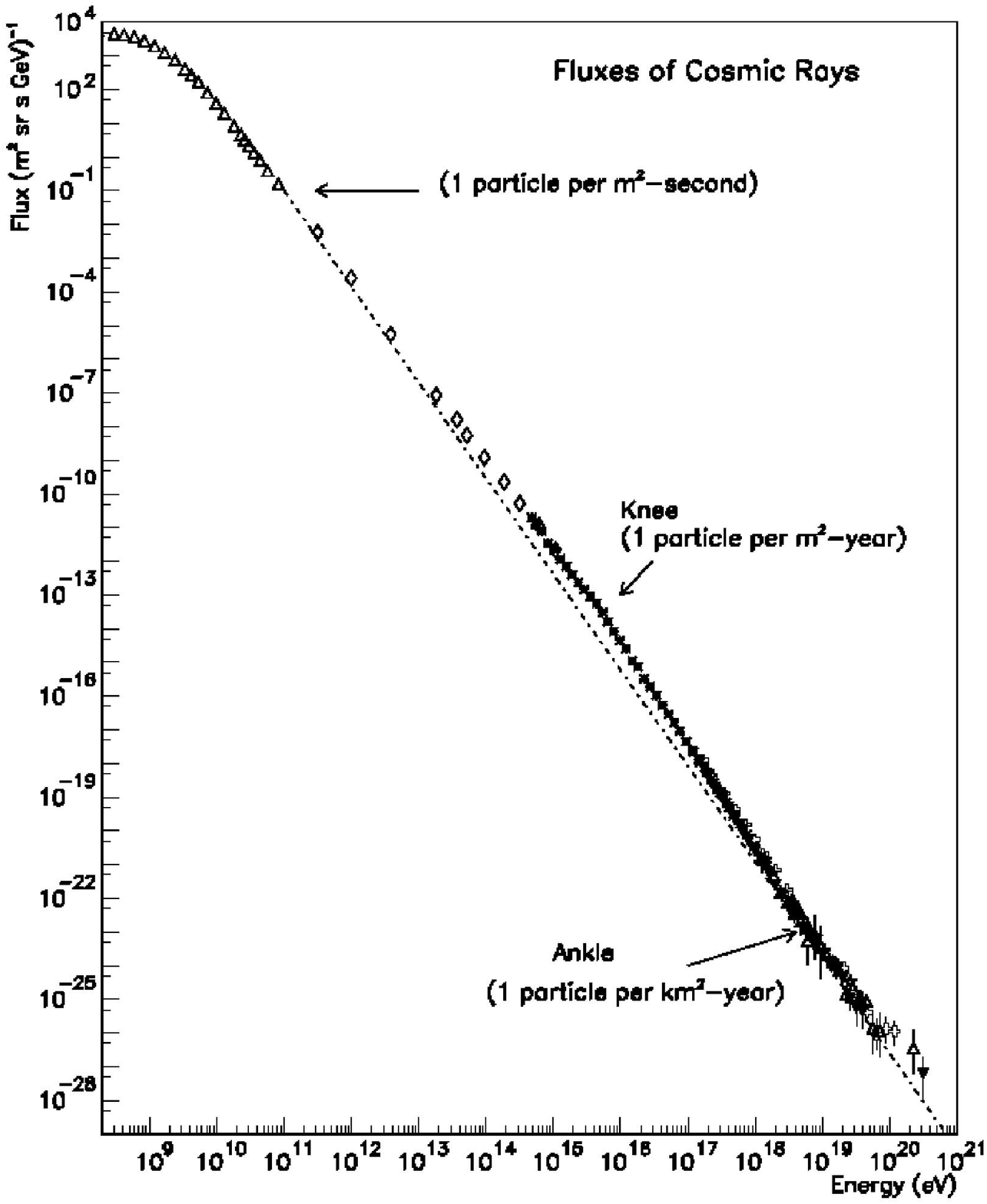}
\caption{$ $ The all particle spectrum of cosmic rays}
\label{fig-crspec}
\end{figure}

The spectrum shown in Figure~\ref{fig-crspec} 
(image credit: S.~Swordy,  University of Chicago, \cite{Swordy2001})
is a compilation of the various measurements. This spectrum
cannot be identified with a single CR source or even multiple CR
sources; in fact, it represents a superposition
of the multitude of Galactic CR sources, integrated over a time
scale of millions of years, and convolved with the
history of their propagation in the Galactic magnetic fields from
all parts of the Galaxy.

Most researchers these days are convinced that the
bulk of the Galactic CRs at least up to the `knee' of the CR spectrum
(i.e., up to the energies of $3\cdot 10^{15}$~eV), 
are produced in astrophysical collisionless shocks
\cite{BE87}. While the details of this process may be uncertain (how
much shocks of individual SNRs contribute to the CR production,
in comparison with SNR shock ensembles in the so-called superbubbles,
how the interstellar dust is involved, etc.), the general
idea is commonly accepted now. 

The process that accelerates the particles to ultra-relativistic
energies in shocks is known as
the first order Fermi process\footnote{There also exists a model
of the second order Fermi process, in which particles are accelerated
by magnetohydrodynamic turbulence in the absence of shocks.}, or diffusive shock
acceleration (DSA).


\section{DSA -- test-particle approximation}

\label{sec_dsa_tp}

Diffusive shock acceleration, DSA, also known as the first order
Fermi process (often abbreviated as Fermi-I)
was first applied to the problem of cosmic ray production
in shocks by several independent groups and researchers
at the end of the 1970s \cite{Bell78a, BO78, Kry77, Eichler79}. The best
simple mechanical analogy to this process is the acceleration of a rubber ball
elastically bouncing back and forth between two massive walls,
as the walls are slowly moved towards each other. 
In a shock, the role of the moving walls is played
by the bulk gas flow: the faster-moving 
unshocked gas and the slower moving shocked gas form an effectively
converging system.

The spectrum of particles accelerated in such a manner may be calculated in various ways,
including a kinetic approach (see, e.g., \cite{Kry77}). 
Consider a one-dimensional shocked flow,
with the shock located at $x=0$, and the flow speed
\begin{equation}
u(x)=\left\{ \begin{array}{l} u_0, \quad x<0, \\ u_2, \quad x>0,\end{array}\right.
\end{equation}
where $u_0$ is the upstream and $u_2 < u_0$ -- the downstream speed,
and let there be a minor admixture of energetic particles that move diffusively 
in the bulk plasma, the diffusion being isotropic in the plasma frame.
Assume that the diffusion coefficient is independent of momentum and
of coordinate (except it may have different constant values upstream and
downstream of the shock):
\begin{equation}
D(x)=\left\{ \begin{array}{l} D_0, \quad x<0, \\ D_2, \quad x>0,\end{array}\right.
\end{equation}
In a steady state, diffusive propagation of 
the energetic particles, as they are being advected downstream by 
the flow, can be described by the equation
\begin{equation}
\label{unmod_prop}
u(x) \frac{\partial f(x,p)}{\partial x} = D(x) \frac{\partial^2 f(x,p)}{\partial^2 x},
\end{equation}
where $f(x,p)$ is the particle distribution function,
such that $f(x,p)dxdydzdp_xdp_ydp_z$ is the number of particles in the
phase space volume $dxdydzdp_xdp_ydp_z$, and that $f(x,p)$ does not
depend on the direction of $\pvector$.
Suppose the incoming energetic particles have a distribution function:
\begin{equation}
\label{bc_minusinf}
\lim_{x\to -\infty}{f(x, p)}=f_0(p)\equiv f_0\frac{1}{p_0^2} \delta_D(p-p_0),
\end{equation}
where $p_0$ is a momentum
such that the corresponding particle speed is much greater than $u_0$,
$p$ is the current particle momentum, and $\delta_D$ is the Dirac
delta-function.  Assume the trivial downstream boundary condition
\begin{equation}
\label{bc_plusinf}
\lim_{x \to +\infty}{f(x,p)} < \infty,
\end{equation}
and define the conditions at the discontinuity point $x=0$:
\begin{eqnarray}
\label{cond0_1}
\lim_{x \to 0-}{f(x, p)} &=& \lim_{x \to 0+}{f(x, p)}, \\
\label{cond0_2}
\lim_{x \to 0-}{\left(-D_0 \frac{\partial f(x,p)}{\partial x}
- \frac{p}{3}\frac{\partial f(x,p)}{\partial p}\right)} &=&
\lim_{x \to 0+}{\left(-D_2 \frac{\partial f(x,p)}{\partial x}
- \frac{p}{3}\frac{\partial f(x,p)}{\partial p}\right)}.
\end{eqnarray}
The first equation expresses the requirement of continuity
of the particle density, and the second -- of particle flux.
The general solution of equation (\ref{unmod_prop}) may be
written as
\begin{equation}
\label{fp_general}
f(x,p) = \left\{ \begin{array}{l}
A(p) \exp{\left(\displaystyle\frac{u_0 x}{D_0}\right)} + B(p),  \quad x < 0 \\
C(p) \exp{\left(\displaystyle\frac{u_2 x}{D_2}\right)} + E(p),  \quad x > 0.
\end{array}\right.
\end{equation}
Substitution of this form into the boundary condition (\ref{bc_minusinf})
results in
\begin{equation}
B(p) = f_0(p),
\end{equation}
and using the boundary condition (\ref{bc_plusinf}) gives
\begin{equation}
C(p) = 0.
\end{equation}
Now we can use the conditions at $x=0$, where the density continuity
equation (\ref{cond0_1})
can help constrain $A(p)$ and $E(p)$ in (\ref{fp_general}):
\begin{equation}
  \label{equap}
  A(p) + f_0(p) = E(p),
\end{equation}
and flux continuity condition (\ref{cond0_2}), rewritten as
\begin{equation}
D_0 \lim_{x \to 0-}{\left( \frac{\partial f(x,p)}{\partial x}\right)}
 - D_2 \lim_{x \to 0+}{\left(\frac{\partial f(x,p)}{\partial x}\right)} =
-\frac{p}{3} \left(\frac{\partial f(0,p)}{\partial p}\right),
\end{equation}
gives
\begin{equation}
  \label{equdp}
  -D_0 A(p) \frac{u_1}{D_0} \exp{(0)} - 0 = -\frac{p}{3}\frac{dE(p)}{dp}\left(u_0 - u_2\right).
\end{equation}
Combining (\ref{equap}) and (\ref{equdp}), we get
\begin{equation}
\frac{p}{3} \frac{dE(p)}{dp} \left(u_0 - u_2\right) + E(p) u_1
= u_1 \frac{f_0}{p_0^2} \delta_D (p-p_0),
\end{equation}
which can easily be integrated, assuming $E(0)=0$, and the solution is
\begin{equation}
  E(p) = E_0
  \left( \frac{p_0}{p} \right)^{s} H(p-p_0),
\end{equation}
where 
\begin{eqnarray}
E_0 &=& \frac{3 N_0 u_0}{p_0^3 \left(u_0 - u_2\right)} \\
\label{s_one}
s &=& \frac{3 u_0}{u_0 - u_2},
\end{eqnarray}
and $H(z)$ is the Heaviside step function:
\begin{equation}
H(z) = \left\{ \begin{array}{l}
0,  \quad x < 0 \\
1,  \quad x \geq 0.
\end{array}\right.
\end{equation}
Finally, the solution of equation (\ref{unmod_prop}) with boundary
conditions (\ref{bc_minusinf}), (\ref{bc_plusinf}) and the 
continuity conditions at the shock (\ref{cond0_1}) and (\ref{cond0_1}) is:
\begin{equation}
\label{tpsolution}
f(p) = \left\{ 
\begin{array}{l}
E_0 
    \left(\displaystyle \frac{p_0}{p}\right)^{s} H(p-p_0) 
    e^{u_0 x/D_0} +
    \displaystyle \frac{f_0}{p_0^2}\delta_D(p-p_0) \left(1 - e^{u_0 x/D_0}\right),
\quad x < 0, \\
E_0
    \left(\displaystyle \frac{p_0}{p}\right)^{s} H(p-p_0) , \quad x > 0.
\end{array}
\right. 
\end{equation}
This is the so-called \textit{test-particle solution} of the problem
of diffusive shock acceleration, meaning that the accelerated energetic
particles are implicitly assumed to be a small admixture in the vast thermal
pool. This assumption is likely to fail for strong collisionless shocks,
leading to serious modifications of the solution, on which the present work 
concentrates.

Let us analyze the basic properties of the test-particle solution (\ref{tpsolution}).
\begin{itemize}
  \item{It requires that some seed particles be introduced, represented by $f_0(p)$,
        but in real shocks these seed particles must be produced from the thermal
        pool (\textit{injected}, as the theorists of the particle acceleration
        field prefer to put it). This model is unable to predict anything about the
        injection of particles, and their number $f_0$ and momentum $p_0$ are free
        parameters of the test-particle model.}
  \item{Once the seed particles are introduced, they form a power-law superthermal
        tail upward of the injection momentum $p_0$, with the index $s$ that depends
        only on the pre-shock and the post-shock speed, as given by
        equation~\ref{s_one}. That equation can be re-written in terms
        of the shock compression ratio $r=u_0/u_2$ as
        \begin{equation}
          \label{pl_index_tp}
          s=\frac{3u_0}{u_0-u_2}=\frac{3r}{r-1}.
        \end{equation}
        For the strongest
        hydrodynamic shocks in a non-relativistic monatomic gas, 
        the compression ratio $r=u_0/u_2$ approaches the
        value of $r=4$ (this well known result can easily be derived
        from the Hugoniot adiabat presented in Section~\ref{sec-rtot}
        in the limit $M_s \to \infty$ with $\gamma=5/3$). Notably, such
        compression ratio corresponds to the power law index of the
        accelerated particle distribution $s=4$.
        A particle distribution $f(p) \propto p^{-4}$ extending to $p \to \infty$
        in unphysical, because the internal energy of such distribution diverges
        logarithmically at $p \to \infty$. This means that, if compression ratios
        of $r=4$ or greater\footnote{For example, relativistic
        gases with a polytropic index $\gamma = 4/3$ allow the strongest shocks to
        have a compression ratio up to $r=7$, which results in a power law index
        $s=3.5$ -- an even more strongly diverging distribution.} 
        are achieved in space, there must be some process
        responsible for limiting the maximum achievable energy. The
        escape of the highest energy particles from the system,
        or a finite time of particle acceleration in a time-dependent
        calculation may determine the high-energy cutoff of the
        particle spectrum. Such processes are not included in this simplistic model.}
  \item{The basic physical assumption that leads to the emergence of the power-law superthermal
        tail of $f(p)$ is that the particles are subject to diffusion isotropic in the plasma
        frame (this is expressed by the equation (\ref{unmod_prop})). This implies
        that we are dealing with a collisionless shock (otherwise the superthermal particles
        would have to thermalize through collisions with their thermal counterparts)
        that has a certain stochastic magnetic field structure (i.e., turbulence),
        responsible for particle scattering.
        The properties of these stochastic fields are, obviously, beyond the scope of this
        model, but they must influence the solution by at least determining the diffusion
        coefficient $D(x,p)$. In fact, as will be shown later, the magnetic turbulence
        that confines the particles to the acceleration site, and allows for the Fermi-I 
        acceleration, is probably produced by the
        accelerated particles themselves, which raises the question of solving the
        particle acceleration problem consistently with the turbulence production process.
        }
\end{itemize}
It turns out that all of these properties of the test-particle make it unable
to explain some observations of interstellar
collisionless shocks (see the next section), which calls for a better model.

\newpage

\section{DSA -- nonlinear regime}

The example from classical mechanics that illustrates the first order Fermi process  --
the rubber ball bouncing between two converging walls -- may also be used to understand
the nonlinear aspects of diffusive shock acceleration. The ball in classical mechanics
gains energy at every collision, but only as long as the walls are much heavier than the 
ball and continue to move inward despite the ball's kicks. But what if the ball
gains enough energy, so the recoil of the walls makes them slow down their convergence?
In this case we would have to account for the feedback of the ball on the
walls. This makes the problem nonlinear. Suppose, we put one ball between
the walls, and its kinetic energy after $N$ cycles becomes $K$.
If we were to put not one, but two balls, the their total energy
after $N$ cycles would be less than $2K$, because the recoil of the
two balls would have slowed the walls down more efficiently than
the recoil of one ball.
Similarly, in shock acceleration, the energy of the bulk plasma flow powers the energetic particle
acceleration, but once the accelerated particles gain enough energy to push back
on the flow, the situation changes dramatically. Such a system is called 
a \textit{nonlinear}, or a \textit{multicomponent} shock wave, and is the subject
of study of nonlinear diffusive shock acceleration theory.

There is a number of reasons to believe that strong shocks in space accelerate particles
very efficiently, thus operating in the nonlinear regime. I outline these reasons below,
and some of them will be elaborated on further in the dissertation.
\begin{enumerate}
\item{Energy considerations. The energy density of Galactic cosmic rays at the
      location of the Solar System is $\varepsilon\approx 0.6$~eV$\cdot$cm$^{-3}$,
      and their characteristic age inferred from the radioactive nuclei in CRs
      is of order $\tau_\mathrm{cr} \approx 10^7$~yr. Assuming that the escape of CRs
      from the Galaxy has the time scale $\tau_\mathrm{cr}$ and that the
      escape is balanced by the CR production one can estimate
      the required power of CR production in the Galaxy as
      \begin{displaymath}
        P_\mathrm{cr} = 
          \frac{\varepsilon V}{\tau_\mathrm{cr}} = 
          3\cdot10^{41}\,\mathrm{erg}\;\mathrm{s}^{-1},
      \end{displaymath}
      where $V$ is the volume of the Galaxy,
      $V = \pi (10^5 \,\mathrm{pc})^2 \cdot 100 \, \mathrm{pc}
      =3\cdot 10^{12}\,\mathrm{pc}^3 \approx 10^{68} \, \mathrm{cm}^3$.
      Assuming that all of these cosmic rays are produced by shocks of SNRs,
      which occur once in $\tau_\mathrm{sn}=100$~yrs and release $E=10^{51}$~erg as the kinetic
      energy of the shock wave, the Galactic energy production in the form of shocks is
      \begin{displaymath}
        P_\mathrm{sk} = \frac{E}{\tau_\mathrm{sn}} = 3\cdot 10^{41}\,\mathrm{erg}\;\mathrm{s}^{-1}.
      \end{displaymath}
      Our estimates of the quantities $P_\mathrm{cr}$ and $P_\mathrm{sk}$
      are comparable, which means that SN shocks may easily be required
      to have an efficiency on the order of tens of percent of converting the bulk
      motion energy into the energy of accelerated particles. See also
      \cite{BE87} for a detailed discussion.
      }
\item{Numerical simulations (e.g., \cite{Ellison85, Malkov97, KJ97, BGV2005})
      predict efficient particle acceleration given the simplest
      physically realistic model of particle injection, thermal leakage. The above mentioned
      models use different techniques, but all of them predict that in a
      strong collisionless shock, the energy density of energetic
      particles becomes comparable to the kinetic energy density of the
      flow, thus making the problem nonlinear.}
\item{Analysis of the morphology of resolved SNRs indicates high
      compression ratios at the forward shock (e.g., \cite{WarrenEtal2005, Gamil2008})
      which is consistent with the predictions of
      nonlinear particle acceleration theories. There is also observational
      evidence of the nonlinearity of DSA that comes from the analysis
      of nonthermal emission spectra (e.g., \cite{RE92, AHS2008})
      and from spacecraft observations of the Earth's bow shock 
      (e.g., \cite{EMP90}).}
\item{Recent observations indicate that magnetic fields in some SNR shocks are
      much stronger than the ambient magnetic fields, which makes many researchers
      believe that magnetic fields are amplified in situ, i.e. in the shocks,
      by the shock-accelerated particles. If that is the case, the energy density of the 
      accelerated particles must be no less than the energy density of
      the amplified magnetic fields, and, according to the observational estimates,
      the latter is a significant fraction of the dynamical pressure
      of the shock flow. This necessitates the nonlinear DSA (see, e.g., \cite{EV2008}).}
\end{enumerate}

The nonlinear DSA theory was developed by various researchers in the
1980s. Although the details of the models may differ, all of them agree
on the following: when the energetic particles gain enough energy
to feed back on the flow, the unshocked plasma slows down and
becomes compressed even
before it reaches the viscous shock (the latter is renamed 
a \textit{subshock} in context of nonlinear DSA), which means
that a \textit{shock precursor} forms in the upstream region ($x<0$);
the maximum particle energy must be limited either by the age, or by the
size of the shock, and if particles of the highest energies
are allowed to leave the shock far upstream, it leads to an increase
in the compression ratio.


\section{Magnetic field amplification in shocks}

Recent observations and modeling of several young supernova remnants
(SNRs) suggest the presence of magnetic fields at the forward shock
(i.e., the outer blast wave) well in excess of what is expected from
simple compression of the ambient circumstellar field, $\Bism$.  These
large fields are inferred from:
\begin{itemize}
  \item{ spectral curvature in radio emission (e.g.,
    \cite{RE92,BKP99}) },
  \item{ broad-band fits of \syn\ emission between
radio and non-thermal X-rays (e.g., \cite{BKV2003,VBK2005a}, see
also \cite{Cowsik80})},
  \item{ sharp X-ray edges (e.g.,
    \cite{VL2003,BambaEtal2003,VBK2005a,EC2005,CHBD2007}), and}
  \item{ rapid variability of nonthermal X-ray emission from bright
    filaments in SNRs (first reported by \cite{Uchiyama07}).}
\end{itemize}
While these methods are all indirect, fields greater than 500~\muG\ are
inferred in the supernova remnant
Cassiopeia A and values of at least several 100 \muG\ are estimated
in Tycho, Kepler, SN1006, and G347.3-0.5.
If $\Bism \sim 3-10$~\muG, amplification factors of 100 or more may be
required to explain the fields immediately behind the forward shocks and
this is likely the result of a nonlinear amplification process
associated with the efficient acceleration of cosmic-ray ions via
diffusive shock acceleration (DSA).
The magnetic field strength is a critical parameter in DSA and also
strongly influences the \syn\ emission from shock accelerated electrons.
Since shocks are expected to accelerate particles in diverse
astrophysical environments and \syn\ emission is often an important
emission process (e.g., radio jets), quantifying the magnetic field
amplification has become an important problem in particle astrophysics
and has relevance beyond cosmic-ray production in SNRs.

These highly amplified magnetic fields are most likely an intrinsic part of
efficient particle acceleration by shocks.
This strong turbulence, which may result from cosmic ray driven instabilities,
both resonant and non-resonant, in the shock precursor, is certain to
play a critical role in \SC, \NL\ models of strong, \CR\ modified shocks.
Although plasma wave instabilities in presence of accelerated
particles have been studied in the context of shock acceleration before
(e.g., \cite{Bell78a, LC83}), it was only recently suggested that
these instabilities may lead to very efficient amplification of
magnetic field fluctuations, $\Delta B \gg B_0$ \cite{BL2001}. Since
then, new models of plasma instabilities possibly responsible for
efficient magnetic field amplification were proposed 
\cite{LB2000, Bell2004, BT2005} 
and studied in context of shock acceleration 
\cite{PLM2006, AB2006, VEB2006, ZP2008}.

All these plasma instabilities are assumed to amplify pre-existing
waves in a plasma in the presence of an underlying uniform magnetic
field $B_0$ parallel to the flow\footnote{For relativistic shocks, 
  which are beyond the scope of the
  present research, the Weibel instability may be important for
  magnetic field generation, as long as the upstream magnetic field is
  low (see, e.g., \cite{ML99, LE2006, Spitkovsky2008b}).}.
The two models of interest that will be applied to the present work
are:
\begin{enumerate}
\item{Resonant CR streaming instability (see \cite{Skilling75b, LC83}
  and \cite{BL2001, AB2006, VEB2006}), in which particles of a
  certain momentum amplify \Alf\ waves with a wavenumber equal to the
  inverse gyroradius of the particle, and}
\item{Nonresonant CR streaming instability of short-wavelength modes
  (suggested by \cite{Bell2004}), which I will sometimes refer to as
  Bell's instability, in which the diffusive electric current of CRs
  amplifies almost purely growing waves with wavenumbers much greater
  than the inverse particle gyroradius. }
\end{enumerate}
I should also mention a nonresonant instability that produces long-wavelengths
modes and may also be important in shocks (see \cite{BT2005} and 
Section~\ref{nonres_long_inst}), which I am
planning to apply to the modeling of shocks in the future,
as well as other possible mechanisms (e.g., \cite{MD2009}).

Amplification of magnetic turbulence has great importance in
the process of DSA. The amplified turbulence provides the stochastic
magnetic fields that scatter the accelerated particles, allowing them to
participate in the Fermi-I process. The properties of the particle
scattering are therefore dependent on the spectrum of stochastic
magnetic fields, yet the latter are produced by the accelerated
particles. This complex connection between particles and waves in
shocks adds to the nonlinear nature of shock acceleration, discussed
in Chapter~\ref{ch_nonlin_dsa}. Therefore, magnetic field amplification
affects the observable nonthermal synchrotron emission from shocks in
two ways: it determines the structure and strength of the
magnetic fields in which the emission occurs, and shapes
the spectrum of the radiating energetic particles.


\section{Turbulence}

Studying strong magnetic field amplification in interstellar shocks
inevitably makes us face the subject of magnetohydrodynamic (MHD)
turbulence. Usually turbulence is defined as chaotic fluid motion,
that is, a motion with a very sensitive dependence on initial
conditions. Chaotic behavior makes turbulent motions effectively
non-deterministic, but they can be studied using statistical
methods. 

Motions of gases and fluids of high Reynolds number
tend to transit to the turbulent regime (see,
e.g., \cite{LL_HYDRO, MoninYaglom}), which is
encountered on a regular basis in areas ranging from
plasma fusion engineering and race car design to
air transport, plumbing, golf and food processing
(e.g., \cite{FO90, SS1983, USPAT_golf, USPAT_homo}).
Driven by the need of applications like
meteorology, climate modeling, aerospace engineering, and others,
turbulence research has been conducted for many decades, and is
a challenging field of mathematics and
physics (e.g., \cite{FS2006}). Conducting fluids (plasmas) easily develop and sustain
magnetic fields, and the MHD turbulence regime, occurring in plasmas,
is even more complicated by the magnetic field interactions than its
hydrodynamic counterpart \cite{Biskamp2003}.

Considering that plasmas constitute a large fraction of all baryonic
matter in space, their properties have pervading importance for
astrophysics. Namely, turbulence in plasmas determines cosmic ray
acceleration and propagation, plays a crucial role for angular
momentum transfer in accreting systems and impacts the properties
of gravitational collapse. The list of astrophysical objects affected
by MHD turbulence is therefore extensive: large scale structure
of the Universe, quasars, accreting binary systems, 
forming stars, supernova remnants, etc. 

The  primary sources of information about MHD
turbulence are spacecraft observations of interplanetary space and 
numerical simulations. The former provide real, but often hard to
interpret data, the bottom line of which is that turbulence often
consists of stochastic perturbations of plasma velocities 
and magnetic fields spanning many decades of the spatial scales.
Oftentimes, the Fourier spectrum of spatial structure of
turbulent fluctuations reveals a power-law distribution of energy
in wavenumber space. The numerical simulations have the advantage
of providing data that is easy to analyze and scale for practically
applicable theories.

A simplified picture of turbulence evolution, based on extensive research,
involves three dominant processes: energy supply, spectral transfer of energy
and dissipation.
Consider a fluid flow in a pipe, where a large
flux of the fluid leads to the development of a hydrodynamic instability that
creates vortices (eddies) breaking the laminar flow. In this way energy is supplied
to the turbulence in the form of large-scale vortices. 
These eddies then break
down into smaller eddies -- this way, spectral energy transfer
(cascade) from large to small scales is realized. As the scale
of the turbulent structures due to cascading becomes smaller,
fluid viscosity plays an increasingly greater role, eventually
leading to the dissipation of the smallest eddies into heat.

The MHD turbulence, as mentioned above, is difficult to describe.
It was originally treated and analyzed as a set of small perturbations
(i.e., plasma waves) moving in the large-scale uniform magnetic field and
weakly interacting with each other (the so-called Iroshnikov-Kraichnan
approach \cite{Iroshnikov1964, Kraichnan1965}).
However, Goldreich and Sridhar \cite{GS95}\footnote{Note that this 
publication is relatively recent, but it has
laid the groundwork for MHD turbulence research from a new
vantage point, possible only with modern
computational resources.} point out that this approach
may be inappropriate for MHD turbulence due to
its inherent anisotropy introduced by the magnetic field \cite{ES2004a}.
The bottom line of their theory and of the subsequent simulations of MHD
is that the magnetic field plays a stabilizing role. 
The cascading takes place mostly for wave vectors perpendicular
to the uniform magnetic field, while the parallel cascade is
suppressed.

A comprehensive source on classical theory of hydrodynamical
turbulence is \cite{MoninYaglom}. Modern advances in the 
study of MHD turbulence is presented in \cite{Biskamp2003}.

\chapter{The Problem of Nonlinear DSA}

\label{ch_nonlin_dsa}

The general problem of nonlinear diffusive shock acceleration 
of charged particles (DSA)
can be formulated as follows: given
a supersonic flow with a speed $u_0$ of a plasma with a number density
$n_0$, temperature $T_0$ and a pre-existing magnetic field ${\bf B}_0$,
and given the location $x=0$ where this flow develops a subshock,
find the distribution of particles 
$f({\bf x},{\bf p},t)$ and electromagnetic fields in the shock
vicinity. This problem is complicated by two facts: a) particle
acceleration occurs due to complex motions of particles in the turbulent
magnetic field, but the magnetic turbulence itself is dependent upon
the motion of the accelerated particles, and, b) if particle
acceleration is efficient, different parts of the particle spectrum
interact with each other (i.e., the accelerated particles push back on
and slow down the flow of the thermal particles).

This problem cannot be practically tackled by particle simulations from first
principles like Maxwell's equations and Lorentz force (see Section~\ref{sec_pic}),
and the most computationally expensive operations must be performed
analytically. Namely, all currently existing models of nonlinear DSA, including
the one discussed in this dissertation, 
assume that the accelerated particles propagate diffusively with some
diffusion coefficient, or mean free path, prescription. This allows the
models to eliminate the need to describe the complex interactions between
particles and waves, and to concentrate on the physical aspects of
particle acceleration.

\newpage

\section{Analytic models}

In successful analytic models, a one-dimensional steady state shock with a nonlinear
precursor is described by the flow speed $u(x,t)$, mass density of the
plasma $\rho(x,t)$ and an isotropic distribution function of energetic particles,
$f_\mathrm{cr}(x,p,t)$. 
The above mentioned macroscopic quantities must be consistent with the
fundamental conservation laws: mass, momentum and energy must be
conserved. These conditions are expressed with the following system of
equations: 
\begin{eqnarray}
  \label{masscons_aa}
     \rho u &=& \mathrm{const}, \\
  \label{momcons_aa}
     \rho u^2 + P_\mathrm{th} + P_\mathrm{cr} + P_\mathrm{mag} &=&
     \mathrm{const}, \\
  \label{engycons_aa}
     \frac12 \rho u^3 +
     \frac{\gamma_\mathrm{th}}{\gamma_\mathrm{th}-1}P_\mathrm{th}u +
     \frac{\gamma_\mathrm{cr}}{\gamma_\mathrm{cr}-1}P_\mathrm{cr}u +
     \frac32 P_\mathrm{mag} u + Q_\mathrm{esc}
     &=&
     \mathrm{const}.
\end{eqnarray}
and the evolution of the particle distribution is governed by the
kinetic equation of CR transport
\begin{equation}
\label{isotr_cr_prop}
\frac{\partial }{\partial x} \left[ 
D(x,p)\frac{\partial}{\partial x}f(x,p)
\right]
-u\frac{\partial f(x,p)}{\partial x} + 
\frac13 \left( \frac{du}{dx} \right)p\frac{\partial f(x,p)}{\partial p}
+ Q_\mathrm{inj} = 0.
\end{equation}
Equations (\ref{masscons_aa}), (\ref{momcons_aa}) and (\ref{engycons_aa}) 
represent conservation of mass, momentum and energy fluxes, respectively.
Equation (\ref{isotr_cr_prop}) is the  kinetic equation describing
propagation of cosmic rays in the diffusion approximation. 
The expressions above are, essentially, a direct generalization of the
test particle model of shock acceleration demonstrated in Section~\ref{sec_dsa_tp},
complemented by the treatment of the flow speed $u(x)$ variability upstream.
Let us use the following notation for the flow speed and other quantities
at points of interest: $u_0$ is the far upstream flow speed, $u_2$ is
the downstream flow speed, and $u_1$ is the flow speed just before the subshock.
Thus, in the upstream region, $x<0$, the flow speed varies from 
$u(x=-\infty)=u_0$ to $u(x=-0)=u_1 < u_0$, and then jumps in a viscous subshock
to $u(x=+0)=u_2 < u_1$. Let us also define the total compression ratio,
$r_\mathrm{tot}=u_0 / u_2$ and the subshock compression ratio 
$r_\mathrm{sub}=u_1 / u_2$.

To close the model, one must describe the evolution of thermal gas pressure,
$P_\mathrm{th}$, define the cosmic ray pressure, $P_\mathrm{cr}$ and
have a model for determining the magnetic field pressure, $P_\mathrm{mag}$.
The term $Q_\mathrm{esc}$ representing the energy escape from 
the system requires a model of particle escape, the term
$Q_\mathrm{inj}$ representing the injection of thermal particles into
the acceleration process calls for a model of particle injection.
The most important parameter of the model, the diffusion coefficient
$D(x,p)$, must be calculated using some simple approximation or using
the assumed spectrum of turbulence.

These models were used in \cite{BE99, Malkov97, KJG2002, BGV2005, Eichler84}.
The major advantage of the analytic models is that they provide a fast
solution of the problem, which can be used in the simulations of objects incorporating
shocks, for example, to calculate the spectrum of electromagnetic
emission from an evolving SNR (see, e.g., \cite{BV2008, MAB2009}).

The computation speed comes at the cost of making some important approximations, as
summarized below.
\begin{enumerate}
\item{Analytic models are limited by the assumptions that go into the analytic
      description of the plasma physics. For example, the diffusion coefficient 
      of charged particles in stochastic magnetic fields can be reliably estimated
      either in the limit of weak turbulence, or in the simplistic Bohm approximation.
      Similarly, the analytic description of the physics of turbulence generation is
      only valid in the quasi-linear regime, i.e., weak turbulence.}
\item{These models adopt a diffusion approximation of particle transport.
      Hand in hand with this approximation goes the assumption that the particle 
      distribution function is isotropic in the plasma frame. Only
      this way can one 
      define the pressures of thermal particles $P_\mathrm{th}(x,t)$
      and cosmic rays $P_\mathrm{cr}(x,t)$ as moments of
      particle distribution function $f(x,p,t)$. The isotropy assumption
      breaks down for relativistic shocks. But even for the non-relativistic shocks
      that we are discussing in this work, the anisotropy of particle distribution
      is important, because it determines the thermal particle injection process. 
      Therefore, analytic models
      require additional parameters or assumptions in order to estimate particle
      injection.}
\item{If a strong uniform magnetic field is present in the shock,
      its strength and orientation may affect particle injection and transport.
      Some SNRs may have an asymmetric appearance due to the
      variation of the obliquity of magnetic field around the rim \cite{Gamil2008}.
      Analytic models are not able to account for this effect due to
      the isotropy assumption.}
\item{Including some important physical processes in the analytic models
      of shock acceleration complicates calculations. These important processes
      include particle escape far upstream, nonlinear processes in turbulence
      generation (such as cascading), modifications of the diffusion regime
      by a specific shape of turbulence spectrum, etc.}
\end{enumerate}

Despite these limitations, analytic models are very helpful for 
making qualitative and quantitative predictions regarding the nonlinear
structure of shocks, and are the current method of choice for modeling
the electromagnetic emission of SNRs, where fast simulations
of shocks are required.

\section{Particle-in-cell (PIC) codes}

\label{sec_pic}

In principle, the problem can be solved completely with few assumptions
and approximations with plasma simulations. Those fall into two
major categories: particle-in-cell (PIC) simulations
(e.g., \cite{Spitkovsky2008,NPS2008}), and
hybrid models that assume that electrons are not dynamically
important (e.g., \cite{WO1996, Giacalone2004})\footnote{
We must also mention the MHD models (e.g., \cite{Bell2004}, \cite{ZPV2008}, 
\cite{ZP2008}) that ignore or treat in a simplified way
the spectral properties of the particle distribution; while they may be important for
describing certain aspects of plasma physics, their application to
nonlinear DSA is limited}.

However, modeling the nonlinear generation of relativistic particles and strong
magnetic turbulence in collisionless shocks is
computationally challenging and PIC simulations will not be able
to fully address this problem in nonrelativistic shocks for some years to come
even though they can provide critical information on the plasma
processes  that can be obtained in no other way.
In this section I outline the requirements that a
PIC simulation must fulfill in order to tackle the problem of efficient
DSA with nonlinear magnetic field amplification (MFA) in SNR shocks. 
The reasoning presented in this section was also laid down in \cite{VBE2008}.

There are two basic reasons why the problem of MFA in nonlinear
diffusive shock acceleration (NL-DSA) is particularly difficult for
particle-in-cell (PIC) simulations. The
first is that PIC simulations must be done fully in three dimensions to
properly account for cross-field diffusion. As Jones 
\cite{JJB98} proved from
first principles, PIC simulations with one or more ignorable dimensions
unphysically prevent particles from crossing magnetic field lines.  In
all but strictly parallel shock geometry,\footnote{Parallel geometry is
where the upstream magnetic field is parallel to the shock normal.} a
condition which never occurs in strong turbulence, cross-field
scattering is expected to contribute importantly to particle injection
and must be fully accounted for if injection from the thermal background
is to be modeled accurately.

The second reason is that, in nonrelativistic shocks, NL-DSA spans large
spatial, temporal, and momentum scales. The range of scales is more
important than might be expected because DSA is intrinsically efficient
and nonlinear effects tend to place a large fraction of the particle pressure
in the highest energy particles.
The highest energy particles, with the
largest diffusion lengths and longest acceleration times, feed back on
the injection of the lowest energy particles with the shortest scales.
The accelerated particles exchange their momentum and energy with the
incoming thermal plasma through the magnetic fluctuations coupled to the
flow. This results in the flow being decelerated and the plasma being
heated. The structure of the shock, including the subshock where fresh
particles are injected, depends critically on the highest energy
particles in the system.

A plasma simulation must resolve the electron skin depth,
$c/\omega_{pe}$, i.e., $L_\mathrm{cell} < c/ \omega_{pe}$, where
$\omega_{pe}=[4 \pi n_e e^2/m_e]^{1/2}$ is the electron plasma
frequency and $L_\mathrm{cell}$ is the simulation cell size. Here, $n_e$ is
the electron number density, $m_e$ is the electron mass and $c$ and
$e$ have their usual meanings (the speed of light and the
elementary charge, respectively). The simulation must
also have a time step small compared to $\omega_{pe}^{-1}$, i.e.,
$t_\mathrm{tstep} < \omega_{pe}^{-1}$.
If one wishes to follow the acceleration of protons in DSA to the TeV
energies present in SNRs, one must have a simulation box that is as large
as the upstream diffusion length of the highest energy protons, i.e.,
$\kappa(\mathrm{E_\mathrm{max}})/u_0 \sim r_g(E_\mathrm{max}) c / (3 u_0)$, where $\kappa$
is the diffusion coefficient, $r_g(E_\mathrm{max})$ is the gyroradius of a relativistic
proton with the energy $E_\mathrm{max}$, $u_0$ is the shock speed, and 
assuming Bohm diffusion.
The simulation must also be able to run for as long as the acceleration
time of the highest energy protons, 
$\tau_\mathrm{acc}(E_\mathrm{max}) \sim E_\mathrm{max} c /(eBu_0^2)$.
Here, $B$ is some average magnetic field.
The spatial condition gives
\begin{equation}
\frac{\kappa(E_\mathrm{max})/u_0}{(c/\omega_{pe})} \sim 6\cdot 10^{11}
\left ( \frac{E_\mathrm{max}}{\mathrm{TeV}} \right )
\left ( \frac{u_0}{1000 \, \mathrm{km \, s}^{-1}} \right )^{-1}
\left ( \frac{B}{\mu \mathrm{G}} \right )^{-1}
\left ( \frac{n_e}{\mathrm{cm}^{-3}}\right )^{1/2}
\left ( \frac{f}{1836}\right )^{1/2}
\ ,
\end{equation}
for the number of cells {\it in one dimension}. The factor $f=m_p/m_e$
is the proton to electron  mass ratio. From the acceleration
time condition, the required number of time steps is,
\begin{equation}
\frac{\tau_\mathrm{acc}(E_\mathrm{max})}{\omega_{pe}^{-1}} \sim 6\cdot 10^{14}
\left ( \frac{E_\mathrm{max}}{\mathrm{TeV}} \right )
\left ( \frac{u_0}{1000\,\mathrm{km\,s}^{-1}}\right )^{-2}
\left ( \frac{B}{\mu \mathrm{G}} \right )^{-1}
\left ( \frac{n_e}{\mathrm{cm}^{-3}}\right )^{1/2}
\left ( \frac{f}{1836}\right )^{1/2}
\ .
\end{equation}
Even with $f=1$ these numbers are obviously far beyond any conceivable
computing capabilities and they show that approximate methods are
essential for studying NL-DSA.

One approximation that is often used is a hybrid PIC simulation where
the electrons are treated as a background fluid. To get the 
estimate of the requirements in this case, we can take the
minimum cell size as the thermal proton gyroradius, 
$r_\mathrm{g0} = c \sqrt{2 m_p E_\mathrm{th}}/(e B)$.
Now, the number of cells, {\it again in one dimension,} is: 
\begin{equation}
\label{eq:cell}
\frac{\kappa(E_\mathrm{max})/u_0}{r_\mathrm{g0}} \sim 7\cdot 10^{7}
\left ( \frac{E_\mathrm{max}}{\mathrm{TeV}} \right )
\left ( \frac{u_0}{1000 \, \mathrm{km \, s}^{-1}} \right)^{-1}
\left ( \frac{E\mathrm{th}}{\mathrm{keV}} \right )^{-1/2}
\ .
\end{equation}
The time step size must be $\tau_\mathrm{step} < \omega_\mathrm{cp}^{-1}$, 
where $\omega_\mathrm{cp}=eB/m_{p}c$ is the thermal proton gyrofrequency.
This gives the number of time steps to reach 1 TeV,
\begin{equation}
\label{eq:step}
\frac{\tau_\mathrm{acc}(E_\mathrm{max})}{\omega_\mathrm{cp}^{-1}} \sim 1\cdot 10^{8}
\left ( \frac{E_\mathrm{max}}{\mathrm{TeV}} \right )
\left ( \frac{u_0}{1000\,\mathrm{km\,s}^{-1}}\right )^{-2}
\ .
\end{equation}
These combined spatial and temporal requirements, even for the most
optimistic case of a hybrid simulation with an unrealistically large
$\tau_\mathrm{step}$, are well beyond existing computing capabilities unless
a maximum energy well below 1 TeV is used.

Since the three-dimensional requirement is fundamental and relaxing it
eliminates cross-field diffusion, restricting the energy range is the
best way to make the problem accessible to hybrid PIC
simulations. However, since producing relativistic particles from 
nonrelativistic ones
is an essential part of the NL problem, the energy range must
comfortably span $m_p c^2$ to be realistic.  If $E_\mathrm{max}=10$\,GeV is used,
with $u_0=5000$\,km~s$^{-1}$, and $E_\mathrm{th}= 10$\,MeV, equation~(\ref{eq:cell})
gives $\sim 1400$ and equation~(\ref{eq:step}) gives $\sim 4 \cdot 10^{4}$.
Now, the computation may be possible, even with the 3\nobreakdash-D requirement, but
the hybrid simulation can't fully investigate MFA since electron return
currents are not modeled.  The exact microscopic description of the
system is not currently feasible.

It's hard to make a comparison in run-time between PIC simulations
and the Monte Carlo technique used here because we are not aware of any published
results of 3\nobreakdash-D PIC simulations of nonrelativistic shocks that follow particles
from fully nonrelativistic to fully relativistic energies. A direct
comparison of 1\nobreakdash-D
hybrid and Monte Carlo codes was given in \cite{EGBS93} for energies
consistent with the acceleration of diffuse ions at the quasi-parallel
Earth bow shock. Three-dimensional hybrid PIC results for nonrelativistic
shocks were presented in \cite{GE2000} and these were barely able to
show injection and acceleration given the computational limits at that
time. As for the Monte Carlo technique, 
a simulation that calculates the nonlinear structure of a
shock with a dynamic range typical for SNRs,
typically takes several hours on 4-10 processors.
Thus, realistic Monte Carlo SNR models are possible with modest
computing resources.

Despite these limitations, PIC simulations are the only way of self-consistently
modeling the plasma physics of collisionless shocks.
In particular, the injection of thermal particles in the large amplitude
waves and time varying structure of the subshock can only only be
determined with PIC simulations (e.g., \cite{NPS2008,Spitkovsky2008}).
Injection is one of the most
important aspects of DSA and one 
where analytic and Monte Carlo techniques have large uncertainties.

\section{Monte Carlo Simulation}

The Monte Carlo method of solving the problem of nonlinear DSA was
developed by Ellison and co-workers (see \cite{EJR90,JE91,VEB2006} and
references therein for more complete details). This method provides an
excellent compromise between the physically realistic, but
computationally limited PIC simulations and fast, but simplified
analytic models. The compromise is achieved by replacing the
solution of coupled equations of particle propagation, of
conservation laws, and of turbulence generation, with a Monte Carlo
simulation of particle transport that incorporates an iterative procedure
that ensures the simultaneous consistency of all assumed laws. The
Monte Carlo method goes beyond the diffusion approximation of particle
transport and allows the calculation of rates of particle injection into
the acceleration process and of energetic particle escape upstream
and/or downstream of the shock.

In this method, particle transport is described as a stochastic process.
Particles move in small time steps, as their local plasma frame momenta
are `scattered' at each step in a random walk process on a sphere in
momentum space. The properties of the random walk are determined by
the assumption of a certain particle mean free path (or diffusion
coefficient), that statistically describes the interactions of particles
with the stochastic magnetic fields. By assuming that such a
description is possible, Monte Carlo methods gets a speed advantage over the
PIC simulations at the cost of relying on theoretical models of
the diffusive properties of the plasma. On the one hand, these models can be
rather advanced and successful, therefore making this approximation
justified. On the other hand, ignoring the spatial structure of
electromagnetic fields and replacing it with a statistical description
is the biggest simplification of this model.

Acceleration of particles takes place naturally in this model, as long
as a shocked flow is described. Some shock heated thermal particles
are injected into the
acceleration process when their history of random scatterings in the
downstream region takes them back upstream.  These particles gain
energy and some continue to be accelerated in the first-order Fermi
mechanism. This form of injection is generally called `thermal
leakage' and was first used in the context of DSA in \cite{EJE1981}
(see also \cite{Ellison82}). 
The number of particles that do this back-crossing, and the energy they gain, are
determined only by the random particle histories; no parameterization of
the injection process is made other than the assumption of the diffusion
coefficient value at various particle energies.

The nonlinearity of the problem is dealt with by employing an
iterative scheme that ensures the conservation of mass, momentum, 
and energy fluxes, thus producing a
self-consistent solution for a steady-state, plane shock, with particle
injection and acceleration coupled to the bulk plasma flow
modification.

An important advantage of the Monte Carlo model is that it was shown
to agree well with spacecraft observations of the
Earth's bow shock \cite{EM87, EMP90}, interplanetary shocks
\cite{BOEF97}, and with 1\nobreakdash-D hybrid PIC simulations
\cite{EGBS93}.

\section{Objectives of this dissertation}

Hopefully, I have convinced the reader of
the far-reaching impact of processes in shocks on many
astrophysical objects. Considering the observations of supernova remnants
that indicate the possibility of strong magnetic field amplifications
at shocks in-situ, I would like to theoretically investigate the
physics of shock acceleration in the presence of strong MHD
turbulence generation by the accelerated particles.

I favor the Monte Carlo approach to this problem, because of the
growing complexity of the models of nonlinear DSA (which makes
the analytic approach less productive), and because I would like to
probe the aspects of NL-DSA that neither the analytic models, nor the
PIC simulations have yet constrained.

The questions that interest me include: 
\begin{itemize}
\item{How efficient can magnetic field amplification be, considering
  the nonlinear effects in the system? What impact does the generated
  strong magnetic turbulence have on the efficiency of particle
  acceleration?}
\item{How do these results depend on the model of turbulence
  generation and on the model for statistical description of 
  particle transport?}
\item{What are the consequences of efficient MFA on the maximum
  momentum of accelerated particles and how do they impact
  the shock structure? }
\item{Highest energy particles must escape upstream of the shock. What
  are the properties of the escaping particles?}
\item{What does the shock precursor look like (scale, structure,
  processes)?}
\item{What is the qualitative and quantitative dependence of some
  observable parameters (i.e., effective magnetic field strength,
  shocked gas temperature, flow compression ratio, etc.) on the
  properties of the shock (shock speed, plasma density and
  magnetization, etc.)? } 
\end{itemize}

Over the past 3 years that the work on this project was being done, we
(I, under the guidance of Prof.~Ellison, and with the help of
Prof.~Bykov's advice)
have successfully developed the model and obtained results
shedding light on most of these
questions. Our results have appeared in several peer-reviewed
journal publications. 

The purpose of this dissertation is to make a record of the process of
developing and testing the Monte Carlo simulation of NL-DSA with MFA,
and to exhibit and summarize our results that have been or will soon
be presented in conferences and in the press.

\chapter{Model}

In the present research, I used the Monte Carlo method developed
by Ellison and co-workers to build a self-consistent model of shock
acceleration of charged particles, now with efficient magnetic field
amplification. 
I wrote the computer code realizing the Monte Carlo model 
from scratch, but making a full use of the formerly
developed procedures. I also contributed some essential
improvements to the Monte Carlo method, that were necessary for
the implementation of magnetic turbulence amplification models.

In this Chapter, I will discuss the model. In Section~\ref{coremc}
I will present the fundamentals of the Monte Carlo simulation
and the tests performed to confirm that my numerical model
reproduces the known analytic results and conforms with the
fundamental laws of physics. Section~\ref{mfa_in_mc} will be devoted
to the state-of-the-art models of magnetic turbulence amplification
discussed these days in the astrophysical literature, and to the
implementation of these models in the Monte Carlo code. This part of
the model is the essence of my research project. Another
original contribution I made to the model in this project is 
the adaptation and incorporation of advanced particle transport
techniques into the simulation, as
discussed in Section~\ref{advanced_transport}. Finally,
in Section~\ref{parallel_computing} I will discss
the realization of parallel computing in the simulation.

\newpage

\section{Core Monte Carlo}

\label{coremc}

This section discusses the techniques that were used in the
Monte Carlo simulation before the incorporation of the magnetic
turbulence amplification. Most of them had been developed before the
author of this dissertation began contributing to the model. However, the
tests of the model demonstrated in this section were performed by the
computer code written by me.

\subsection{Overview}

The simulation of nonlinear particle acceleration with the Monte Carlo
particle transport starts by assuming an unmodified shocked flow
[$u(x<0)=u_0$, $u(x >0) = u_2$]. One must also assume
some scattering properties of the medium, i.e., assign a mean
free path $\lambda(x,p)$ to the whole particle energy range,
at every point in space. 

Then thermal particles are introduced
far upstream, and the code propagates these particles
until they cross the subshock at $x=0$. Particle
propagation is diffusive, according to the
chosen mean free path $\lambda(x,p)$, and it is performed
as described in Section~\ref{subsec-pascatt}. Some of these particles
will be advected downstream with the flow $u_2$, and once
the code finds any particle many diffusion lengths downstream
of the shock, its propagation may be terminated. However,
the downstream flow speed must by definition be smaller than 
the downstream speed of sound (i.e., a thermal particle speed),
so a small fraction of the downstream thermal particles
may, in the random walk process, find themselves upstream.
If this happens to a particle, it is said to have been {\it injected}
into the acceleration process and becomes a CR particle 
(as opposed to having been a thermal one). An injected
particle is much more likely than a thermal one
to cross the shock again and again, and eventually 
gain a relativistic energy in this process\footnote{Note that in 
the collisionless shocks discussed in this dissertation,
particles with superthermal energies
do not lose energy in particle-particle collisions and can
therefore be easily accelerated once they are injected.}. The action
of the advection with the flow combined with particle
diffusion in a non-uniform flow is described in 
Section~\ref{subsec-motion}.

The particles that have been injected will, due to the flow
speed difference across the shock, find themselves moving 
at a high speed with respect to the plasma, at least 
at the speed $v=u_0-u_2$. This completes the first cycle
of the Fermi-I process. In a short time, 
the accelerated particles will
again be advected downstream, but having a greater
energy, they will find it easier to return upstream again
and get accelerated a second time. As this process goes on,
a few particles may achieve, in principle, unlimited
energy.

The acceleration process in reality must have an upper
limit for particle energy. The two possible causes of
such a limit are {\it a)} time-limited acceleration: the
particles only gain as much energy as they can in the 
amount of time that the shock has been in existence, and
{\it b)} size-limited acceleration: particles have enough
time to achieve such high energies that their scattering
length becomes comparable to the size of the accelerating
system, and they escape. The Monte Carlo method can, in principle,
model both situations by terminating the acceleration
of every given particle either after it had been in the
process for a certain amount of time, or once it has
reached a certain boundary. I favor the second scenario,
due to its inherent consistency with the assumption
of a steady state solution,
and throughout this work it will be assumed that the acceleration
is size-limited. For the model it means assuming a
boundary, let us call it the free escape boundary (FEB) located
at $\xfeb<0$, such that any particle that crosses this boundary
while moving against the flow
leaves the system forever. The mean free path of
the accelerated particles generally increases with energy,
therefore only the highest energy particles can
reach the distant boundary at $\xfeb$. This way,
the acceleration has an upper energy determined by the
location of the free escape boundary. In reality,
the distance $\xfeb$ must be comparable to,
or be a fraction of, the radius of the SNR shell shock.

After all the accelerated particles either had been advected
downstream, or had escaped through the free escape boundary,
the Monte Carlo transport process finishes.
During the transport, the model was calculating the contributions
of the particles, both thermal and CRs, to the particle distribution
at every point. This process is described
in Section~\ref{subsec-calcfp}. The fluxes of mass,
momentum, and energy of the particles (and the accompanying
magnetic fields) are the important moments of the
particle distribution function, that can be calculated directly
from particle trajectories.
`Balancing the books' after the Monte Carlo iteration, one may
either find that the above mentioned fluxes were constant
throughout the shock, or, if the particle injection was
efficient, one may find that the calculated fluxes
deviated from their upstream value. The latter case
means that the accelerated particles gained too much
energy from the bulk flow to be just a small admixture.
Indeed, the Fermi-I process powers particle acceleration
by giving a fraction of the bulk flow energy to the
particle scattering in it. There is only a finite
amount of energy available for the accelerated particles,
and as soon as they borrow a significant amount of it,
the energy-bearing flow must change. This is where 
the nonlinearity of efficient particle acceleration
comes into play.

In order to obtain a steady state solution consistent
with the fundamental conservation laws (i.e., conservation
of mass, momentum and energy), the simulation invokes
an iterative procedure. Using the calculated, non-equilibrium,
fluxes of mass, momentum and energy, the simulation adjusts
the flow speed in the precursor, $u(x<0)$, making it decrease 
slightly towards the subshock, as outlined
in Section~\ref{subsec_smoothing}. It will be referred to as
precursor smoothing.
The value $u(x=-0)\equiv u_1$
is the flow speed just before the subshock.
Then a second Monte Carlo
iteration of particle propagation may be run. Thermal
particles are introduced far upstream, where the flow
is yet unmodified, $u(x)=u_0$, and allowed to propagate,
get injected and accelerated. However, this time, the flow
speed difference between the downstream region
and the upstream region is smaller due to the slowing
down of the incoming flow in the precursor. This means
that particle acceleration will borrow less energy from
the flow than in the previous iteration.

Continuing the iterative process of particle propagation
followed by the flow speed adjustment, one may obtain
a self-consistent solution, in which particles gain just
the right amount of energy from the flow, to conserve
mass, momentum and energy of the sum of the
slowed down upstream flow and the accelerated particle
distribution. It turns out that, in the presence of
highly relativistic particles that change the compressibility
of the plasma, and due to the assumption of the upstream
particle escape at $\xfeb$, one must also adjust the
downstream flow speed, $u_2$, in order to conserve the
fundamental fluxes across the subshock as well.

The procedure outlined above is the method of solving
the problem of nonlinear particle acceleration developed
by Ellison and co-workers.

In order to implement efficient magnetic field amplification
and self-consistent particle transport, the author made
adjustments to the procedure. The underlying theory and
practical details are explained in the following two
Sections, \ref{mfa_in_mc} and \ref{advanced_transport}.

First of all, in addition to the flow speed, $u(x)$ being
an unknown function derived by the iterative procedure,
we now seek for turbulence spectrum, $W(x,k)$,
in a similar way (see Section~\ref{mfa_in_mc}).
Starting with some initial guess
for $W(x,k)$, the simulation runs the diffusion module, that analyzes
the spectrum $W(x,k)$ and calculates the corresponding
mean free path at all energies, $\lambda(x,p)$
(Section~\ref{advanced_transport} is devoted to this
calculation).
Then the Monte Carlo transport module is executed, that
simulates particle acceleration in the given
structure of the shock with the scattering properties
determined by $\lambda$. After that, collecting the information
about mass, momentum, and energy fluxes, the model estimates
the smoothing of the precursor required for the next iteration.
Additionally, it collects the information about particle streaming
during the acceleration process [i.e., the diffusive
current of CRs, $j_d(x)$ or the CR pressure, $\Pcr(x,p)$].
Using this information, the code runs the magnetic field amplification
module that calculates the turbulence generation by the
particle streaming and adjusts the turbulence spectrum
for the next iteration, $W(x,k)$. The latter is then used
to improve the guess on the particle mean free path,
$\lambda(x,p)$, and the next Monte Carlo transport
iteration starts. Similarly, this process continues until
a self-consistent solution is derived, one that preserves the 
mass, momentum and energy fluxes, and in which particle acceleration
produces the spectrum of accelerated particles that
generates precisely the magnetic turbulence spectrum used
for simulating the particle acceleration.

\subsection{Particle propagation, pitch angle scattering}

\label{subsec-pascatt}

\subsubsection{Theory}

Propagation of particles is performed using the methods developed and
presented in \cite{EJR90}. The bottom line of the reasoning provided
in this work is the following procedure. If the particle has a mean
free path $\lambda$ and a corresponding collision time
$t_c=\lambda/v$, where $v$ is the particle speed, then the scheme
allows the particle to travel a finite time, $\Delta t \ll t_c$, in a straight line,
and then rotates the particle's momentum. The momentum is rotated by
an angle $\delta \theta$, which is chosen randomly as follows:
\begin{equation}
\cos{\delta \theta} = 1 - \mathcal{X}\left(1-\cos{\Delta
  \theta_\mathrm{max}}\right). 
\end{equation}
Here $\mathcal{X}$ is a random number with a uniform distribution
between $0$ and $1$, and 
\begin{equation}
\label{deltathetamax}
  \Delta \theta_\mathrm{max}= \sqrt{\frac{6 \Delta t}{t_c}}
\end{equation}
is the maximal scattering angle. The scheme works consistently when
this angle is small. After choosing the polar angle of scattering,
$\delta \theta$, one must choose the azimuthal direction of
scattering, $\delta \phi$. Assuming the scattering is isotropic,
\begin{equation}
\delta \phi = 2 \pi \mathcal{Y} - \pi,
\end{equation}
where $\mathcal{Y}$ is another random number uniformly distributed
between $0$ and $1$. To rotate the momentum, let's define spherical
coordinates in the momentum space so that the azimuthal angle,
$\theta$, is measured from the positive $p_x$ axis. Given the
spherical angle of the original momentum,  $\theta$,
such that $\cos{\theta} = p_x / p$, 
we can calculate the spherical angle of the scattered momentum,
$\theta'$, from:
\begin{equation}
  \cos{\theta'} = \cos{\theta}\cos{\delta \theta} + 
                  \sin{\theta}\sin{\delta \theta} \cos{\delta \phi}
\end{equation}
The spherical angles $\phi$ and $\phi'$ don't matter in our
1\nobreakdash-dimensional, axially-symmetric model. 

\begin{figure}[htbp]
\centering
\vskip 0.5in
\includegraphics[width=4.0in, clip=true, trim=2.0in 1.5in 2.0in 1.0in]{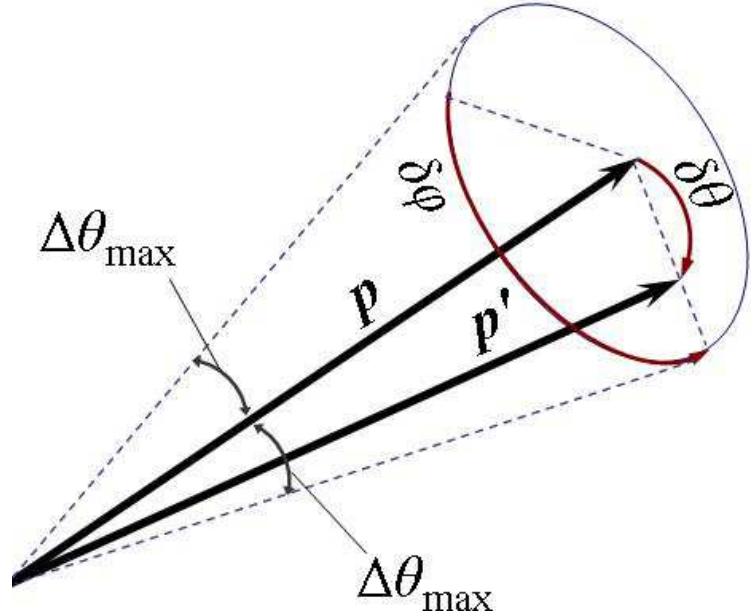}
\caption{$ $ Pitch angle scattering diagram.}
\label{fig-pas}
\end{figure}

The diagram illustrating the process is shown in Figure~\ref{fig-pas}.
The initial momentum vector $\pvector$ and the final (scattered) 
momentum vector $\pvector'$ are shown with thick arrows,
and the dotted lines crossing at the tails of $\pvector$
and $\pvector'$ show the maximal scattering cone with
the half-opening angle of $\Delta \theta_\mathrm{max}$.
The vector $\pvector'$ is obtained from $\pvector$ by rotating
the latter by a random angle $\delta \theta$
(such that $0 < \delta \theta < \Delta \theta_\mathrm{max}$)
about the tail of $\pvector$, and then by turning it
by a random angle $\delta \phi$ (such that
$0 < \delta \phi < 2 \pi$) about the axis of $\pvector$.

Note that the term `pitch angle' in plasma physics means
the angle between the magnetic field vector and the
particle momentum vector. This term was used in the name of this
procedure because initially it was assumed that a uniform
magnetic field, ${\bf B}_0$, dominates the magnetization, with small
fluctuations of this field providing the scattering. In
the case of strong turbulence, however, the uniform pre-existing field will
be overwhelmed by the self-generated fluctuations
$\Delta B  \gg B_0$, so the term `pitch angle scattering'
is actually a misnomer. By the pitch angle in this scheme
I simply mean the angle between the momentum vector
$\pvector$ and the positive direction of the $x$-axis
(the direction of upstream plasma flow).

\subsubsection{Tests}

\begin{figure}[htbp]
\centering
\vskip 0.5in
\includegraphics[angle=-90, width=4.0in, clip=true, trim=0.7in 0.0in 0.0in 0.0in]{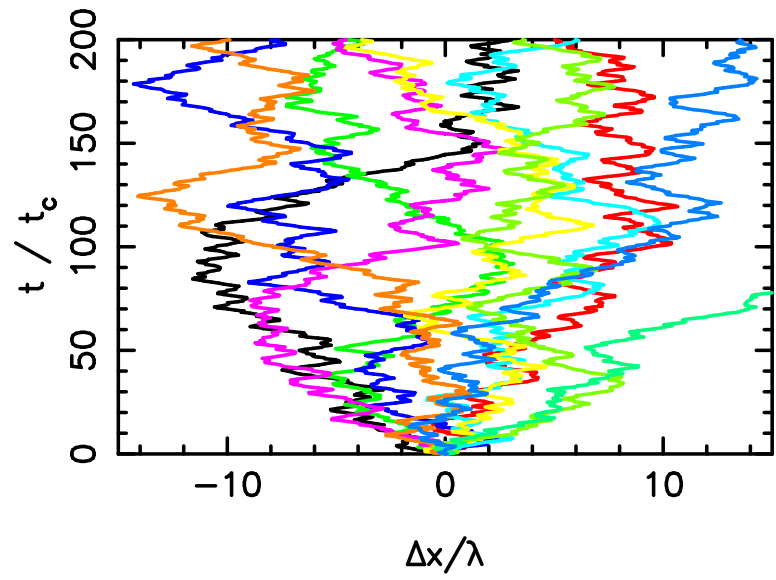}
\caption{$ $ Particle trajectories calculated in the Monte Carlo code.}
\label{fig-dxvst}
\end{figure}

We illustrate the trajectories of the particles subject to the pitch
angle scattering in Figure~\ref{fig-dxvst}. For about 10 particles
injected at some position at $t=0$, the program recorded their deviations from the
original position, $\Delta x$, versus time, $t$. It can be
seen that the particles frequently change the direction of motion,
which results in a stochastic transport. In the plot, the time $t$ is
normalized to the collision time, $t_c$, which is defined as the ratio
of the mean free path, $\lambda$, to the particle speed, $v$.

\begin{figure}[htbp]
\centering
\vskip 0.5in
\includegraphics[angle=-90, width=4.0in, clip=true, trim=0.8in 0.0in 0.0in 0.0in]{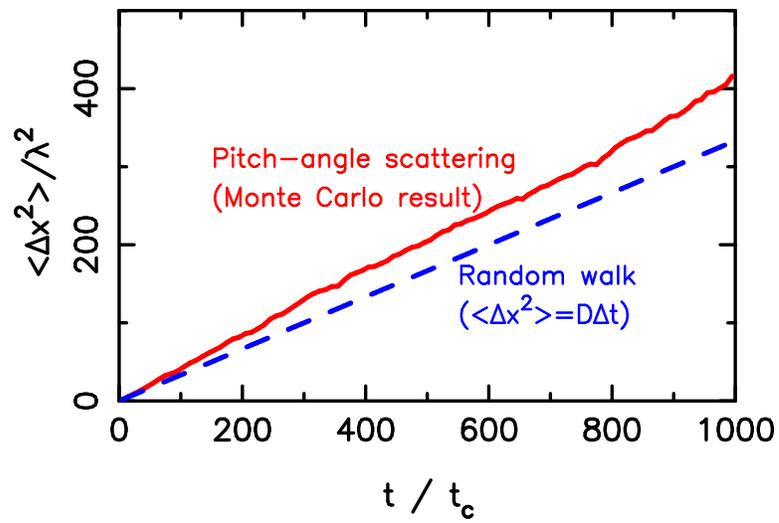}
\caption{$ $ Particle diffusion in the Monte Carlo code.}
\label{fig-dx2vst}
\end{figure}

In order to verify that the properties of this particle transport
correspond to diffusion, I plotted in Figure~\ref{fig-dx2vst} the
mean square of deviation of particle coordinates, 
$\left<\Delta x^2\right>$, as a function of time. The
random walk process (i.e., such particle propagation that
every step has the length $\lambda$ and is taken in a
random direction) is a well-known textbook problem, and
the solution predicts that the displacement 
$\sqrt{ \left< \Delta r^2 \right>} = \sqrt{N} \lambda$, where $N$ is
the number of the steps, and coordinate
$r^2=x^2 + y^2 + z^2$. Therefore $\left<\Delta x^2\right> = D t$,
where $D=v \lambda/3$, and $t=N t_c$. This dependence is shown in
Figure~\ref{fig-dx2vst} with the dashed line labelled ``Theory'',
while the result of the Monte Carlo pitch-angle scattering simulation, 
averaged over 1000 particles, is shown with the solid line.

One can clearly see that there is a linear dependence of the mean
square displacement, $\left<\Delta x^2\right>$, on time, $t$. However,
the slope of the solid line is some $20$\% greater than
in the random walk theory. This discrepancy is due to the definition of the mean
free path made in \cite{EJR90} in the derivation of the pitch angle 
scattering scheme. This definition is slightly different from the
definition of the MFP in the random-walk model, where particles make
discrete steps of length $\lambda$ in a completely random direction.
However, a $20$\% difference in the value of the diffusion
coefficient is a small factor compared to the uncertainties in the
models of particle transport, and one can conclude that
both models (random walk and pitch-angle scattering) 
describe the process of particle diffusion reasonably
well.

It must be explained why the authors of the Monte Carlo transport
simulation chose the pitch angle scattering
scheme instead of a simpler version of a random walk process
with the step size being equal to the particle mean free path.
The solution may, in principle, have sharp gradients of flow speed,
magnetic turbulence, or particle pressure, in which
case the momentum distribution of some particles will be anisotropic.
For example, the thermal particles first crossing the subshock,
and finding themselves downstream, initially have a strongly
anisotropic distribution of momentum. The random walk process assumes 
isotropy of particles in $p$-space, which may limit the
physical veracity of the model, while pitch angle
scattering accounts for the anisotropies exactly. In this sense,
the Monte Carlo model goes one step beyond the diffusion approximation
adopted in the simpler analytic models. This allows us to calculate the
particle injection rate in the thermal leakage model, where the
downstream thermal particles get injected by returning
upstream in the process of their stochastic propagation,
while some simpler analytic models resort to introducing an
additional free parameter to fix particle injection rate
(e.g., \cite{BGV2005}).

\subsection{Motion of the scattering medium}

\label{subsec-motion}

\subsubsection{Theory}

With the diffusive transport of particles developed and tested, one
needs to incorporate the advection of the plasma. The paradigm of the Monte
Carlo model (as well as of other approximations of the NL DSA problem)
is that there is a bulk flow of thermal plasma with respect to the
shock, with magnetic fields frozen into it. Therefore, the scattering
is elastic and isotropic in the plasma reference frame (elastic,
because the scattering represents the action of the magnetic force,
which doesn't perform work on the particle, and isotropic due to the
assumption of strong developed turbulence adopted in the model; in
some cases, anisotropic scattering is a better approximation). If
there is a spatial variation in the speed of the bulk flow, 
then two subsequent particle scatterings may take place in two different reference
frames, which may increase the energy of the particle (resulting eventually in
the Fermi-I process).

It is important to properly account for all effects of special
relativity in order to model relativistic particle propagation.
In order to model particle propagation in the moving plasma
along with the diffusive transport,  let us assume the following
process. Suppose the particle is at the location $x$ with 
the local plasma flow speed $u(x)$. The code will allow this
particle to travel a certain time in the plasma frame,
and accordingly change its coordinate in the shock frame.
Therefore, one may introduce an effective speed
\begin{equation}
v_{x,\;\mathrm{eff}} =
\frac{ v_x + u(x) }{\sqrt{1-\displaystyle\frac{u^2(x)}{c^2}}} .
\end{equation}
Here $v_x$ is the $x$-component of the particle's velocity
in the plasma frame, and the physical meaning of
$v_{x,\;\mathrm{eff}}$ is that, given a time $\Delta t$ in 
the plasma frame, the particle's displacement in the
shock frame will be $\Delta x = v_{x,\;\mathrm{eff}} \Delta t$.
Then the program chooses the time the particle will be allowed
to travel in the plasma frame, $\Delta t$, so that
the resulting $\Delta \theta_\mathrm{max}$ will be small enough,
as given by equation~(\ref{deltathetamax}). After that,
we change the particle's coordinate in the shock frame by
$\Delta x = v_{x,\;\mathrm{eff}} \Delta t$ and proceed
with the pitch angle scattering routine.

An important new aspect of this process is that in a model
with efficient magnetic field amplification,  one expects large
gradients of the mean free path as well as of $u(x)$. This means
that when $\Delta x$ is large enough for the particle to cross
one or more numerical grid planes at which all
the physical quantities are defined (see Section~\ref{subsec-calcfp}),
care must be taken to account for
the changing properties of the medium. The simulation deals with this in
the following manner. I choose a single numerical
value of the angle $\Delta \Theta_\mathrm{max}$
for all particles throughout the simulation.
Then the code allows each particle to propagate just
enough time, so that the `accumulated' maximal scattering
angle is exactly $\Delta \Theta_\mathrm{max}$. If a particle is to 
cross several grid planes in the course of this time, then
I define a cumulative maximal scattering angle as
\begin{equation}
\Delta \Theta_\mathrm{max, \; cml}^{2} = 6 \sum_{i}{\frac{\Delta t_{i}}{\tau_\mathrm{coll, \; i}}},
\end{equation}
where the summation index $i$ runs over all the spatial bins
that the particle had crossed, each bin with a different
value of $\taucoll$, and $\Delta t_{i}$ is the amount
of time that this particle spent in bin $i$. Only after 
$\Delta \Theta_\mathrm{max, \; cml} =  \Delta \Theta_\mathrm{max}$,
does the Monte Carlo routine scatter the particle. This makes the results
independent of the choice of grid plane locations, as long as the
separation between them is small enough.

\subsubsection{Test}

Let us perform a test of the advection superimposed on diffusion
in the Monte Carlo code. With a bulk plasma flow of speed $u_0$ set
up, I introduce the test particles at $x=0$ with their plasma
frame velocity in the positive $x$ direction, and let them
propagate. Particles are assumed to move according to the pitch 
angle scattering scheme described in the previous section;
between the scatterings, the motion of the particles is
ballistic in the plasma frame, which moves at the speed $u_0$ 
with respect to the stationary frame.
The scatterings are isotropic and
elastic in the plasma frame, with the corresponding momenta in the
stationary frame Lorentz transformed.

\begin{figure}[htbp]
\centering
\vskip 0.5in
\includegraphics[angle=-90, width=4in, clip=true, trim=0.7in 0.0in 0.0in 0.0in]{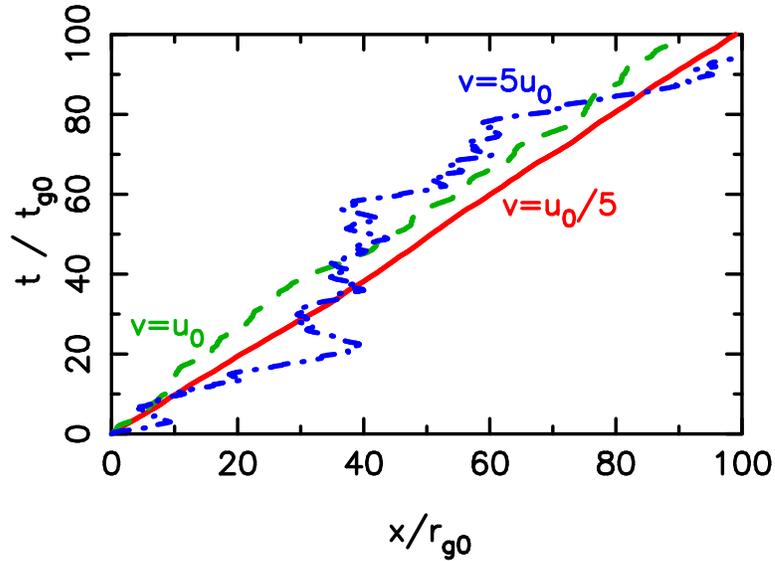}
\caption{$ $ Advection with diffusion.}
\label{fig-adv}
\end{figure}

In Figure~\ref{fig-adv}, I illustrate the results of the test.
I introduced 3 particles with different
speeds in the plasma frame: a slow particle with $v=u_0/5$, a
moderately fast particle with $v=u_0$, and a fast particle with
$v=5u_0$ (here and below, the letter $u$ will denote the speed of the
flow, and the letter $v$ will refer to the speed of a particle,
measured either in the stationary, or in the plasma reference
frame). The solid line represents the motion of the first, the dashed
line -- of the second, and the dash-dotted line -- of the third
particle, respectively. The spatial coordinate, $x$, is measured in
the units of $\rgzero$, the latter being the mean free path of the
particle with speed $v=u_0$, and time is measured in units of
$t_\mathrm{g0}=\rgzero / u_0$. It was assumed that the mean free path
is proportional to the speed of the particle in the plasma frame.
One can see from the Figure~\ref{fig-adv} that the slow particle moves
almost synchronously with the flow (the solid line is very close to
$x=u_0 t$). The faster particle, with $v=u_0$, on the average moves
along with the plasma, but sometimes slower, and sometimes faster, 
because its x-component of velocity in the stationary frame,
$v_x+u_0$, varies between $0$ and $2u_0$, depending on the
orientation of the particle's momentum. The fastest particle, $v>u_0$,
can move backwards, as the dash-dotted line shows. However, its motion
is still affected by the flow, shifting the average location at the
advection speed $u_0$.

Because the purpose of the model under development is to model nonlinear shock
acceleration, where the flow speed $u(x)$ can vary with distance,
smoothly (in the shock precursor) or discontinuously (across the
subshock), one must see how the code treats such varying flow
speeds. Let us study three cases: a uniform flow speed,
a smoothly varying flow speed, and a discontinuity in the flow speed
(representing a shock). In a separate test,
I will introduce a slow particle ($v=0.3u_0$)
at $x=-100 \; \rgzero$, where $u(x)=u_0$, and trace it as it propagates
in the flow. The code will record the positions of the particle, measured in
$\rgzero$ (the latter is, again, the mean free path of a particle with
$v=u_0$, and the mean free path for any other particle energy is
proportional to the particle speed), and the $x$-components of the
particle's velocity in the plasma frame, $v_x$.

\begin{figure}[hbtp]
\centering
\vskip 0.5in
\includegraphics[height=7in]{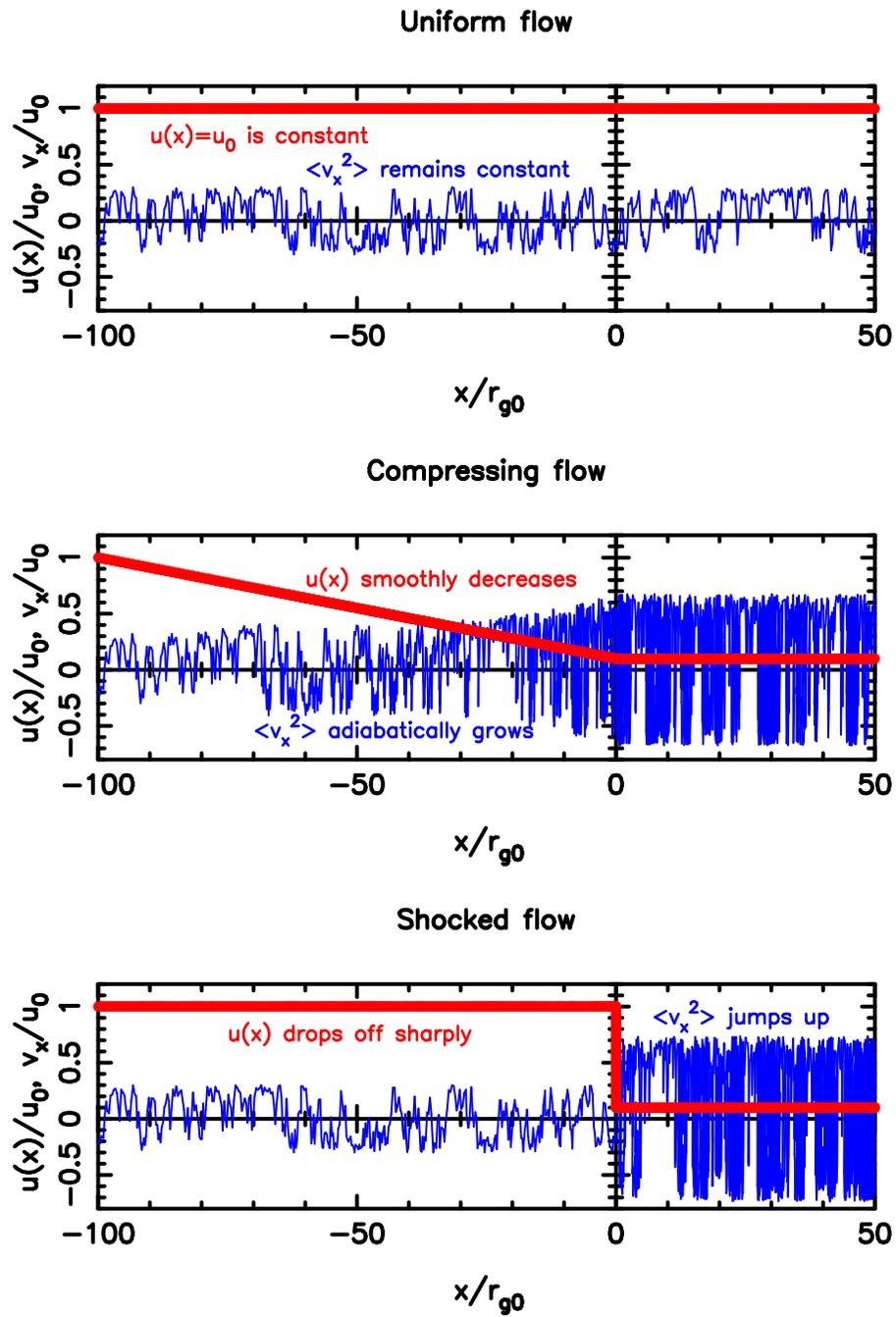}
\caption{$ $ Particle heating in a compressing flow.}
\label{fig-compr}
\end{figure}

Figure~\ref{fig-compr} shows the results for these three cases. In the
first case (top panel), with constant flow speed, $v_x$ varies with time
in the range $-0.3 u_0 < v_x < +0.3u_0$, with the average $\left<v_x\right>=0$,
but the dispersion $\left<v_x^2\right>$ remains constant. This case is similar to
the situation studied in the previous test. In the second case (middle
panel of Figure~\ref{fig-compr}), where
the flow speed linearly drops from $u(x=-100\;\rgzero)=u_0$ to
$u(x=0)=u_0/10$ and then remains constant, we see a different
behavior. The average particle motion is still locked with the
plasma ($\left<v_x\right>=0$), but the dispersion $\left<v_x^2\right>$ increases as $u(x)$
drops. This means that the particle's energy in the plasma frame
grows. Such energizing of the particles is, in
fact, the adiabatic heating of a gas put in a slowly shrinking
volume (see Appendix~B of \cite{VBE2008} or Section~1.2 in \cite{Wentzel74}).
In the third case (bottom panel of Figure~\ref{fig-compr}),
where the flow speed is constant at $x<0$,
but drops abruptly from $u(x=-0)=u_0$ to $u(x=+0)=u_0/10$, similarly
to what it looks like in a shock, the particle is energized
significantly at the shock crossing. I, actually, had to introduce a
reflecting boundary at $x=0$ that doesn't allow the particles to cross
the shock backwards, from $x>0$ to $x<0$, in order to show a concise
plot. Such crossing back becomes possible because the speed of the
particle in the plasma frame is greater than the shock speed.

\subsection{Calculating particle distribution and its moments}

\label{subsec-calcfp}

To calculate the particle distribution in the simulation, the simulation registers particles
crossing certain locations, that we hereafter refer to as `grid planes',
because these locations also define the spatial grid, at the nodes of which
all the quantities: $u(x)$, $W(x,k)$, etc., are defined. One may think of this
as detection of particles by imaginary detectors placed at
discrete locations upstream and downstream of the shock. This calculation may
seem a bit tricky because each crossing of a detector by a particle
contributes not to the density, but to the flux of particles, so in order
to extract the particle distribution information, one needs to properly weight
the detected information. However, this weighting is a standard procedure
for simulations like ours, and below I demonstrate the reasoning leading to
it and examples of the scheme at work.
 
The Monte Carlo simulation does not populate the whole space with particles; 
instead, it introduces $N_p$ particles upstream and propagates them one by one,
until each leaves the system; yet, it is simulating a steady state 
solution with this process. This means that if one wants to calculate 
the particle distribution function $f(x,\pvector)$ and its moments 
(momentum and energy fluxes) at some spatial locations, one must collect 
the information about the particles in such a way that all the data collected 
in the course of one iteration (i.e., during the propagation of all 
the $N_p$ simulation particles) represents the information that would be 
collected by particle detectors placed in the plasma {\it in a unit time} 
(e.g., in one second).

The simulation only registers the particles' contribution to 
$f(x, \pvector)$ when they cross one of the `detectors'. Consider an infinite plane detector 
in a spatially uniform plasma with no bulk motion ($u_0=0$) and an isotropic distribution 
of particle momenta; assume also that all particles have the same speed $v$. Clearly, 
in a unit time such a detector will register more particles incident normally onto 
it than tangentially to its plane. It is easy to understand that the number of 
detected particles with a certain $v_x$ is proportional to $P(v_x) \propto | v_x |$, 
where $v_x$  is the x-component of the particle's velocity in the rest frame of the 
detector and $P(v_x)$ is the probability density of its detection in a unit time. 

%
When a `detector' in the simulation registers a particle with small $| v_x |$, 
we must interpret it as that there are many similar particles at this location, 
but, because of an unfavorable direction of motion, only a few reach this 
parallel-plane detector in a unit time that one iteration represents. 

Quantitatively, if one wants to calculate the number density of particles at the
location of the detector, the code must compute the following sum:
\begin{equation}
\label{dens_calc}
n(x_i) = \sum_{j} {w_j w_p} = \sum_{j} {\left| \frac{u_0}{v_{x,\;j}} \right| w_p }
\end{equation}
where $i$ is the number of the grid plane, and $x_i$ -- its coordinate,
$n(x_i)$ is the number density of particles at $x_i$, and the summation index $j$
runs over all the events of a particle crossing 
this detector in the course of an iteration.
The weight $w_j$ is the statistical weight of the $j$-th event,
and $w_p$ is the statistical weight of the particle participating in the event.
The statistical weight of a particle, if $N_p$ particles are introduced
upstream, representing a plasma density $n_0$, is simply $w_p = n_0 / N_p$.
The statistical weight of the event is expressed by the ratio
$| u_0 / v_{x,\;j} |$, where the denominator $v_{x,\;j}$ is the $x$-component
of the particle velocity measured in the rest frame of the detector,
at the $j$-th event, and $u_0$ acts as the appropriate normalization factor.

If one wants to calculate the particle distribution function, i.e., the
number of particles in a unit phase space volume $dxdydzdp_xdp_ydp_z$, then the
summation must be restricted to the events corresponding to that phase
space volume. In practice, one may be interested in the distribution 
function $f(x, \pvector)$ such that 
\begin{equation}
  \label{norm_of_fp}
  \int{f(x_i, \pvector)\; d^3 p} = \int{f(x_i,\pvector) p^2 dp d\Omega}= n(x),
\end{equation}
where $d\Omega$ represents the infinitesimal spherical angle 
corresponding to the momentum space volume $d^3p=dp_xdp_ydp_z$.
In the simulation,
given a phase space binned so that $\Delta p_k$ is the width of the
$k$-th momentum bin centered at the momentum value $p_k$, 
and averaging over the angles, one gets
\begin{equation}
  \label{calculation_of_fp}
  \bar{f}(x_i,p_k) = \frac{1}{4\pi}\int{f(x_i, \pvector)\;d\Omega} = 
  \frac{1}{4\pi p_k^2 \Delta p_k}
    \sum_{p_j \in \Delta p_k} \left| \frac{u_0}{v_{x, \; j}}\right| w_p,
\end{equation}
where $d\Omega$ is the differential of the solid angle in $p$-space,
and the index $j$ runs over all particles whose momentum falls into the
$k$-th momentum bin.
Indeed, with $\bar{f}(x,p)$ defined this way,
\begin{eqnarray}
  \nonumber
  \int{f(x_i, \pvector)\; d^3 p} &\equiv& \int{f(x_i,\pvector) p^2 dp d\Omega} = \\
  \nonumber
    &&=\sum_{k}  \bar{f}(x_i,p_k) 4\pi p_k^2 \Delta p_k \\
  \nonumber
    &&=\sum_{k}  \frac{1}{4\pi p_k^2 \Delta p_k} \sum_{p_{j} \in \Delta p_k} {\left| \frac{u_0}{v_{x,\;j}}\right| w_p } 
                                                  \cdot 4\pi p_k^2\Delta p_k=\\
  \nonumber
    &&=\sum_{k}  \sum_{p_{j} \in \Delta p_k} {\left| \frac{u_0}{v_{x,\;j}}\right| w_p }=\\
  \label{eq_finddens}
    &&= \sum_{j} {\left| \frac{u_0}{v_{x,\;j}}\right| w_p } = n(x_i),
\end{eqnarray}
as expected (in the summation, the index $k$ runs over all the momentum
bins defined in the simulation, and in the last summation, $j$ runs over
all the events of particle crossing of the detector).
In the future, the plots showing $f(x,p)$ actually show
$\bar{f}(x,p)$, but I omit the bar representing the angular averaging
for simplicity.

Calculating the angle-averaged distribution function may be informative,
but for practical purposes one needs the moments of the distribution
function (i.e., the mass, momentum and density fluxes)
that require the information about the angular dependence. From
the above reasoning (namely, Equation~\ref{eq_finddens}) 
one may conclude that the correspondence
between the integration over the momentum space and summation
over particle detection events is as follows:
\begin{equation}
f(x, \pvector) d^3p = 
       \sum_{p_j \in \Delta p_k} \left| \frac{u_0}{v_{x,\;j}} \right| w_p,
\end{equation}
Therefore, for any function of coordinate and momenta $\gfunc(x, \pvector)$,
its expectation value may be calculated directly in the simulation as
\begin{equation}
\int{ \gfunc(x_i, \pvector) f(x_i, \pvector) \;d^3p} = 
       \sum_{j} \gfunc(x, \pvector_j)\left| \frac{u_0}{v_{x,\;j}} \right| w_p.
\end{equation}
Namely, for $\gfunc(x, \pvector)=1$ one finds the particle density,
for $\gfunc(x, \pvector) = m_p v_x$ -- the mass flux in the $x$-direction,
for $\gfunc(x, \pvector) = p_x v_x$ -- the flux of the
$x$-component of momentum in the $x$-direction, and for
$\gfunc(x, \pvector) = K(p) v_x$ -- the energy flux in
the $x$-direction ($K(p)$ is the relativistic
kinetic energy corresponding to the momentum $p=|\pvector|$).
The expectation value of the function $\gfunc(x, \pvector) = e \vvector$ 
is the diffusive current $j_d(x)$, and $\gfunc(x, \pvector) = \frac13 p v$
in the case of isotropic momentum distribution gives the pressure.
\begin{eqnarray}
n(x_i) &=& \int{ f(x, \pvector) \; d^3p} =
       \sum_{j} \left| \frac{u_0}{v_{x,\;j}} \right| w_p, \\
\Phi_\mathrm{M, \, p}(x_i) &=& \int{ m_p v_x f(x, \pvector) \; d^3p} = 
       \sum_{j} m_p v_{x,\;j} \left| \frac{u_0}{v_{x,\;j}} \right| w_p, \\
\Phi_\mathrm{P, \, p}(x_i) &=& \int{ p_x v_x f(x, \pvector) \; d^3p} = 
       \sum_{j} p_{x,\;j} v_{x, \; j} \left| \frac{u_0}{v_{x,\;j}} \right| w_p, \\
\Phi_\mathrm{E, \, p}(x_i) &=& \int{ K v_x f(x, \pvector) \; d^3p} = 
       \sum_{j} K_j v_{x, \; j} \left| \frac{u_0}{v_{x,\;j}} \right| w_p, \\
j_d(x_i) &=& \int{ e v_x f(x, \pvector) \; d^3p} = 
       \sum_{j} e v_{x,\;j} \left| \frac{u_0}{v_{x,\;j}} \right| w_p, \\
\Pth(x_i) &=& \int\limits_\mathrm{th}{ \frac13 v p f(x, \pvector) \; d^3p} = 
       \sum_{j \in \mathrm{th}} \frac13 v_j p_j \left| \frac{u_0}{v_{x,\;j}} \right| w_p, \\
\Pcr(x_i) &=& \int\limits_\mathrm{cr}{ \frac13 v p f(x, \pvector) \; d^3p} = 
       \sum_{j \in \mathrm{cr}} \frac13 v_j p_j \left| \frac{u_0}{v_{x,\;j}} \right| w_p,
\end{eqnarray}
etc. In the last two equations, the integration was limited to the thermal or to CR
particles only. In this way I define the thermal pressure
and the CR pressure. I must remind the reader here that
a peculiarity of the approach I adopted is that in order to separate the CR
particles from the thermal ones, the code uses their history, and not their
energy.  By my definition, a thermal particle is one that had been
introduced into the simulation upstream with a random thermal energy and
that may have crossed the subshock going downstream, but has never crossed
it back. Once a particle crosses the subshock (the coordinate $x=0$, to
be more precise) in the upstream direction, it by my definition
is injected and becomes a CR particle.

Let us note here that, assuming that the particle distribution is
isotropic, the fluxes of energy and momentum can be expressed via
gas pressure as
\begin{eqnarray}
\label{momentumviapressures}
\int p_x v_x f(x,\pvector)d^3p &=&
\rho(x) u^2(x) + P_p(x), \\
\label{energyviapressures}
\int K v_x f(x,\pvector)d^3p &=&
\frac{1}{2} \rho(x) u^3(x) + w_p(x)u(x) \ ,
\end{eqnarray}
where $P_p(x)$ is the pressure and $w_p(x)$ is the enthalpy
of the particles. These are well known results
of the kinetic theory of gases.

\subsection{Introducing particles into the simulation}

\label{subsec_particleintro}

\subsubsection{Theory}

In order to start the Monte Carlo simulation of particle transport,
we must introduce thermal particles far upstream, or close
to the subshock (the latter method is described in 
Appendix~B of \cite{VBE2008}).
This task has two components: generating a particle population
with the proper energy distribution, and choosing the appropriate
angular distribution for these particles. 

We normally assume that the unshocked plasma is thermal
and has a certain temperature $T_0$ (a typical cold
interstellar plasma has $T_0 \approx 10^4$~K).
In order to generate a thermal population,
we randomly choose the momentum of every particle introduced into
the simulation so that the resulting distribution
is Maxwellian:
\begin{equation}
\label{eq_maxwellian}
f(p) = n_0\left( \frac{1}{2 \pi m k_B T_0} \right)^{3/2}
  \exp{\left({-\frac{p^2}{2 m k_B T_0}}\right)}.
\end{equation}
In order to accomplish this, consider the function 
$F(p)$ such that 
$F(p)\Delta p$ is the fraction of the thermal particles with
momenta in the interval $[p-\Delta p/2 ; p+\Delta p/2]$, i.e.,
\begin{equation}
  F(p)=\frac{4\pi p^2f(p)}{n_0} = \frac{4}{\sqrt{\pi}} \left( \frac{1}{2 m k_B T_0} \right)^{3/2}
  p^2 \exp{\left({-\frac{p^2}{2 m k_B T_0}}\right)}.
\end{equation}
By substitution 
\begin{equation}
y = \frac{p^2}{2 m k_B T_0},
\end{equation}
using the identity $G(y)dy=F(p)dp$, we find that $y$ has the following distribution function:
\begin{equation}
  G(y) = \frac{2}{\sqrt{\pi}}y^{\frac{1}{2}}e^{-y}
\end{equation}
(here $G(y)\Delta y$ is the fraction of particles with momenta
corresponding to the interval $[y-\Delta y/2; y+\Delta y/2]$).
The latter is a Gamma distribution with parameter $a = 3/2$. To generate
a quantity $y$ with the above distribution, one can use the following recipe:
\begin{equation}
  \label{maxwinj}
  y= \frac12 \mathcal{Z}^2 - \ln \mathcal{X}.
\end{equation}
Here $\mathcal{X}$ is a random deviate with a uniform distribution in $(0;\;1]$, and
$\mathcal{Z}$ is a deviate the the normal (Gaussian) distribution 
with the mean equal to $0$ and the dispersion equal to $1$. The above
method was adopted from \cite{Devroye86}.

\subsubsection{Test}

\begin{figure}[htbp]
\centering
\vskip 0.5in
\includegraphics[angle=-90, width=4in]{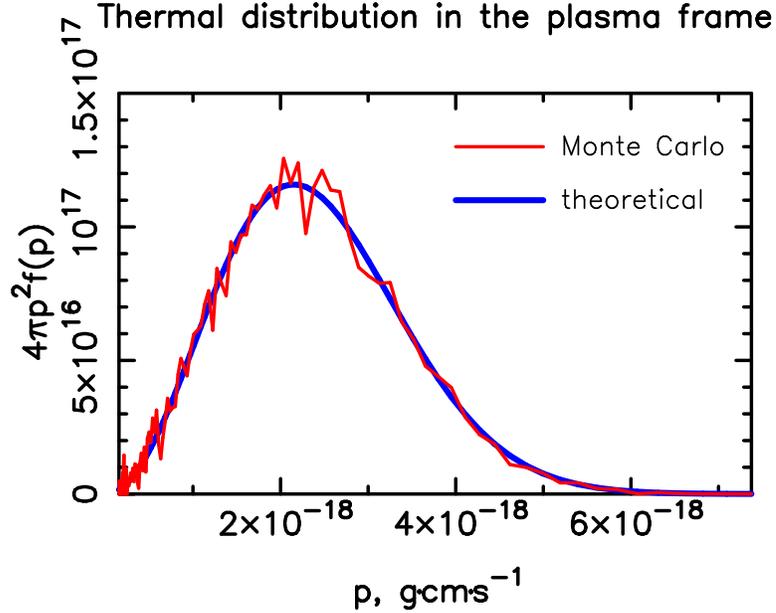}
\caption{$ $ Generation and detection of thermal particle population.}
\label{fig-pdftest}
\end{figure}

I tested the implementation of this procedure, and verified the
distribution function in the Monte Carlo
simulation, and the test
results are shown in Figure~\ref{fig-pdftest}.

The code introduced $N_p = 10^4$ particles into the simulation, assuming
that their temperature is $T_0=10^4$~K, and the density
$n_0=0.3$~\pcc. The thick line shows the desired Maxwellian
distribution calculated according to Equation~(\ref{eq_maxwellian}).
The thin line shows the result of the detection of the introduced
particles at some grid plane, as described in
Section~\ref{subsec-calcfp}, according to 
Equation~(\ref{calculation_of_fp}). The local deviations of the Monte 
Carlo result from the theoretical curve are statistical fluctuations,
and these decrease for greater number of particles. Otherwise,
the match is excellent, demonstrating the correct implementation
of the introduction and the detection of particles in my code.

The angular distribution of momenta of the
introduced particles is a major issue of concern for a simulation
like ours, because it determines the rate of particle
injection into the acceleration process. When the simulation introduces particles at the coordinate $x$,
it is replacing the dynamics of these particles upstream of $x$
with an analytic description, consequently it must  distribute
particles in $p$-space at $x$ the way they would be distributed
having traveled from far upstream and reaching $x$ {\it for the first
time}. This is equivalent to calculating a $p$-space distribution of
particles incident on a {\it fully absorbing boundary} at $x$ after
scattering in a non-uniform flow $u(x)$. This is easy to do analytically
if all particles have a plasma frame speed $v$ less than the flow speed
$u(x)$ (because then all particles crossing position $x$ do it for
the first and the last time), and fairly complicated otherwise
(see Appendix~B). Let us
assume $v<u(x)$ in further reasoning, which is justified as long as
the local sonic Mach number at the introduction position is large.

The problem is now reduced to the following. We know how to designate
an introduced particle's momentum, $p$. But how do we choose
its direction, identified by the angle $\mu$ such that
$\cos{\mu}=p_x/p$? As was stated earlier, it is assumed that the angular distribution of
momenta of the introduced thermal particles is isotropic in the plasma
frame, and there is an overall drift speed $u$ superimposed over
this motion in the plasma. Therefore, it may seem natural (but is
incorrect) to just  choose $p_x$ isotropically in the plasma frame, 
and then transform them into the shock frame.
The correct solution must account for the fact that
when these particles cross a grid plane, 
their flux must be `flux-weighted' as seen in Equation~(\ref{dens_calc}),
because the
number of particles arriving at $x$ in a unit time is proportional to
the cosine of the angle that their shock frame velocity
$\vvector_\mathrm{sf}$ makes with the $x$-axis.
This can be  done by assuming a probability density of
of $v_\mathrm{sf,\:x}$ as
\begin{equation}
F(v_\mathrm{sf,\:x}) = \left\{ 
\begin{array}{l r} 
   A v_\mathrm{sf,\:x}, & \quad v_\mathrm{min} < v_\mathrm{sf,\:x} < v_\mathrm{max}, \\ 
   0,     & \mathrm{otherwise}. \end{array} \right.
\end{equation}
Here $v_\mathrm{min} = u - v$, $v_\mathrm{max} = u + v$, and $v$ is the particle
speed in the plasma frame chosen using a random number generator
according to~(\ref{maxwinj}). The constant $A$ can be found from
the normalization condition
\begin{equation}
\int\limits_{v_\mathrm{min}}^{v_\mathrm{max}} F(v_\mathrm{sf,\:x}) dv_\mathrm{sf,\:x} = 1
\end{equation}
as
\begin{equation}
A = \frac{2}{v_\mathrm{max}^2 - v_\mathrm{min}^2},
\end{equation}
so
\begin{equation}
\label{fastpushslowdistr}
F(v_\mathrm{sf,\:x}) = \left\{ 
\begin{array}{l r} 
   \displaystyle\frac{2 v_\mathrm{sf,\:x}}{v_\mathrm{max}^2 - v_\mathrm{min}^2}, & \quad v_\mathrm{min} < v_\mathrm{sf,\:x} < v_\mathrm{max}, \\ 
   0,     & \mathrm{otherwise} \end{array} \right.
\end{equation}
In order to generate such a particle distribution,
we use a random number $\mathcal{Z}$ uniformly distributed between 
$0$ and $1$ and calculate $v_\mathrm{sf,\:x}$
as a function of $\mathcal{Z}$. The identity
\begin{equation}
F(v'_x) dv'_x = H(z') dz',
\end{equation}
and substitution of the uniform distribution
\begin{equation}
H(z') = \left\{ \begin{array}{l r} 1, & \quad 0 < z' < 1, \\ 0, & \quad \mathrm{otherwise} \end{array} \right. 
\end{equation}
lead to:
\begin{equation}
\label{vxvsx}
\int\limits_0^{v_\mathrm{sf,\:x}} F(v'_x) dv'_x = \int\limits_{0}^{z} dz' = z,
\end{equation}
which is where we can derive $v_\mathrm{sf,\:x}(\mathcal{Z})$ from.
Substituting (\ref{fastpushslowdistr}) into (\ref{vxvsx}), we get
\begin{equation}
\label{slowanginj}
v_\mathrm{sf,\:x} = \sqrt{(v_\mathrm{max}^2 - v_\mathrm{min}^2)\mathcal{Z} + v_\mathrm{min}^2}.
\end{equation}

Note that, if we did not account for the flux weighting and just
prepared an isotropic distribution of particles in the plasma frame and
then transformed it into the shock frame, then instead of 
the prescription~(\ref{slowanginj}) we would use
\begin{equation}
\label{wrongslowinj}
v_\mathrm{sf,\:x} = (2\mathcal{Z}-1)v + u,
\end{equation}
which is incorrect, as I show below.

If properly implemented, the introduced particle population
as detected by the grid plane detectors should have a uniform
distribution of $\mu$ in the plasma frame, $g(\mu)=1/2$. This
corresponds to isotropy of $f(\pvector)$. Let us perform
two tests to confirm that the procedure~(\ref{slowanginj})
gives the correct particle distribution isotropic in the plasma
frame.

\begin{figure}[htbp]
\centering
\vskip 0.5in
\includegraphics[angle=-90, width=4in, clip=true, trim=0.8in 0.0in 0.0in 0.0in]{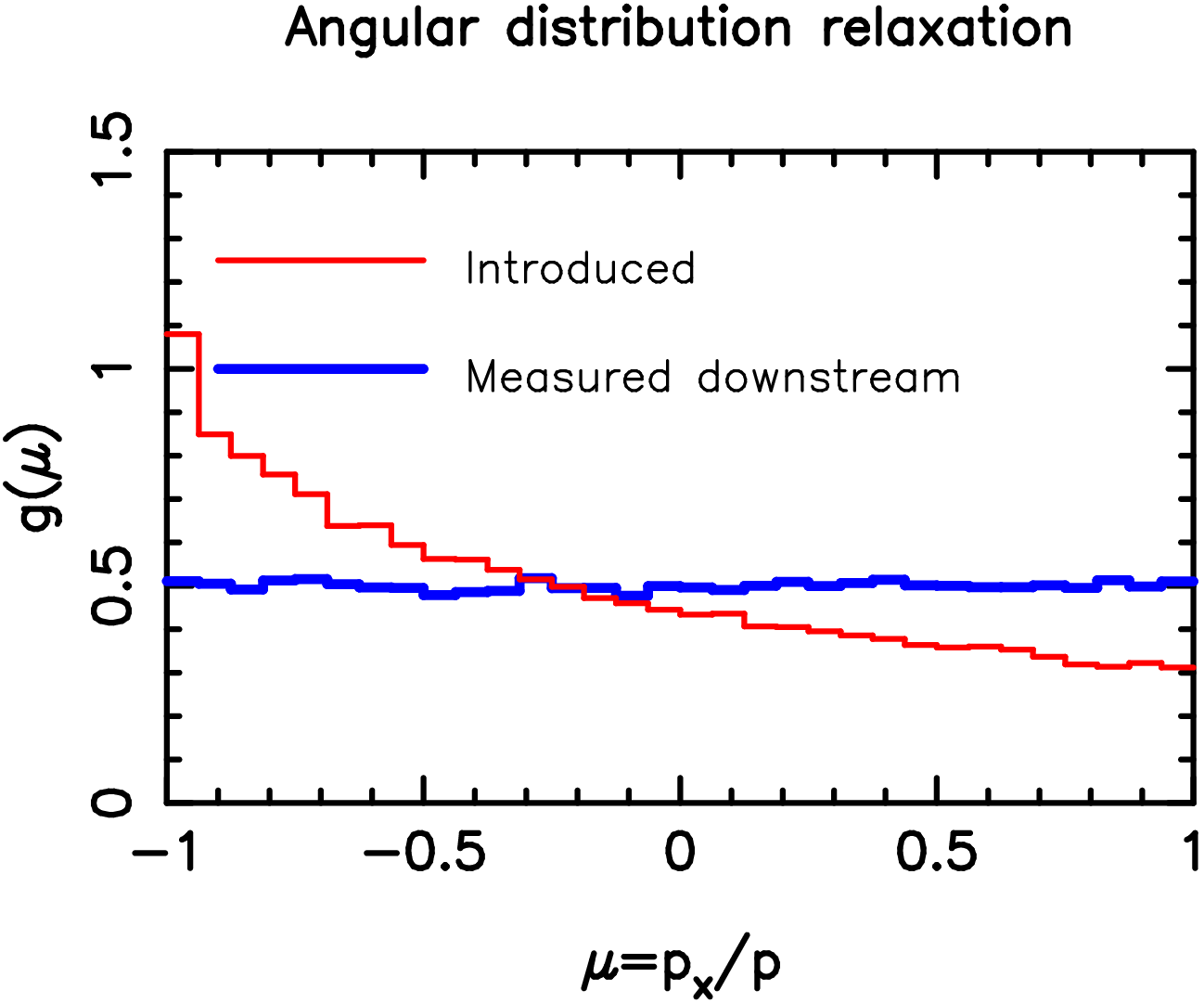}
\caption{$ $ Relaxation of particle distribution to isotropy.}
\label{fig-angtest}
%
\centering
\vskip 0.5in
\includegraphics[angle=-90, width=4in, clip=true, trim=0.8in 0.0in 0.0in 0.0in]{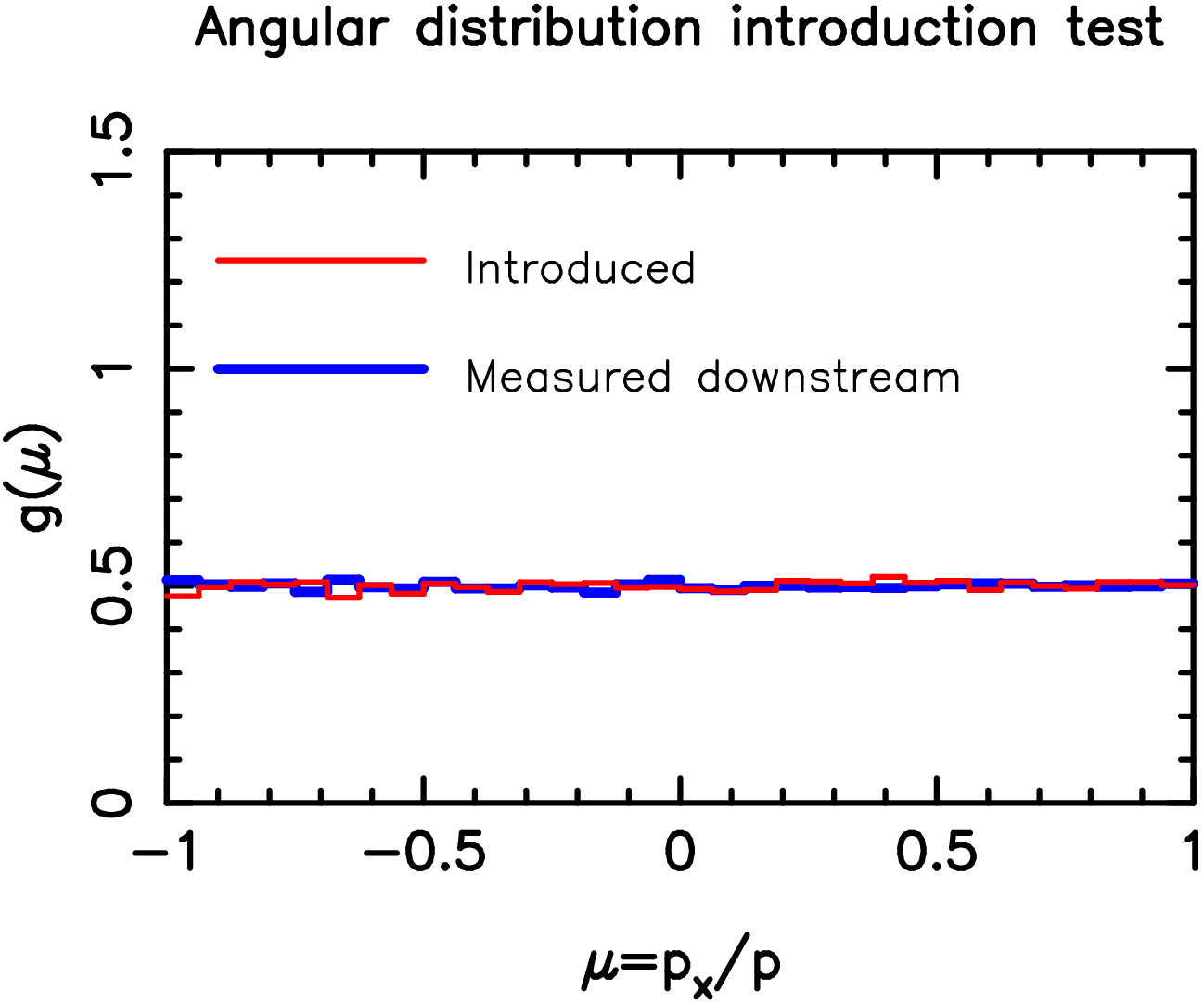}
\caption{$ $ Test of introduced particle distribution isotropy.}
\label{fig-angtest2}
\end{figure}

In the first test, let us introduce $N_p=10^5$ particles into the simulation
with a supersonic flow (sonic Mach number $M_s =2.5$)
using the incorrect recipe~(\ref{wrongslowinj}). At the grid plane
very close to the introduction position, the model will measure the angular
distribution of particles, $g(\mu)$. Several 
(approximately 20) diffusion lengths downstream of the introduction
position, it will measure $g(\mu)$ again. By the time of the second
measurement, the particles must have scattered 
enough to assume an isotropic velocity distribution in 
the plasma frame corresponding to $g(\mu)=1/2$.
The results of the test
are shown in Figure~\ref{fig-angtest}.

As expected, the thick line in Figure~\ref{fig-angtest} is
$g(\mu)=1/2$, meaning that after many scatterings, particles 
isotropize their velocities, and also confirming that I implemented
correctly the calculation of particle distribution and the
pitch angle scattering routine. At the same time, the tilt of the
thin line tells us that the introduced particle distribution was
not isotropic in the plasma frame, thus the
recipe~(\ref{wrongslowinj}) is incorrect.

In the second test, $N_p=10^5$ particles will be introduced,
now using the recipe (\ref{slowanginj}) to choose their angular
distribution. Let the model measure the angular distribution immediately,
and some distance downstream of the introduction position.
The results are shown in Figure~\ref{fig-angtest2}.
The two angular distributions match exactly, which means that
the introduced angular distribution was isotropic, and remained
such after many scatterings.  

These tests conclude the verification of particle propagation
methods of Monte Carlo, and now we can get to testing the
particle acceleration properties.

\subsection{Test particle case of DSA}

\label{tp_dsa_test}

A crucial test that the nonlinear simulation of DSA must pass is the
production of a power-law spectrum of accelerated particle in the
test particle case. Similarly to the solution shown in
Section~\ref{sec_dsa_tp}, a shock characterized by a sharp
transition at $x=0$, in which the flow speed jumps from the
$u_0$ to $u_2$, with a compression ratio $r=u_0/u_2 > 1$ 
and particle injection taking place at the shock via
thermal leakage, must produce a power-law spectrum of 
accelerated particles with the power law index $s$
given by Equation~(\ref{pl_index_tp}). Verifying the power
law index is a strong argument for the correctness
of the model.

We ran 3 simulations with a discontinuous flow speed. In
each simulation, $u_0 = 10^4$~\kmps, $T_0=10^4$~K,
$n_0=0.3$~\pcc,
and to define the mean free path, the model assumed Bohm scattering in 
a magnetic field $B_0=3\cdot 10^{-6}$~G (see 
Section~\ref{sec_bohm}). The difference between the 3
models was the compression ratio, $r$. I chose
$r=7.0$ for the first model, $r=4.0$ for the second
and $r=3.0$ for the third. Note that I did not
require consistency of these compression ratios with the laws
of hydrodynamics, and was only interested in particle
acceleration and its properties. Namely, according
to Equation~(\ref{pl_index_tp}), one expects to get
power-law distributions of accelerated particles downstream
with the indices $s=3.5$, $s=4.0$ and $s=4.5$, for
the first, second and the third model, respectively.
In these runs I took advantage of the procedure
of particle introduction developed in 
Section~\ref{subsec_particleintro} and introduced particles
close to the shock instead of far upstream. This allowed us to
speed up the calculation significantly, and thus I put the free
escape boundary rather far upstream, at $\xfeb=-10^6\;\rgzero$.

\begin{figure}[htbp]
\centering
\vskip 0.5in
\includegraphics[angle=-90, width=4in, clip=true, trim=0.8in 0.0in 0.0in 0.0in]{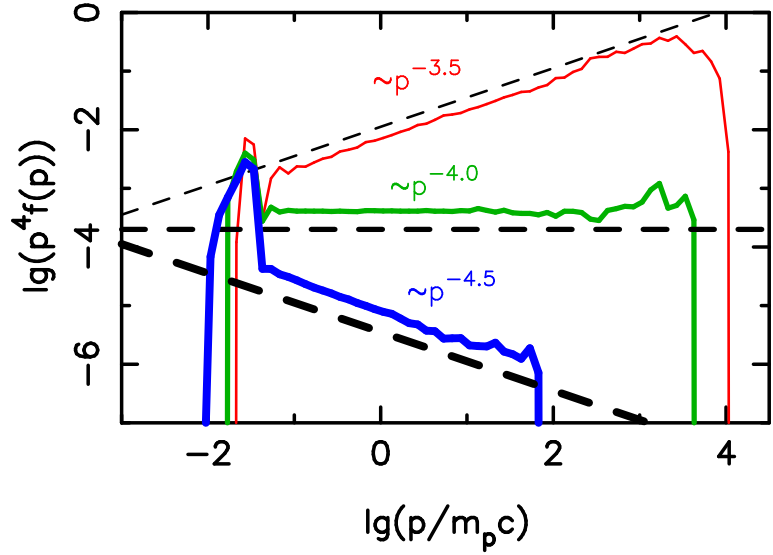}
\caption{$ $ Test particle case of DSA.}
\label{fig-tptest}
\end{figure}

The results of the test -- the particle distributions
measured in the shock rest frame -- are shown
in Figure~\ref{fig-tptest}.
The solid lines are the particle distributions in the shock frame
measured in the downstream region. The $x$-axis shows proton momentum
in units of $m_p c$, and the $y$-axis -- the distribution function
$f(p)$ multiplied by $p^4$ for convenience. This multiplication factor
makes the $r=4.0$ case with the corresponding $s=4.0$ appear
as a horizontal line, which is
a desirable feature of this plot. In the future, all particle
distribution functions shown will have this multiplication factor.
The thin solid line shows the detected distribution function for
$r=7.0$, the medium thickness solid line -- for $r=4.0$ and the thick 
solid line -- for $r=3.0$ case. The dashed lines are added for a comparison.
They have the slopes predicted by Equation~(\ref{pl_index_tp}),
and for the correct result, the dashed lines must be parallel
to the high-energy parts of the distribution functions,
which is certainly the case in the presented results.

\begin{figure}[hbtp]
\centering
\vskip 0.5in
\includegraphics[angle=0, height=7in]{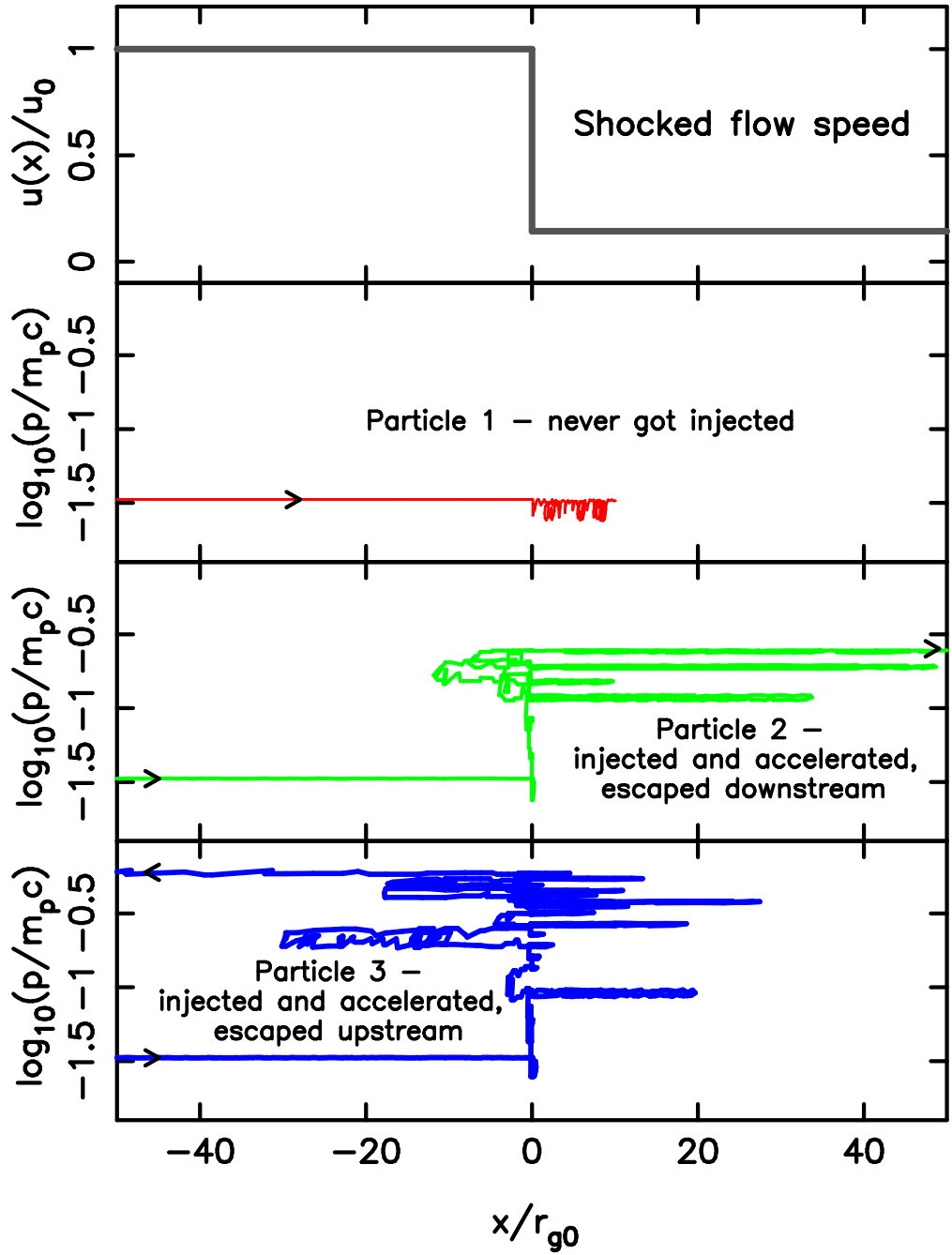}
\caption{$ $ Particle trajectories in DSA.}
\label{fig-dsamech}
\end{figure}

I would also like to illustrate the physical process that leads to
particle acceleration in these tests. In a separate simulation
similar to the $r=7.0$ case, but with a free escape boundary
close to the shock, at $\xfeb=-50\;\rgzero$, the code traced several
particles, and I show their motion in the phase space in
Figure~\ref{fig-dsamech}

The top panel of Figure~\ref{fig-dsamech} shows the shocked flow speed.
Upstream, at $x<0$, the flow is uniform and fast, $u_0=10^4$~\kmps,
and at the shock located at $x=0$, the flow speed drops
down to $u_2 = u_0/r$. This flow speed is measured in the system
where the shock is stationary, so the flow of matter is
directed from the left end to the right end of the plot,
and the shock is directed to the left.

The second panel shows a particle introduced
far upstream, that crossed the shock, got heated, but
never returned upstream and escaped trapped in the downstream
flow (the particle's motion is from the left to the right end
of the plot). This is what happens to all particles in
collisional shocks that are usually observed on Earth,
and no particle acceleration occurs.

The third panel shows a particle initially introduced
far upstream as a thermal particle, but its random motion in the
stochastic magnetic fields, induced by the pitch angle scattering,
lead the particle
back upstream, and it crossed the shock from the $x>0$ region
into the $x<0$ region. In several such subsequent crossings,
the particle gained energy due to the flow speed difference
across $x=0$ (the trajectory moved up). Eventually this particle's
energy became so large that it was able to protrude quite
far upstream, to $x\approx -10\;\rgzero$, but eventually
it was advected downstream with the flow.

The fourth (bottom) panel in Figure~\ref{fig-dsamech} shows
the phase space trajectory of a `lucky' particle that
not only got injected (crossed the shock against the flow),
but gained enough energy to find itself very far upstream,
to the left of the free escape boundary located
at $x=-50\;\rgzero$. There such particles were assumed to escape
from the system in the upstream direction -- this is how
the finite size of the accelerator is modeled in the Monte 
Carlo simulation.
The energy of this particle at the moment of escape was 
close to the maximum achievable particle energy in this shock.
The particle's speed greatly exceeded the shock speed at this
moment, with the particle actually being mildly relativistic
$p\approx 0.6 \; m_p c$. In a model representing a real
SNR shock, the free escape
boundary would have been located much farther upstream,
and particles would be able to make many more crossings before
escaping upstream, and the maximum particle momenta could
be ultrarelativistic, $p \approx 10^6 \; m_p c$.

\subsection{Shock compression ratio in nonlinear DSA}

\label{sec-rtot}

\subsubsection{Theory}

In a standard steady state hydrodynamic shock, the relationship between the
pre-shock macroscopic quantities: $u_0$, $n_0$, $T_0$ and their
post-shock values: $u_2$, $n_2$, $T_2$ is determined by
the sonic Mach number of the shock, and can be derived
from the Rankine-Hugoniot equations:
\begin{eqnarray}
\rho_2 u_2 &=& \rho_0 u_u, \\
\rho_2 u_2^2 + P_2 &=& \rho_0 u_0^2 + P_0, \\
\frac12 \rho_2 u_2^3 + w_2 u_2 &=& \frac12 \rho_0 u_0^3 + w_0 u_0,
\end{eqnarray}
which express the conservation of mass, momentum and energy
fluxes, respectively. Here $P$ is the gas pressure, and 
$w$ -- its enthalpy, $w=\epsilon + P$, and the
internal energy $\epsilon$ of the gas is proportional to
the pressure $P$. For an adiabatic gas with the ratio of specific
heats $\gamma$, one can write $w = \gamma P/(\gamma-1)$,
and the upstream gas pressure $P_0$ can be related to the
sonic Mach number $M_s$ as 
$M_s^2 = (u_0 / c_s)^2 = \rho_0 u_0^2 / (\gamma P_0)$.
This leads to a solution of the Rankine-Hugoniot equations,
relating the pre-shock and the post-shock flow speed and temperature.
That solution is called the Hugoniot adiabat:
\begin{eqnarray}
\label{hugoniot_u}
\frac{u_0}{u_2} &=& \frac{\gamma + 1}{\gamma + 2 / M_s^2 - 1}, \\
\label{hugoniot_T}
\frac{T_2}{T_0} &=& 
   \frac{\left( 2 \gamma M_s^2 - (\gamma-1)\right)
         \left( 2 / M_s^2 + (\gamma-1)\right)}
        {(\gamma + 1)^2},
\end{eqnarray}
where $u_0 / u_2 = \rho_2 / \rho_0 \equiv \rtot$ is the
compression ratio\footnote{Hereafter let us replace the notation of
the total shock compression ratio.
Instead of $r$, we will now denote it as $\rtot\equiv u_0/u_2$, 
to distinguish it from the subshock compression 
$\rsub \equiv u_1/u_2$.}.

In a nonlinear shock with magnetic field amplification,
the situation is complicated by the contributions of cosmic ray
pressure and magnetic turbulence pressure, and by particle
escape far upstream. The procedure of the search
of the self-consistent compression ratio, developed in \cite{Ellison85},
is based on the requirement that in a steady-state system,
mass, momentum and energy fluxes must be constant in space. 
I generalized this procedure 
for the problem of magnetic field amplification by including
the contributions of magnetic turbulence to the
conservation relations, and
developed an iterative procedure for an automated search
of the self-consistent $\rtot$. Consider the conservation relations
\begin{eqnarray}
\label{fluxmass}
\rho(x) u(x) &=& \rho_0 u_0\\
\label{fluxmomentum}
\momentumflux(x) &=&  \upstreammomentumflux,\\
\label{fluxenergy}
\energyflux(x) + \Qesc(x) &=& \upstreamenergyflux.
\end{eqnarray}
Here $\rho$ and $u$ are the mass density and the flow speed,
$\momentumflux(x)$ is the flux of the $x$-component of momentum in the
$x$-direction including the contributions from particles and turbulence,
and $\upstreammomentumflux$ is the far upstream value of momentum flux, i.e.,
\begin{equation}
\upstreammomentumflux  =
   \rho_0 u_0^2 + \Pthzero + \Pwzero
\ .
\end{equation}
The quantity $\momentumflux$ is defined as
\begin{equation}
  \label{mfasmoment}
  \momentumflux(x) = \int p_x v_x f(x, \pvector) d^3p + P_w(x),
\end{equation}
where $p_x$ and $v_x$ are the $x$-components of momentum and
velocity of particles, and $f(x,\pvector)$ is their distribution
function, all measured in the shock frame.
The quantity $\energyflux(x)$ is the energy flux of particles and turbulence in
the $x$-direction, $\Qesc$ is the energy flux of escaping particles
at the FEB,\footnote{Particle escape at an upstream FEB also
causes the mass and momentum fluxes to change but these changes are
negligible as long as $u_0 \ll c$ (see \cite{Ellison85}).} and the far
upstream value of the energy flux is
\begin{equation}
\upstreamenergyflux = \frac{1}{2} \rho_0 u_0^3 +
\frac{\gamma}{\gamma-1}\Pthzero u_0  + \Fwzero.
\end{equation}
The quantity $\energyflux(x)$ is defined as
\begin{equation}
\label{efasmoment}
\energyflux(x) = \int K v_x f(x,\pvector) d^3 p + F_w(x),
\end{equation}
$K$ being the kinetic energy of a particle with momentum $p$ measured in
the shock frame. $P_w$ and $F_w$ are the momentum and energy fluxes of the turbulence
defined in Section~\ref{turb_fluxes}.

Writing equations (\ref{fluxmass}), (\ref{fluxmomentum}) and
(\ref{fluxenergy}) for a point downstream of the shock, sufficiently
far from it that the distribution of particle momenta is isotropic,
and the approximations~(\ref{momentumviapressures}) and
(\ref{energyviapressures}) are valid, and
denoting the corresponding quantities by index `2', we get the
equivalent of the Rankine-Hugoniot relations, that
accounts for particle acceleration and escape, and for the presence
of magnetic turbulence:
\begin{eqnarray}
\label{cons1}
\rho_2 u_2 &=& \rho_0 u_0, \\
\label{cons2}
\rho_2 u_2^2 + \Pptwo + \Pwtwo &=& \rho_0 u_0^2 + \Ppzero + \Pwzero\equiv\upstreammomentumflux, \\
\label{cons3}
\frac12 \rho_2 u_2^3 + \wptwo u_2 + \Fwtwo + \Qesc&=&
     \frac12 \rho_0 u_0^3 + \wpzero u_0 + \Fwzero\equiv\upstreamenergyflux.
\end{eqnarray}
The particle gas enthalpy $w_p$ is $w_p=\epsilon_p + P_p$, and the
internal energy $\epsilon_p$ of gas is proportional to
the pressure $P_p$. Introducing the quantity $\gammabar$ so that
$\epsilon_p = P_p/(\gammabar-1)$, one can write
\begin{equation}
\label{fppp}
w_p u=\frac{\gammabar}{\gammabar-1}P_p u
\end{equation}
The value of $\gammabar$ is averaged over the whole particle
spectrum, and it ranges between $5/3$ for a nonrelativistic
and $4/3$ for an ultra-relativistic gas.
The local value of $\gammabar$ can be easily calculated in our
code from the particle distribution, along with
$P_p$ and $\epsilon_p$, as $\gammabar=1 + P_p / \epsilon_p$.
Similarly, one can define $\deltabar=F_w/(u P_w)$ and calculate a local
value of $\deltabar$ anywhere in the code in order to express
\begin{equation}
\label{fwpw}
F_w=\deltabar \cdot P_w u \ .
\end{equation}
The value of $\deltabar$ depends on the nature of the turbulence.
For instance, in \Alf ic turbulence, one expects $\deltabar
\approx 3$, [see Equation~(\ref{fwdef})].

Substituting (\ref{fppp}) and (\ref{fwpw}) into the above equations
and introducing $\rtot=u_0 / u_2$, we can eliminate $\rho_2$ using
(\ref{cons1}) and $\Pptwo$ using (\ref{cons2}), which allows us to
express from (\ref{cons3}) the quantity $\qesc \equiv \Qesc /
\upstreamenergyflux$ as
\begin{equation}
\label{qescrtot}
\qesc = 1 + \frac{A/\rtot^2 - B/\rtot}{C},
\end{equation}
where
\begin{eqnarray}
A&=& \frac{\gammabar_2 + 1}{\gammabar_2 - 1} ,\\
B&=& \frac{2 \gammabar_2}{\gammabar_2 - 1}
     \left( 1 + \frac{\Ppzero + \Pwzero - \Pwtwo}{\rho_0 u_0^2}\right) +
            \frac{2 \deltabar_2 \Pwtwo}{ \rho_0 u_0^2 },\\
C&=& 1 + \frac{2 \gammabar_0}{\gammabar_0 - 1}\frac{\Ppzero}{\rho_0 u_0^2} +
            \frac{2 \deltabar_0 \Pwzero}{\rho_0 u_0^2}.
\end{eqnarray}
Note that $\rho_0 u_0^2 / \Ppzero = \gammabar_0 M_s^2$, where
$\gammabar_0=\gamma=5/3$ due to the absence of CRs far upstream.
The pressure of stochastic magnetic fields
$\Pwzero$ can be found from the spectrum of seed turbulence far upstream
(see Section~\ref{mfa_in_mc}).

The quantity $\qesc$ is readily available in the simulation after
the end of any iteration.
Comparing it to the value predicted by (\ref{qescrtot}), one may
evaluate the self-consistency of the solution and make the correction to
$\rtot$, if necessary, for further iterations. For making these
corrections it is helpful to use in the simulation the inverse of
(\ref{qescrtot}), the physically relevant branch of which is
\begin{equation}
\label{rtotqesc}
\rtot = \frac{2 A}{B - \sqrt{B^2 - 4 A C (1-\qesc)}}.
\end{equation}

It is important to emphasize here that an iterative procedure
is required to find the compression ratio $\rtot$ of a non-linearly
modified shock, because quantities $\qesc$, $\Pwtwo$ and $\gammabar_2$
depend on $\rtot$, so (\ref{rtotqesc}) only provides a practical
way to perform the iterations. The code employs the following procedure:
\begin{eqnarray}
\label{iteration_rtot}
  \rtot'= (1-\zeta) \rtot + \zeta \frac{2 A}{B - \sqrt{B^2 - 4 A C (1-\qesc)}},
\end{eqnarray}
where $\zeta$ is a small number (typically $\zeta = 0.01 \dots 0.1$). Here
$\rtot$ is the compression ratio assumed for the last iteration,
and $\rtot'$ -- the compression ratio chosen for the following
iteration. The weighting using the parameter $\zeta$ is chosen so that,
when the deviation of the nonlinear structure of the shock from
the self-consistent solution is large, this iterative procedure
would not overcompensate the discrepancy, which may lead to the
breakdown of the model (for example, $\rtot <1$ is unphysical,
and $\rtot$ too high may stall the particle transport procedure).

Another complication that may arise with the procedure described
by~(\ref{iteration_rtot}) is that instead of converging to a
solution that satisfies the conservation relations
(\ref{cons1}), (\ref{cons2}) and (\ref{cons3}), the procedure
may find an attracting cycle around the self-consistent
value of $\rtot$. For example, $\rtot$ may be too low in one 
iteration, underestimating particle acceleration efficiency,
which leads to a shock with a high $\rtot'$ predicted
by~(\ref{iteration_rtot}). The latter, in turn, overestimates
particle acceleration, leading to a low compensatory $\rtot$ prediction again.
These cycles may be merely a mathematical consequence of our
numerical model of particle accelerating shocks. On the other
hand, it is conceivable that a real shock, instead of
evolving into a steady-state system, may behave periodically 
or even chaotically, turning particle acceleration on and off.
However, these effects are beyond the scope of the present research,
because we look for the steady-state structure of collisionless shocks.
The simulation avoids the attractors other than the self-consistent solution
by randomizing the value of $\zeta$, so that it varies between
a finite value and $0$ in every iteration. This way, any attracting cycle
that the system~(\ref{iteration_rtot}) may have with a constant
value of $\zeta$ will eventually be broken, 
but if $\rtot$ is at its self-consistent value ($\rtot'\approx\rtot$),
the randomization of $\zeta$ will not take the solution away
from this point. The latter is expected as long as statistical fluctuations of
quantities $A$, $B$, $C$ and $\qesc$ keep $\rtot'$ in the attracting
domain. This only requires that a high enough number
of particles is used in the Monte Carlo routine.

\newpage

\subsubsection{Test of the implementation}

We test the procedure of the iterative estimation of 
the compression ratio~(\ref{iteration_rtot}) by confirming
that it reproduces the solid predictions of the Hugoniot adiabat
(\ref{hugoniot_u}) and (\ref{hugoniot_T}). In 10 runs
described below, the shocks propagate in a gas with density
$n_0=0.3$~\pcc, temperature $T_0 = 7.3\cdot 10^3$~K (corresponding to
a sound speed $c_s = 10$~\kmps), and magnetic field $B_0 = 10^{-9}$~G
determining the Bohm diffusion (this magnetic field is too small
to influence the momentum and energy balance, therefore
the Hugoniot adiabat should apply). I performed 9 runs,
in which the flow speed, $u_0$, varied from $5$~\kmps\ 
to $1000$~\kmps, corresponding to the sonic Mach number,
$M_s$, varying from $0.5$ to $100$. In these simulations,
I artificially eliminated particle acceleration by assuming
that the subshock is fully reflective for particles trying
to cross it from the downstream into the upstream region.
Additionally, I ran simulation number 10, for which the subshock
is assumed fully transparent, and particle acceleration 
occurs (limited by a free escape boundary at $x=-80\;\rgzero$).
We start the simulations off by assuming a flow with the
speed $u_0$, which at the point $x=0$ abruptly slows
down by $0.1\%$ to $u_2 = u_0/1.001$. This tiny flow speed
jump heats the particles a little, just enough to make
the procedure (\ref{iteration_rtot}) start converging
to a self-consistent $\rtot$. Setting the parameter
$\zeta=0.3$ and randomizing it between $0$ and $0.3$, in $30$
iterations the simulation obtains the self-consistent value of $\rtot$.
I averaged the $\rtot$ prediction over the last $10$ iterations
out of $30$, and showed the results in Table~\ref{search_rtot}. Additionally,
the simulation measured the downstream gas temperature, $T_2$, by
detecting thermal particle pressure $\Pthtwo$ and relating it to the
temperature by the ideal gas law.


\begin{table}[b]
\caption{$ $ Test of iterative search of the compression ratio, $\rtot$}
\label{search_rtot}
\begin{center}
\begin{scriptsizetabular}{lccccccc}
\hline
Model & $u_0$, km/s & $M_s$ & Accel. & 
            $\left(\rtot\right)_\mathrm{HA}$ & 
            $\left(\rtot\right)_\mathrm{MC}$ & 
            $\left(T_2/T_0\right)_\mathrm{HA}$ & 
            $\left(T_2/T_0\right)_\mathrm{MC}$ -- MC\\
\hline
 1 &   5 & 0.5 & no  & 1.0*  & 1.022$\pm$ 0.003& 1.0* & 0.99 $\pm$ 0.08 \\
 2 &  10 & 1.0 & no  & 1.0   & 1.05 $\pm$ 0.01 & 1.0  & 1.02 $\pm$ 0.05 \\
 3 &  12 & 1.2 & no  & 1.30  & 1.31 $\pm$ 0.02 & 1.19 & 1.19 $\pm$ 0.06 \\
 4 &  15 & 1.5 & no  & 1.71  & 1.73 $\pm$ 0.01 & 1.49 & 1.48 $\pm$ 0.06 \\
 5 &  20 & 2.0 & no  & 2.29  & 2.29 $\pm$ 0.01 & 2.08 & 2.0  $\pm$ 0.1 \\
 6 &  30 & 3.0 & no  & 3.00  & 2.99 $\pm$ 0.01 & 3.67 & 3.7  $\pm$ 0.1 \\
 7 &  50 & 5.0 & no  & 3.57  & 3.58 $\pm$ 0.02 & 8.68 & 8.4  $\pm$ 0.3 \\
 8 & 100 &  10 & no  & 3.88  & 3.90 $\pm$ 0.01 & 32.1 & 31   $\pm$ 1 \\
 9 & 300 &  30 & no  & 3.99  & 4.00 $\pm$ 0.01 & 282  & 290  $\pm$ 10 \\
10 & 300 &  30 & yes & 3.99**& 3.11 $\pm$ 0.09 & 282**& 230  $\pm$ 10 \\
\hline
\end{scriptsizetabular}
\end{center}
\vskip -0.2in
\end{table}

Column `Model' in Table~\ref{search_rtot} shows the number of the
model, $u_0$ is the upstream flow speed, $M_s$ -- the corresponding
Mach number, `Accel.' shows whether the Fermi-I acceleration 
was allowed, as described above. The columns 
`$\left(\rtot\right)_\mathrm{HA}$'
and 
`$\left(T_2/T_0\right)_\mathrm{HA}$' are the predictions of the Hugoniot adiabat
(\ref{hugoniot_u}) and (\ref{hugoniot_T}) corresponding
to the Mach number $M_s$ and the adiabatic index of a non-relativistic
ideal gas $\gamma=5/3$. The columns 
`$\left(\rtot\right)_\mathrm{MC}$'
and 
`$\left(T_2/T_0\right)_\mathrm{MC}$' show the result of the simulation, i.e.,
the average of the last $10$ of the $30$ iterations, along
with the $1 \sigma$ standard deviation. For $M_s=0.5$, the
Hugoniot adiabat doesn't apply, because the flow is subsonic,
and no shock should form, corresponding to $\rtot = 1$ and 
$T_2 = T_1$ (values marked with the asterisk `*').

It is clear that, within statistical errors, the simulation
n Models 1-9 reproduced the Hugoniot adiabat results for nonrelativistic
hydrodynamic shocks. Note that the physics that the simulation
is based on at this point involves only the isotropic
particle scattering, the Lorentz transformations between
the reference frame of the flow and that of the shock, 
and the requirement that the calculated momentum, energy and
mass fluxes be conserved. 

\begin{figure}[htbp]
\centering
\vskip 0.5in
\includegraphics[angle=-90, width=4.0in, clip=true, trim=1.0in 0.0in 0.0in 0.0in]{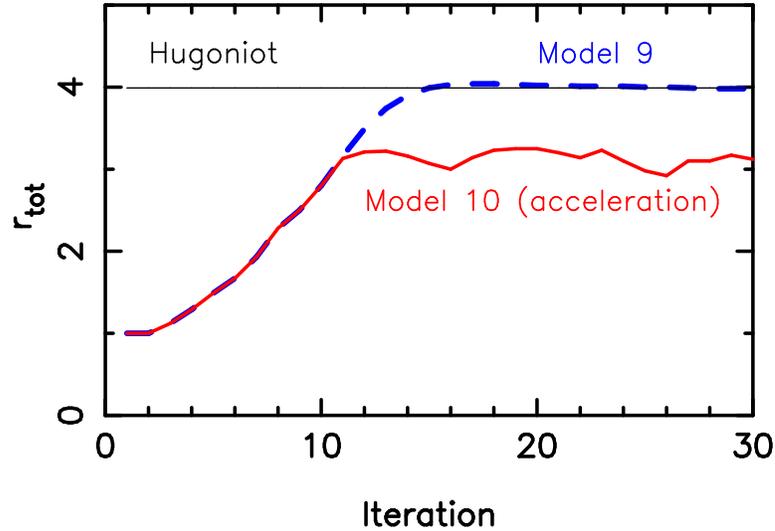}
\caption{$ $ Iterative estimation of compression ratio.}
\label{fig-rtot_iter}
\end{figure}

Figure~\ref{fig-rtot_iter} illustrates how the prediction
of the self-consistent compression ratio for Model 9
starts off with a trivial initial guess of $\rtot=1.001$,
rises up to the value predicted by the Hugoniot
adiabat in 15 iterations, and stays there.

\begin{figure}[htbp]
\centering
\vskip 0.5in
\includegraphics[height=7.00in]{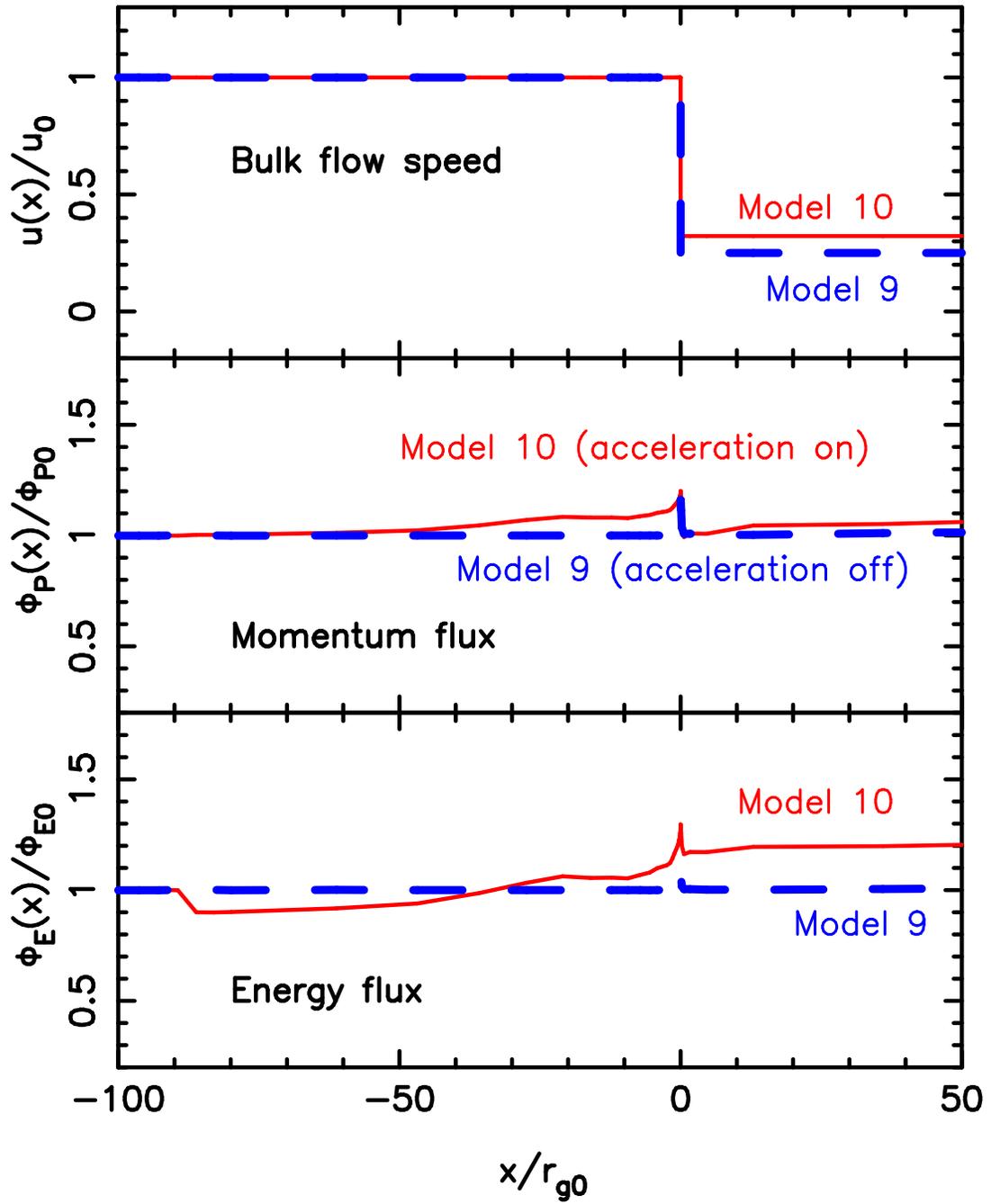}
\caption{$ $ Momentum and energy conservation illustration.}
\label{fig-sharpshock_pe}
\end{figure}

Figure~\ref{fig-sharpshock_pe} confirms the conservation of
momentum and energy across the shock in Model 9. The top panel
shows the flow speed with the compression ratio $\rtot=4.0$,
the middle panel shows the momentum flux measured in the
units of the upstream flux, and the bottom panel shows
the energy flux. These fluxes remain constant throughout
the shock for Model 9 (the dashed thick line).

Now let's analyze the results for Model 10, which is the
same as Model 9, but with the shock transparent to the
particles, so that the process of diffusive shock acceleration
can occur. The Hugoniot adiabate is not applicable in this
case, and the corresponding values in Table~\ref{search_rtot}
are marked with a double asterisk `**'.
Indeed, the values of the compression ratio
and the downstream temperature determined 
by the simulation and 
shown in Table~\ref{search_rtot} are $\rtot \approx 3.09$,
and $T_2/T_0 \approx 230$. These values are
is lower than the prediction of the
Hugoniot adiabat for a shock with the sonic Mach number $M_s=30$. 
In Figure~\ref{fig-rtot_iter},
the thick solid line demonstrates that the iterative
procedure given by Equation~(\ref{iteration_rtot})
has converged, but the fluxes shown with the
solid lines in Figure~\ref{fig-sharpshock_pe}
are not constant, meaning that the derived solution
is not physical. This is the effect of particle acceleration
studied by Ellison and co-workers using the Monte Carlo model,
and the commonly accepted solution is that a shock
precursor must form, i.e., the upstream flow speed $u(x<0)$ 
must decrease towards the shock. The procedure
that models it is demonstrated in Section~\ref{subsec_smoothing}.

\subsection{Nonlinear structure of the shock precursor}

\label{subsec_smoothing}

\subsubsection{Theory}

In the test-particle limit,
the aftermath of particle acceleration by a shock with
the compression ratio $r$ is the power law spectrum
of accelerated particles every point in space,
$f(x,p)\propto p^{-s}$. The power law index of the spectrum
for a strong shock with $r=4.0$ is $s=4.0$, which
has the unphysical property that, if it
were to stretch from $p=0$ to $p=\infty$, the pressure
(and the internal energy) of the particles with such a
spectrum, $P \propto \int p f(p) p^2 \;dp \propto \int p^{-1}\;dp$,
would logarithmically diverge at the high momentum end. In reality,
of course, the spectrum is limited
by the maximum momentum determined by the size of the shock
or the acceleration time. However, the logarithmic
divergence of pressure at high energies means that
there may be a large amount of energy in the highest
momentum particles. Of course, the actual amount of energy is
determined by the rate of particle injection.

Initially, the shock in the simulation doesn't have a \SC\
structure because it starts with an unmodified shock and
$\momentumflux(x)$ is overestimated at all locations where
accelerated particles are present (see, e.g., the plots
for Model 10 in Figure~\ref{fig-sharpshock_pe}).
Therefore, it must choose $u(x)$
to reduce the mismatch between the local momentum
flux and the far upstream value of it $\upstreammomentumflux$
for $x<0$, as described by Equation~(\ref{fluxmomentum}). 
It can be done by calculating
\begin{eqnarray}
\label{iteration_ux}
  u'(x)=u(x) + \zeta \cdot \frac{\momentumflux(x) - \upstreammomentumflux}{\rho_0 u_0},
\end{eqnarray}
where $u'(x)$ is the predicted flow speed for the
next iteration, and $\zeta$ is a small positive number (typically around
0.1), characterizing the pace of the iterative procedure. The
value of the parameter $\zeta$ is randomly chosen between $0$ and a finite
value for reasons similar to those described in
Section~\ref{sec-rtot}. If magnetic field amplification is invoked,
then at this point the simulation also refines its
estimate for the particle diffusion coefficient (see 
Section~\ref{advanced_transport}).

The predicted $u(x)$ and $D(x,p)$ are then used in a new iteration where
particles are injected and propagated. The calculated CR pressure, momentum
flux, etc. are then used to refine the guesses for $u(x)$ and $D(x,p)$
for the next iteration, along with the guess for the compression
ration $\rtot$. This procedure is continued until all
quantities converge.

\subsubsection{Test of implementation}

\begin{figure}[htbp]
\centering
\vskip 0.5in
\includegraphics[angle=-90, width=4.0in, clip=true, trim=0.8in 0.0in 0.0in 0.0in]{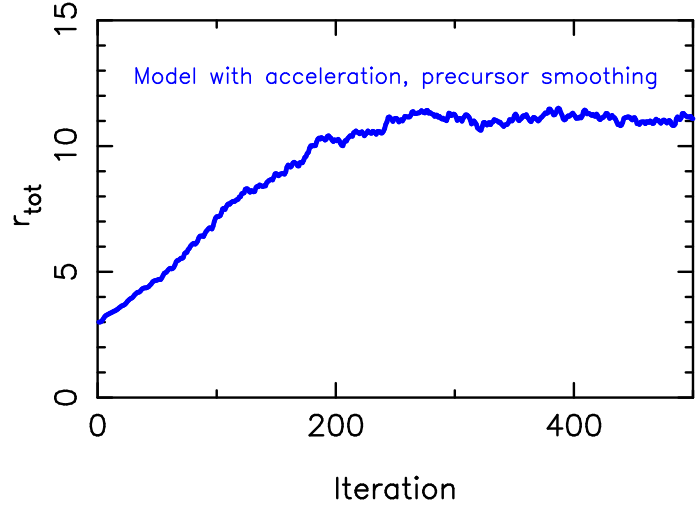}
\caption{$ $ Search for the self-consistent compression ratio.}
\label{fig-rtot_iter_smoothing}
\end{figure}

In order to test the implementation and the effects of the precursor
smoothing, I complemented Model~10 with the
procedure~(\ref{iteration_ux}). I had to reduce the maximum
value of $\zeta$ from $0.3$ to $0.1$ and make many more iterations
in order to restrict the self-consistent compression ratio
$\rtot$. As Figure~\ref{fig-rtot_iter_smoothing} shows,
the iterative procedure converged to a value $\rtot\approx 11$,
much higher than the Hugoniot adiabat predicted\footnote{This
and other nonlinear effects are discussed in \cite{Ellison85}.}.

\begin{figure}[htbp]
\centering
\vskip 0.5in
\includegraphics[height=7.0in]{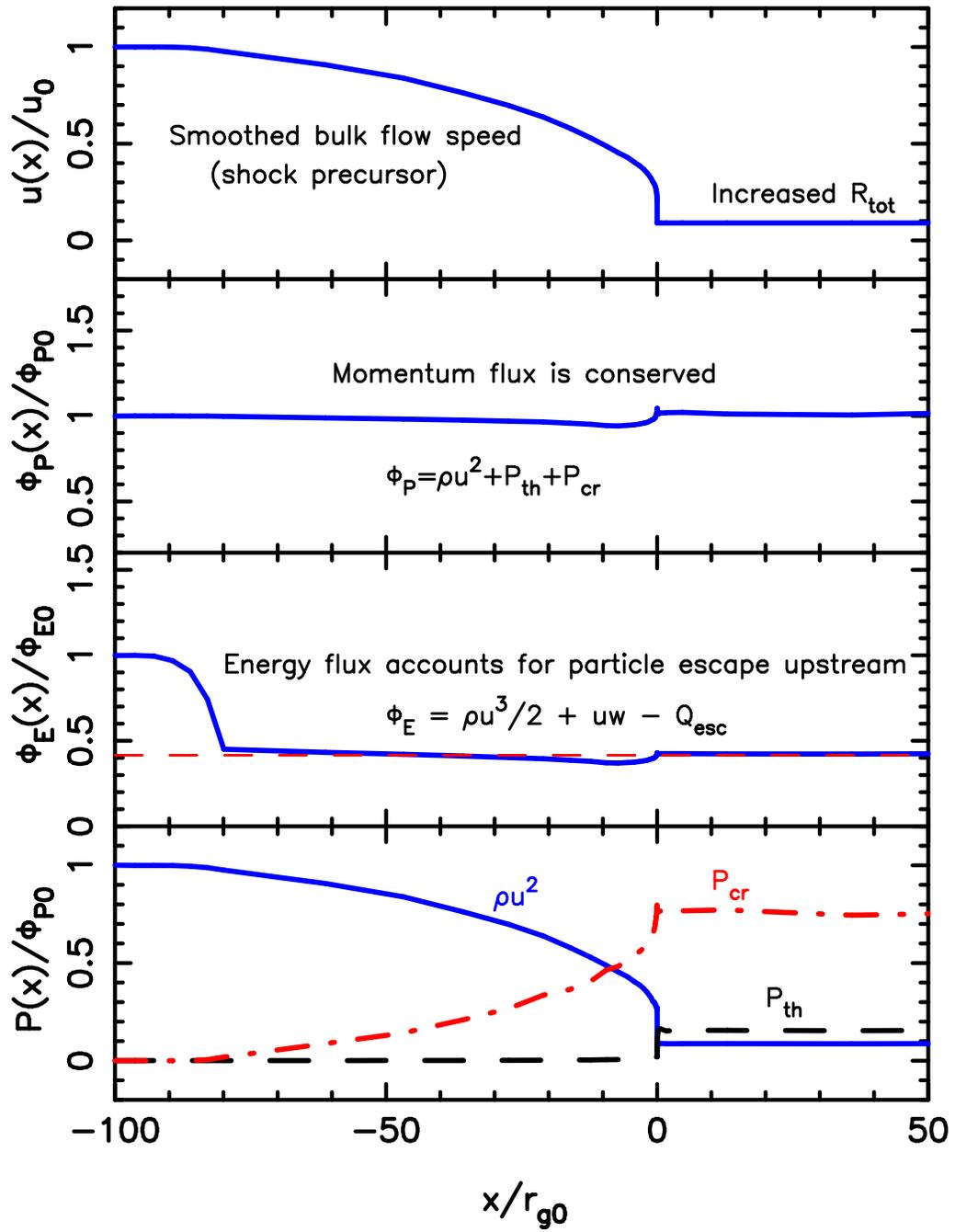}
\caption{$ $ Precursor smoothing for momentum and energy conservation.}
\label{fig-smoothshock_pe}
\end{figure}

In Figure~\ref{fig-smoothshock_pe}, I show the spatial structure
of the smoothed shock. In the top panel, a reduction of the flow
speed from $u_0$ in the upstream region, $x<0$, is apparent. Let us
refer to the smoothed region as a shock precursor. A subshock
at $x=0$ has a compression ratio, 
$\rsub = u(x=-0)/u(x=+0) \approx 2.5$. Momentum flux shown in the
second panel is conserved within a few percent, and its deviation
from conservation is statistical (the shown plots are an average of
14 iterations at the end of the 500 iterations leading to a 
consistent solution). The energy flux (third panel from top)
drops at the upstream free escape boundary, $x=-80\;\rgzero$,
by 60\%, which is explained by particle escape. Downstream of
the free escape boundary, the energy flux is almost constant
(within statistical deviations). The thin dashed line in
this panel shows the self-consistent value of energy flux
accounting for particle escape, as determined by
equation~(\ref{qescrtot}); as one can see, the actual value
of $\energyflux$ is in excellent agreement with this quantity.
The bottom panel shows the constituents of the momentum flux
$\momentumflux$, the dynamic pressure $\rho u^2$, the thermal
particle pressure $\Pth$ and the cosmic ray pressure $\Pcr$.
One can easily see that the shock is dominated by the accelerated
particles in this case.

\newpage

\subsection{Summary}

I would like to conclude this lengthy section with a summary.
The simulation of nonlinear shock acceleration presented here 
features a Monte Carlo code of particle transport and an iterative
procedure for deriving a self-consistent shock structure.

The tests I presented here confirm that:
\begin{itemize}
\item{Particle propagation
simulated by the Monte Carlo method is diffusive, and that it
reproduces qualitatively and quantitatively the expected behavior
of such particle transport;}
\item{The simulation reproduces the well-known results for 
hydrodynamic shocks (i.e., the Hugoniot adiabat), if particle
acceleration is artificially blocked;}
\item{Accelerated particle spectra predicted by the simulation
in the test-particle regime agree with the solid predictions
of analytic models of test-particle acceleration;}
\item{When the feedback of accelerated particles in the nonlinear
regime is accounted for, the simulation obtains a stable solution that 
conforms with mass, momentum and energy conservation laws.}
\end{itemize}

The methods of the Monte Carlo simulation of particle transport and the 
idea of iterative derivation of nonlinear shock structure presented
here are not original to this dissertation, they were developed
by Ellison and co-workers. I will not elaborate any more on the subject
of nonlinear shock acceleration and refer the reader
to the literature for more information (e.g., \cite{JE91}
and \cite{MD2001}).

The original part of the model -- magnetic field amplification
and self-consistent particle transport -- will be presented
in the rest of this Chapter.
The reason I presented and tested the Monte Carlo part
of the model in such detail is that the actual computer code
used in the research was written by the author of this dissertation
and had to be tested to confirm that it reproduces the well
known, previously established results. Besides that,
some details of the currently employed methods did not appear in
our publications, and I would like to make a record of these
details here.

\newpage

\section{Magnetic Field Amplification}

\label{mfa_in_mc}

In this section I will describe two theoretical models of magnetic
turbulence amplification by streaming particles available in the
modern scientific literature. Then I will proceed with a generalization
of these models for the problem of nonlinear DSA, and present
the analytic description of magnetic field amplification 
adopted for the model, and its numerical solution.
I will conclude by describing the feedback of the amplified magnetic
turbulence on the plasma flow.

The physical conditions in which the  instabilities take place
are representative of the conditions in a collisionless shock precursor. A
fully ionized plasma is moving at a speed $u(x)$ from the far upstream, unshocked
region ($x\to -\infty$) towards the subshock located at $x=0$,
where it gets non-adiabatically compressed.
We assume that a pre-existing uniform magnetic field
$\bf{B}_0$ parallel to the plasma flow fills the space.
The instabilities are induced by the accelerated
particles produced by diffusive shock acceleration, 
and described by $f(x,\bf{p})$.
These particles are subject to diffusion in the plasma and advection along with
it. Therefore, in the reference frame locally co-moving with the
plasma, the CRs appear to move against the bulk flow, away from the
subshock. The CR density and pressure increase from $0$ at $x\to
-\infty$ to a finite value at $x=0$, and so does the diffusive current
of CR measured in the plasma reference frame\footnote{Speaking of CR
  current and CRs streaming, I will always mean the apparent drift
of CRs in the plasma reference frame.}. 

We describe the fluctuations of magnetic field in the plasma by the
energy spectrum of turbulence, $W(x,k)$. The latter is a quantity such
that $W(x,k)\Delta k$ is the volume density
of turbulent energy (i.e., the energy of the waves, including the
magnetic field energy and that of the associated stochastic plasma
motions) in the waveband $\Delta k$. This means that, instead
of the the spatial structure of the turbulent magnetic fields,
the simulation only follows the evolution of its Fourier transform,
locally calculated over a large enough volume, and averaged
over a large enough time interval.
This approach relieves our model from the computational expenditure
of PIC plasma simulations, because the latter have to have a spatial
resolution finer than the size of the smallest turbulent
vortex, while our simulation gets away with a resolution that is 
more coarse than the largest turbulent harmonic by describing
the processes on smaller scales statistically. Obviously, this
advantage is gained at the cost of having to rely on theoretical
models of processes in plasmas instead of performing a numerical experiment
based on more fundamental physical principles.

Magnetic turbulence amplification is modeled under the assumption
of a steady state situation, i.e., the time derivatives 
$\partial/\partial t$ in the respective equations are set to $0$.
It is also postulated that
in the interstellar medium (i.e., far upstream), there exists seed 
turbulence, which is expressed by the boundary condition
\begin{equation}
\label{eq_seed_turb}
  W(- \infty, k) = 
  \left\{
  \begin{array}{l}
    \displaystyle\frac{\left(\Bseed\right)^2}{4\pi} 
      \displaystyle\frac{k^{-1}}{\ln{\left(\kmax/\kmin\right)}},
    \; \mathrm{if} \; \kmin < k < \kmax, \\
    0, \; \mathrm{otherwise}.
  \end{array}
  \right.
\end{equation}
Expression~(\ref{eq_seed_turb}) describes a far
upstream  seed turbulence with a power law spectrum,
$W \propto k^{-1}$, normalized so that $\Beff(x \to -\infty) = \Bseed$,
where $\Bseed$ is a parameter of the model
representing the assumed effective magnetic field of the seed turbulence.
The values $\kmin$ and $\kmax$ limiting the wavenumber range of the
seed turbulence spectrum are also parameters of the model. Normally, the code
will choose them so that for particles of all energies found in the
simulation, the resonant wavenumber (defined later) is
between $\kmin$ and $\kmax$.

The effective magnetic field $\Beff(x)$
at any point is a quantity that I define as
\begin{equation}
\label{eq_beff}
\frac{\Beff^2(x)}{8 \pi} = \frac12 \int_{0}^{\infty} W(x,k) \: dk.
\end{equation}
The fraction $1/2$ before the integral in the right-hand side of
Equation~(\ref{eq_beff}) expresses the assumption that one half of the
turbulence energy is contained in the magnetic field
fluctutaions, and the other half is carried by the stochastic
fluctuations of the plasma velocity associated with the waves. The factor
$1/2$ is exactly correct when the turbulence is purely \alf ic
(see Section~\ref{turb_fluxes}). The expression~(\ref{eq_beff}) is therefore
a good approximation if the turbulence is generated by the resonant
streaming instability (see Section~\ref{res_desc}),
but does not apply, for example, to
the waves generated by the nonresonant Bell's 
instability (see Section~\ref{bells_nonres_desc}).
For the latter case, the relationship between the plasma velocity
fluctuations, $\delta u$, and the magnetic field fluctuations, $\delta B$,
can be inferred, for example, from Equation~(17) in
\cite{Bell2004}. It can be shown that for wavelengths at the peak of the
amplification rate, $k=k_c/2$ (defined in Section~\ref{bells_nonres_desc})
magnetic field contains $3/4$, and velocity fluctuations --
 $1/4$ of the total energy density $W(x,k)$. However, I use the
`50/50' distribution of the turbulent energy between magnetic and kinetic
fluctuations, expressed by the factor $1/2$ in
Equation~(\ref{eq_beff}), for turbulence produced by any
source. I do so for the following
reasons. First, considering the
much larger uncertainties in some other factors of the model (e.g.,
the diffusion coefficient), an adjustment of the expression
(\ref{eq_beff}) to account for the nature of the turbulence
would be a minor correction. Second, nonlinear turbulent processes like 
dissipation and cascading may change the energy distribution between
magnetic and velocity fluctuations, and there exists no analytic
description of this process adequate for the strong turbulence
considered in this model.

\subsection{Resonant cosmic ray streaming instability}

\label{res_desc}

Charged particles streaming along a uniform magnetic field are able to
resonantly amplify \Alf\ waves traveling along the same field
(\cite{Kulsrud2005, Tsytovich1966, Wentzel74, Bell78a}, etc.). 
The mechanism of \Alf\ wave amplification is
similar to that used to amplify electromagnetic waves in the
electronic device known as the {\it traveling-wave tube}
\cite{Gittins65}. 
\Alf\ waves exchange energy with fast particles of resonant
momenta. If the particle distribution is anisotropic, 
\Alf\ waves traveling in one direction get amplified, and
waves traveling in the opposite direction get dampened by
the interactions with the particles close to the resonance. I will
not describe this instability in detail and refer the reader
to the sources cited above.

Assuming a steady state and that the particle distribution is
controlled by advection of the flow and resonant turbulent diffusion
(see Section~\ref{subsec_resonant_mfp}), and that the generated
waves are a weak perturbation of the uniform magnetic field,
$\Delta B \ll B_0$, the evolution of 
$W(x,k)$ may be described (see \cite{LC83} and references therein) as
\begin{equation}
\label{eq_turb_evol_res}
  u(x)\frac{\partial W(x,k)}{\partial x} = \resgrowthrate(x,k) W(x,k),
\end{equation}
if all other effects accompanying wave generation are ignored. Here
the growth rate
\begin{equation}
\label{resonant_increment}
  \resgrowthrate(x,k) = v_A 
  \frac{\partial \Pcr(x,\pres)}{\partial x} 
  \left| \frac{d\pres}{dk} \right|
  \frac{1}{W(x,k)},
\end{equation}
and 
\begin{equation}
  \frac{c \pres}{e B_0} k = 1.
\end{equation}
defines the resonant momentum, $\pres$.
Here $v_A$ is the \Alf\ wave speed, and $\Pcr(x,p)$ is the
spectrum of particle pressure, normalized so that $\Pcr \Delta p$
is the pressure of accelerated particles in the momentum
range $\Delta p$.

It was usually assumed that the fluctuations grow until
$\Delta B \approx B_0$, after which the instability
saturates (e.g., \cite{MV82}). Recently, Bell and Lucek
\cite{BL2001} suggested that, if the instability
can grow beyond this point, to $\Delta B \gg B_0$, the observations
of large magnetic fields in some SNRs can be explained by
generation of magnetic fields in the process of DSA.
Simulations done by the same authors \cite{LB2000}
support such a possibility. Here, I adopt
Bell and Lucek's idea and assume that the streaming of CRs
is able to produce strong magnetic field fluctuations.

As the perturbations grow and reach $\Delta B \gtrsim B_0$, however,
it is likely
that waves with wave vectors ${\bf k}$ not aligned with ${\bf B}_0$ will
be generated, due to local CR pressure gradients along the total ${\bf
B}={\bf B}_0 + \Delta{\bf B}$.  With $\Delta B \gtrsim B_0$, it becomes
impossible to predict the average value of the transverse pressure
gradients and the resulting magnetic field structure without knowing the
relative phases of different wave
harmonics.
The problem is further complicated by the fact that this longitudinal,
compressible turbulence may produce a strong 2nd order Fermi particle
acceleration effect which, in turn, can damp the longitudinal
fluctuations (see, for example, \cite{SchlickeiserEtal1993}). 

These complications place a precise description of plasma turbulence
beyond current analytic capabilities.
%
%
However, valuable conclusions about MFA in efficient DSA
can be made by considering the two limiting cases
of the resonant instability development in the nonlinear
regime.
The first assumes there is no
longitudinal turbulence, 
in which case the wave growth rate is determined by the \Alf\ speed in
the non-amplified field $B_0$.
This gives a lower bound to the growth rate. 
The upper limit
assumes that the turbulence is isotropic, in which case the growth rate
is determined by the \Alf\ speed in the much larger
amplified field $\Beff$ (defined in Equation~(\ref{eq_beff})).  
The real situation should lie
between these two cases, and while I consider these limits, I do not
explicitly include second-order Fermi acceleration in the calculations.
Section~\ref{res2006} describes the parameterization of
the growth rate of the resonant instability that encompasses
the minimum and the maximum growth rate in the regime of strong
fluctuations.

\newpage

\subsection{Bell's nonresonant instability}

\label{bells_nonres_desc}

A nonresonant instability, theoretically described by Bell in 2004 
(\cite{Bell2004}),
occurs when a strong external electric current of CRs is put through the
plasma. The instability develops because the thermal plasma
must provide a current in response to the external current of streaming
CRs, in order to maintain quasi-neutrality. This current makes
certain MHD modes unstable; these modes can be described
as driven circularly polarized \Alf\ waves. Again,
I will not discuss the details of the instability; the reader may
find more information in \cite{Bell2004, ZPV2008, AB2009}.

In the linear regime ($\Delta B  \ll B_0$) the growth of
the unstable modes can be described by the following equation:
\begin{equation}
\label{eq_turb_evol_nr}
u(x)\frac{\partial W(x,k)}{\partial x} = \nrgrowthrate(x,k) W(x,k).
\end{equation}

The dispersion relation of waves subject to Bell's instability is
\begin{equation}
\label{bell_dispersion}
\omega^2 - v_A^2 k^2 \pm \frac{B_0 k j_d}{c \rho} = 0,
\end{equation}
where $\omega$ and $k$ are the frequency and the wavenumber of 
the generated waves, and $j_d$ is the diffusive current of CRs
directed along the magnetic field ${\bf B_0}$. The frequency
$\omega$ has an imaginary part when
\begin{equation}
\label{bell_cond_kcrit}
k < k_c \equiv \frac{B_0 j_d}{c \rho v_A^2},
\end{equation}
and the reasoning leading to (\ref{bell_dispersion}) is applicable when the
wavelengths of the generated waves are shorter than the smallest
energetic particle gyroradius in the system, $\rgone$:
\begin{equation}
\label{bell_cond_shortwave}
\frac{1}{\rgone} < k.
\end{equation}
Along with its applicability conditions (\ref{bell_cond_kcrit}) and (\ref{bell_cond_shortwave}),
the growth rate of the energy of the waves $\nrgrowthrate(k) \equiv 2\Im{\omega(k)}$ is
\begin{equation}
\label{bell_increment}
\nrgrowthrate = \left\{
\begin{array}{l}
2 v_A k\sqrt{\displaystyle\frac{k_c}{k}-1}, \quad 
  \mathrm{if} \quad 1/\rgone < k < k_c,\\
0, \quad \mathrm{otherwise}.
\end{array}
\right.
\end{equation}
Here $v_A=B_0/\sqrt{4 \pi \rho}$ is the
\Alf\ speed, and  the critical wavenumber $k_c$ is
\begin{equation}
\label{bell_kcrit}
k_c = \frac{B_0 j_d}{c \rho v_A^2}.
\end{equation}

This instability has been studied theoretically and using MHD and PIC
simulations (\cite{Bell2004, ZPV2008, ABG2008, AB2009, RS2008}), which
show that it is capable of generating large magnetic fields,
and may even dominate the resonant CR streaming instability in
young shocks of SNRs \cite{PLM2006}.

It is informative for future reasoning to point out the dependence of
$\nrgrowthrate$ on $k$: the rate $\nrgrowthrate$ becomes non-zero at large
wavelengths at $k=1/\rgone$, and then grows as $k^{1/2}$ towards the
smaller wavelengths, until it peaks at $k=k_c/2$, and then rapidly
falls off down to zero at $k=k_c$. 

\subsection{Nonresonant long-wavelength instability}

\label{nonres_long_inst}

Other instabilities possibly leading to magnetic field amplification
may exist in an interstellar plasma in vicinity of a particle
accelerating shock. Bykov and Toptygin \cite{BT2005} suggested a model
in which streaming cosmic rays may amplify waves in plasma with
wavelengths much larger than the gyroradii of the particles.

The model presented in \cite{BT2005} requires the presence of a neutral
component in the plasma (i.e., unionized hydrogen atoms) that
suppresses the transverse conductivity. The Balmer series
lines of neutral hydrogen 
have been observed in the emission spectra of forward shocks
of Type Ia SNRs (e.g., in SN~1006 and Tycho's SNR \cite{KWC1987}. See
also the references in \cite{GRSH2001}). 
It is also possible that short-scale turbulence may act similarly
to neutral plasma component at suppressing the transverse conductivity,
which will make the model of nonresonant long-wavelength instability
applicable to fully ionized plasmas as well (A.~M.~Bykov, in private
communication, and \cite{BOT2008}, in press). 

I have not incorporated this effect into the simulation, because the
details of this model are being developed. However, if this
instability operates in the precursor of a collisionless shock,
it may have a very strong impact on the maximum momentum of
the accelerated particles due to its long-wavelength nature.

\newpage

\subsection{Evolution of turbulence in a nonlinear shock -- other effects}

\label{turb_effects}

\subsubsection{Flow compression}

Equations of the form (\ref{eq_turb_evol_res}) and
(\ref{eq_turb_evol_nr}) generally apply to a uniform
flow $u(x)=const$, and do not take into account the possibility
of a changing flow speed in the precursor. In the geometric optics
approximation, the propagation of waves in a medium with changing
properties may lead to modulation of the wavelengths and of the
amplitudes of the waves. Different instabilities generate
different types of waves, which may evolve differently.
Namely, the resonant CR streaming instability in the
quasi-linear approximation $\Delta B \ll B_0$ generates
\Alf\ waves propagating against the flow, and Bell's
nonresonant instability generates almost purely growing
harmonics that can be described as driven \Alf\ waves \cite{Bell2004}.
Considering that compression ratios in strong SNR
may reach $\rtot\approx 5-15$ (see, e.g,
observational arguments \cite{WarrenEtal2005, Gamil2008},
and theoretical predictions \cite{JE91, MD2001}),
these effects may change the wavelengths and amplitudes
of the generated waves by a large factor, and should be
considered.

I propose to include the effects of flow compression in the
model by adding the corresponding terms to the equation of turbulence
evolution.  Consider the equation
\begin{equation}
u \frac{\partial W}{\partial x} + \parcompamp W \frac{du}{dx} 
- \parcompwav \frac{\partial}{\partial k} \left( k W \frac{du}{dx} \right) = 0.
\end{equation}
Given a boundary condition $W(x_0, k) = W_0(k)$, one can readily
solve this equation for $x>x_0$ using the method of characteristics:
see Equation~(\ref{wxk_nocasc_noamp}).
The ratio $u(x_0)/u(x)$ in this equation is the factor by which the plasma
is compressed, because due to Equation~(\ref{fluxmass}),
$\rho(x)/\rho(x_0) = u(x_0) / u(x)$. 
Equation (\ref{wxk_nocasc_noamp}) shows that the parameter $\alpha$ determines how
the amplitudes of the waves react to compression, namely,
$W\propto \rho^{\alpha}$. 
For \Alf\ waves, the correct
value of this parameter is $\alpha = 1.5$ (see
the wave generation equations and the corresponding
explanation in \cite{VEB2006}).
The parameter $\beta$ describes what happens to the wavenumber
of the waves as they propagate in the compressing medium,
that is, $k \propto \rho^{\beta}$. 

\subsubsection{Dissipation}

The amplified turbulence may be dissipated through collisional and/or
collisionless mechanisms and these include:
\newlistroman
\listromanDE linear and \NL\ Landau damping 
(e.g., \cite{AB86,Kulsr78,VBT93,Z2000}),
\listromanDE particle trapping (e.g., \cite{Medvedev99}),
and
\listromanDE ion-neutral wave damping (e.g, \cite{DDK96,BT2005}).  
Existing analytic descriptions of MHD wave damping
rely on the quasi-linear approximation $\Delta B \ll B_0$, which is
inapplicable for strong turbulence, and numerical models
with varying ranges of applicability have been proposed which offer a
compromise between completeness and speed
(e.g., \cite{Bell2004,AB2006,VEB2006,ZPV2008}).
Because no consistent analytic description of magnetic turbulence
generation with $\Delta B \gtrsim B_0$ exists,
and because an numerical (MHD or PIC) description of this process
in the framework of non-linear DSA is very computationally
expensive, we propose a
parameterization of the turbulence damping rate.
In doing this, we are pursuing two goals.
First, we make some predictions connecting cosmic
ray spectra, turbulent magnetic fields
and plasma temperatures, which, in principle, can be tested
against high resolution X-ray observations in order to estimate the
heating of the thermal gas by turbulence dissipation.
And second, once heating is included in our simulation in a
parameterized fashion, we will be ready to implement more realistic
models of turbulence generation and dissipation as they are developed.

The heating
of the precursor plasma by dissipation
modifies the subshock Mach number (e.g., \cite{EBB2000, VBE2008}) and this in
turn modifies injection.
The overall acceleration efficiency and, of particular importance for X-ray
observations, the temperature of the shocked plasma
(e.g., \cite{DEB2000,HRD2000,EDB2004}) will depend on wave dissipation.

The generalized way of including the dissipation of turbulence 
into the simulation is introducing a corresponding term into the
equation of turbulence evolution:
\begin{equation}
u\frac{\partial W}{\partial x} = G - L,
\end{equation}
where $G$ stands for the rate of growth of the instability
driving the turbulence (i.e., $G=\Gamma W$), and $L=L(x,k)$ is the
rate of turbulence dissipation (measured in
ergs$\cdot$cm$^{-3}$s$^{-1}$). In the simulation, I allow for one of
two prescriptions for the dissipation rate to be realized.

In the absence of a better model, one may assume that the dissipation
rate $L$ is proportional to the amplification rate $G$, i.e.
\begin{equation}
\label{diss_param}
L_F = \heatpar G,
\end{equation}
where $\heatpar$ is a number between $0$ and
$1$. Equation~(\ref{diss_param}) is a mere parameterization of the
dissipation rate, in which $\heatpar$ is the fraction of the
instability generation rate that is assumed to go directly into
particle heating rather than magnetic fields (in $L_F$, the subscript
$F$ stands for `fraction'). In particular, $\heatpar = 0$ corresponds to no
dissipation, and $\heatpar = 1$ describes a situation where all the
turbulence generated by a particle streaming instability immediately
gets dissipated and transformed into heat. 

Another prescription for the rate of dissipation is
\begin{equation}
\label{diss_ksq}
L_V = \frac{v_A}{\kdiss}k^2 W.
\end{equation}
The wavenumber at which the dissipation begins to dominate, $\kdiss$,
is identified with the inverse of a thermal proton gyroradius:
\begin{equation}
\label{kdiss_def}
  \kdiss = \frac{eB_0}{c\sqrt{m_pk_BT}},
\end{equation}
where $m_p$ is the proton mass, $k_B$ is the
Boltzmann constant and $T=T(x)$ is the local gas temperature.
This prescription is a relaxation-time approximation,
defined in such a way that at $k=\kdiss$, the dissipation
time equals the period of an \Alf\ wave with the wavenumber $k$,
and the $k$-dependence is $L\propto k^2$. The $k^2$ dependence
of the dissipation rate is based on the assumption
that viscosity (i.e., magnetic viscosity in this case)
drives the dissipation (see, e.g., \cite{VBT93}).

\subsubsection{Influence of turbulence dissipation on thermal plasma heating}

Dissipation of turbulence acts as an energy sink, in which the
magnetic and kinetic energy of turbulent fluctuations are transformed
into the internal energy of the thermal particle gas. This means that,
in order to conserve energy, the appearance of the term $L$ in the
equation of turbulence evolution must be accompanied by the
corresponding correction to the equations of motion of the thermal plasma.
The way to incorporate the thermal plasma heating due to 
turbulence dissipation was shown in~\cite{MV82}, who derived
the equation of thermal pressure evolution in the shock precursor:
\begin{equation}
\label{pressuregrowth}
\frac{ u\rho^{\gamma}}{\gamma-1}
\frac{d}{dx}\left(\Pth \rho^{-\gamma} \right) = L
\ .
\end{equation}
Here the ratio of specific
heats of an ideal nonrelativistic gas is $\gamma=5/3$.
For $L=0$, equation (\ref{pressuregrowth})
reduces to the adiabatic heating law,
$\Pth \sim \rho^{\gamma}$ and, for a non-zero $L$, it describes
the heating of the thermal plasma in the shock precursor due to the
dissipation of magnetic turbulence.
The fluid description of heating given by
equation~(\ref{pressuregrowth}), while it doesn't include details of
individual particle scattering, can be used in the \mc\ simulation to
replace particle scattering and
determine heating in the shock precursor.  This merging of analytic and
\mc\ techniques, or \FP\ (\FPshort), is described in detail
in Appendix~B of \cite{VBE2008}, and briefly summarized below.

When the heating rate, $L$, becomes available from the solution
of the turbulence growth equation~(\ref{eq-genmfa}),
the code solves (\ref{pressuregrowth}) and substitutes the
solution, $\Pth(x)$, for the thermal pressure calculated
from particle trajectories. It is done in the upstream 
region up to the point $\xFP$, at which thermal particles are
subsequently introduced into the simulation for the
next iteration.
In order to include the effects of heating in the model, we must
introduce thermal particles at $\xFP$ as if they were heated in the
precursor, i.e., their temperature $T(\xFP)$ must be determined
by~(\ref{pressuregrowth}) and the ideal gas law:
\begin{equation}
\label{idealgaslaw}
T(x)=\frac{\Pth(x)}{k_B n_0 (u_0/u(x))},
\end{equation}
The simulation therefore chooses the magnitude of every introduced
particle's momentum $p$ in the local plasma frame
distributed according to Maxwell-Boltzmann distribution with 
temperature $T$ determined by (\ref{idealgaslaw})
at $x=\xFP$. As long as the local sonic Mach number
at this location is large (i.e., $\Msone > 3$),
it can be done using the results of the section~\ref{subsec_particleintro}
for the angular distribution of the introduced particles.
If the $\Msone < 3$ (which does not usually happen
in a self consistent solution), then the results
described in Appendix~B may be applicable.

\subsubsection{Spectral energy transfer}

Observations of turbulence, including the MHD turbulence in the
interplanetary plasma, often report spectra that look like
power law functions of $k$ over many decades. 
This phenomenon can successfully be explained by spectral energy
transfer (cascading). After the energy has been generated
by an instability on some dominant
spatial scale,  nonlinear motions in the turbulent fluid
cause splitting and merging of the turbulent vortices
(i.e., a cascade), leading to a
re-distribution of energy between different scales. This way,
turbulence initiated by large-scale vortex formation due to an external
power source can cascade into smaller vortices, producing a
power-law distribution of energy in the so-called inertial
range (i.e., the interval in $k$-space where the turbulence
spectrum is populated by cascading rather than directly
by the instability). This cascade continues until the size of the vortices
is small enough so that dissipation [e.g., (\ref{diss_ksq})]
terminates it by converting the energy of motion into heat
in the so-called dissipative region of $k$-space
(e.g., \cite{Biskamp2003}).

There are various ways to describe spectral energy transfer
(see, e.g., \cite{MoninYaglom}). One of the simplest methods, listed
in \cite{MoninYaglom} as the Kovazhny hypothesis, involves a dimensional
analysis argument. If one writes the equation of turbulence evolution
in the inertial range as
\begin{equation}
\frac{dW}{dt} - \frac{\partial}{\partial k}\Pi = 0,
\end{equation}
then by the physical meaning, $\Pi$ is the flux of energy through
$k$-space towards larger $k$. Assuming that $\Pi$ is a product of
powers of the minimum set of relevant quantities, one can find the
simplest form of the corresponding cascading rate. That is,
if $\Pi = W^{a} k^{b} \rho ^{c}$, then the only combination of
$a$, $b$ and $c$ that gives $\Pi$ the correct units is
\begin{equation}
\label{pi_kolmogorov}
\Pi_{K} = W^{3/2} k^{5/3} \rho^{-1/2},
\end{equation}
As we will see later, this cascading rate gives a stationary solution
$W \propto k^{-5/3}$, which is known as the Kolmogorov spectrum,
and the corresponding cascade will be referred to as Kolmogorov-type
cascade (here denoted by the subscript `K').

When MHD turbulence is considered, the simple dimensional argument
shown above does not work because the magnetic field is another
relevant quantity. There are two approaches to describing nonlinear
effects (spectral energy transfer) in MHD turbulence. One, proposed
by Iroshnikov and, independently, by Kraichnan 
\cite{Iroshnikov1964, Kraichnan1965}, treats the MHD
turbulence as weakly interacting plasma waves that can undergo mergers
and splitting. The bottom line of this approach is that a stationary
spectrum $W \propto k^{-3/2}$ is predicted. Because $5/3$ and $3/2$
are so close, it is difficult to distinguish between the two indices in
the analysis of observations. Goldreich and Sridhar \cite{GS95}
point out that the MHD turbulence is inherently anisotropic
(even if there is no mean magnetic field, the effective field of the
large scale harmonics can play its role for the processes
in the inertial interval),
and the weak-turbulence approach is not applicable.
These authors proposed another theoretical approach: they suggested
a certain anisotropic damping rate and postulated a critical
balanced state, which allowed them to derive an anisotropic turbulence
spectrum.  Their results predict that harmonics with wavevectors transverse
to the uniform magnetic field experience a Kolmogorov-like
cascade, while the cascade in wavevectors parallel to the field
is suppressed. The waves generated with streaming instabilities
are transverse; therefore the diffusion coefficient for particle 
transport parallel to the flow depends on the wavenumbers parallel to
the magnetic field. Biskamp \cite{Biskamp2003}
shows that the Goldreich-Sridhar spectrum for parallel wavenumbers
may be expressed as $W \propto k_{\parallel}^{-5/2}$.

We can find the corresponding cascading rate, such that
$\Pi = W^{a} k^{\frac52 a} \rho^{b} v_A^{c}$, which would
lead to a steady state spectrum with $W \propto
k^{-5/2}$. From the dimensional argument,
\begin{equation}
\label{pi_gs}
\Pi_\mathrm{GS} = W^{2/3} k^{5/3} \rho^{1/3} v_A^{5/3}.
\end{equation}
One may do a simple estimate and compare the Kolmogorov and
the Goldreich-Sridhar cascading rates:
\begin{equation}
\frac{\Pi_\mathrm{GS}}{\Pi_K} = 
\frac{W^{2/3} k^{5/3} \rho^{1/3} v_A^{5/3}}{W^{3/2} k^{5/2}
  \rho^{-1/2}}=\left( \frac{B_0^2}{4 \pi k W}\right)^{5/6}.
\end{equation}

\subsubsection{Transition to turbulence}

One may pose a relevant question: at what point do the
linear plasma waves acquire the nonlinear behavior that
leads to their cascade and dissipation at short wavelengths? 
We assume that it happens when some of the waves reach
strong amplitudes, i.e., $\Delta B(k) \approx B_0$.
In terms of the quantities that we use to describe the turbulence
spectrum, I postulate that if, at the coordinate $x$,
there is a wavenumber $k$ such that
\begin{equation}
\frac12 kW(x,k) \geq \frac{B_0^2}{8\pi},
\end{equation}
then downstream of this coordinate, the turbulent
cascade and dissipation start (i.e., a transition
to turbulence occurs). In accordance with
that, upstream of this coordinate, the energy transport
$\Pi$ and the dissipation rate $L$ are both set to zero.

\subsubsection{Anisotropy relaxation}

The resonant streaming instability of \Alf\ waves amplifies
the waves traveling in the direction of the diffusive particle 
stream (i.e., in the upstream direction) and damps the
waves traveling in the opposite direction. The distribution of
energy between the upstream and downstream traveling
waves may be important for some applications:
for example, the mean speed of scattering centers, if
it is not negligible compared to the flow speed, 
calls for the appropriate
reference frame transformations for particle scattering.
This may be important for low \Alf\ Mach number shocks.
In the strong, fast shocks of young SNRs,
the generated waves predominantly travel upstream,
but for older shocks, nonlinear interactions between
upstream and downstream traveling structures may lead
to the appearance of downstream traveling waves
(e.g., \cite{BL2001}).

Following \cite{BL2001}, one can notionally separate the
turbulence spectrum as in the plasma frame  
\begin{equation}
W(x,k) = U_-(x,k) + U_+(x,k),
\end{equation}
where $U_-$ is the spectral energy density of waves
traveling upstream, and $U_+$ -- that of the downstream-directed
waves. The equation of turbulence growth due to 
a streaming instability, accounting only for the wave
advection, growth and the nonlinear interactions between
the waves traveling in different directions can be
written (see also \cite{VEB2006}) as 
\begin{eqnarray}
\label{uminuskp_int}
[u(x) - v_A]\frac{\partial}{\partial x}U_- &=& \resgrowthrate U_- 
        -v_A k \left( U_- - U_+ \right)
\ ; \\
\label{upluskp_int}
[u(x) + v_A]\frac{\partial}{\partial x}U_+ &=& -\resgrowthrate U_+ 
        +v_A k \left( U_- - U_+ \right)
\ ,
\end{eqnarray}
where $r_g = cp / (e B_0)$ and $v_A$ is the \Alf\ speed.
The factor $u\pm v_A$ represents the fact that the considered
waves travel at a velocity $v_A$ with respect to the plasma
along the magnetic field. The term proportional to
$U_- - U_+$ in the right-hand side describes nonlinear
interactions between the oppositely directed waves that
lead to isotropization of the wave spectrum (i.e.,
to $U_- = U_+$) with a relaxation time of about the
\Alf\ wave period. This effect may be important
for weaker shocks.

\newpage

\subsection{Generalized model of magnetic turbulence amplification}

\label{generalized_mfa}

Considering the effects described above (except for the interactions
with the waves traveling downstream), let us write the equation of
turbulence spectrum evolution in the following parameterized form:
\begin{equation}
\label{eq-genmfa}
u\frac{\partial W}{\partial x}
  +\parcompamp W \frac{du}{dx}
  -\parcompwav \frac{\partial}{\partial k} \left( k W \frac{du}{dx} \right) 
  -\paramplif G
  +\parcasc \frac{\partial}{\partial k} \Pi
  +\pardiss L
  = 0,
\end{equation}
and assume that a boundary condition is given at the
coordinate $x=x_0$ in the form
\begin{equation}
\label{bc-genmfa}
W(x_0, k) = W_0(k).
\end{equation}
The coordinate $x_0$ is typically located far upstream
of the shock, and the function $W_0(k)$ describes the seed turbulence
spectrum that, we assume, exists in the unshocked interstellar medium.

In Equation~(\ref{eq-genmfa}), the first term describes the advection
of turbulence with the flow. In the Lagrangian view, one may think of the
turbulence amplification process as the evolution of a matter element
advected towards and across the subshock, compressed and penetrated by
cosmic ray flux on the way, which leads to a buildup of stochastic
magnetic fields in this element. In the Eulerian perspective, this term 
represents the full derivative of $W$ with respect to
time, $d/dt = \partial/\partial t + u\partial/\partial x$, with the
local derivative $\partial / \partial t$ set to zero to model the
steady-state solution.

The second term, proportional to $\parcompamp$, represents the effect 
of plasma compression on the amplitude of the waves, as described in
the previous section. The parameter $\parcompamp$ measures the degree of
this effect. With all other terms set to zero,
equation~(\ref{eq-genmfa}) has the solution $W\propto
u^{-\parcompamp}$, i.e., the amplitude of the waves grows
proportionally to the power $\parcompamp/2$ of the plasma density.
For \Alf\ waves, $\alpha = 1.5$.

The third term, containing $\parcompwav$, describes the effect of
compression on the wavenumber of the waves. With all other effects
inactive, (\ref{eq-genmfa}) has the solution 
$W(k)\propto W(k u^{\beta})u^{\beta}$, which means that the
spectrum $W(k)$ shifts in $\log{k}$ space, while
preserving the normalization: $\int{W(k)} dk = const$.
Setting $\beta$ to 0 may be
used to `turn off' this effect in the model.

The term that contains $\paramplif$ is the driving term of instability
growth. The function $G$ is $G=\Gamma W$, where the growth rate $\Gamma$
can take on values: $\resgrowthrate$ from (\ref{resonant_increment}),
or $\nrgrowthrate$ defined by (\ref{bell_increment}), or the sum
of the two, depending on which instability one wishes to consider in the
model. The value $\paramplif = 1$ can be used to `turn on',
and $\paramplif = 0$ -- to `turn off' the turbulence amplification for
purposes of testing the code or making predictions relevant 
to the physics of shock acceleration.

The parameter $\parcasc$ in the fourth term of (\ref{eq-genmfa}) controls
the rate of cascading. For the energy flux $\Pi=\Pi_K$ given 
by~(\ref{pi_kolmogorov}), the quantity $\parcasc$ is essentially
the Kolmogorov constant, a factor that complements the dimensional
analysis leading to the derivation of (\ref{pi_kolmogorov}), and
that should be taken from experiments or numerical simulations.
There seems to be a universal value of the Kolmogorov
constant (see \cite{Sreenivasan1995} for
a review of experiments (note that this article has a different definition
of the constant) and \cite{GF2001}
for simulation results), $\parcasc = 1.6 - 1.7$. As was mentioned
earlier, MHD turbulence may have cascade properties different
from those of hydrodynamic turbulence, and $\Pi$ may assume
different forms, for example $\Pi_\mathrm{GS}$ from 
(\ref{pi_gs}). For lack of better knowledge, I will use the
value $\parcasc=1$ to include turbulent cascading and
$\parcasc=0$ to omit it from the model.

Dissipation of turbulence is controlled by the last term in
(\ref{eq-genmfa}), and the parameter $\pardiss$ can be set
to $1$ or $0$ to include or omit the dissipation. The function
$L$ can assume the parameterized form
$L_F$ from (\ref{diss_param}) or in the form of viscous
dissipation $L_V$ defined in (\ref{diss_ksq}).

\newpage

\subsection{Analytic solutions for turbulence spectrum}

The expression~(\ref{eq-genmfa}) is a nonlinear partial differential
equation of first order for a function of two variables, $W(x,k)$. 
Its particular solution is determined by the initial 
conditions~(\ref{eq_seed_turb}) and by the nonlinear
dynamics of the system that couples $W(x,k)$ to particle
propagation, and the latter to the driving term $G$ in (\ref{eq-genmfa})
and to the flow speed $u(x)$ determined with the iterative
procedures~(\ref{iteration_rtot}) and
(\ref{iteration_ux}). Therefore, the solver of (\ref{eq-genmfa}) has
to be run after every Monte Carlo iteration to advance the solution
towards self-consistency.

A powerful tool for tackling first order nonlinear equations is the
method of characteristics \cite{Lopez99}. In fact, in some simple
cases, i.e. when the terms in (\ref{eq-genmfa}) assume a simple
form, analytic solution is possible. Although these simple cases are
not directly applicable to the physical system we are studying,
I would like to derive these solutions below, because not only do they
reveal the influence of various effects on the solution, but they will
also be used for testing of the numerical solver.

\subsubsection{Parametric form}

In order to apply the method of characteristics to Equation
(\ref{eq-genmfa}), let us re-write it, collecting the terms containing
the partial derivatives of $W$, and assuming that $\Pi$ is given by
(\ref{pi_kolmogorov}): 
\begin{displaymath}
  u\frac{\partial W}{\partial x} +
  \left(-\parcompwav  k \frac{du}{dx} + \frac32 \parcasc \frac{W^{1/2}k^{5/2}}{\rho^{1/2}}\right)
    \frac{\partial W}{\partial k} =  \qquad \qquad
\end{displaymath}
\begin{equation}
\qquad\qquad = 
    (\parcompwav-\parcompamp) \frac{du}{dx} W -
  \frac52 \parcasc \frac{W^{3/2}k^{3/2}}{\rho^{1/2}} +
  \paramplif G - \pardiss L.
\end{equation}
In the spirit of the method of characteristics, this equation can be
written in a parametric form \cite{Lopez99}, describing $x$, $k$ and
$W$ as functions of a new parameter $t$:
\begin{eqnarray}
\frac{dx}{dt}&=& u, \\
\label{dkdt_param}
\frac{dk}{dt}&=& -\parcompwav  k \frac{du}{dx} + \frac32 \parcasc \frac{W^{1/2}k^{5/2}}{\rho^{1/2}}, \\
\frac{dW}{dt}&=& (\parcompwav-\parcompamp) \frac{du}{dx} W -
  \frac52 \parcasc \frac{W^{3/2}k^{3/2}}{\rho^{1/2}} + 
   \paramplif G - \pardiss L
\end{eqnarray}
The other two parametric equations form a system of nonlinear mutually dependent
equations, which cannot be solved in a closed form, but the solution
can be expressed in a form suitable for analysis. Consider the substitution
$p(t)=W(x(t), k(t))u^{\parcompamp-\parcompwav}(x(t))$, $q(t)=k(t)u^{\parcompwav}(x(t))$, and notice
that $(d/dx)=u^{-1}(x(t))(d/dt)$. Then the above system of equations can
be written as
\begin{eqnarray}
  \label{xoft}
  \frac{dx}{dt}&=& u, \\
  \label{qoft}
  \frac{dq}{dt} &=& 
  \frac32 \parcasc \frac{p^{1/2}q^{5/2}}{\rho^{1/2}}u^{-\parcompamp/2-\parcompwav},\\
  \label{poft}
  \frac{dp}{dt}&=& 
  - \frac52 \parcasc
  \frac{p^{3/2}q^{3/2}}{\rho^{1/2}}u^{-\parcompamp/2-\parcompwav} +
  \left(\paramplif G - \pardiss L\right) u^{\parcompamp-\parcompwav} .
\end{eqnarray}
Equations (\ref{xoft}), (\ref{qoft}) and (\ref{poft}) are the desired
parametric form of the generalized equation of turbulence evolution
(\ref{eq-genmfa}). This form will be used in the numerical solver.
The physical meaning of the parameter $t$ is obvious from 
equation~(\ref{xoft}): it is the time elapsed since a particular
harmonic at $k=k_0$ started evolving at $x=x_0$ (corresponding to $t=0$).

I must point out that this system is strongly nonlinear, because
the quantities $G$ and $L$ depend not only on the coordinate $x$ and
the wavenumber $k$, but also on the values of $p$ and $q$,
and not only locally, but also on the integrals of $W\propto p$ with respect
to $x$ and $q$, via the transport and particle acceleration properties of the 
turbulence.

However, assuming simplified expressions for $G$ and $L$, we may
obtain analytic solutions, as shown in the next two sections.

\subsubsection{Solution without cascades}

In the absence of cascading ($\parcasc=0$), equations~(\ref{xoft}),
(\ref{qoft}) and (\ref{poft}) are not coupled, and have the obvious
solution 
\begin{eqnarray}
  t    &=& \int\limits_{x_0}^{x}\frac{dx'}{u(x')}, \\
  q(t) &=& q_0, \\
  p(t) &=& p_0 + 
   \int\limits_{0}^{t}
   \left(\paramplif G - \pardiss L\right)u^{\parcompamp-\parcompwav}\;dt'.
\end{eqnarray}
or, in terms of $k$ and $W$,
\begin{eqnarray}
  t    &=& \int\limits_{x_0}^{x}\frac{dx'}{u(x')}, \\
   k(t)\left[u(x(t))\right]^{\parcompwav} &=& k_0 \left[u(x_0)\right]^{\parcompwav}, \\
   W(t)\left[u(x(t))\right]^{\parcompamp-\parcompwav} &=& 
                    W(x_0,k_0)\left[u(x_0)\right]^{\parcompamp-\parcompwav}+ \\
  & & +
        \int\limits_{t_0}^{t}
        \left(\paramplif G(x',k') 
            - \pardiss L(x',k') \right)
           \left[u(x')\right]^{\parcompamp-\parcompwav}\;dt',
\end{eqnarray}
where $x'\equiv x(t')$, $k'\equiv k(t')$. Finally, for the spectrum
in terms of the original variables, $W(x,k)$, we may write
\begin{displaymath}
W(x,k) = W_0\left(k\left(\frac{u(x)}{u(x_0)}\right)^{\parcompwav}\right)
       \left[ \frac{u(x_0)}{u(x)} \right]^{\parcompamp-\parcompwav} +
       \qquad 
\end{displaymath}
\begin{equation}
\label{wxk_nocasc}
\qquad + \int\limits_{x_0}^{x}
    \left\{ 
      \paramplif G\left(x',k\left[\frac{u(x)}{u(x')}\right]^{\parcompwav}\right)-
      \pardiss   L\left(x',k\left[\frac{u(x)}{u(x')}\right]^{\parcompwav}\right)
      \right\}
    \left[\frac{u(x')}{u(x)}\right]^{\parcompamp-\parcompwav}
    \;\frac{dx'}{u(x')}
\end{equation}
Expression (\ref{wxk_nocasc}) is the solution of equation 
(\ref{eq-genmfa}) with the boundary condition (\ref{bc-genmfa})
for $\parcasc=0$ (no cascading).

In particular, setting $\parcompamp=\parcompwav=0$ in
(\ref{wxk_nocasc}), we get the
solution describing the turbulence evolution with only
amplification $G$ and dissipation $L$ accounted for:
\begin{equation}
\label{wxk_onlyamp}
W(x,k) = W_0(k) + \int\limits_{x_0}^{x} 
\left\{\paramplif G(x',k) - \pardiss L(x',k)\right\}\frac{dx'}{u(x')}.
\end{equation}

Assuming the opposite, $\paramplif=\pardiss=0$, but 
$\parcompamp\neq 0$ and $\parcompwav \neq 0$, one obtains
\begin{equation}
\label{wxk_nocasc_noamp}
W(x,k) = W_0\left(k\left(\frac{u(x)}{u(x_0)}\right)^{\parcompwav}\right)
       \left[ \frac{u(x_0)}{u(x)} \right]^{\parcompamp-\parcompwav},
\end{equation}
which describes the effect of compression on the plasma turbulence:
the energy density increases in proportion to $\rho^{\parcompamp}$,
and the wavenumber grows as $\rho^{\parcompwav}$ (see also
Section~\ref{turb_effects}).

\subsubsection{Integral form with cascades}

Now let us return to the parametric form of the turbulence evolution
equation given by (\ref{xoft}), (\ref{qoft}) and (\ref{poft}).
This time let us not set $\parcasc=0$, but try to derive a solution
that accounts for cascading. The last two equations are
coupled via the cascading terms, so in order to
get an integral form of the solution, let us express $p$ as a function of
$q$ by dividing equation (\ref{poft}) by equation (\ref{qoft}),
which is a correct operation because $\parcasc \neq 0$. This 
leads to:
\begin{equation}
  \frac{dp}{dq} = 
  -\frac53 \frac{p}{q} + 
   \frac23 \frac{\rho^{1/2}}{p^{1/2}q^{5/2}} 
   \frac{1}{\parcasc} (\paramplif G - \pardiss L) u^{3 \parcompamp/2},
\end{equation}
or
\begin{equation}
  \frac{d}{dq} \left[ \left(pq^{5/3}\right)^{3/2} \right]= 
  \rho^{1/2} \frac{1}{\parcasc}(\paramplif G - \pardiss L) u^{3\parcompamp/2},
\end{equation}
which gives the dependence of $p$ on $q$ in the following form:
\begin{equation}
  \label{pofq}
  p=q^{-5/3}
  \left[\left(p_0 q_0^{5/3}\right)^{3/2} +
    \frac{1}{\parcasc}
     \int_{q_0}^{q} \rho^{1/2} 
         (\paramplif G(x', k') - \pardiss L(x', k'))
               u^{3\alpha/2}(x') dq'
  \right]^{2/3},
\end{equation}
where $x' \equiv x(t')$, $k' \equiv k(t')$, and $t'$ is the moment in time 
corresponding to $q(t')=q'$. Constants $p_0$
and $q_0$ define the characteristic curve by its initial conditions as:
$p_0=W(x_0, k_0)u^{\alpha-\beta}(x_0)$, and $q_0=k_0u^{\beta}(x_0)$. 
Substituting (\ref{pofq}) into (\ref{qoft}) gives:
\begin{displaymath}
  \frac{d}{dt}\left[ q^{-2/3} \right] = 
  - \parcasc
  u^{-\parcompamp/2 - \parcompwav} \rho^{-1/2} \times \qquad\qquad
\end{displaymath}
\begin{equation}
  \label{qoftsol}
\qquad \times  \left[ 
 \left( p_0 q_0^{5/3} \right)^{3/2} + 
    \frac{1}{\parcasc} \int_{q_0}^{q} 
       \rho^{\frac 12}(x') 
       (\paramplif G(x',k') - \pardiss L(x',k'))
       u^{3\alpha/2}(x') dq'
  \right]^{1/3}.
\end{equation}
Expressions (\ref{xoft}), (\ref{pofq}) and (\ref{qoftsol}) are almost a solution
to equation (\ref{eq-genmfa}). We collect them below:
\begin{eqnarray}
\label{xoftsol_c}
  \frac{dx}{dt}&=& u, \\
\nonumber
  \frac{d}{dt}\left[ q^{-2/3} \right] &=&
            - \parcasc  u^{-\parcompamp/2 - \parcompwav} \rho^{-1/2} \left[ 
    \left( p_0 q_0^{5/3} \right)^{3/2} + \right. \\
  \label{qoftsol_c}
  & &\quad \left. + 
    \frac{1}{\parcasc} \int_{q_0}^{q} 
       \rho^{\frac 12}(x') 
       (\paramplif G(x',k') - \pardiss L(x',k'))
       u^{3\alpha/2}(x') dq'
  \right]^{1/3}, \\
  \nonumber
  p&=&q^{-5/3}
  \left[\left(p_0 q_0^{5/3}\right)^{3/2}\right. + \\
  \label{poftsol_c}
  & & \quad \left.
    \frac{1}{\parcasc}
     \int_{q_0}^{q} \rho^{1/2} 
         (\paramplif G(x', k') - \pardiss L(x', k'))
               u^{3\alpha/2}(x') dq'
  \right]^{2/3}
\end{eqnarray}
The system (\ref{xoftsol_c}), (\ref{qoftsol_c}) and (\ref{poftsol_c}),
the integral (integro-differential) form of the solution
to equation (\ref{eq-genmfa}), is rather unsightly for a physicist, so 
we will simplify it by setting $\parcompamp=\parcompwav=\pardiss=0$,
$\paramplif=\parcasc = 1$,
and assuming that $u(x)=u_0$, $\rho(x)=\rho_0$,
and $G(x,k) = G_0 \delta_D(k-k_c)$, where $\delta_D$ is the
Dirac delta function. This corresponds to the case where energy
is supplied to the turbulence at a wavenumber $k_c$ throughout
the spatial extent of the system. In addition, let us assume
that there is no seed turbulence, i.e., $W_0(k)=0$.  Then
(\ref{qoftsol_c}) and (\ref{poftsol_c}) yield:
\begin{eqnarray}
k(t) &=& 
   \left\{ \begin{array}{l}
     \displaystyle
     \left[k_0^{-2/3} - F_0^{1/3}t\right]^{-3/2}, \quad t<t_c, \\
     \displaystyle
     \left[k_0^{-2/3} - \left(
            F_0 + \frac{G_0}{\rho_0}\right)^{1/3}t\right]^{-3/2}, \quad t \geq t_c;
   \end{array} \right.\\
W(x(t), k(t)) &=& 
    \left\{ \begin{array}{l}
    \displaystyle
    \left[k_0^{-2/3} - F_0^{1/3}t\right]^{5/2}
    \rho_0 F_0^{2/3},  \quad t<t_c, \\
    \displaystyle
    \left[k_0^{-2/3} - \left( F_0 + \frac{G_0}{\rho_0}\right)^{1/3}t\right]^{5/2}
    \rho_0 \left( F_0 + \frac{G_0}{\rho_0} \right)^{2/3}, \quad t \geq t_c.
    \end{array} \right. 
\end{eqnarray}
Here
\begin{equation}
F_0 = \left( \frac{W(x_0,k_0)k_0^{5/3}}{\rho_0} \right)^{3/2} 
\end{equation}
and
\begin{equation}
t_c = \frac{k_0^{-2/3} - k_c^{-2/3}}{F_0^{1/3}}.
\end{equation}
Assuming $G_0/\rho_0 \gg F_0$ (that is, the generation of the 
turbulence at the wavenumber $k_c$ overpowers the seed turbulence),
the solution for $t>t_c$ (in other words, for $k>k_c$) is:
\begin{eqnarray}
k(t) &=& 
     \left[k_0^{-2/3} -
       \left(\frac{G_0}{\rho_0}\right)^{1/3}t\right]^{-3/2},  \\
\label{amp_with_cascades}
W(x(t), k(t)) &=& 
    \left[k_0^{-2/3} - \left( \frac{G_0}{\rho_0}\right)^{1/3}t\right]^{5/2}
    \rho_0 \left(\frac{G_0}{\rho_0} \right)^{2/3}=k^{-5/3}(t)\rho_0 \left(\frac{G_0}{\rho_0} \right)^{2/3}.
\end{eqnarray}
The expression for $W(x,k)$ does not depend on $k_0$, $x_0$ or $W(x_0, k_0)$,
and therefore it describes explicitly the turbulence spectrum at $k>k_0$.
Namely, shortward of $k_c$, the effect of cascading leads
to a formation of a power-law spectrum of turbulence $W(k) \propto k^{-5/3}$, which
is the Kolmogorov spectrum, as discussed in
Section~\ref{turb_effects}. This result can directly be used
for testing of the numerical routine solving the equation
(\ref{eq-genmfa}).

\newpage

\subsection{Development of the numerical integrator}

In order to calculate the spectrum of MHD turbulence
produced by the instabilities of the precursor plasma in the presence
of the accelerated particle stream, the model solves
equation~(\ref{eq-genmfa}). The driving term in this
equation, $G$, is calculated using the information about particle
streaming simulated in the Monte Carlo transport module.
The numerical procedure that will be run in the simulation
must solve Equation~(\ref{eq-genmfa}) with arbitrary driving term $G$
and with or without all the other terms in this
equation, parameterized by $\parcompamp$, $\parcompwav$,
$\paramplif$, $\parcasc$ and $\pardiss$.
I have developed such an integrator, and the algorithm
of integration is presented in this section.

In brief, equation~(\ref{eq-genmfa}) is solved by integrating
the system of coupled first-order ordinary differential
equations: (\ref{xoft}), (\ref{qoft}) and (\ref{poft}). This system
is derived using the method of characteristics,
and its solutions for different values of $k_0$ are the characteristic
curves. The numerical method used for integration is a 
finite differencing scheme (based on the implicit Gauss's
method), with an adaptive step size in $x$-space and adaptive
mesh refinement in $k$-space.  The implicit nature of Gauss's
method is beneficial for the stability of the results,
and is achieved with an iterative procedure.

Here is the outline of the procedure.
Integrating from $x=-\infty$ to $x>0$, the scheme will make 
$N_x$ steps, $N_x$ being the number of grid planes. For every $k$-bin,
every spatial step from $x_{(i-1)}$ to $x_{(i)}$ will consist of
$N_{sub}$ substeps, enumerated by the index $l$, in which 
the code will propagate $k$ and $W$ from 
$x_{(i-1)}$ to $x_{(i)}$; the size of each substep will be adaptively chosen
to ensure the stability of Gauss's method.
After all $k$-bins have been propagated from $x_{(i-1)}$ to $x_{(i)}$,
the program will use the $k$-grid modified by compression and cascading
to project the amplified $W$ onto the fixed $k$-grid of the simulation at
$x_{(i)}$, and then proceed with the step to the next grid plane.
If the code finds that the evolved $k$-grid has too large a spacing between
some nodes, it will refine the problematic regions of the
$k$-grid at $x_{(i-1)}$ and repeat the integration of the equation.
The scheme will keep refining the $k$-grid at $x_{(i-1)}$
until the resulting $k$-grid at $x_{(i)}$ is satisfactory (i.e., fine enough).

\subsubsection{Notation for this section}

An index in round parentheses, as in $x_{(i)}$,
enumerates the $x$-grid plane, and one in square parentheses,
as in $k_{[j]}$, indicates the number of $k$-space bin.
Subscripts without parentheses (e.g., in $p_l$) mean
the number of the substep between $x_{(i-1)}$
and $x_{(i)}$, and superscripts in parentheses (e.g.,
$q_l^{(m)}$) are reserved for the number of the cycle in the
iteration used to achieve the implicitness of the method.

For brevity, we re-write equations 
(\ref{xoft}), (\ref{qoft}) and (\ref{poft}) as 
\begin{eqnarray}
  \label{casc_par_x}
  \frac{dx}{dt}&=& u, \\
  \label{casc_par_q}
  \frac{dq}{dt} &=& \parcasc C q,\\
  \label{casc_par_p}
  \frac{dp}{dt}&=& 
  - \frac53 \parcasc C p +
  (\paramplif G - \pardiss L)u^{\parcompamp-\parcompwav} ,
\end{eqnarray}
where
\begin{equation}
C = \frac32 \frac{p^{1/2}q^{3/2}}{u^{\parcompamp/2 + \parcompwav} \rho^{1/2}}.
\end{equation}

\subsubsection{Making a substep}

The substeps will be enumerated by the index $l$, so that $q_l$ and $p_l$
are the quantities $q$ and $p$ at the end of the $l$-th substep.
The code starts making the $l$ substeps by initializing the following quantities:
\begin{eqnarray}
  x_0 &=& x_{(i-1)}, \\
  t_0 &=& 0, \\
  q_0 &=& k_{[j]}(x_{(i-1)}) u^{\parcompwav}(x_{(i-1)}), \\
  p_0 &=& W_{[j]}(x_{(i-1)}) u^{\parcompamp-\parcompwav}(x_{(i-1)}).
\end{eqnarray}
To make the $l$-th substep, let us first assign
the following quantities:
\begin{eqnarray}
  x_{l} &=& x_{l-1} + \Delta x_{l},\\
  u_{l} &=& u(x_{l}), \\
  \rho_{l} &=& \rho(x_{l}).
\end{eqnarray}
The step width $\Delta x_{l}$ will be initially (for $l=1$) set as
\begin{equation}
  \Delta x_{1} = x_{(i)} - x_{(i-1)},
\end{equation}
and if this attempted substep succeeds, there will be only one substep
($l=1$), after which the scheme will move on to the next grid plane $i$.
If the scheme finds this substep too large, it
will choose a smaller substep. For the subsequent substeps we will set
\begin{equation}
  \Delta x_l = X_l \cdot \Delta x_{l-1},
\end{equation}
where $X_l$ is a number, greater or smaller than 1, depending on whether
the previous substep was estimated as too short or too long,
as discussed later.
The program can integrate (\ref{casc_par_x}) to get:
\begin{eqnarray}
  \Delta t_{l} &=& \frac{\Delta x_{l}}{u_l}, \\
  t_{l} &=& t_{l-1} + \Delta t_{l}.
\end{eqnarray}
To derive $q_{l}$ from $q_{l-1}$ and $p_{l}$ from $p_{l-1}$,
the code will need to use an iterative procedure in order to 
implement an implicit finite differencing scheme for 
solving (\ref{casc_par_q}) and (\ref{casc_par_p}).
The superscript $(m)$ will denote the cycle of iteration, and it will
run from $0$ as far as it takes for convergence, restarting as $m=0$ with
each new $l$. The initial step in this iteration will be
\begin{eqnarray}
q_{l}^{(0)}&=&q_{l-1},\\
p_{l}^{(0)}&=&p_{l-1}.
\end{eqnarray}
and the subsequent iterations will be derived from
\begin{eqnarray}
  \label{qlmiter}
  q_{l}^{(m)} &=& q_{l-1} \exp\left( \left[\frac{d \ln{q}}{dt}\right]_{l}^{(m-1)} \Delta t_{l}\right), \\
  \label{plmiter}
  p_{l}^{(m)} &=& 
        (p_{l-1} - p_{\star}) \exp\left( \left[\frac{d \ln{p}}{dt}\right]_{l}^{(m-1)} \Delta t_{l}\right) + p_{\star},
\end{eqnarray}
The values of the above mentioned derivatives and of the quantity $p_{\star}$ are discussed later.
Before making the $(m)$-th iteration, the code must check whether the substep size $\Delta x_l$ was small enough.
It does so by comparing the arguments of the above mentioned exponentials to a pre-set number $\eta$.
The value $\eta=0.01$ seems to work well as the target step size. If at any step the arguments of the
exponentials are greater than $\eta$, the $(m)$ iteration terminates,
the code chooses a proportionally
lower $\Delta x_{l}$ by setting $X_l<1$, and tries making the $l$-th substep again. 
If the value of the arguments of the exponentials in (\ref{qlmiter}) and (\ref{plmiter}) are
by a factor of a few smaller than $\eta$ in
all $(m)$ iterations, then for the $(l+1)$-th substep the code chooses
$X_{l+1}>1$ in order to speed up the integration. Choosing the spatial step size this
way makes the scheme adaptive in $x$-space.

The value $p_{\star}$ is used to tend to a nasty property of our
equations: the cascading and dissipation terms eventually drive the
solution to $p(t \to \infty)\to 0$, which happens to be the boundary
of the range of definition of some of the functions in the equations. 
For the analytic solution, it is not a problem, 
because at $p=0$ processes further decreasing $p$ (cascading
and dissipation) naturally cease. But in a numerical solution, 
there is a danger of marginally running into the $p<0$ region, if
$p$ is evolved with a finite differencing scheme, which will
cause an error, because the factor $p^{1/2}$ in some of the functions
is not defined for a negative $p$.
I eliminate the possibility of getting $p<0$ by evolving 
$\ln{p}$ instead of $p$ with the finite differencing method.
However, when $p\to 0$, the program risks dividing by zero.
To avoid zero values of $p$, I re-define the point at which
the processes decreasing $p$ stop: 
from $p=0$ to $p=p_{\star}$. The solution in each bin
subject to dissipation then converges to $p=p_{\star}$
instead of $p=0$. The value of $p_{\star}$
is chosen small enough so that it doesn't affect the physical solution,
but large enough to be treated numerically without problems.

One danger possible with an iteration on $q_l^{(m)}$ and $p_l^{(m)}$ like 
(\ref{qlmiter}) and (\ref{plmiter})
is that the solution may find an attractor cycle around the equilibrium
point instead of converging to it, in which case we may find ourselves stuck in
an infinite cycle (the equilibrium point 
is the point at which $q_{l}^{(m)}=q_{l}^{(m-1)}$ and $p_{l}^{(m)}=p_{l}^{(m-1)}$). 
Theory suggests that there is a finite domain of attraction around the attracting
equilibria of this system, so all we have to do
to ensure convergence in the end is perturb the solution occasionally. If the
iteration is stuck in an attractor cycle,
with the perturbation it usually jumps into the domain of attraction of the equilibrium
point and converges (or finds another cycle, which it will be
driven out of with a later perturbation).
In practice, the code perturbs the solution whenever $m$ equals a multiple of a large integer,
for example, 1000. Then it adjusts $q_{l}^{(m)}$ and $p_{l}^{(m)}$ only 
half way from $q_{l}^{(m-1)}$ and $p_{l}^{(m-1)}$ to what 
(\ref{qlmiter}) and (\ref{plmiter}) suggest (this going half way is the perturbation). Experience
shows that this procedure successfully finds the equilibrium points
of the above system of equations, thus yielding the implicit
Gauss's integration scheme.

The iteration deriving $q_l^{(m)}$ from $q_l^{(m-1)}$ 
and $p_l^{(m)}$ from $p_{l}^{(m-1)}$ will continue until
it converges, that is, the relative difference between the 
values obtained at the previous and the current
step becomes small enough. Suppose it happens at step $m=N_m$. Then
the code will assign
\begin{eqnarray}
q_l &=& q_l^{(N_m)},\\
p_l &=& p_l^{(N_m)}.
\end{eqnarray}
and increment $l$. As soon as the last substep is completed
($x_l = x_{(i)}$), the program names $N_{sub}=l$ and assigns
\begin{eqnarray}
q(t_{fin}) &=& q_{N_{sub}}, \\
p(t_{fin}) &=& p_{N_{sub}}.
\end{eqnarray}
Having $q(t_{fin})$ and $p(t_{fin})$ allows one to revert back to the physical 
quantities and assign
\begin{eqnarray}
k_{[j]}(x_{i}) &=& q(t_{fin}) u(x_{i})^{-\parcompwav}, \\
W_{[j]}(x_{i}) &=& p(t_{fin}) u(x_{i})^{-\parcompamp+\parcompwav}.
\end{eqnarray}

\subsubsection{Calculating the derivatives}

In equations (\ref{qlmiter}) and (\ref{plmiter}), the derivatives of
$\ln{q}$ and $\ln{p}$ are calculated, according to (\ref{casc_par_q}) and
(\ref{casc_par_p}), as:
\begin{eqnarray}
\left[\frac{d \ln{q}}{dt}\right]_{l}^{(m-1)} &=& \parcasc C_{l}^{(m-1)},\\
\left[\frac{d \ln{p}}{dt}\right]_{l}^{(m-1)} &=& -\frac53 \parcasc C_{l}^{(m-1)} + 
        \left(\paramplif G_{l}^{(m-1)} - \pardiss L_{l}^{(m-1)}\right)\frac{u^{\parcompamp-\parcompwav}}{p_l^{(m-1)}},
\end{eqnarray}
where
\begin{eqnarray}
C_{l}^{(m-1)} &=& \frac32 
      \frac{\left(p_l^{(m-1)} - p_{\star}\right)^{\frac12} \left(q_l^{(m-1)}\right)^{\frac32}}
           {u_l^{\parcompamp/2+\parcompwav} \rho_l^{1/2}}, \\
\label{gl_num}
G_{l}^{(m-1)} &=& V_{G,\,l} \left[
                      \left( \frac{d\Pcr}{dx} \right) 
                               \left|\frac{dp}{dk}\right|
                           \right]
     \left| \begin{array}{l}\\ \\\left(x_{i,\,l},k\right)\end{array}\right., \\
\label{ll_num}
L_{l}^{(m-1)} &=& \frac{\left(p_{l}^{(m-1)}-p_{\star}\right)u_l^{-\parcompamp+\parcompwav}}
            {\tau_D\left(x_{i,\,l},k\right)}
             H\left(x_{i,\,l},k\right)
\end{eqnarray}
More details on evaluating quantities from (\ref{gl_num}) and (\ref{ll_num})
are given in the next subsection.

\subsubsection{Details of the the growth and damping rate calculations}

In expressions (\ref{gl_num}) and (\ref{ll_num}) the following notation
is used:
\begin{eqnarray}
k &\equiv& k_l^{(m-1)} = q_l^{(m-1)} u^{-\parcompwav}_l, \\
V_{G, \, l} &=& \frac{B_0}{\sqrt{4 \pi \rho_l}}.
\end{eqnarray}

The instability growth term, $G$, in the resonant case is
determined by the gradient of CR
pressure at the resonant momentum. The quantity $\Pcr$ is the
pressure per unit interval of particle momentum, thus the
factor $|dp/dk|$ in (\ref{gl_num}). To calculate the pressure gradient in such
a way that the discontinuity of $\Pcr$ in $p$-space
doesn't lead to a discontinuity of $G$ in $k$-space,
I chose to average the pressure over a finite wavenumber interval
$\Delta k$. The code sets $\Delta k = 0.05 k$ and defines
\begin{eqnarray}
  k^{left}(k) &=& k - \frac12 \Delta k, \\
  k^{right}(k)&=& k + \frac12 \Delta k,
\end{eqnarray}
after which it can
calculate the corresponding range of the particle momenta that
interact with the current bin:
\begin{eqnarray}
  p^{high}(k) &=& \frac{e B_0}{c k^{left}(k)}, \\
  p^{low}(k)  &=& \frac{e B_0}{c k^{right}(k)}.
\end{eqnarray}
Then the instantaneous gradient of the CR pressure that powers the
instability (to the best of one's knowledge at the $(m)$-th iteration
of the $l$-th substep from $x_{i-1}$ to $x_{i}$) can be estimated as
\begin{equation}
\left[
\left( \frac{d\Pcr}{dx} \right) 
\left| \frac{dp}{dk} \right|
\right]
\left| \begin{array}{l}\\ \\(x_{l},k) \end{array}\right. 
=
\frac{1}{\Delta x_{(i)}}
\left(
P_{(i)}\left(k\right)    \left|\frac{dp}{dk}\right|- 
P_{(i-1)}\left( k \right) \left|\frac{dp}{dk}\right|
\right),
\end{equation}
where
\begin{eqnarray}
  P_{(i-1)}\left( k \right) \left|\frac{dp}{dk}\right| &=&
  \frac{1}{\Delta k}
  \int\limits_{p^{low}(k)}^{p^{high}(k)} \Pcr(x_{(i-1)}, p) dp, \\
  P_{(i)}\left( k \right) \left|\frac{dp}{dk}\right| &=&
  \frac{1}{\Delta k}
  \int\limits_{p^{low}(k)}^{p^{high}(k)} \Pcr(x_{(i)}, p) dp.
\end{eqnarray}
This allows us to calculate (\ref{gl_num}). Note that I used
the grid nodes $x_{(i-1)}$ and $x_{(i)}$ as reference points
for calculating the gradient. That is done because the CR pressure
is evaluated in the Monte Carlo simulation directly at these 
locations.

In the dissipation rate $L$, calculated in (\ref{ll_num}),
\begin{eqnarray}
  \tau_D\left(x_{l},k\right) &=& 
     \frac{k^{-1}}{V_{G,\,l}},\\
  H\left(x_{l},k\right) &=& 
     \frac{1}{1+\kdiss\left(x_{l}\right)/k}, \\
  \kdiss\left(x_{l}\right) &=& 
     \frac{e B_0}{c \sqrt{m_p k_B T\left(x_{l}\right)}}.
\end{eqnarray}

\subsubsection{Adaptive $k$-grid}

I use the parametric form of the turbulence growth equation,
in which the $k$-grid evolves in time. It may happen that two
$k$-grid nodes that were adjacent far upstream will move 
apart significantly by the time the turbulence advects downstream.
In practice, starting off at $x=-\infty$
with 80 $k$-grid nodes equally spaced in 
$\log {k}$ space and spanning 10 orders of magnitude of $k$, 
we are likely to get two adjacent nodes that move apart
by several orders of magnitude (!) at $x>0$. This makes it 
problematic to interpolate the wave spectrum $W(x,k)$ between
these two nodes, and, in fact, a lot of information about
this $k$-region is missing from the solution. An attempt to boost
the $k$-resolution by 
increasing the density of $k$-grid nodes uniformly throughout
$k$-space leads to a significant increase of computation time
and to the need to have tens of thousands of $k$-nodes.

To solve this problem, I use an iterative approach to the refinement
of the $k$-grid. After the system is integrated to $x_{(i)}$, the code evaluates
the $k$-grid at this final point.  If it finds two $k$-nodes that
are too far apart at $x_{(i)}$ (by too far apart
I usually mean $\Delta \ln{k}\equiv\ln(k_{[j]}/k_{[j-1]}) \geq 0.5$), 
it inserts a number of new nodes into the
integration grid at $x_{(i-1)}$ and interpolates the seed turbulence
spectrum into these nodes to repeat the calculation. Several
(less than 10) iterations like that allow to get enough resolution
in $k$-space throughout the system with minimal time 
(in practice, the whole computation takes a few seconds) and 
minimal memory (I usually have to have only a few hundred $k$-bins).

\subsubsection{Turning over in $k$-space}

Equation (\ref{dkdt_param}), for $\parcompwav=0$,
shows that cascading leads to 
the motion of a harmonic with wavenumber $k$ at a speed
of $V_k=1.5 W^{1/2}k^{5/2}\rho^{-1/2}$ (for $\parcasc=1$) in 
$k$-space. This dispersion relation has an interesting feature:
if the spectrum $W(k)$ has a power-law shape, $W \propto k^{s}$, 
then $V_k$ is an increasing function of $k$ for $s>-5$, but
a decreasing function of $k$ for $s<-5$. That is, for
the parts of the spectrum in which it rapidly drops off
with $k$ (more quickly than $k^{-5}$), the lower $k$ harmonics
increase their $k$ {\it faster} than the greater $k$.
In this situation the fast-moving low-$k$ harmonics may
catch up and overrun the slow-moving high-$k$ harmonics.

This situation is common for waves in gases and fluids, where
the phase speed of waves increases with density or wave height.
It leads to waves turning over in water, and to shocks in fluid dynamics,
when viscosity is accounted for. Obviously, in this model
the turning over of waves in $k$-space doesn't have a physical
meaning and simply reflects the limited applicability of the 
Kolmogorov cascade to steep wave spectra. However, straightforward
application of this model to non-linear particle accelerating shocks
does lead to the turning over in $k$-space in the numerical
solution.

The place where turning over is most likely to occur is the dissipative
region of the spectrum. There the turbulence dissipation term, $L$, 
makes the wave spectrum drop off exponentially, creating the situation
in which turning over is likely to happen. Another possibility is
turning over in the inertial region, if the generation of waves occurs
on top of a `seed' spectrum, and the generated waves cascade faster
than the seed waves.

I ignore the wave turnover in  $k$-space
in the dissipative region, assuming that it will not affect 
the energetics of the process very much. 
As for the inertial spectrum, the physical
solution for a steady-state nonlinearly modified shock must not have
wave turnover there, if the model is self consistent (otherwise
we must conclude that the Kolmogorov cascade is not a good approximation
for the plasma physics of self-generated turbulence). There is
a natural property of the accelerated particle distribution in shock
precursors that seems to help the situation, if resonant amplification of
waves is assumed. Very far upstream,
only the highest energy particles resonantly generate the smallest $k$ waves.
These waves start to cascade and would outrun the higher $k$ seed waves,
but as the plasma advances toward the subshock, it encounters lower energy
particles, whose pressure builds up exponentially with time. These lower
energy particles should energize the higher $k$ waves, facilitating their
escape from the lower $k$ waves pre-amplified farther upstream. This
way wave turnover in $k$-space may be avoided naturally 
due to the properties of particle accelerating shocks.

\subsection{Tests of the numerical integrator}

In this section I will present the tests of the integrator 
which compare the results of the numerical solution
to the analytic solutions described above.
All these test involve introducing a seed turbulence
spectrum upstream, at $x=x_0<0$, and numerically integrating 
Equation~(\ref{eq-genmfa}) from $x=x_0$ to $x=0$.
In order to test and understand the effects of different
processes parameterized by $\parcompamp$,
$\parcompwav$, $\paramplif$, $\parcasc$ and $\pardiss$,
I executed several runs, in which some of these
parameters were set to finite values, while the other
were set to zero.

\begin{figure}[htbp]
\centering
\vskip 0.5in
\includegraphics[angle=-90, width=4.0in, clip=true, trim=0.8in 0.0in 0.0in 0.0in]{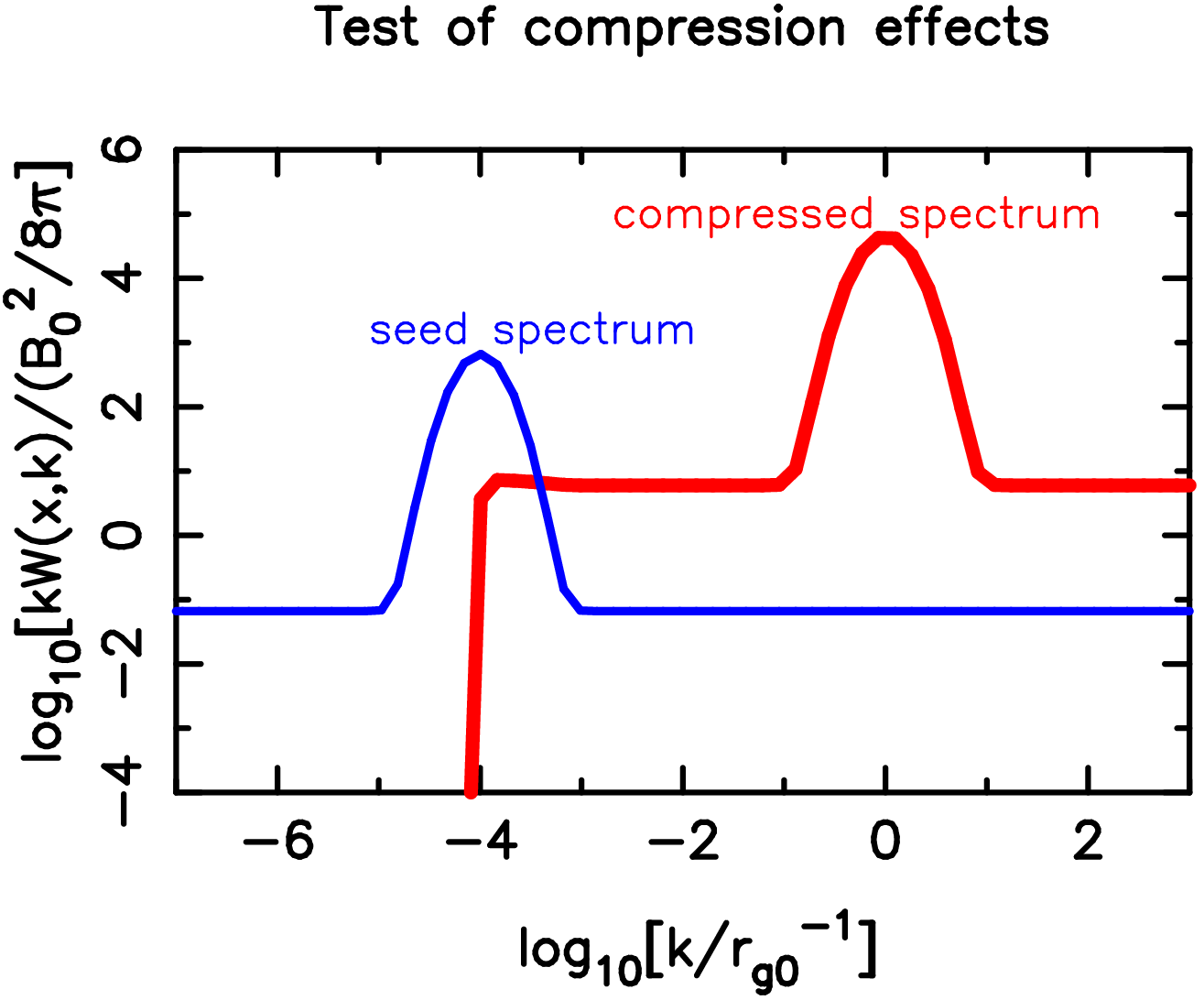}
\caption{$ $ Effect of flow compression on turbulence spectrum.}
\label{fig-turb_test_compr}
%
\vskip 0.5in
\centering
\includegraphics[angle=-90, width=4.0in, clip=true, trim=0.8in 0.0in 0.0in 0.0in]{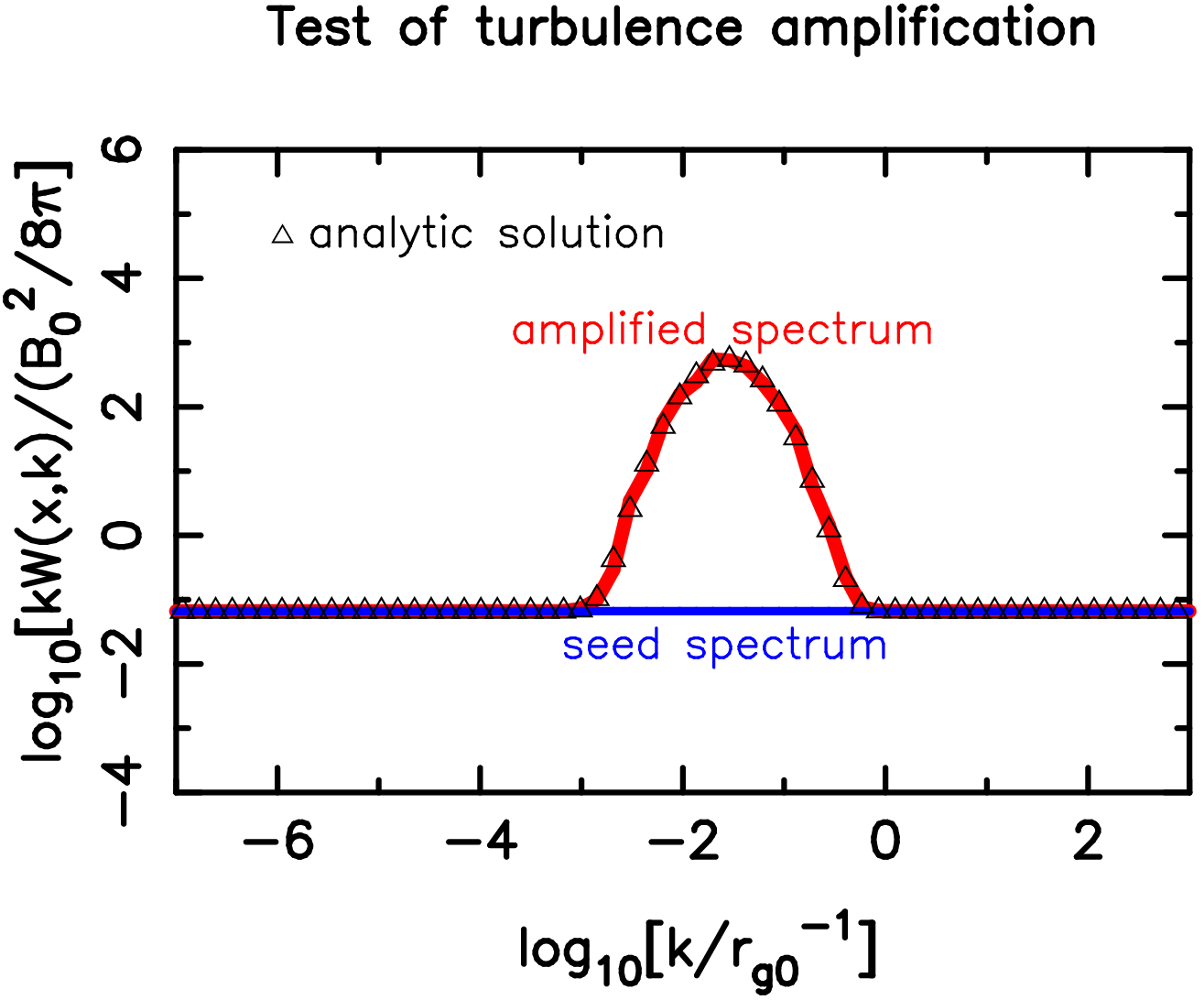}
\caption{$ $ Amplification of turbulence spectrum.}
\label{fig-turb_test_amp}
\end{figure}

First, I tested the effects of the compression of the
flow: the increase in the amplitude and the wavenumber of
the harmonics. At $x=x_0$ I introduced a Bohm seed spectrum with
a Gaussian feature on top of it, located at $k=10^{-4}\,\rgzero^{-1}$
(see the thin line in Figure~\ref{fig-turb_test_compr}). I imposed
a flow speed that drops by a factor of $r=10^2$ from
$x=x_0$ to $x=0$. Then the code solved Equation~(\ref{eq-genmfa})
using $\parcompamp=1.0$ and $\parcompwav=2.0$ (these 
values were used just for testing; physically justified
values are discussed in Section~\ref{turb_effects}). The resulting
spectrum at $x=0$, shown with the thick line in
Figure~\ref{fig-turb_test_compr}, agrees with one's expectation
based on the analytic solution~(\ref{wxk_nocasc_noamp}):
the feature moved to the right, towards greater $k$ by
a factor of $r^{\parcompwav}=10^4$ and upward, to greater
amplitudes, by a factor of $r^{\parcompamp}=10^2$.
Note that in the plots, the spectrum $W(x,k)$ is multiplied
by $k$, so a horizontal line represents
the seed spectrum, $W\propto k^{-1}$. 

The second test, illustrated in Figure~\ref{fig-turb_test_amp},
confirms that the amplification term proportional
to $\paramplif$ in (\ref{eq-genmfa}) is handled correctly
by the numerical solver. The introduced seed spectrum
(shown with the thin line) is the boundary condition
at $x=x_0$ for (\ref{eq-genmfa}), in which
$\parcompamp=\parcompwav=\parcasc=\pardiss=0$,
and $\paramplif=1$. The growth term $G$ is modeled
using the assumption that the resonant instability
operates, i.e.,  $G=\resgrowthrate W$
[see Equation~(\ref{resonant_increment})], where
an artificial CR pressure spectrum was imposed, described
by the expression
\begin{equation}
\Pcr(x,p) = 0.5 \rho_0 u_0^2 \frac{1}{p_0}
       e^{-(\ln{p}-\ln{p_0})^2}
         \exp\left( -\frac{x}{x_0}\frac{p_0}{p} \right).
\end{equation}
(this pressure was simulated and binned into the 
momentum and spatial grids in order to emulate the actual
run, where the pressure $\Pcr$ is calculated by the Monte
Carlo particle transport routine). The corresponding
solution given by~(\ref{wxk_onlyamp}) is:
\begin{displaymath}
W(0,k) = W(x_0,k) + \int_{x_0}^{0} v_A 
  \frac{\partial}{\partial x}\left[
        0.5 \rho_0 u_0^2 \frac{1}{p_0}
       e^{-(\ln{p}-\ln{p_0})^2}
         \exp\left( -\frac{x'}{x_0}\frac{p_0}{p} \right)\right]
         \frac{p}{k}\frac{dx'}{u_0}=
\end{displaymath}
\begin{equation}
\label{test_amp_sol}
\qquad\qquad =
W(x_0,k) + 0.5 \rho_0 u_0^2 \frac{v_A}{u_0} \frac{k_0}{k^2}
     e^{-(\ln{k}-\ln{k_0})^2}
      \left[1 - \exp\left(-\frac{k}{k_0}\right)\right],
\end{equation}
where $k_0 = eB_0/cp_0 = (mu_0/p_0) \, \rgzero^{-1}$,
and $p_0 = m_p c$. The result
of the numerical integration, shown with the solid thick
line, coincides perfectly with the analytic
solution~(\ref{test_amp_sol}) shown with the triangular markers.

The cascading term, proportional to $\parcasc$, along
with the viscous dissipation in the term proportional
to $\pardiss$,  are tested in the following two
runs. 

\begin{figure}[htbp]
\vskip 0.5in
\centering
\includegraphics[angle=-90, width=4.0in, clip=true, trim=0.8in 0.0in 0.0in 0.0in]{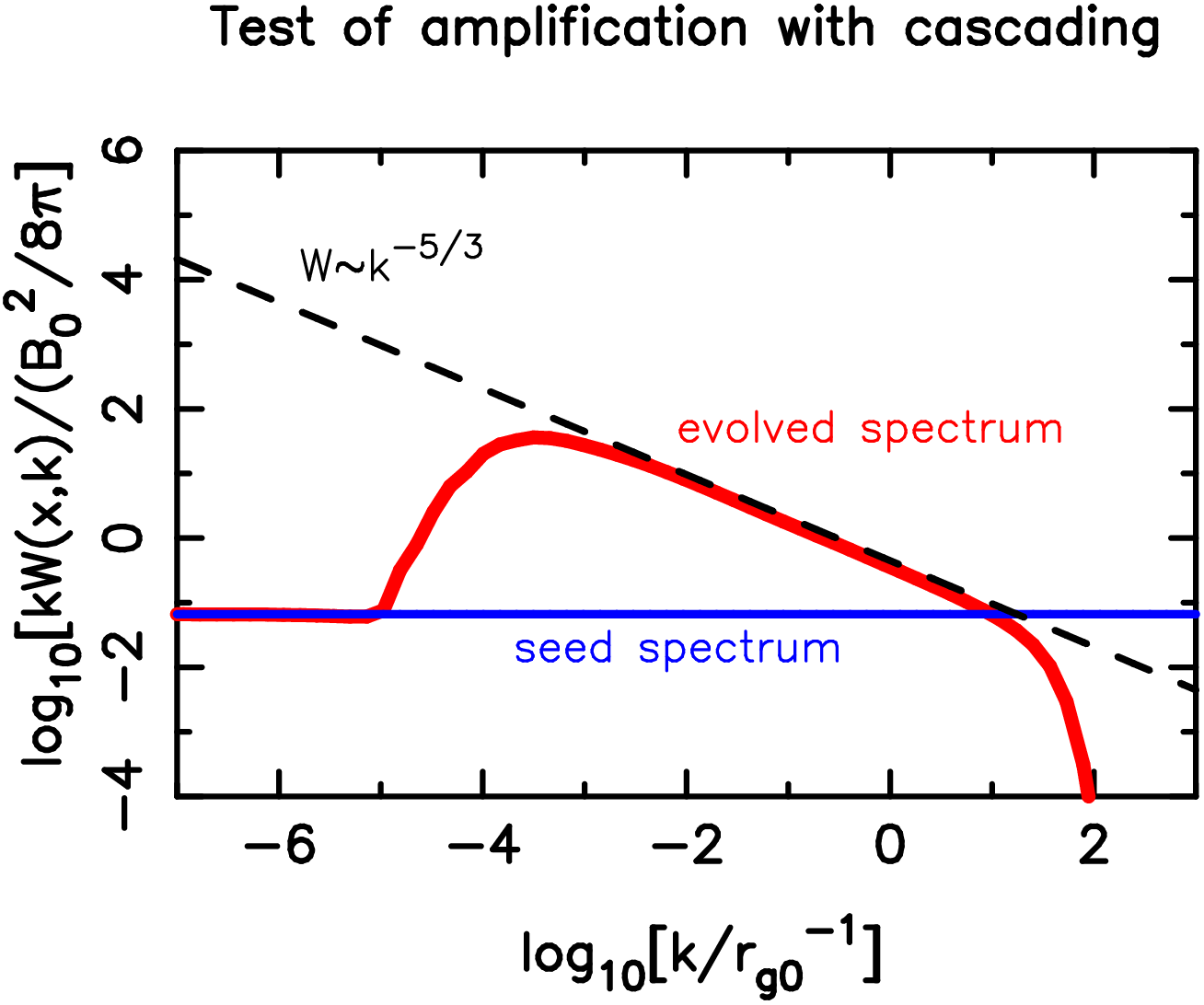}
\caption{$ $ Amplification and cascading of turbulence.}
\label{fig-turb_test_ampcasc}
%
\vskip 0.5in
\centering
\includegraphics[angle=-90, width=4.0in, clip=true, trim=0.8in 0.0in 0.0in 0.0in]{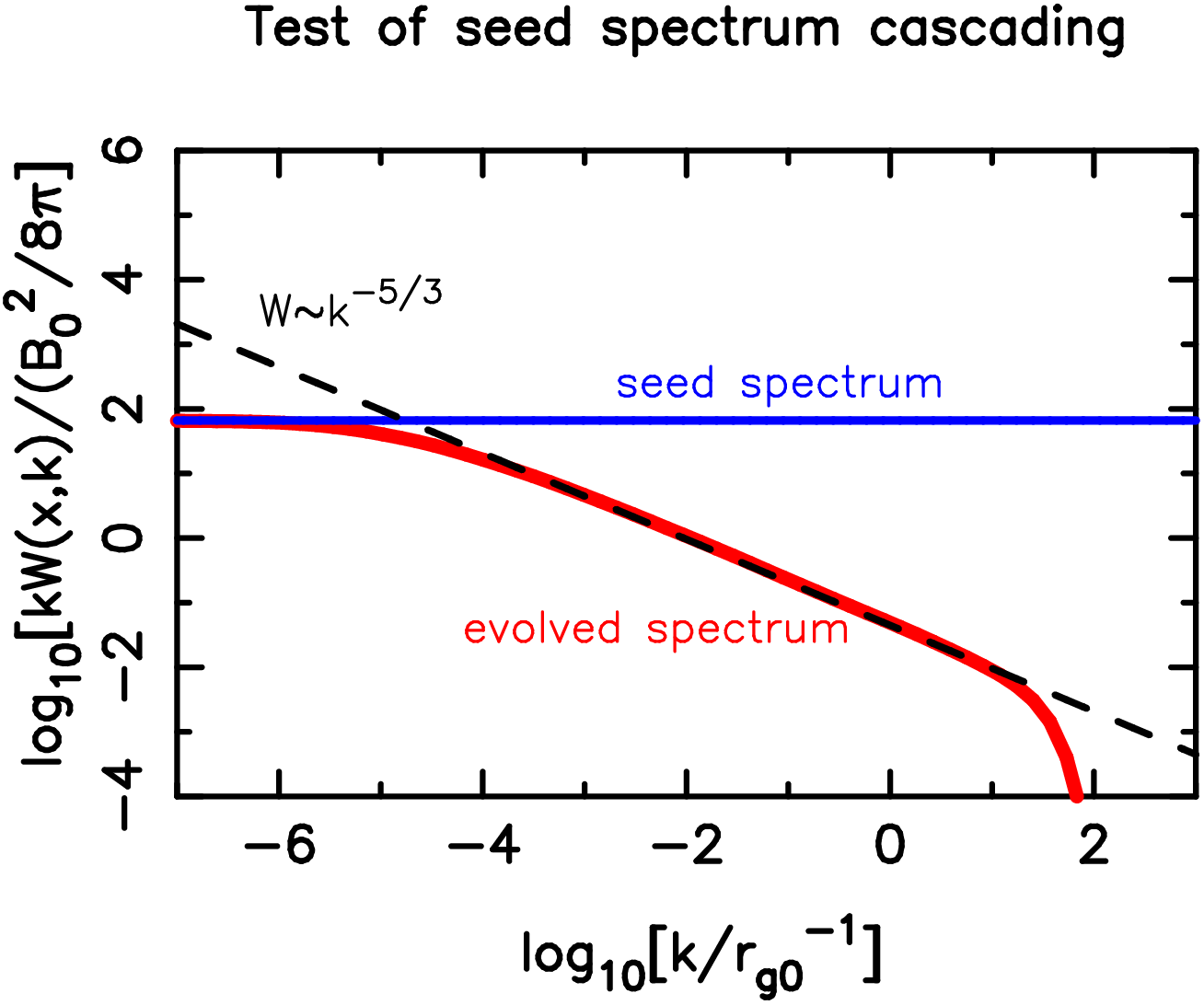}
\caption{$ $ Cascading of seed power law spectrum of turbulence.}
\label{fig-turb_test_casc}
\end{figure}

In Figure~\ref{fig-turb_test_ampcasc}, I illustrate
the third test -- the solution of~(\ref{eq-genmfa}) with 
$\parcompamp=\parcompwav=0$ and 
$\paramplif=\parcasc=\pardiss=1$; the growth rate,
$G$, was chosen similarly to the previous example, but with $p=10^2
m_pc$; the cascading rate, $\Pi$, was taken in the form
(\ref{pi_kolmogorov}); and I chose the viscous dissipation
model described by~(\ref{diss_ksq}) with 
$\kdiss \approx 1.1\cdot 10^{3}\,\rgzero$, corresponding
to a temperature $T_0 = 10^4$~K in~(\ref{kdiss_def}).
The resulting turbulence spectrum
is consistent with the predictions of the Kolmogorov theory.
The energy-containing interval of wavenumbers
is around $k_0=eB_0/cp_0\approx 3\cdot 10^{-4}\,\rgzero^{-1}$, 
where the turbulence amplification takes place
(see the previous example for the amplified spectrum
not modified by cascading). Then
follows the inertial interval, where the energy
is carried from small $k$ to the greater $k$ by cascading;
the power law index of the spectrum matches very well
the Kolmogorov's $k^{-5/3}$ law described by the analytic
solution~(\ref{amp_with_cascades}). Finally, at short
wavelengths, the dissipative interval is marked by the
spectrum turning down exponentially due to
the effect of viscous dissipation, $L$. It happens
at $k\approx 0.1 \, \kdiss$.

In another test of cascading, I confirm that, if the seed
turbulence has a power-law form, and is not amplified,
the cascading leads to the formation of an inertial
interval with $W\propto k^{-5/3}$ followed by the dissipative
interval, where the spectrum turns down exponentially.
The setup of the run shown in Figure~\ref{fig-turb_test_casc}
is similar to that of the previous example, but $\paramplif=0$,
and the seed turbulence spectrum contains more energy
by a factor $10^3$ (the solid thin line). The evolution
with cascading leads to the formation of the spectrum
shown with the thick solid line. Its slope is in agreement
with the Kolmogorov's law indicated with the
dashed line.

The tests presented above are only a few of the multitude
of tests that I performed in order to confirm that my
major contribution to the model, the magnetic field
amplification module, adequately solves 
equation~(\ref{eq-genmfa}) and calculates the effects
of turbulence generation and dissipation on the flow.
These effects are: plasma heating due to the turbulence 
dissipation [see equation~(\ref{pressuregrowth} and
the text explaining it], the contribution of turbulence
to the momentum and energy balance, which affects
the plasma flow (see
Section~\ref{turb_fluxes}), and the determination
of particle transport by the spectrum $W(x,k)$
(Section~\ref{advanced_transport}).
I should note that in several publications we used a 
model for magnetic field amplification that included the
generation of waves traveling in both directions, but
did not include cascading. This model and the
corresponding numerical integrator are described
in Appendix~A.

\subsection{Turbulence and equations of motion}

\label{turb_fluxes}

The fundamental idea on which the Monte Carlo model,
as well as simpler analytic models, is based, is that
the dynamics of matter, particles and magnetic fields 
are described on scales much larger than the scale of
turbulent
fluctuations. That is, the model does not contain and
describe the information about the spatial structure of
stochastic flows and magnetic fields, substituting
an averaged statistical description. This is
expressed in the following approximations:
\begin{itemize}
\item{Instead of a field
of turbulent fluctuations of the plasma velocities, 
the model has the averaged flow speed $u(x)$;}
\item{Instead of the spatial structure of magnetic
fields ${\bf B}({\bf r}, t)$, the Fourier spectrum of 
fluctuations, averaged over a large enough volume surrounding a
coordinate $x$, is used, denoted as $W(x,k)$;}
\item{Instead of describing particle
transport using the equations of motion based
on the Lorentz force, the model employs a diffusion model,
in which the mean free paths depend on $W(x,k)$.
This diffusion approach applies on scales on which
the particles `lose memory' of their 
initial direction of motion, and these
scales must be greater than the size of the
turbulent structures scattering the particles.}
\end{itemize}

The above approximations mean that the equations of motion describing $u(x)$
must contain the properly averaged contributions of
the turbulence to the fluxes of mass, momentum and energy.
In this section, we present and explain these contributions.
The equations and reasoning shown here are pertinent
to the discussions in Sections~\ref{sec-rtot} and 
\ref{subsec_smoothing}.

One may calculate the flux of momentum and energy,
accounting for the turbulent contribution,
using the general expression for
the energy density $W_t$, 
the stress tensor $T_{ik}$,
and the energy flux ${\bf q}$
(e.g., equations (2.48), (2.49) and (2.67) in \cite{VBT93})
\begin{eqnarray}
\label{totenergy}
W_t &=&
 \rho\left( \frac12 u^2 + \frac{\epsilon}{\rho} \right) + \frac{B^2}{8\pi},\\
\label{totstresstens}
T_{ik} &=& 
    P \delta_{ik} + \rho u_i u_k + \frac{B^2}{8\pi}\delta_{ik} - 
     \frac{B_i B_k} {4\pi} \\
\label{totengyflux}
  {\bf q} &=& \rho {\bf u}
     \left(\frac12 u^2 + \frac{\epsilon}{\rho} + \frac{P}{\rho}\right)
    + \frac{{\bf B} \times ({\bf u} \times {\bf B})}{4\pi}.
\end{eqnarray}
Here $\delta_{ik}$ is the Kronecker delta-symbol, and the
index `t' in $W_t$ indicates that this is the total energy
density of the bulk flow, accelerated particles,
and turbulence.

For simplicity (also see the comment at the end of this section), let
us assume that the spectrum of turbulence, $W(x,k)$,
is a power spectrum of \Alf\ waves traveling along
the magnetic field ${\bf B}_0$ in a plasma
moving at a constant speed ${\bf u}_0$ with
mass density $\rho_0$.
Such waves induce perturbations of the matter velocity and
magnetic field, and the total flow velocity ${\bf u}$ and
total magnetic field ${\bf B}$ can be written as:
\begin{eqnarray}
\label{ualf}
{\bf u} &=& {\bf u}_0 + 
      \delta{\bf u}_m \exp{\left[ i k (x - (u_x \mp v_A))\omega t
          \right]} = {\bf u}_0 + \delta{\bf u}\\
\label{balf}
{\bf B} &=& {\bf B}_0 + 
      \delta{\bf B}_m \exp{\left[ i k (x - (u_x \mp v_A))\omega t
          \right]} = {\bf B}_0 + \delta{\bf B},
\end{eqnarray}
where $\delta{\bf B}$ and $\delta{\bf u}$
are the time and coordinate-dependent values of the fluctuations
of the magnetic field and the plasma velocity in the wave, 
$\delta{\bf B}_m$ and $\delta{\bf u}_m$ are the amplitudes 
of these fluctuations,
and $u_x$ is the average x-component of the flow velocity, 
also denoted throughout this work as $u$. The $\pm$
signs correspond to different polarization (i.e., directions of motion).
For \Alf\ waves, the following properties must be listed:
$\delta{\bf u} = \pm\delta{\bf B}/\sqrt{4\pi\rho_0}$, 
${\bf B}_0 \parallel {\bf u_0}$,
$\delta{\bf B} \perp {\bf B}_0$, $\delta{\bf u} \perp {\bf u}_0$.
Also, because \Alf\ waves are an incompressible motion of plasma, 
one may add these conditions: $\rho = \rho_0$, 
$\epsilon = \epsilon_0$ and $P = P_0$ 
(here $\epsilon$ is the internal energy, and $P$ -- the
pressure of the gas).

Substituting the expressions
for ${\bf u}$ and ${\bf B}$ from (\ref{ualf})
and (\ref{balf}) into (\ref{totenergy}), (\ref{totstresstens})
and (\ref{totengyflux}), one may derive the quantities
that interest us in the 1\nobreakdash-D simulation: $W(x,k)$,
$T_{xx}$ and $q_x$. Note that these quantities have also
been denoted above as $\momentumflux$ and $\energyflux$.
\begin{eqnarray}
W_t &=& \rho_0 \left(\frac12 ({\bf u_0} + \delta{\bf u})^2 
               + \frac{\epsilon_0}{\rho_0} \right) + 
              \frac{({\bf B}_0 + \delta{\bf B})^2}{8\pi},\\
\momentumflux \equiv T_{xx} &=&
          P_0 + \rho u_0^2 + 
          \frac{({\bf B} + \delta{\bf B})^2}{8\pi} - 
          \frac{B_0^2}{4\pi},\\
\nonumber
\energyflux \equiv q_x &=&
  \rho_0 u_0\left(
              \frac12 ({\bf u}_0 + \delta{\bf u})^2 
              + \frac{\epsilon_0}{\rho_0}
              + \frac{P_0}{\rho_0}
            \right) + \\
 & & 
   + \frac{ ({\bf B}_0 + \delta{\bf B})\times 
          [ ({\bf u}_0 + \delta{\bf u})\times
            ({\bf B}_0 + \delta{\bf B}) ]}{4\pi}.
\end{eqnarray}
Simplifying the vector operations and averaging over
many wavelengths in $x$ and many cycles in $t$ 
(this leads to
$\left< \delta{\bf B}^2 \right>=\delta{\bf B}_m^2/2$
and
$\left< \delta{\bf u}^2 \right>=\delta u_m^2/2$),
one gets:
\begin{eqnarray}
\left< W_t \right> &=& \frac12 \rho_0 u_0^2 + \epsilon_0 + \frac{B_0^2}{8\pi} + 
        \left( \frac12 \rho_0\delta u_m^2 +
              \frac{\delta B_m^2}{8\pi}\right)/2,\\  
\left<\momentumflux\right> \equiv \left< T_{xx} \right> &=&
          \rho u_0^2 + P_0 - \frac{B_0^2}{8\pi} + 
          \left(\frac{\delta B_m^2}{8\pi}\right)/2,\\
\nonumber
\left<\energyflux\right> \equiv \left< q_x \right> &=&
   \frac12  \rho_0 u_0^3 +
     (P_0 + \epsilon_0)u_0  + \\
 & & 
   + \left(\frac12 \rho_0 u_0 \delta u_m^2 + 
    \frac{u_0 \delta B_m^2 \mp B_0 \delta u_m^2}{4\pi}
   \right)/2.
\end{eqnarray}
In the following, we omit the averaging signs 
$\left<\right.$~$\left.\right>$.
Associating the last terms in the above equations with
the contributions of turbulence
we have:
\begin{eqnarray}
W &=& \left( \frac12 \rho_0\delta u_m^2 +
              \frac{\delta B_m^2}{8\pi}\right)/2,\\  
P_w &=& \left(\frac{\delta B_m^2}{8\pi}\right)/2,\\
F_w &=& \left(\frac12 \rho_0 u_0 \delta u_m^2 + 
    \frac{u_0 \delta B_m^2 \mp B_0 \delta B_m \delta u_m}{4\pi}
   \right)/2.
\end{eqnarray}
Now, using the `equipartition' characteristic of \Alf\ waves,
i.e., the identity $\delta u_m^2 = \delta B_m^2 / (4\pi\rho_0)$,
and the definition of \Alf\ velocity $v_A = B_0/\sqrt{4\pi\rho_0}$,
we arrive at:
\begin{eqnarray}
W &=& W_k +  W_m = 
       \frac12 \frac{\rho_0 \delta u_m^2}{2} +
       \frac12 \frac{B_m^2}{8\pi}, \\
\label{pwdef}
P_w &=& \frac12 W, \\
\label{fwdef}
F_w &=& \frac32 (u_0 \mp v_A) W.
\end{eqnarray}
In these equations, $W_k=W_m$ are the energy densities
of, respectively, kinetic and magnetic turbulent
fluctuations. Equations (\ref{pwdef}) and (\ref{fwdef})
define the `pressure' (i.e., flux of the $x$-component of
momentum in the $x$-direction) and the energy flux
(in the $x$-direction) of turbulence. These quantities should
be added to the corresponding fluxes of particles 
in order to account for turbulence in the momentum 
and energy balance; in other words, in order to account
for turbulence in the equations of averaged motion.

Let us discuss the equations (\ref{pwdef}) and (\ref{fwdef})
defined above. First of all, they only strictly apply to
\Alf\ waves (but, thankfully, of arbitrary amplitude).
Nonlinear interactions between high amplitude waves and
particles may, as explained in earlier sections, lead to the 
turbulent behavior characterized by cascading and by
significant changes in the geometry of magnetic fields
and random plasma velocities, invalidating
(\ref{pwdef}) and (\ref{fwdef}). Also, even without the transition
to turbulence, these equations do not rigorously
apply to any waves other than \Alf. For instance,
the short-wavelength harmonics generated by Bell's
instability are not \Alf ic; one may show that
for waves at $k=k_c/2$ (the peak of the growth rate),
the balance between the kinetic and magnetic energy
density of these waves, $W_k$ and $W_m$, is
$W_m = 3 W_k$, as opposed to $W_k = W_m$ for
\Alf\ waves. 

In the absence of a more detailed model of turbulence evolution
that describes the geometry and dynamics of stochastic motions
and fields in the plasma, one cannot expect to significantly improve
the calculation of $P_w$ and $F_w$. However,
I argue that, as shown by the example of Bell's harmonics,
different geometry or dynamics of turbulence may just
lead to changes in the factors such as $1/2$  and $3/2$
in equations~(\ref{pwdef}) and (\ref{fwdef}).
One may hope that this would be a minor
change, where by `minor' I mean a change by a factor
of a few. This is as much certainty as one may expect to achieve
without describing the spatial structure of turbulence
with a PIC or MHD model. That approach, as we
saw earlier, is extremely computationally expensive,
especially for nonlinear shocks that require
a large spatial and temporal dynamic range,
and I accept the equations (\ref{pwdef}) and (\ref{fwdef})
in the model for the sake of achieving the designated goal
of this work: studying the nonlinear structure of shocks
undergoing efficient particle acceleration and strong
magnetic field amplification.

\newpage

\section{Particle transport}

\label{advanced_transport}

The problem of diffusive transport of charged particles in
magnetized plasmas is fundamental for plasma physics. In collisionless
plasmas typically found in astrophysics, this transport is
generally turbulent diffusion as particles propagate in stochastic magnetic fields
and the associated stochastic plasma motions. The question
usually asked is, given the spectrum (or a more complete description
-- correlation tensors) of turbulence, find the diffusion coefficient
of a particle with a certain momentum $\pvector$. In this
work I used several approximations of diffusion coefficients,
as described below. Each of these approximations has a
its own domain of applicability.

\subsection{Bohm diffusion limit}

\label{sec_bohm}

Bohm diffusion was first observed for electrons in a magnetized
laboratory plasma \cite{Kaufman90}, but the Bohm diffusion model is often applied 
in astrophysics due to its simplicity. The principal assumption is
that the plasma is magnetized and turbulent, so that a particle's
mean free path between strong deflections is equal to its gyroradius,
\begin{equation}
\label{mfp_bohm}
\lbohm = \frac{c p}{e B}.
\end{equation}
Here $p$ is the momentum of the particle, and $B$ is the
magnetic field in the plasma.
The corresponding diffusion coefficient, assuming isotropic diffusion, is
\begin{equation}
\label{dif_bohm}
\Dbohm = \frac{\lbohm v}{3},
\end{equation}
where $v$ is the speed of the particle corresponding to momentum 
$\pvector$. Note that for non-relativistic particles ($p=mv \ll mc$),
$\Dbohm \propto p^2$, and for ultra-relativistic ones ($p \gg mc$, 
$v \equiv c$), the scaling is $\Dbohm \propto p$.

The Bohm approximation is clear and intuitive. It features two most
important dependencies: the diffusion coefficient increases with
the particle momentum, $p$, and decreases with the magnetic field $B$.
This diffusion model rests on the assumption that $B$ is rather
strong: it confines the particle gyromotion to scales on which the
field itself varies significantly (so that the diffusive character of
motion is effectuated).

\subsection{Resonant scattering by \Alf\ waves}

\label{subsec_resonant_mfp}

When a uniform field, $B_0$, exists in a plasma on scales much
larger than the sizes of particle gyroradii and turbulent harmonics,
and a train of low amplitude $\Delta B \ll B_0$
\Alf\ waves travels along this field, the mean free
path of an energetic particle along the uniform field
can be estimated as
\begin{equation}
  \label{mfp_resonant}
  \lres = \frac{4}{\pi} \frac{cp_{\perp}/eB_0}{\mathcal{F}},
\end{equation}
where
\begin{equation}
  \label{F_quantity}
  \mathcal{F} = \frac{\kres W(\kres)}{B_0^2 / 8\pi}.
\end{equation}
and
\begin{equation}
\label{kres_in_b0}
\kres = \frac{1}{cp_{\parallel}/eB_0}
\end{equation}
(see \cite{Wentzel74} or \cite{LC83}).
In expression (\ref{mfp_resonant}), the numerator of the second
fraction is the gyroradius of the particle ($p_{\perp}$ is the component
of the particle's momentum transverse to the field $\bf{B}_0$), and the
denominator $\mathcal{F}$ is, within a factor, the energy density of \Alf\ waves
(per unit logarithmic waveband $d\ln{k}=1$) normalized to the energy
density of the underlying uniform field. The energy density $W(k)$ in
(\ref{F_quantity}) is
taken at the resonant wavenumber $\kres$ defined by
(\ref{kres_in_b0}). When $\mathcal{F}$ approaches $1$, the mean free path
shrinks down to the particle gyroradius, and the Bohm limit is
realized. Increasing $\mathcal{F}$ further takes this theory beyond its
applicability limits.

\subsection{Diffusion in short scale turbulent fluctuations}

If the bulk of the turbulence energy is in small-scale harmonics
with respect to the particle mean free path, then the motion of
the particle is nearly ballistic, with frequent and small deflections
from the stochastic Lorentz force. The collision length of such motion
can be expressed (\cite{DT68}, see also \cite{Toptygin1985} and \cite{Jokipii1971}) as:
\begin{equation}
\label{mfp_shortscale_1}
\lss(x,p) = \frac{4}{\pi}\frac{p^2 c^2}{e^2} 
   \left[4\pi\int\limits_{0}^{\infty}\frac{W(x,k)}{k}\;dk\right]^{-1}.
\end{equation}
This corresponds to a mean free path in the small-scale field, 
$\lss$, given by the expression 
$\lss=\rss^2/\lcor$, where $\rss=cp/e\Bss$ is the gyroradius of the particle
with momentum $p$ in the effective small-scale field, $\Bss$, and
$\lcor$ is equal to the correlation length of the small-scale magnetic field 
(see below for exact definitions). This relationship is easy to
understand. Consider a thought experiment: an energetic particle with
momentum $p$ is propagating through a medium consisting of regions of
scale $\lcor$, each of which contains a magnetic field with
magnitude $\Bss$, pointing in a different random direction in each
region. 
In the course of the path $\lss \gg \lcor$, the particle
encounters $N=\lss/\lcor \gg 1$ such regions, and in each of them its
momentum gets a random scattering in the amount 
$\Delta p_\mathrm{v} \approx F\Delta t = e \Bss \lcor / c$ (here $F$ is the magnitude
of Lorentz force, and $\Delta t$ is the time of the particle crossing
the region). Considering this process a random walk in $p$-space,
the mean square deflection of momentum along the path $\lss$ is
$\left<\Delta p\right>^2=N (\Delta p_\mathrm{v})^2=\lss / \lcor (e \Bss \lcor /c)^2$,
and setting $\left<\Delta p\right>^2 = p^2$, corresponding to $\lss$ being the
mean free path, one can solve this equation to find
$\lss=(cp/e\Bss)^2 / \lcor=\rss^2 / \lcor$. This mean free path
depends on the particle momentum as $\lss \propto p^2$, as opposed
to the Bohm behavior $\propto p$, which is a significant difference.

\subsection{Low energy particle trapping by turbulent vortices}

Suppose the turbulence has a power law spectrum that
contains a significant fraction of energy in the smallest scales
(such a spectrum may be produced by cascading 
as described in Section~\ref{turb_effects}).
A particle with a low enough energy will be effectively confined
by resonant scattering on the small scale turbulence
fluctuations. But its transport on scales greater than the
correlation length of the turbulence (i.e., greater than the
largest turbulent harmonics), which is of interest
for the Monte Carlo code, may be significantly different
from the directly applied model of resonant scattering transport. 
The efficient resonant scattering effectively confines
the particles to the large-scale turbulent structures,
and their diffusion on large scales is determined by the
motions of the turbulence rather than the particles' own
motion. A theoretical description of such transport is
described by Bykov and Toptygin 
in \cite{BT92}, \cite{BT93} and \cite{VBT93}. A rough approximation
of their result is that, if the mean free path
of a low energy particle due to resonant scattering
is $\lres \ll \lcor$, where $\lcor$ is the correlation length of
the turbulence, then the diffusion coefficient of such
particle on scales greater than $\lcor$ is on the order of
\begin{equation}
\label{diff_largescale_trapped}
D \approx u_c \lcor,
\end{equation}
where $u_c$ is the typical speed of turbulent motions
with correlation length $\lcor$. This applies
when $D \gg v \lres$, where $v$ is the speed
of the particle, and $\lres$ is its mean free path
between the resonant scatterings, meaning that
the `convective' diffusion coefficient (\ref{diff_largescale_trapped})
is much greater than the resonant scattering coefficient.
This situation
is analogous to the convective diffusion of cream in
a coffee cup. Pour the cream into the coffee and,
even without stirring, it will spread through the cup
in minutes. If one naively assumes molecular diffusion
and estimates the time it takes  the cream to diffuse
from one end of the cup to another,
this time will be much longer, on the order of hours. 
The discrepancy is successfully explained with a model similar to
(\ref{diff_largescale_trapped}): molecules of the admixture
are confined to the turbulent vortices in the medium
(in the coffee cup, those are induced by the temperature
difference between the top and the bottom, and by
the energy introduced during the pouring of
the coffee into the cup and of the cream into the coffee),
and the propagation of the admixture is determined
by the motion of these vortices rather than of the
admixture with respect to the vortices.

Another possibility of particle trapping in turbulent
structures is when there is no short-scale turbulence to
produce effective resonant scattering, but a particle has a low
enough energy so that its gyroradius in the large scale turbulent
magnetic field, $r_g$, is small compared to the correlation length
of the turbulence, $\lcor$. Then the particle will gyrate
around the turbulent magnetic fields, losing the memory
of its initial direction of motion on the length comparable
to $\lcor$. If the particle's speed $v \gg u_c$ 
(so that the turbulence is essentially stationary for
the particle), then one may estimate the coefficient
of diffusion of the particle on scales greater than $\lcor$
as
\begin{equation}
D \approx v \lcor,
\end{equation}
or the effective mean free path of the particle as
\begin{equation}
\label{diffusion_trapping}
\lambda \approx \lcor.
\end{equation}
This means that particles trapped in the turbulent
vortices by gyration in the turbulent
large-scale magnetic fields have a mean free path
nearly independent of the particle energy and
equal to the size of the turbulent vortices.
More realistic models of this process may be necessary,
because effects such as drifts in magnetic fields 
and time dependence of the vortex structure
may change the dependence of the mean free path on
the particle energy.

The above approximation applies to particle transport
on scales greater than the turbulence correlation length.
The transport of very low energy particles on smaller scales
depends on geometry and evolution of the turbulent structures,
which is beyond the reach of our model. The motion 
of magnetic field lines (sometimes called
magnetic field line wandering) may be 
non-diffusive on small spatial scales, resulting in CR
transport that cannot be described as diffusion
(e.g., \cite{Ragot99}).

\subsection{Implementation of diffusion models in the Monte Carlo code}

Based on the theoretical models of particle transport outlined above,
I implemented the corresponding mean free path prescriptions into
the Monte Carlo code. When the model is run, the user can specify
which prescription is to be used in the simulation. It allows the
application of
transport models of various degrees of physical accuracy and
applicability to study their effects on the self-consistent shock structure.

\subsubsection{Bohm diffusion}

If the user specifies the Bohm regime of diffusion in the
simulation, then given the momentum of the particle, $p$, measured in
the plasma frame, the code will calculate the mean free path $\lbohm$ using
(\ref{mfp_bohm}), where for $B$ it substitutes the effective local
magnetic field, $\Beff$, defined in~(\ref{eq_beff}).

This is the simplest method of describing diffusion in the presence of
efficient MFA. It should give an accurate (within an order of
magnitude) estimate of the collision mean free paths for moderate energy
cosmic rays. For the highest energy cosmic rays, with gyroradii
greater than the magnetic field correlation length, the
turbulence acts as small-scale magnetic fluctuations, an Bohm diffusion is an
overestimate of the confinement strength. For the lowest energy CRs
and thermal particles, Bohm diffusion is also not a good
approximation, because the particles may be trapped in magnetic
structures, in which case their diffusion is determined by the evolution of the
small scale turbulence rather than their own motion.

\subsubsection{Resonant scattering}

I have the option of describing the particle scattering with a
form similar to
(\ref{mfp_resonant}) in the simulation. If this model is adopted,
it calculates the resonant wavenumber as given by
(\ref{kres_in_b0}), except that it uses the total momentum $p$ instead
of $p_{\parallel}$, and to calculate the mean free path, it uses
(\ref{mfp_resonant}), but with $p$ instead of $p_{\perp}$. This
replacement of the components of the particle momentum with its
magnitude is done in order to account for the strong nature of the
turbulence. Actually, if $\Delta B \gg B_0$, then (\ref{mfp_resonant})
is not applicable in all rigorousness, but I use this theory in order
to grasp the most important qualitative behavior of the turbulent
transport: the stronger the turbulent structures of scales comparable
to the particle gyroradius, the more efficient is particle scattering.

If the turbulence spectrum has the shape $W = W_0 (k/k_0)^{-1}$, which
will hereafter be called the Bohm spectrum, then 
(\ref{mfp_resonant}) gives a mean free path similar to the Bohm
prescription (\ref{mfp_bohm}). Namely, when 
$\sqrt{4 \pi W_0 k_0}=B_0^2$, and $\mathcal{F}=1$, the two models match within a factor of
$4/\pi$. The latter condition is equivalent to the condition that a
unit logarithmic waveband $d\ln{k}=1$ contains the same amount of
turbulent energy as the underlying magnetic field $B_0$. 

Thus, for relativistic particles, $\lres \propto p$ in a Bohm spectrum
$W(k) \propto k^{-1}$. Steeper spectra of turbulence ($W \propto
k^{-q}$ for $q>1$) give weaker dependencies of $\lres$ on $p$. The
spectrum $W(k) \propto k^{-2}$ gives a constant $\lres(p)$.

\subsubsection{Hybrid model of diffusion in strong turbulence}

\label{subsec_hybr_diff}

It is useful to re-write equation (\ref{mfp_shortscale_1}) as
\begin{equation}
\label{mfp_shortscale}
\lss(x,p) = \frac{\rss^2}{\lcor},
\end{equation}
where $\rss$ is the particle gyroradius in the effective
magnetic fields of the short-scale 
magnetic perturbations. In the model, I adopt the prescription~%
(\ref{mfp_shortscale}), and generalize it with two assumptions,
as outlined below, so it can be applied to particles of lower energies
as well. The first assumption is that for a particle of momentum $p$,
the local turbulence spectrum can be divided into the large-scale
and the short-scale part, the wavenumber $\kstar$ being the
boundary between them. The effective large-scale magnetic field 
is then
\begin{equation}
\frac{\Bls^2(x, \kstar)}{8\pi} =
  \frac{B_0^2}{8\pi} + 
     \frac12 \int\limits_{0}^{\kstar} W(x,k') \; dk',
\end{equation}
the effective small-scale field is
\begin{equation}
\frac{\Bss^2(x, \kstar)}{8\pi} =
     \frac12 \int\limits_{\kstar}^{\infty} W(x,k') \; dk',
\end{equation}
and the correlation length of short-scale field $\lcor$ can be 
estimated as
\begin{equation}
\lcor = \frac{\int_{\kstar}^{\infty} W(x,k')/k'\;dk'}
             {\int_{\kstar}^{\infty} W(x,k')   \;dk'}.
\end{equation}
I define $\kstar$ using the condition $r_g(\Bls)\kstar=1$, 
where $r_g(\Bls)=cp/e\Bls$ is the gyroradius
of the particle in the large-scale magnetic field $\Bls$.
The latter is dependent on $\kstar$, therefore a nonlinear
equation must be solved at every point in space for every
particle momentum in order to determine $\kstar$.
The second assumption is that the calculated $\lss(p)$
does not increase as momentum $p$ decreases.

Let us comment on the physics behind the assumptions outlined
above. Equation~(\ref{mfp_shortscale}) applies if the turbulence is
predominantly short scale. However, if the turbulence spectrum
incorporates a wide range of wavenumbers
(for example, the assumed upstream spectrum $W \propto k^{-1}$)
then a good quasi-linear approximation to the particle transport
properties is the resonant scattering prescription
(\ref{mfp_resonant}) (see, e.g., \cite{AB2006, VEB2006} and references therein).
However, for a spectrum similar to (\ref{eq_seed_turb})
and  $\kstar \ll \kmax$, the mean free path (\ref{mfp_shortscale})
can be represented after some simple mathematical transformations
as
\begin{equation}
\label{diff_res_eff}
\lambda = \frac{cp}{e \Bls} \frac{\Bls^2}{4 \pi k W(x,k)},
\end{equation}
where $k = e\Bls/(cp)$. For $\Bls \approx B_0$ (weak turbulence case),
this is  precisely the resonant scattering mean free path 
(\ref{mfp_resonant}), and
for $\Bls > B_0$ (strong perturbations), it may be a good
generalization of the latter. Therefore, dividing
the turbulence spectrum at $\kstar$ allows one to correctly
describe the mean free path of intermediate-energy particles
using (\ref{mfp_shortscale}), along with the high energy particles.

The second assumption, that of monotonic behavior of $\lambda(x,p)$
with respect to $p$, doesn't influence the case of a power-law
turbulence spectrum, but affects the diffusive transport
of low energy particles in case of turbulence with a marked
concentration of energy around a wavenumber $\kwhirl$, i.e. 
containing strong vortices of size $1/\kwhirl$. Indeed, assume 
for simplicity a Gaussian spectrum
$W(x,k) \propto \exp{\left[(k-\kwhirl)^2/(2\sigma^2)\right]}$, 
where  $\sigma$ is the width of the spectrum.
If the particle momentum $p$ is large enough so that 
$\kstar < \kwhirl$, and $\rss \gg \lcor$ holds, then 
the particle is scattered by frequent deflections in the
short-scale magnetic field of the vortices, and
equation~(\ref{mfp_shortscale}) applies unconditionally. 
However, a particle with a low enough momentum so that
$\kstar \gg \kwhirl$ will find itself trapped in the
large-scale magnetic fields of the vortices, and one may assume
that its transport on scales larger that $1/\kwhirl$ is
diffusive (see equation~(\ref{diffusion_trapping}), 
with the effective mean free path
\begin{equation}
\label{diff_whirl}
\lambda \approx 1/\kwhirl.
\end{equation}
Now consider the above Gaussian spectrum.
The prescription~(\ref{mfp_shortscale}), with $\rss$ and $\lcor$
determined using the $\kstar$ formalism, does not describe
the trapping of the low energy particles. However, at such
momentum $\ptr$ that
$\kstar \approx \kwhirl$ for this momentum,
the magnetic field $\Bls \approx \Bss \gg B_0$,
(assuming strong turbulence), and lowering the value of $p$
will lead to an exponentially rapid decrease of $\Bss$, and an equally rapid
increase in $\rss$, which will make $\lambda=\rss^2/\lcor$ 
unphysically increase for smaller $p$. The monotonicity assumption will
correct this unphysical behavior by fixing $\lambda(x,p)$
at the value $\lambda(x,\ptr)$ for $p<\ptr$. And this value
will approximately be (\ref{diff_whirl}),
because $\ptr$ corresponds to $\rss \approx 1/\kwhirl \approx \lcor$.

Summarizing, I state that I choose the mean free path of particles
with momentum $p$ according to (\ref{mfp_shortscale}), where $\rss$ and $\lcor$ are calculated
for the short-scale part of the magnetic field, $k>\kstar$,
and force this prescription to be monotonic in $p$
for low momenta. The reasoning provided above shows that our
prescription properly describes particle transport
a) for high $p$ particles in short-scale field, as per the derivation
of (\ref{mfp_shortscale}); b) for intermediate to low $p$ in a power-law turbulence
spectrum, assuming resonant scattering, and c) for low energy
particles in large-scale turbulent vortices, assuming particle
trapping. In between these important regimes, the prescription
provides an interpolation.

\newpage

\section{Parallel computing with MPI}

\label{parallel_computing}

The code used for this dissertation is written for parallel
processing using
the MPI (Message Passing Interface) protocol. In this section I will
summarize the parallelization algorithm, outline its advantages and
drawbacks, and present a performance test.

By far the most time consuming part of the simulation is the Monte
Carlo transport of particles that simulates the Fermi-I acceleration
process. This procedure is intrinsically very well suited for
parallel computing, because particles are propagated one after
another, and each particle's motion within an iteration is completely
independent of any other particle's history. Multiple particles are
only required in order to decrease random deviations of the results,
i.e. to `improve statistics'. This also means that 
the quality of random numbers is not a major
issue of concern for this Monte Carlo code, because even if the random
numbers are correlated within a sequence, correlated between different
processors, or do not continuously fill their range of definition,
it does not affect the quality of results. That is because
the trajectory of each particle depends not only on the latest
scattering outcome, but also on the previous history
of acceleration of this particle, which effectively diminishes
any possible correlations in particle histories due to the
imperfections of the random numbers used by the Monte
Carlo code\footnote{The random number generator used in the code
is an excellent match for the single-processor version of
the Monte Carlo simulation, but was not specifically designed 
for parallel processing. This, however, turns out not to be a problem
for the reasons stated in the text.}.

I implemented the following algorithm of parallelization
of the calculations. First, a `master' processor
divides the user-specified number of particles equally between
the available processors, including itself. 
Then each processor (the `slaves' and the `master') performs one
iteration, i.e. propagates the particles it is responsible for
until they reach the highest achievable energies, and
the iteration terminates. After a processor completes
its iteration, it returns the output, (the particle
distribution function $f(\pvector)$ and its moments)
to the `master' processor (which also performs its iteration
equally with the other processors, and returns
the collected information to itself). 
The `master' processor then averages the incoming results
(which improves the statistical certainty of the calculated
particle distribution, of momentum and energy fluxes,
and of the increments of field-amplifying
instabilities) and uses them to
calculate magnetic field amplification and precursor
smoothing. These procedures are not easily parallelized,
but they take relatively little time, and I chose to
leave them to just one processor. After that,
the `master' processor gives the other processors
the updated flow speed $u(x)$, the re-iterated magnetic turbulence
$W(x,k)$, and each processor computes the
corresponding mean free path prescription $\lambda(x,p)$
and performs another iteration. This cycle continues
until the self-consistent solution is derived.

The primary advantage of this procedure is its ultimate simplicity.
In terms accepted in the parallel computing field, this is
an `embarrassingly parallel' code, which means that the
interactions between processors take place very infrequently
(in practice, they exchange several megabytes of data
once every several minutes). Another advantage is that
one processor's runtime performance does
not affect another processor's particle history. It is
a welcome feature of the method, because it makes
it easier to debug, if problems arise: the results, including
the run-time errors, are reproducible.
We must note here that the sequences of random numbers generated by
the code are, in fact, deterministic: in two identical
runs executed at different times, the random number sequences
and the final results will be identical. The same
applies to the version of the code with parallel
computing performed as described above.

A  disadvantage of this method
is that there may be situations when most processors
had finished their iterations, but must wait for one
processor working on a particle with an `unfortunate',
long history of acceleration. Computing time is lost
in this case, because the duration of every Monte Carlo iteration
is as long as the worst processor's performance.
This is not a major issue of concern when the number of particles
per processor is large, but when many processors
are available, and each gets only a few particles,
the deviation of the worst processor's performance
from the average performance may be significant.

In Table~\ref{test_parallel} I listed the results of
a simulation similar to that done in 
Section~\ref{subsec_smoothing}. I executed
5 runs, with identical input parameters,
but with different numbers of
processors: 1, 2, 4, 8 and 16.
Each run obtained a self-consistent
shock structure, and the results were identical
in all runs within small statistical deviations.

\begin{table}
\caption{$ $ Test of performance boost with parallel computing}
\label{test_parallel}
\begin{center}
\begin{tabular}{lrrr}
\hline
$N_\mathrm{proc}$ & $N_p/N_\mathrm{proc}$ & Time, s & Speedup \\
\hline
 1  & 160 & 7622 & 1.0 \\
 2  &  80 & 5614 & 1.4 \\
 4  &  40 & 3049 & 2.5 \\
 8  &  20 & 2092 & 3.6 \\
 16 &  10 & 1688 & 4.5 \\
\hline
\end{tabular}
\end{center}
\end{table}

Column `$N_\mathrm{proc}$' lists the number of processors
used in the run. There were a total of $N_p=160$ particles 
in each run\footnote{This is not the number of thermal
particles. A numerical procedure called `particle splitting'
is used in the Monte Carlo model, which allows to maintain
nearly equal number of particles at any energy --  a
necessary condition to simulate rapidly decreasing particle
spectra over many decades of the energy. I did not
describe the `particle splitting' in this dissertation,
because it is purely technical, and was explained in the
literature (e.g., \cite{JE91}). }, and they were equally divided between 
the processors, as shown in column `$N_p/N_\mathrm{proc}$'.
The time that the each simulation took is listed
in seconds, and the column `Speedup' shows the ratio
of the time of the iteration with 1 processor
(i.e., without parallel computing) to that of the parallelized
run.

The results show that the parallelization does lead to 
an increase in the speed of the calculations, but 
the speedup is proportional to, approximately, 
$(N_\mathrm{proc})^{0.6}$, which is not very efficient.
An alternative parallelization algorithm
that may improve the efficiency is  reserving
the `master' processor for dispatching particles between the
`slaves' in the run-time, i.e., each `slave' gets a new particle
from the `master' as soon as it finishes with the previous
one. This way the situation when many processors await 
a few `unlucky' ones to finish with their iterations will
not last as long.

However, because this procedure is incompatible with
the reproducibility of results (unless each particle
has its own random number sequence, that is
stored and passed between the processors; care
must be taken in this case to ensure that
correlations between particle histories do
not occur when particle splitting is performed).
This makes it difficult
to debug the code (see above), I chose to stay
with the currently implemented,
less efficient, but more predictable scheme.
Despite the less than perfect scaling of performance
with the number of processors, 
it allows me to achieve reasonable computation times 
even with the available modest computational 
resources (8-16 processors per run).
Typical run times are seconds to minutes for test runs, and 
around one day for a self-consistent
simulation with a realistic dynamic range and small enough statistical
deviations.

\chapter{Applications of the model}

\label{ch-applications}

In this chapter I will show the basic results of the model of
magnetic field amplification in collisionless shocks based on the
Monte Carlo simulation of DSA. Some of these results have appeared in
peer-reviewed publications (Sections~\ref{res2006}, \ref{res2007} and
\ref{res2008}), and some will soon be submitted for publication 
(section~\ref{res2009}).

For the published material, in this Chapter I only
provide a condensed version of the articles. For the work
that has not yet appeared in press
(Sections~\ref{res_scalings}, \ref{res_angular}), the reader will find
an outline of the proposed direction of research, 
a presentation of the preliminary results and a discussion
of the applicability to astrophysical research.

\newpage

\section{Turbulence growth rate and self-consistent solutions}

\label{res2006}

In \cite{VEB2006}\footnote{The results presented here first appeared in \cite{VEB2006}
and largely are reproduced from this publication.},
we introduced a \MC\ model of \NL\ diffusive shock acceleration allowing
for the generation of large-amplitude magnetic turbulence, i.e., $\Delta
B \gg B_0$, where $B_0$ is the ambient magnetic field.  The model is the
first to include strong wave generation, efficient particle acceleration
to \rel\ energies in \nonrel\ shocks, and thermal particle injection in
an internally \SC\ manner.  
In order to describe the field growth rate in the regime of strong fluctuations,
we use a parameterization that is consistent
with the resonant quasi-linear growth rate in the weak turbulence limit.
We believe our parameterization spans the range between maximum and
minimum rates of fluctuation growth.

We find that the upstream magnetic field
$B_0$ can be amplified by large factors and show that this amplification
depends strongly on the ambient \alf\ Mach number. We also show that in
the nonlinear model large increases in $B$ do not necessarily translate
into a large increase in the maximum particle momentum a particular
shock can produce. The most direct application of our
results will be to estimate magnetic fields amplified by strong
cosmic-ray modified shocks in supernova remnants.

\subsection{Model}

In \cite{VEB2006}, we described the amplification of
magnetic turbulence by the following set of equations:
\begin{displaymath}
[u(x) - V_G]\frac{\partial}{\partial x}U_- + U_-\frac{d}{dx} \left(
\frac32 u(x) - V_G \right) = 
\qquad\qquad\qquad\qquad\qquad
\end{displaymath}
\begin{equation}
\label{uminuskp}
\qquad\qquad\qquad\qquad\qquad \frac{U_-}{U_+ + U_-} V_G \frac{\partial
    \Pp(x,p)}{\partial x}\left|\frac{dp}{dk}\right|- \frac{V_G}{\rgzero}
    \left( U_- - U_+ \right)
\ ;
\end{equation}
\begin{displaymath}
[u(x) + V_G]\frac{\partial}{\partial x}U_+ + U_+\frac{d}{dx} \left(
\frac32 u(x) + V_G \right) = 
\qquad\qquad\qquad\qquad\qquad
\end{displaymath}
\begin{equation}
\label{upluskp}
\qquad\qquad\qquad\qquad\qquad -\frac{U_+}{U_+ + U_-} V_G
\frac{\partial \Pp(x,p)}{\partial x}\left|\frac{dp}{dk}\right|+
\frac{V_G}{\rgzero} \left( U_- - U_+ \right)
\ ,
\end{equation}
which were solved iteratively in the MC simulation.
This system describes the development of the resonant
cosmic ray streaming instability of \Alf\ waves along
with the processes of wave amplitude increase due to
the plasma compression, and of interactions between
waves traveling in opposite directions
(see Section~\ref{turb_effects} and Appendix~A). 
For the growth of \alf\ waves in \Qlin\ theory, $V_G=\Valf$, where
$\Valf = B_0/\sqrt{4\pi \rho(x)}$ is the \alf\ speed calculated with the
non-amplified field and $\rho(x)$ is the matter density  at position $x$.
This choice of $V_G$
provides a lower limit on the amplification rate for
the nonlinear regime, $\Delta B \gg B_0$, and was used in
\cite{AB2006}.  If, on the contrary, we define $V_G$ using the
amplified field, i.e., $V_G = \Beff(x)/\sqrt{4 \pi \rho(x)}$, it
reflects the situation where the growth rate is determined by the
maximum gradient of $\Pp(x,p)$ along the fluctuating field lines. This
provides an upper limit on the wave growth rate and was used in
\cite{BL2001}.
The real situation should lie between the two extremes for $V_G$.
For this preliminary work, we vary
$V_G$ between the two limits, i.e., we introduce a parameter, $0 \le
\fValf \le 1$, such that
\begin{equation}
\label{VG}
V_G = \Valf \left \{ 1 + \left [\frac{\Beff(x)}{B_0} -1 \right ] 
\fValf \right \}
\ ,
\end{equation}
and $V_G$ varies linearly between $\Valf$ (for $\fValf=0$) and
$\Beff/\sqrt{4 \pi \rho(x)}$ (for $\fValf=1$).

Finally, we assume a Bohm model for diffusion. The mean free path of a
particle with momentum $p$ at position $x$ is taken to be equal to the
gyroradius of this particle in the amplified field, i.e., $\lambda(x,p)
= r_g(x,p) = pc/[q \Beff(x)]$, and the diffusion coefficient is then
$\Diff = \lambda v/3$, where $v$ is the particle speed,
and $\Beff$ was defined according to (\ref{eq_beff}).

The Monte Carlo code used here was the original simulation 
developed by Ellison and co-workers, not the version developed
by the author of this dissertation for the problem of magnetic field
amplification. I have confirmed that the latter model reproduces
the results presented here.

\subsection{Results}

In all of the following examples we set the shock speed
$u_0=5000$\,\kmps, the unshocked proton number density $n_{p0}=1$\,\pcc,
and the unshocked proton temperature $T_0=10^6$\,K. For simplicity, the
electron temperature is set to zero and the electron contribution to
the jump conditions is ignored. With these parameters, the sonic Mach
number $\MS \simeq 43$ and the \alf\ Mach number $\Malf \simeq 2300
(1\mu \mathrm{G}/B_0)$.

\subsubsection{With and Without Magnetic Field Amplification}

\begin{figure}[htbp]
\centering
\vskip 0.5in
\includegraphics[height=7.0in]{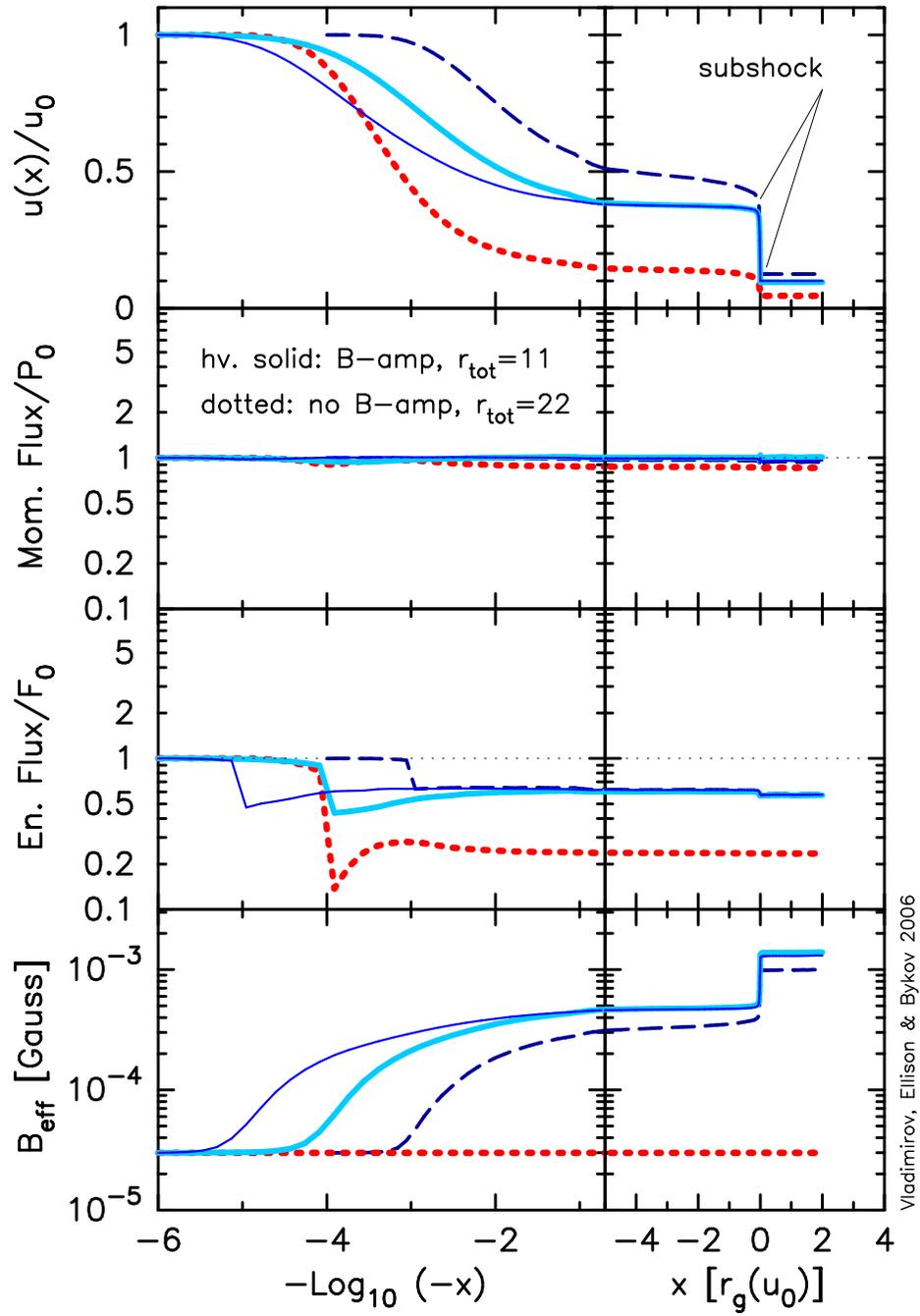}
\caption{$ $ Shock structure with and without MFA\label{BL_noBL}}
\end{figure}

Figure~\ref{BL_noBL} shows self-consistent
solutions for four shocks, obtained with $\fValf=0$.
Note that the horizontal scale has units of $\rg(u_0)=m_pu_0/(e B_0)$ and
is divided at $x=-5\rg(u_0)$ between a linear and log scale.  The
the heavy dotted curves show results without amplification and
all other curves are with amplification.  
The heavy solid and dotted curves have
$\Dfeb=-10^4\,\rg(u_0)$, the dashed curve has $\Dfeb=-1000\,\rg(u_0)$, 
the light solid curve has $\Dfeb=-10^5\,\rg(u_0)$.

First, we compare the results shown with heavy-weight solid curves to
those shown with heavy-weight dotted curves.  The heavy solid curves
were determined with $B$-field amplification while the dotted curves
were determined with a constant $\Beff(x)=B_0$. All other input
parameters were the same for these two models, 
and an upstream free escape boundary was placed at
$\Dfeb=-10^4\,\rg(u_0)$, where $\rg(u_0) \equiv m_p u_0 c/(eB_0)$.
The most striking aspect of this comparison is the increase in
$\Beff(x)$ when field amplification is included (bottom panels). The
magnetic field goes from $\Beff(x\to -\infty) = 30$\,\muG, to $\Beff >
1000$\,\muG\ for $x > 0$, and this factor of $>30$ increase in $B$ will
influence the shock structure and the particle distributions in
important ways.
The solution without $B$-field amplification (dotted curves) has a
considerably larger $\Rtot$ than the one with amplification, i.e., for
no $B$-field amplification, $\Rtot \simeq 22$, and with $B$-field
amplification (heavy solid curves), $\Rtot \simeq 11$.\footnote{See
\cite{BE99} for a discussion of how very large $\Rtot$'s can result in
high Mach number shocks if only adiabatic heating is included in the
precursor. The uncertainty on the compression ratios for the examples in
this paper is typically $\pm 10\%$.}
This  difference in overall compression results because the wave pressure
$\Pwtot$ is much larger in the field
amplified case making the plasma less compressible.   
A large $\Rtot$
means that high energy particles with long diffusion lengths get accelerated
very efficiently and, therefore, the fraction of particles injected must
decrease accordingly to conserve energy. The shock structure adjusts so
weakened injection (i.e., a small $\Rsub$) just balances the more efficient
acceleration produced by a large $\Rtot$. Since $\Rsub$ largely
determines the plasma heating, the more efficiently a shock accelerates
particles causing $\Rtot$ to increase, the less efficiently the plasma
is heated.

\begin{figure}[htbp]
\centering
\vskip 0.5in
\includegraphics[width=5.0in]{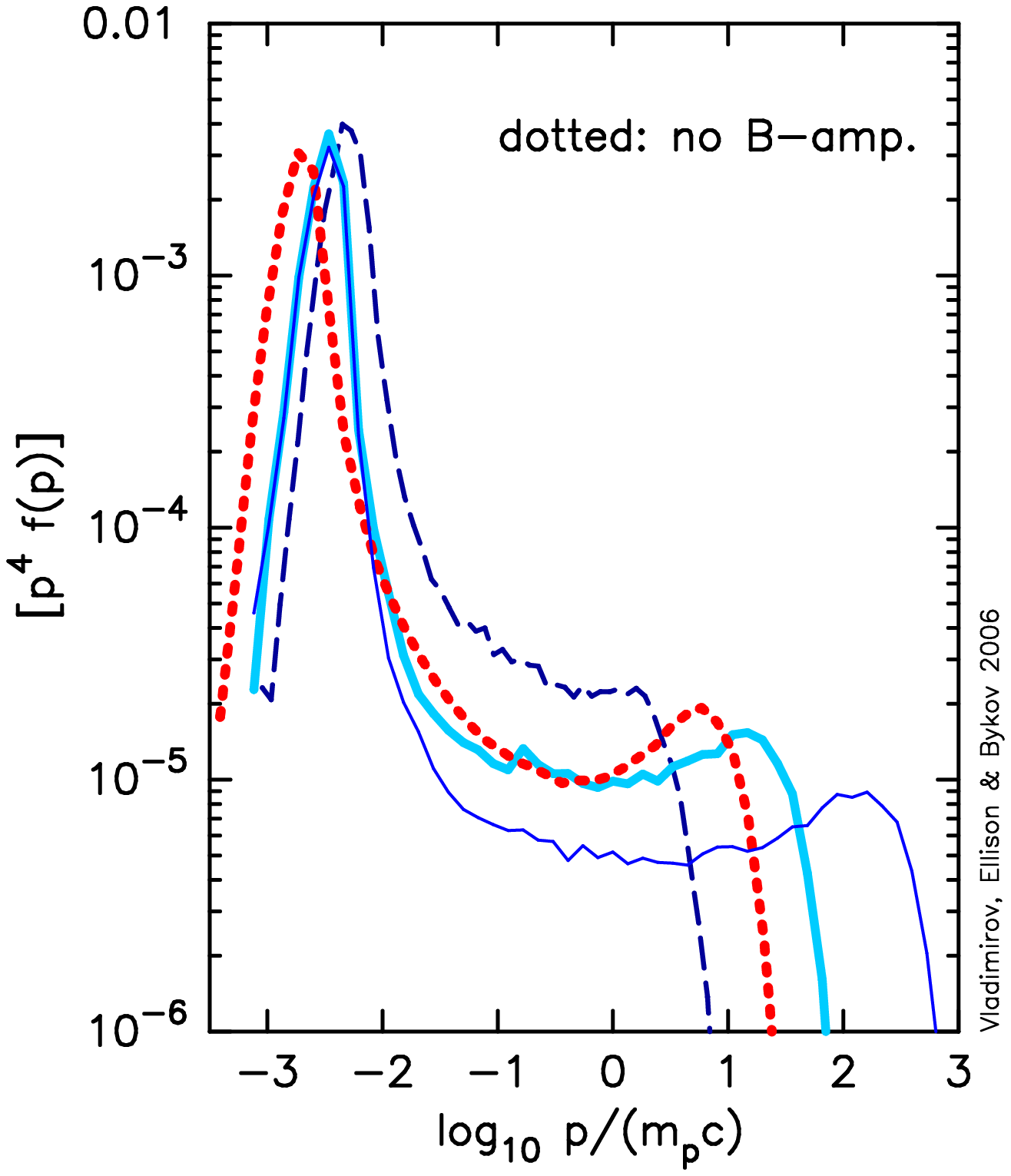}
\caption{$ $ Phase space distributions with and without MFA\label{BL_noBL_fp}}
\end{figure}

In Figure~\ref{BL_noBL_fp} we show the phase space distributions, $f(p)$,
for the shocks shown in Figure~\ref{BL_noBL}.  
These spectra are multiplied by $[p/(m_pc)]^4$ and are calculated
downstream from the shock in the shock rest frame.
For the two cases with the
same parameters except field amplification, we note that the amplified
field case (heavy solid curve) obtains a higher $\Pmax$ and has a higher
shocked temperature (indicated by the shift of the ``thermal'' peak and
caused by the larger $\Rsub$) than the case with no field amplification
(heavy dotted curves).
It is significant that the increase in $\Pmax$ is modest even though $B$
increases by more than a factor or 30 with field amplification.
We emphasize that $\Pmax$ as such is not a free parameter in this model;
$\Pmax$ is determined \SCly\ once the size of the shock system, i.e.,
$\Dfeb$, and the other environmental parameters are set.

In order to show the effect of changing $\Dfeb$, we include in
Figs.~\ref{BL_noBL} and \ref{BL_noBL_fp} field amplification shocks with
the same parameters except that $\Dfeb$ is changed to $-1000\,\rg(u_0)$
(dashed curves) and $-10^5\,\rg(u_0)$ (light-weight solid curves).
From Figure~\ref{BL_noBL_fp}, it's clear that $\Pmax$ scales approximately
as $\Dfeb$ and that the concave nature of $f(p)$ is more pronounced for
larger $\Pmax$.  The field amplification also increases with $\Pmax$,
but the increase between the $\Dfeb=-1000\,\rg(u_0)$ and $\Dfeb =
-10^5\,\rg(u_0)$ cases is less than a factor of two 
(bottom panels of Figure~\ref{BL_noBL}).

\begin{figure}[htbp]
\centering
\vskip 0.5in
\includegraphics[height=7.0in]{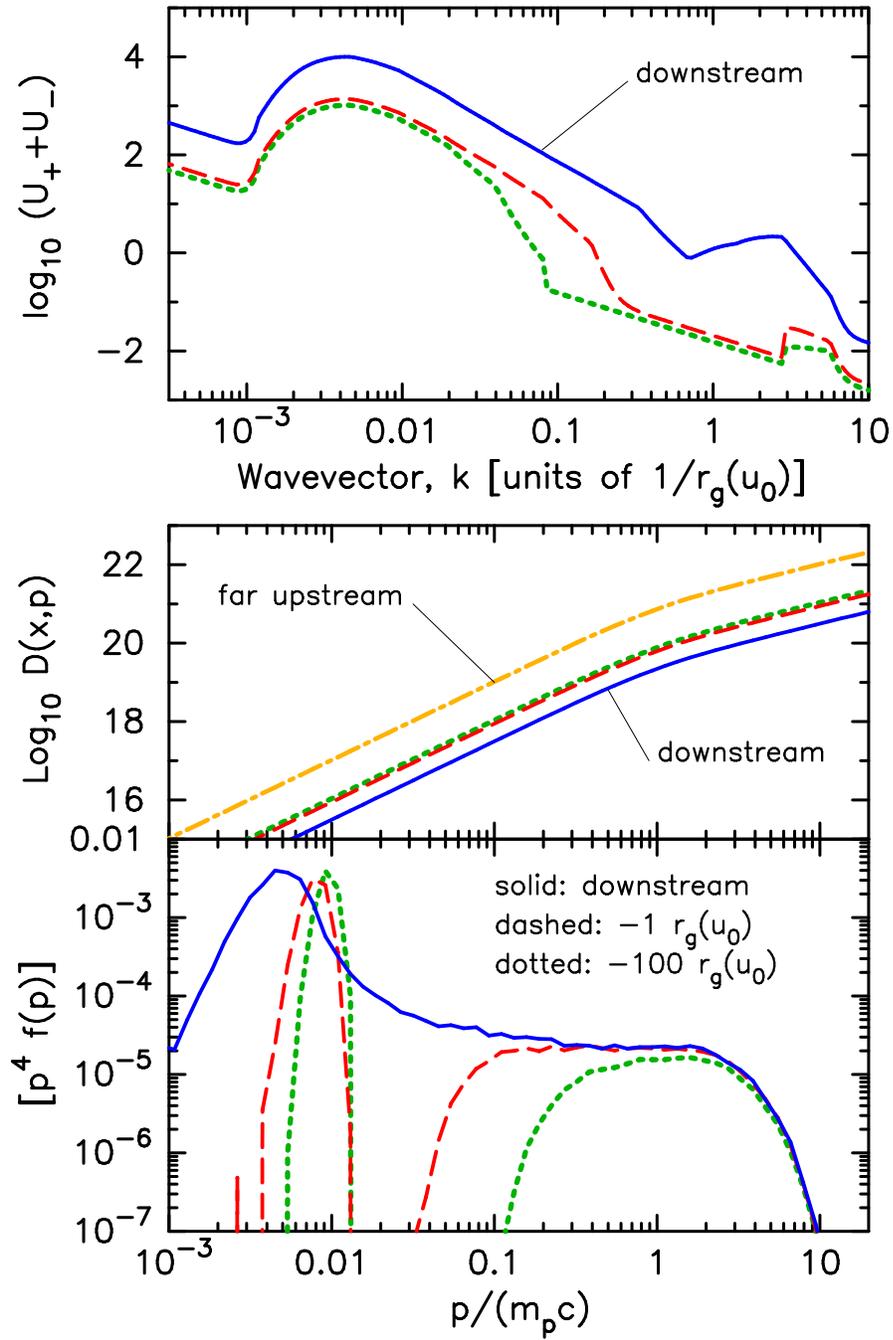}  
\caption{$ $ Turbulence and particle spectra with MFA\label{U_D_fp}}
\end{figure}

\begin{figure}[hbtp]
\centering
\vskip 0.5in
\includegraphics[width=5.0in]{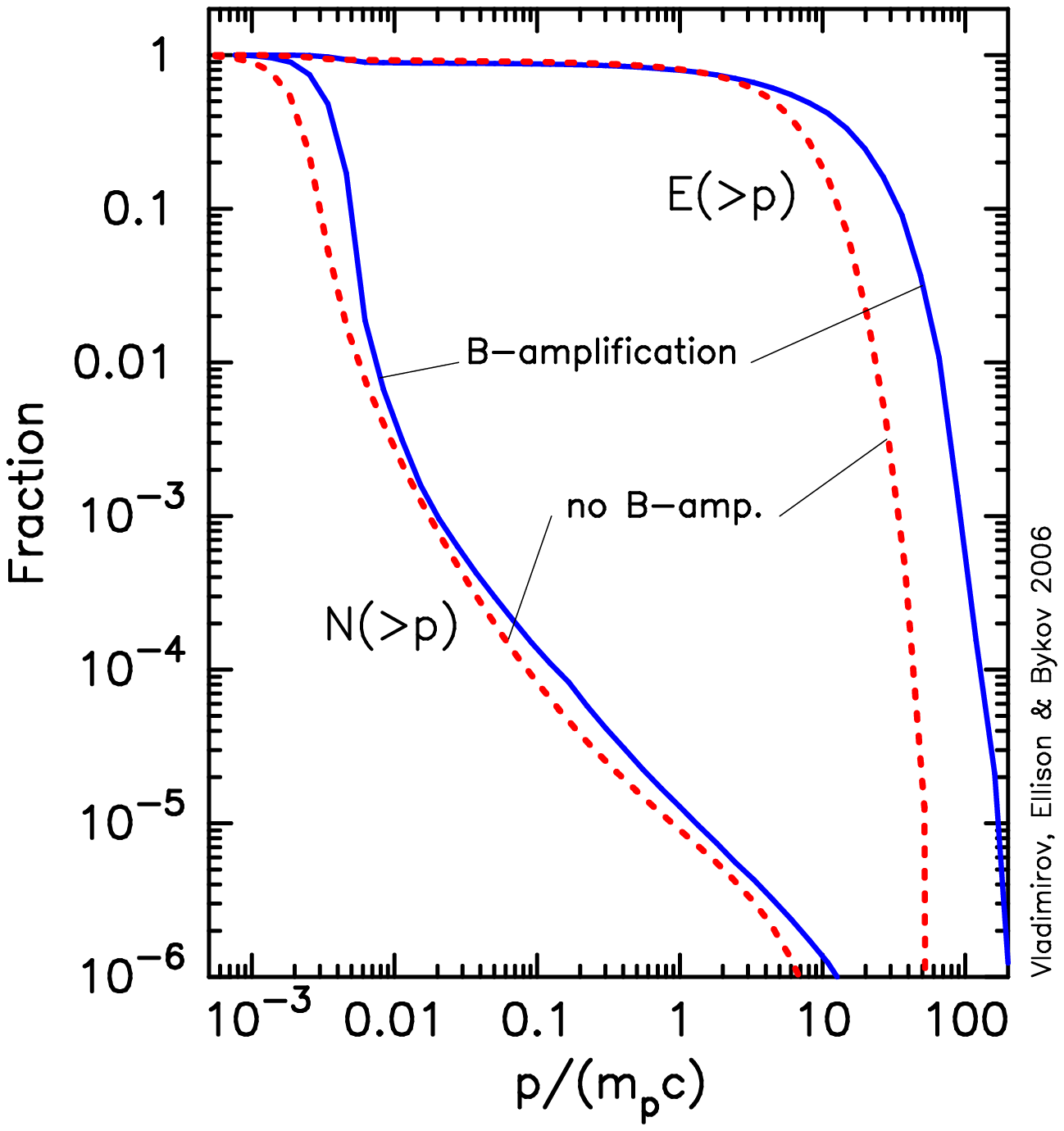}
\caption{$ $ Acceleration efficiency with and without MFA\label{inj_eff}}
\end{figure}

In Figure~\ref{U_D_fp} we show the energy density in magnetic turbulence,
$U_+(x,k) + U_-(x,k)$, the diffusion coefficient, $D(x,p)$, 
and particle distributions as functions of $k$ and $p$ at three different
positions in the shock.  All of these plots
are for the example shown with dashed curves in Figs.~\ref{BL_noBL}
and \ref{BL_noBL_fp} (i.e., with $\Dfeb=-1000\,\rg(u_0)$).  
The solid curve is calculated downstream from the shock, the dashed
curve is calculated at $x = - \rg(u_0)$ upstream from the subshock, and
the dotted curve is calculated at $x = -100\rg(u_0)$ upstream from the
subshock.

The efficiency of the shock acceleration process can be inferred from
Figure~\ref{inj_eff}. It shows the number density of
particles with momentum greater than $p$, i.e.,  $N(>p)$, 
the energy density in particles with momentum greater
than $p$, i.e.,  $E(>p)$,  for the shocks shown in Figs.~\ref{BL_noBL}
and \ref{BL_noBL_fp} with heavy solid and dotted curves.
The plots in Figure~\ref{inj_eff} indicate that the shocks are extremely
efficient accelerators with $>50\%$ of the energy density in $f(p)$
placed in \rel\ particles (i.e., $p \ge m_pc$). 
The actual energy
efficiencies are considerably higher since the escaping particles carry
away a larger fraction of the total energy than is placed in magnetic
turbulence. 
With $\Qesc$ included, well over 50\% of the
total shock energy is placed in \rel\ particles. Despite this high
energy efficiency, the fraction of total particles that become \rel\ is
small, i.e., $N(>p=m_pc) \sim 10^{-5}$ in both cases.

The effect of magnetic field amplification on the number of particles
injected is evident in the left-hand curves. The larger $\Rsub$ (solid
curve) results in more downstream particles being injected into the
Fermi mechanism with amplification than without. 
While it is hard to see from Figure~\ref{inj_eff}, when the escaping
energy flux is included, the shock with \BFA\ puts a considerably
smaller fraction of energy in \rel\ particles than the shock without
amplification.  Again, injection depends in a \NL\ fashion on the shock
parameters and the subshock strength will adjust to ensure that just the
right amount of injection occurs so that momentum and energy are
conserved.

\subsubsection{Alfv\'en Mach Number Dependence}
\begin{figure}[htbp]
\centering
\vskip 0.5in
\includegraphics[height=7.0in]{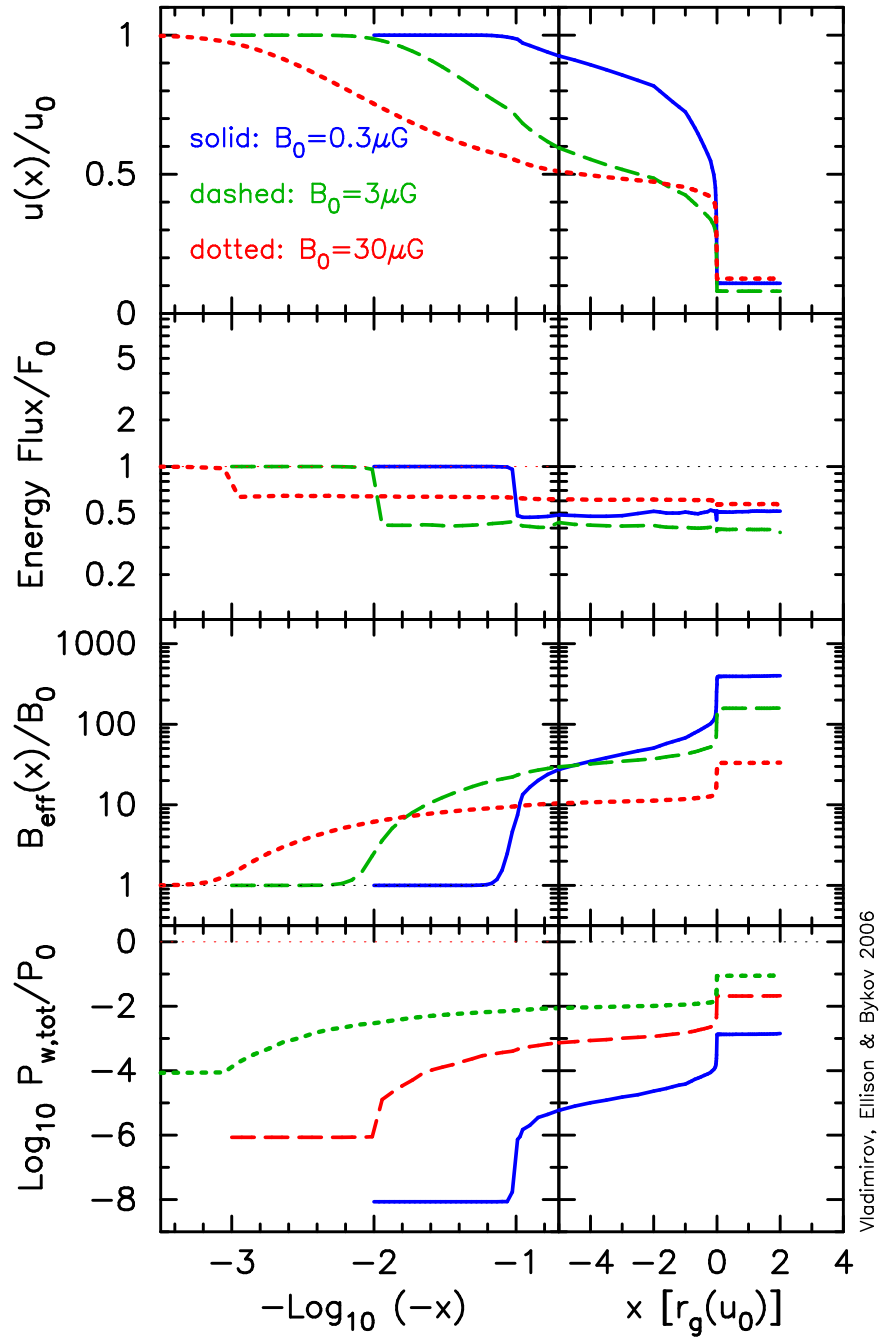}
\caption{$ $ Comparison of shocks with different far upstream fields
  $B_0$. \label{vary_Bz}}
\end{figure}

In Figure~\ref{vary_Bz} we show three examples where $B_0$ was varied
and all other input parameters were kept constant.

In all panels, the solid curves are for $B_0=0.3$\,\muG, the
dashed curves are for $B_0=3$\,\muG, and the dotted curves are for
$B_0=30$\,\muG. The FEB is placed at the same physical distance in all
cases with $\Dfeb=-1.7\xx{10}$\,m. 
Note that $\Beff$ increases most
strongly for $B_0=0.3$\,\muG, but that the pressure in magnetic
turbulence (bottom panel) 
does not gets above $\sim 10$\% of the total pressure,
which but still contains a significant fraction of the total pressure.

The resulting overall compression ratios are: 
$\Rtot \simeq 9$ for $B_0=0.3$\,\muG,    
$\Rtot \simeq 12$ for $B_0=3$\,\muG,    
$\Rtot \simeq 8$ for $B_0=30$\,\muG, values consistent, within
  statistical errors, with $\Qesc$, as indicated in the energy flux
  panels.
As for the self-consistent amplified magnetic fields in these
examples,
$B_2/B_0 \simeq 400$ for $B_0=0.3$~\muG,
$B_2/B_0 \simeq 150$ for $B_0=3$~\muG, and
$B_2/B_0 \simeq 30$ for $B_0=30$~\muG.

\subsubsection{Wave Amplification factor, $\fValf$}
All of the examples shown so far have used the minimum amplification
factor $\fValf=0$ (equation~\ref{VG}). We now investigate the effects of
varying $\fValf$ between 0 and 1 so that $V_G$ varies between $v_a(x)$
and $\Beff(x)/\sqrt{4 \pi \rho(x)}$. The other shock parameters are the
same as used for the dashed curves in Figure~\ref{BL_noBL}, i.e.,
$u_0=5000$\,\kmps, $B_0=30$\,\muG, and $\Dfeb=-1000\,\rg(u_0)$.

\begin{figure}[hbtp]
\centering
\vskip 0.5in
\includegraphics[width=5.0in]{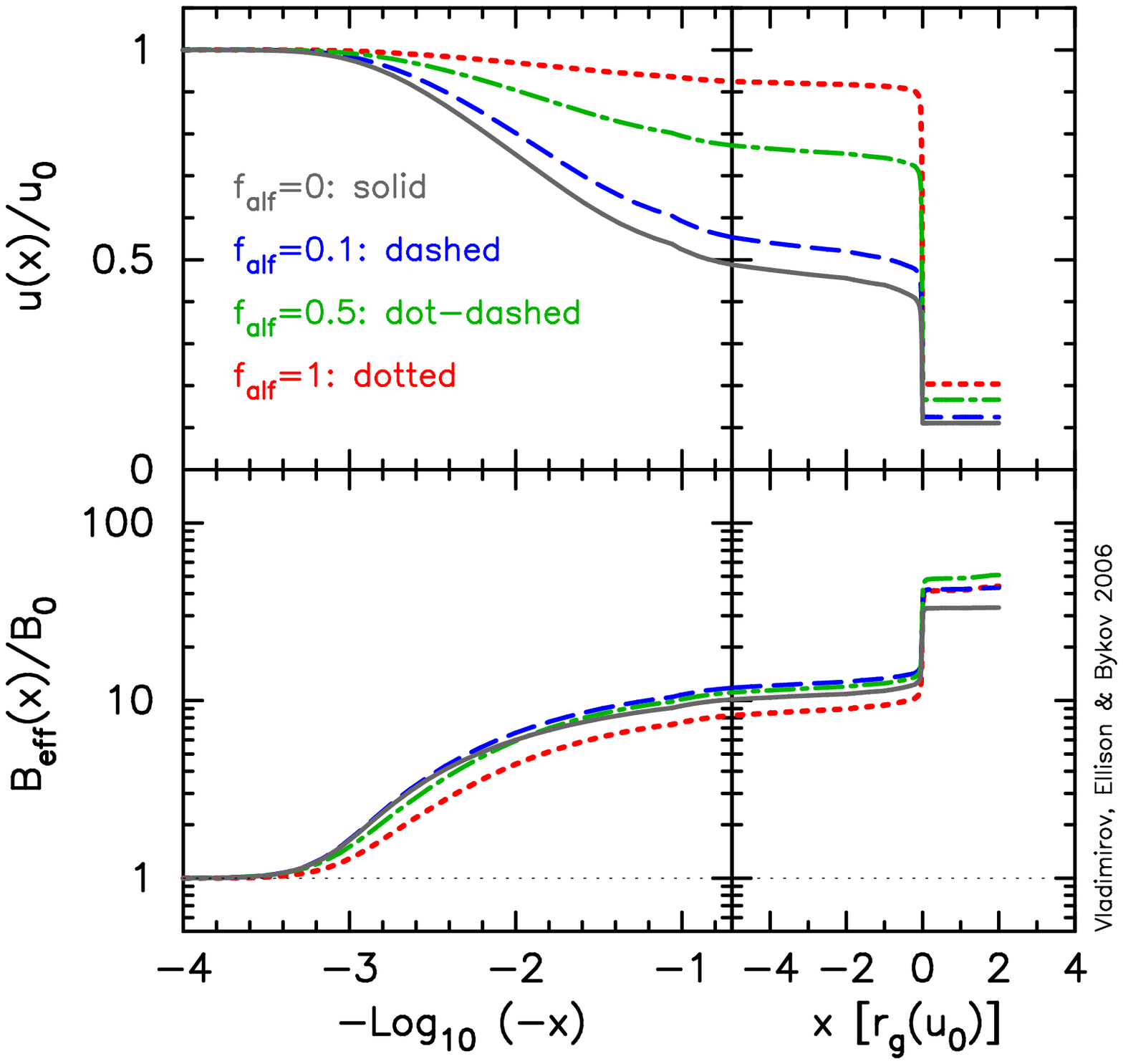}
\caption{$ $ Shocks with varying $\fValf$\label{amp_grid}}
\end{figure}

\begin{figure}[htbp]
\centering
\vskip 0.5in
\includegraphics[width=5.0in]{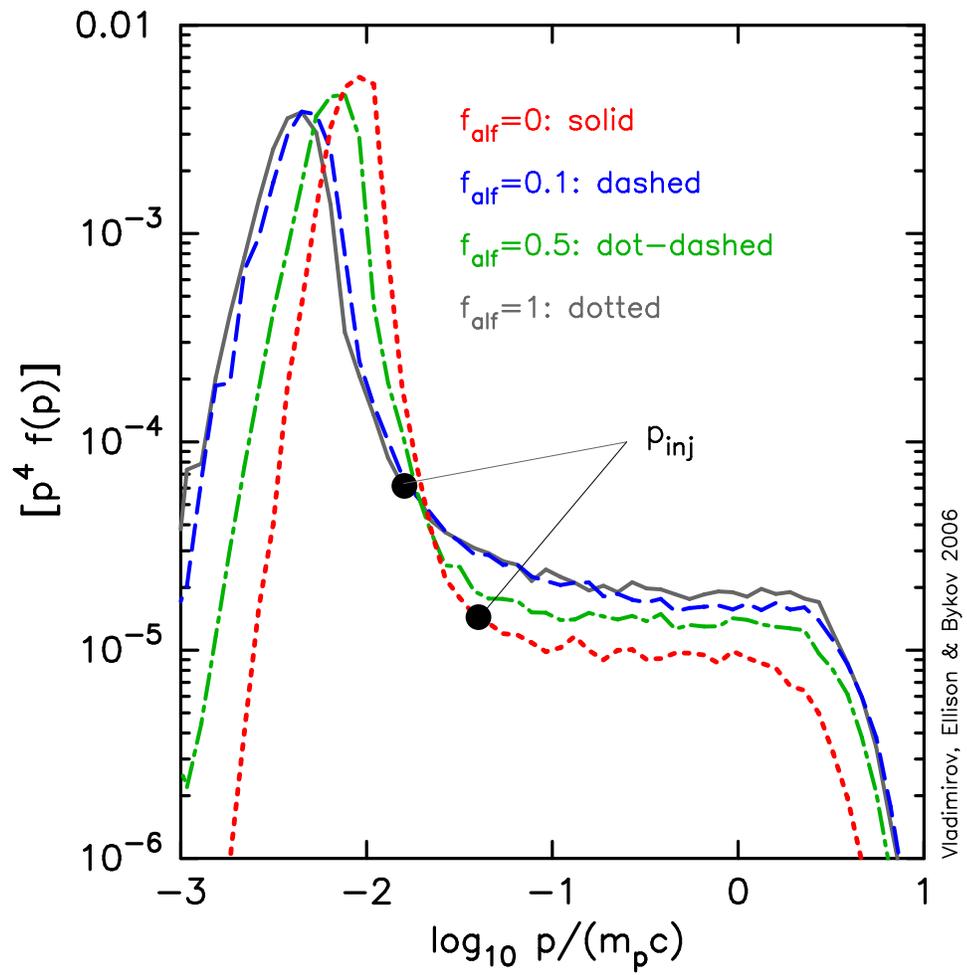}  
\caption{$ $ Distribution functions for the shocks shown in
Figure~\ref{amp_grid}. \label{amp_fp}}
\end{figure}

Figure~\ref{amp_grid} shows $u(x)/u_0$ and $\Beff(x)/B_0$ for 
$\fValf = 0$, $0.1$, $0.5$, and 1 as indicated.  
In all cases, $u_0=5000$\,\kmps, $B_0=30$\,\muG, and $\Dfeb=-1000\,\rg(u_0)$.
The top panels show that
increasing the growth rate (increasing $\fValf$ and therefore $V_G$)
produces a large change in the shock structure and causes the overall
shock compression ratio, $\Rtot$, to decrease. 
The decrease in $\Rtot$
signifies a decrease in the acceleration efficiency 
and a decrease in the fraction of energy that escapes at the FEB, and
the subshock compression adjusts to ensure conservation of momentum and
energy. It is interesting to note that $\Rsub$ increases as $\Rtot$ decreases and
becomes greater than 4 for $\fValf \gtrsim 0.5$. 
In contrast to the strong modification of $u(x)$, there is little
difference in $\Beff(x)/B_0$ (bottom panels of Figure~\ref{amp_grid}) and
little change in $\Pmax$ (Figure~\ref{amp_fp}), between these examples.
The fact that increasing the wave growth rate decreases the acceleration
efficiency shows the \NL\ nature of the wave generation process.
The most important reason for this is that the magnetic
pressure $\Pwtot$, becomes
significant compared to $\rho(x)u^2(x)$ when $\fValf \rightarrow 1$. 
The wave pressure causes the shock to be less compressive overall and
forces $\Rtot$ down.  

In Figure~\ref{amp_fp} we show the distribution functions for the four
examples of Figure~\ref{amp_grid} and note that the low momentum peaks
shift upward significantly with increasing $\fValf$.
As we have emphasized, the injection efficiency, i.e., the fraction of
particles that enter the Fermi process, must adjust to conserve momentum
and energy and the low momentum peaks shift as a result of this.
The solid dots in Figure~\ref{amp_fp} roughly indicate
the injection point separating ``thermal''
and superthermal particles for the two extreme cases of $\fValf = 0$ and 1. 
The first thing to note is that this injection point is not well
defined, a consequence of the fact that the MC model
doesn't distinguish between ``thermal'' and ``nonthermal''
particles. Once the shock has become smooth, the injection process is
smooth and the superthermal population smoothly emerges from the
quasi-thermal population.\footnote{We note that the smooth emergence of
a superthermal tail has been seen in spacecraft observations of the
quasi-parallel Earth bow shock (i.e., \cite{EMP90}) and at
interplanetary traveling shocks (i.e., \cite{BOEF97}).} 
Nevertheless, the approximate momentum where the superthermal population
develops, $\pinj$, can be estimated and we mark this position with solid
dots for $\fValf=0$ and $1$.
What is illustrated by this is that the injection point shifts,
{\it relative to the post-shock distribution}, when $\fValf$ is varied.
This implies that, if injection is parameterized, the parameterization
must somehow be connected to modifications in the shock structure.

\subsection{Discussion}
We have introduced a model of diffusive shock acceleration which couples
thermal particle injection, \NL\ shock structure, \MFA, and the \SC\
determination of the maximum particle momentum. This is a first step
toward a more complete solution and, in this preliminary work, we make
a number of approximations dealing mainly with the plasma physics of
wave growth. 
Keeping in mind that our results are subject to the validity of our
approximations, we reach a number of interesting conclusions.

First, our calculations find that efficient shock acceleration can
amplify ambient magnetic fields by large factors and are generally
consistent with the large fields believed to exist at blast waves in
young SNRs, although we have not attempted a detailed fit to SNR
observations in this paper.
More specifically, we find that the amplification, in terms of the
downstream to far upstream field ratio $B_2/B_0$, is a strong function of
\alf\ Mach number, with weak ambient fields being amplified more than
strong ones. For the range of examples shown in Figure~\ref{vary_Bz},
$B_2/B_0 \sim 30$ for $\Malf \sim 80$ and $B_2/B_0 \sim 400$ for $\Malf
\sim 8000$. 
Qualitatively, a strong correlation between amplification and $\Malf$
should not depend strongly on our approximations and may have important
consequences.
Considering that evidence for radio emission at
reverse shocks in SNRs has been reported (see \cite{GotthelfEtal2001},
for example) and the strong amplification of low fields we
see here, it may be possible for reverse shocks in young SNRs to
accelerate electrons to \rel\ energies and produce radio \syn\ emission.
If similar effects occur in \rel\ shocks, these large amplification
factors will be critical for the internal shocks presumed to exist in
\gamray\ bursts (GRBs). Even if large \BFA\ is confined to \nonrel\
shocks, amplification will be important for understanding GRB
afterglows, in the stages when the expanding fireball 
has slowed down.

As expected, amplifying the magnetic field leads to a greater maximum
particle momentum, $\Pmax$, a given shock can produce. Quantifying
$\Pmax$ is one of the outstanding problems in shock physics because of
the difficulty in obtaining parameters for typical SNRs that allow the
production of cosmic rays to energies at and above the CR knee near
$10^{15}$\,eV. 
Assuming that acceleration is truncated by the size of the shock system,
we determine $\Pmax$ from a physical constraint: the relevant parameter
is the distance to the free escape boundary in diffusion lengths. 
Our results show that $\Pmax$ does increase when field amplification is
included, but the increase is considerably less than the amplification
factor at the shock $B_2/B_0$ (compare the heavy dotted and heavy solid
curves in Figure~\ref{BL_noBL_fp}).
The main reason for this is that high momentum particles have long
diffusion lengths, and the weak precursor magnetic field well upstream from
the subshock determines $\Pmax$. 
If the shock size, in our case $\Dfeb$,
limits acceleration, $\Pmax$ will be considerably less than crude
estimates using a spatially independent $B_2$ (see also Section~\ref{res2007}).
On the other hand, particles spend a large fraction of their time
downstream from the shock where the field is high and collision times
are short. If shock age limits acceleration rather than size, we expect
the increase in $\Pmax$ from the amplified field to be closer to the
amplification factor, $B_2/B_0$. 

Finally, it is well known that DSA is inherently efficient. Field
amplification reduces the fraction of shock ram kinetic energy that is
placed in \rel\ particles but, at least for the limited examples we show
here, the overall acceleration process remains extremely efficient. Even
with large increases in $\Beff(x)$, well over 50\% of the shock energy can go
into \rel\ particles (Figure~\ref{inj_eff}).  As in all \SC\ calculations,
the injection efficiency must adjust to conserve momentum and energy. In
comparing shocks with and without field amplification, we find that
field amplification lowers $\Rtot$ and, therefore, individual energetic
particles are, on average, accelerated less efficiently. In order to
conserve momentum and energy, this means that more thermal particles
must be injected when amplification occurs. The shock accomplishes this
by establishing a strong subshock which not only injects a larger
fraction of particles, but also more strongly heats the downstream
plasma. This establishes a \NL\ connection between the field
amplification, the production of cosmic rays, and the X-ray emission
from the shocked heated plasma.

\newpage

\section{Impact of MFA on the maximum particle energy}

\label{res2007}

Evidence is accumulating\footnote{This section presents, 
in a condensed form, our publication \cite{EV2008}.} 
suggesting that collisionless shocks in
supernova remnants (SNRs) can amplify the interstellar magnetic field to
hundreds of microgauss or even milligauss levels, as recently claimed
for \SNRJ~ \cite{Uchiyama07}.
Ironically, the evidence for large magnetic fields and, therefore,
nonlinear MFA is obtained exclusively from radiation emitted by \rel\
electrons, while the nonlinear processes responsible for MFA are driven
by the efficient acceleration of \rel\ ions, mainly protons.

Here we address a single question: Can, as asserted by Uchiyama et
al. \cite{Uchiyama07}, the large amplified fields inferred for {\it electrons} from
radiation losses in a \NL\ shock also determine the maximum {\it proton}
energy produced in the SNR shock?  We find the answer to be no because
the inevitable \NL\ shock modification (due to efficient DSA) and the
magnetic field variation in the shock precursor (due to MFA) make the
maximum proton energy smaller than what is expected without accounting
for these effects.  

Our result is similar to that found by
\cite{BAC2007} in a time-dependent calculation of DSA where the
acceleration is limited by the age of the shock rather than the size, an
indication that the \NL\ effects we discuss are robust.
 
\subsection{Model}

In a size limited shock, the proton maximum energy, $\Epmax$, will be
determined when the upstream diffusion length of the most energetic
protons becomes comparable to the confinement size of the shock,
typically some fraction of the shock radius.  We model the confinement
size with a free escape boundary (FEB) at a distance $\Lfeb$ in front of
the shock. Protons that reach this position stream freely away from the
shock without producing any more magnetic turbulence. Therefore,
for Bohm diffusion (see \cite{EV2008} for details), 
\begin{equation}
\Epmax \propto \Lfeb  \usk \Bsk,
\label{Emax}
\end{equation}
where $\usk$ is the upstream flow speed.
For a quasi-parallel, \Unmod\ (UM) shock with no MFA,
$\usk \Bsk = u_0 B_0 = u_0 B_2$, $u_0$ being the shock speed and
$B_2$ being the downstream magnetic field derivable from synchrotron
emission of accelerated electrons.  However, for a \NL\ (NL) CR modified
shock of the same physical confinement size, $\Lfeb$, 
the maximum proton energy $\EpmaxNL$ will be
determined by some mean value $\left<u(x)B(x)\right>$, giving
\begin{equation}
    \frac{\EpmaxNL}{\EpmaxUM}=\frac{\left<u(x) B(x) \right>}{u_0 B_2}
\ .
\end{equation}
For a strongly modified shock, $\left < u(x) B(x)\right > \ll u_0
B_2$, and in the following we determine $\left <
u(x) B(x)\right >/ (u_0 B_2)$ using the Monte Carlo model
described in detail in \cite{VEB2006}.

The \mc\ model we use (see \cite{VEB2006} for full details) calculates
NL DSA and the magnetic turbulence produced in a steady-state,
plane-parallel shock precursor by the CR streaming instability. 
We \SCly\ determine the \NL\ shock
structure [i.e., $u(x)$ vs. $x$], the MFA [$\Beff(x)$ vs. $x$], and the
thermal particle injection.\footnote{Note that the \mc\ model ignores
the dynamic effects of electrons and the NL shock structure is
determined solely from the pressure of the accelerated protons and 
of the amplified magnetic fields. While electron acceleration can be modeled
(e.g., \cite{BaringEtal99}), we only show proton spectra here.}

The NL results we investigate do not depend qualitatively on the
particular shock parameters as long as the sonic Mach number is large
enough to result in efficient DSA.  Here, we use a
shock speed = $u_0 = 3000$\,\kmps,
sonic Mach number $M_s \approx 30$, plasma density 
$n_\mathrm{ISM} = 1$ protons cm$^{-3}$, 
and $B_0=\Bism=10$\,\muG, yielding 
an \alf\ Mach number $M_A \approx 140$.
To these parameters we add a FEB boundary at $\Lfeb \sim 0.1$\,pc,
corresponding to $10^8\, \rgzero$, where $\rgzero \equiv m_p u_0 c /(e B_0)$.
This size is comparable to the hot spots in \SNRJ\ 
and produces a proton energy $\sim 10^{15}$\,eV in our \Unmod\ shock
approximation.
Using the above parameters, we simulate two cases: a \NL\ solution,
where $B$ is amplified from an upstream value $B_0=10$\,\muG\ to a
downstream value $B_2=450$\,\muG\ (obtained self-consistently by our
model), and an \Unmod\ solution with a magnetic field set equal
everywhere to $B_2=450$\,\muG.  In these two cases we look at $\Epmax$
to see how the prediction of the NL model, conserving momentum and
energy, compares to the prediction of the UM model, implicitly assumed
by \cite{Uchiyama07}. The information about the maximum energy of
electrons (which are not included in our calculations) can be inferred
graphically from the plot of the acceleration time (see Fig.~\ref{fp}).

\subsection{Results}

\begin{figure}[htbp]
\centering
\vskip 0.5in
\includegraphics[height=7.0in]{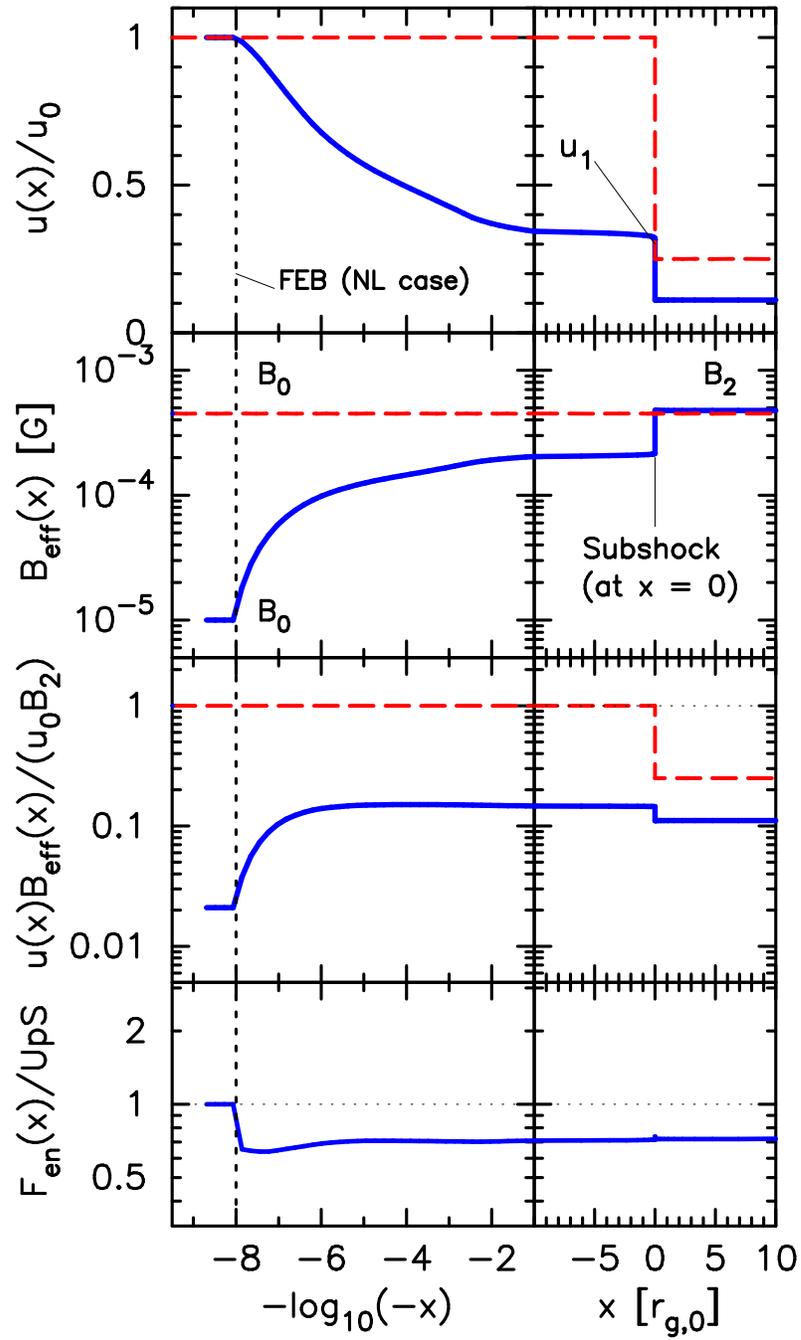}
\caption{$ $ Flow structure: unmodified versus \NL.\label{shock_profile}}
\end{figure}

Figure~\ref{shock_profile} shows the shock structure, $u(x)$, the
effective magnetic field after amplification, $\Beff(x)$, and $u(x)
\Beff(x)/(u_0 B_2)$, for the \Unmod\ case (dashed lines), and the \NL\
case (solid lines). The bottom panel shows the energy flux, normalized
to the far upstream value, for the NL case.
The smoothing of $u(x)$, the weak subshock ($\Rsub \simeq
2.9$), and the increase in $\Rtot$ above 4 ($\Rtot \simeq 9$) are
clearly present in the top panel for the NL case. These three effects
must occur to conserve momentum and energy if CRs are efficiently
accelerated.  The quantity $u(x) \Beff(x)/(u_0 B_2) \sim 0.1$ over
most of the precursor in the NL case.

\begin{figure}[htbp]
\centering
\vskip 0.5in
\includegraphics[width=5.0in]{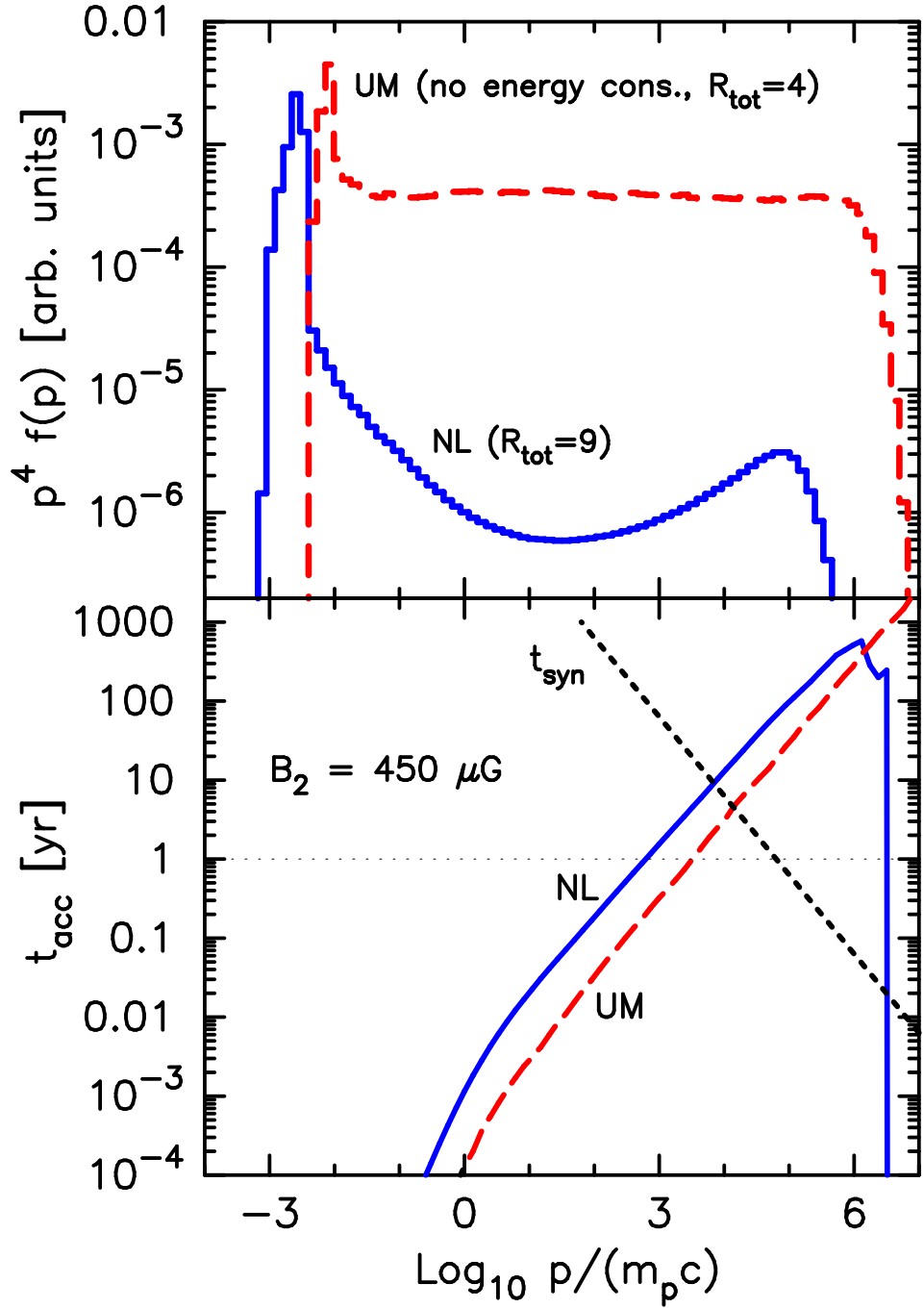}  
\caption{$ $ Proton spectra and acceleration times: unmodified versus
  \NL.\label{fp}}
\end{figure}

In Figure~\ref{fp} we show the momentum distributions
functions, $f(p)$ (multiplied by $p^4$), and the acceleration time,
$\tacc$.  The NL effects evident in Fig.~\ref{shock_profile} result in,
\begin{equation}
\pNL / \pUM \lesssim 0.1
\ ,
\end{equation}
and a longer $\tacc$ to a given momentum. Here, $\pNL = \EpmaxNL/c$ and
$\pUM = \EpmaxUM/c$.
We have not attempted a detailed fit to \SNRJ, but note that
the concave shape of our proton spectrum, above the thermal peak, is
similar to that obtained by \cite{BV2006} who find a good fit to the
data, including the HESS TeV observations
\cite{AharonianJ1713_2006}.

The UM result is obtained from the \mc\ simulation assuming the same
``thermal leakage'' model for injection as in the NL result
(e.g., \cite{JE91}).  This injection scheme works \SCly\ with
modifications in the shock structure and overall compression ratio,
$\Rtot$, to conserve momentum and energy in the NL case. In the UM case,
the shock structure and $\Rtot$ are {\it not} adjusted and the thermal
leakage model produces far too many injected particles to conserve
momentum and energy.
For
the UM shock to become a \TP\ shock with energy conservation, far
fewer particles would need to be injected so that the normalization
of the superthermal $f(p) \propto p^{-4}$ power law becomes low enough,
relative to the thermal peak, so that in contains an insignificant
fraction of the total shock ram kinetic energy.
Since we are only interested in comparing $\ppmax$ in the two cases, the
normalization of the unmodified power law is unimportant since $\pUM$
only depends on $\Lfeb$.

\subsection{Discussion}

The possibility of strong MFA in SNR shocks has been strengthened by the
recent observations of rapid time variability in hot spots in \SNRJ\ by
\cite{Uchiyama07}.
If we accept the conclusions of \cite{Uchiyama07}, the $\sim 1$~yr
variations in X-ray emission in some hot spots stem from radiation
losses for electrons and indicate magnetic fields on the order of
1\,mG. Such large fields would almost certainly be caused by MFA
occurring simultaneously with the efficient production of CR ions in
DSA.

While a number of other interpretations of the X-ray and broadband
emission in \SNRJ\ have concluded that the magnetic field present in the
particle acceleration site is considerably less than 1\,mG
(e.g., \cite{ESG2001,Lazendic2004,BV2006,PMS2006}), we have shown 
that even if the magnetic field inferred from electron
radiation losses is as high as \cite{Uchiyama07} claim, the underlying
physics of MFA in DSA shows that this field cannot be simply applied to
protons to estimate their maximum energy.

The essential point is that, if MFA to milligauss levels is occurring as
part of DSA, the acceleration must be efficient and the system is
strongly \NL. The accelerated particles and the pressure from the
amplified field feedback on the shock structure
(Fig.~\ref{shock_profile}) and this feedback makes the precursor less
confining [i.e., $\uBwt \ll u_0 B_2$]. Therefore, a shock of a given
physical size will not be able to accelerate protons to an energy as
large as estimated ignoring NL effects.

Despite the reduction in $\Epmax$ compared to \TP\ predictions that
our results imply, a remnant such as \SNRJ\ might still produce CRs up
to the knee.  The NL example we have presented with $B_2 \simeq
450$\,\muG\ produces protons up to $\sim 100$\,TeV in $\sim
100$\,yr in a confinement region of $\sim 0.1$\,pc.  If instead we had
taken $\Lfeb = 1$\,pc, a size comparable to the western shell of \SNRJ,
our NL model would produce $\sim 1$\,PeV protons in $\sim 1000$\,yr.
Protons of this energy are consistent with
the $\sim 30$\,TeV \gamrays\ observed from \SNRJ\
\cite{AharonianJ1713_2006} and when the acceleration of heavy ions
such as Fe$^{+26}$ is considered, the maximum particle energy extends to
$> 10^{16}$\,eV.

As a final comment we emphasize a point also made by \cite{BAC2007}. If
MFA is occurring and the system is highly NL, it may not be possible to
explain temporal variations in nonthermal X-ray emission
simply as a radiation loss time.
There cannot be variations in X-ray emission on
short time scales unless the accelerator changes in some fashion on
these time scales, otherwise the radiation would be steady, or varying
on the shock dynamic timescale, regardless of
how short the radiation loss time was.
Since the injection and acceleration of protons and electrons is
nonlinearly connected to the amplified magnetic field,
changes in the
electron particle distribution and changes in the field producing
the \syn\ emission, will go together and it may be difficult to
unambiguously determine the field strength from temporal variations.

\newpage

\section{Turbulence dissipation in shock precursor}

\label{res2008}

Here we present the results of our model regarding the effects
of dissipation of turbulence upstream of the shock and the subsequent
precursor plasma heating\footnote{Results presented here were a part 
of our article \cite{VBE2008}.}.

The magnetic turbulence generated by the instability is
assumed to dissipate at a rate proportional to the turbulence generation
rate, and the dissipated energy is pumped directly into the thermal
particle pool (i.e., the model described by
Equation~(\ref{diss_param}) is assumed).
An iterative scheme is employed to ensure the
conservation of mass, momentum, and energy fluxes, thus producing a
self-consistent solution of a steady-state, plane shock, with particle
injection and acceleration coupled to the bulk plasma flow
modification and to the magnetic field amplification and damping.

Our results show that even a small rate of turbulence dissipation
can significantly increase the precursor temperature and that this, in
turn, can increase the rate of injection of thermal particles. The
\NL\ feedback of these changes on the shock structure, however, tend
to cancel so that the spectrum of high energy particles is only
modestly affected.

\subsection{Model}

We model
the evolution of the turbulence, as it is being advected with the plasma
and amplified, with the following equations:
\begin{equation}
\label{ampeq}
E_{\pm}[U] = (1-\heatpar)G_{\pm}[U] + I_{\pm}[U].
\end{equation}
Here, for readability, we abbreviated as $E$ the evolution operator,
as $G$ the growth operator and as $I$ the wave-wave interactions
operator, acting on the spectrum of turbulence energy density
$U=\{U_-(x,k), U_+(x,k)\}$. These quantities are defined as follows:
\begin{eqnarray}
  E_{\pm}[U] & = &
     \left(u \pm V_G\right)\frac{\partial}{\partial x} U_{\pm} + 
      U_{\pm}\frac{d}{dx}\left(\frac32 u \pm V_G\right), \\
  G_{\pm}[U] & = &
     \mp\frac{U_\pm}{U_+ + U_-} V_G \times 
     \frac{\partial \Pcrhat(x,p)}{\partial x}
     \left| \frac{dp}{dk} \right|, \\
  I_{\pm}[U] & = &
     {\pm} \frac{V_G}{\rgzero}\left(U_- - U_+\right).
\end{eqnarray}
The parameter $\heatpar$ describes the turbulence dissipation rate, and
for $\heatpar=0$, the equations~(\ref{ampeq}) becomes exactly
the system of equations (\ref{uminuskp}) and (\ref{upluskp}).
In this system $u=u(x)$ is the flow speed and
$V_G=V_G(x)$ is the parameter defining the turbulence growth rate
and the wave speed\footnote{As explained in \cite{VEB2006},
in the quasi-linear case, $\Delta B \ll B_0$,
the wave speed and the speed determining turbulence growth rate are
both equal to the \Alf\ speed,
$V_G (x) = v_A = B_0 / \sqrt{4 \pi \rho(x)}$.
In the case of strong turbulence, $\Delta B \gtrsim B_0$,
we hypothesize that the resonant streaming instability can
still be described by equations~(\ref{ampeq}) with $V_G$
being a free parameter ranging from $B_0 / \sqrt{4 \pi \rho(x)}$
to $\Beff / \sqrt{4 \pi \rho(x)}$.}.
The parameter  $\heatpar$ enters the
equations of turbulence evolution (\ref{ampeq}) through the factor
$(1-\heatpar)$, which represents the assumption that at all
wavelengths only a fraction $(1 -\heatpar)$ of the
instability growth rate goes into the magnetic turbulence, and the
remaining fraction $\heatpar$ is lost in the dissipation process.
See Equation~(\ref{diss_param}) and the corresponding part
of Section~\ref{turb_effects}.

Equation (\ref{pressuregrowth}) is used to account for the precursor
plasma heating by the dissipated turbulence.
For $\Lbar(x)=0$, equation (\ref{pressuregrowth})
reduces to the adiabatic heating law,
$\Pth \sim \rho^{\gamma}$ and, for a non-zero $\Lbar(x)$, it describes
the heating of the thermal plasma in the shock precursor due to the
dissipation of magnetic turbulence.

The main effects of turbulence dissipation in our model are:
\newlistroman\listromanDE a decrease in the value of the amplified field
$\Beff(x)$, which determines the diffusion coefficient, $D(x,p)$;
\listromanDE an increase in the temperature of particles just upstream of
the subshock, which influences the injection of particles into the
acceleration process, and
\listromanDE
an increase in the thermal particle pressure $\Pth(x<0)$,
and a decrease in the turbulence pressure $P_w(x)$,
which enter the conservation equations
described in Section \ref{sec-rtot} and \ref{subsec_smoothing}.
Since all of these processes are coupled, a change in dissipation
influences the overall structure of the shock.

\subsection{Results}

\subsubsection{Particle Injection in Unmodified Shocks (Subshock Modeling)}
\label{injectionvsdissipation}

In order to isolate the effects of plasma heating on particle injection,
we first show results for unmodified shocks, i.e., $u(x<0)=u_0$ and
$u(x>0)=u_0/\rtot$, with fixed $\rtot$. In these models particle
acceleration, \MFA\ and turbulence damping are included consistently
with each other, but we do not obtain fully \SC\ solutions
conserving momentum and energy, since this requires the shock
to be smoothed, while we intentionally fix $u(x)$.

\begin{figure}[htbp]
\centering
\vskip 0.5in
\includegraphics[height=7.0in]{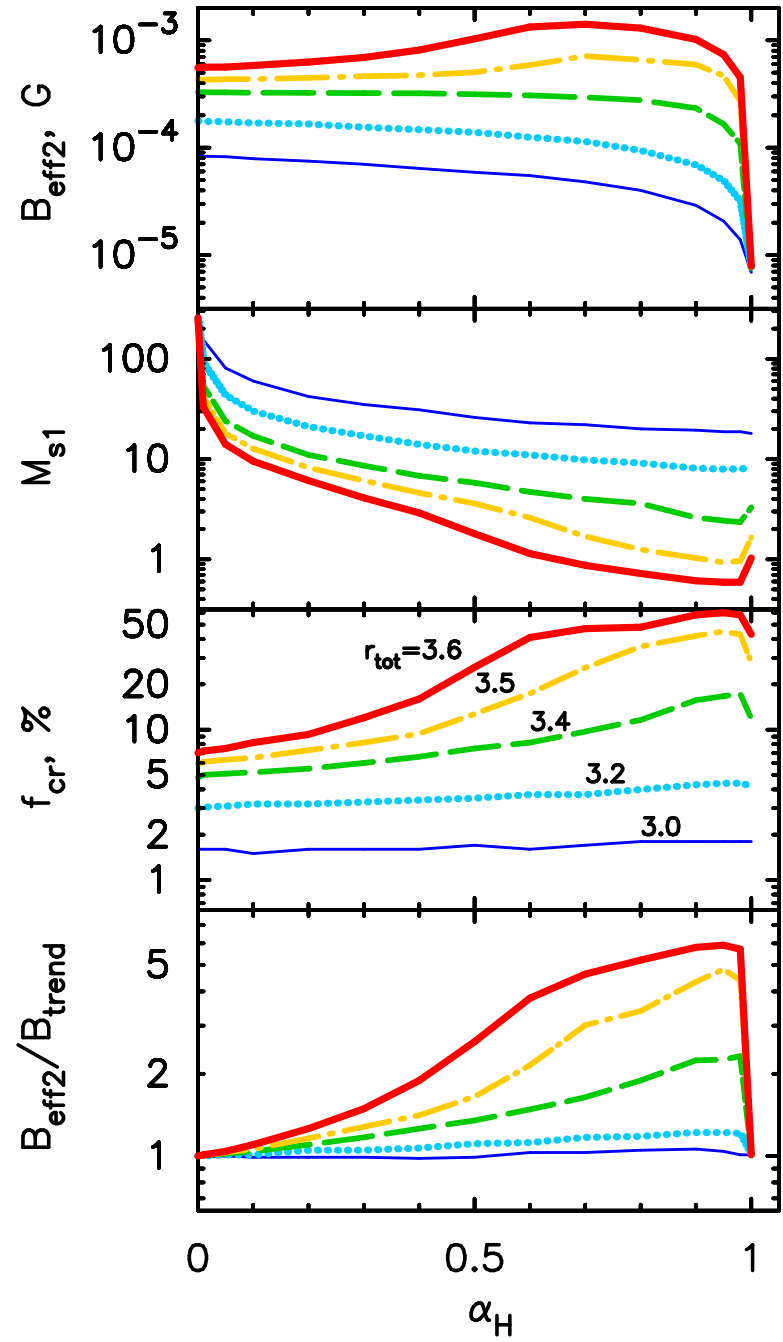}
\caption{$ $ Dissipation effects in unmodified shocks  \label{fig_unmod_inj}}
\end{figure}

\begin{figure}[htbp]
\centering
\vskip 0.5in
\includegraphics[height=7.0in]{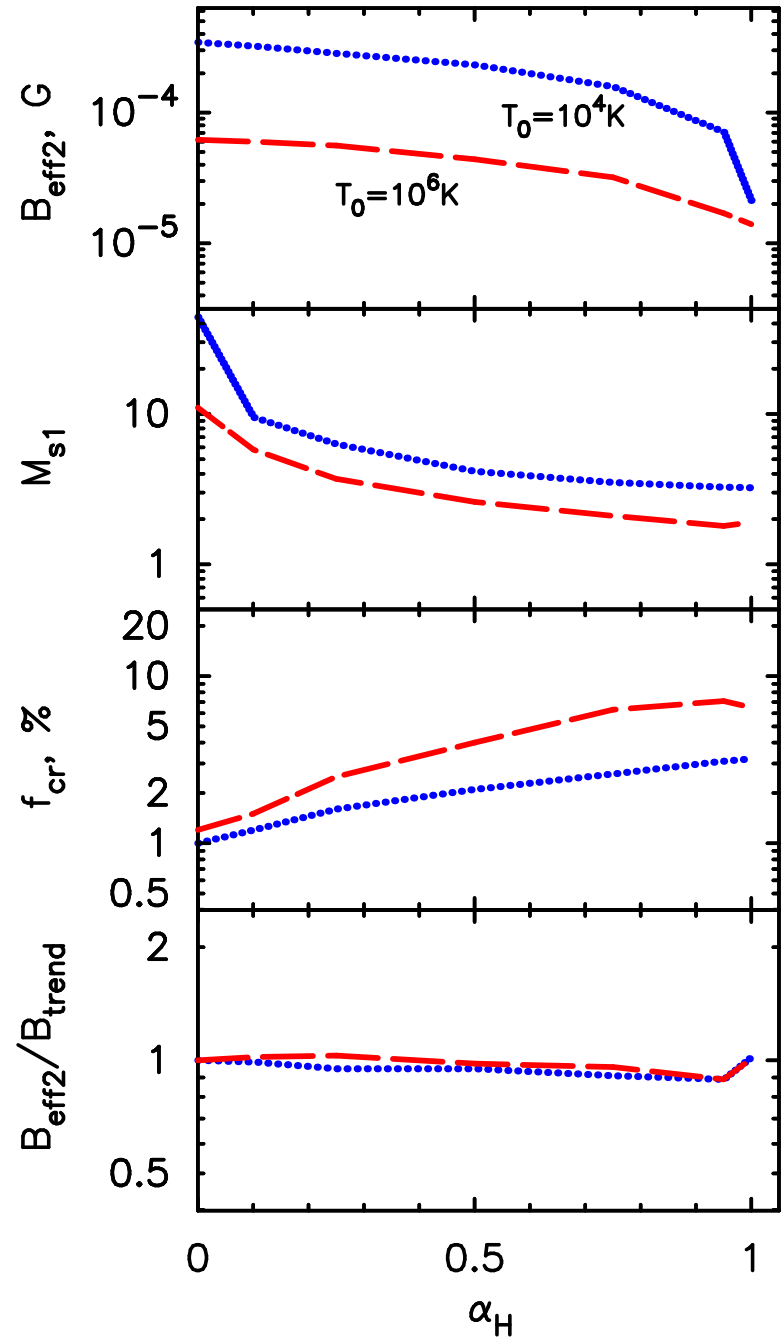}
\caption{$ $ Dissipation effects in nonlinear shocks   \label{fignlsum}}
\end{figure}

In Fig.~\ref{fig_unmod_inj} we show results where the compression ratio is
varied between $\Rtot=3$ and $3.6$ as indicated. In all models, $u_0 =
3000$ \kmps, $T_0 = 10^4$ K, $n_0 = 0.3$\,\pcc\ and $B_0 = 3$\,\muG\
(the corresponding sonic and \Alf\ Mach numbers are $\Mszero \approx
\Mazero \approx 250$). The FEB was set at $\xfeb = -3\cdot 10^4\:
\rgzero$ (our spatial scale unit $\rgzero=m_p u_0 c / (eB_0)$),
and for each $\Rtot$ we obtained results for different values
of $\heatpar$ between 0 and 1. The values plotted in the
top three panels of Fig.~\ref{fig_unmod_inj}
are the amplified magnetic field downstream, $\Befftwo$, the Mach number
right before the shock, $\Msone$ (this is not equal to $\Mszero$ because
of the plasma heating due to turbulence dissipation), and the fraction
of thermal particles in the simulation that crossed the shock in the
upstream direction at least once (i.e., got injected), $\fcr$. The bottom
panel shows the ratio of the calculated downstream effective magnetic
field $\Befftwo$ to trend values $\Btrend(\heatpar)$;
what is meant by ``trend'' is the equation (\ref{btwotrend}) explained
in \cite{VBE2008}:
\begin{equation}
  \label{btwotrend}
  \Btrend^2(\heatpar) =
     \left(B_0 \rtot^{3/4}\right)^2 + (1-\heatpar) 
  \left[ \left. \Befftwo^2 \Big.\right|_{\heatpar=0}
    - \left(B_0 \rtot^{3/4}\right)^2 \right].
\end{equation}

Looking at the curve for $\Befftwo$ in the $\rtot=3.0$ and $\rtot=3.2$
models, one sees
an easy to explain behavior: as the magnetic turbulence dissipation
rate, $\heatpar$,
increases, the value of the amplified magnetic field decreases, going
down to $B_0 \rtot^{3/4}$ (the upstream field compressed at the shock)
for $\heatpar=1$. Increasing $\heatpar$ simply causes more energy to be
removed from magnetic turbulence and put into thermal particles, thus
decreasing the value of $\Befftwo$.

The plots for $\rtot \gtrsim 3.4$ present a
qualitatively different behavior from those with $\rtot \lesssim
3.2$. The downstream magnetic field $\Befftwo$ does
decrease with increasing $\heatpar$, but not as rapidly as in the
previous two cases, and there is a switching point at $\heatpar \approx
0.95$ in the curves for $\Msone$ and $\fcr$.  The bottom panel of
Fig.~\ref{fig_unmod_inj} shows a deviation of $\Befftwo$ from the trend
(\ref{btwotrend}) by a large factor in the $\rtot = 3.4$ case.  This
effect becomes even more dramatic for $\rtot=3.5$ and $\rtot=3.6$ where
$\Befftwo$, contrary to expectations, {\bf increases} with $\heatpar$ before
$\heatpar\rightarrow 1$.  The fact that the final energy in turbulence
can increase as more energy is transferred from the turbulence to heat
indicates the \NL\ behavior of the system and shows how sensitive the
acceleration is to precursor heating.

It is worth mentioning that the observed increase of particle
injection due to the precursor plasma heating is a consequence of the
thermal leakage model of particle injection adopted here (see
\cite{VBE2008} for the explanation of this connection). In this model,
a downstream particle, thermal or otherwise, with plasma frame speed
$v>u_2$, has a probability to return upstream which increases with $v$
(see \cite{Bell78a} for a discussion of the probability of returning
particles).  An alternative
model of injection (see, for example, \cite{BGV2005}) is one where
only particles with a gyroradius greater than the shock thickness can
get injected. 
In the \cite{BGV2005} model the fraction of
injected particles may be insensitive to the precursor heating if
the parameter controlling the injection in that model, $\xi$, is fixed.
While both models are highly simplified
descriptions of the complex subshock
(see, e.g., \cite{Malkov98, GE2000}), they offer two scenarios for
grasping a qualitatively correct behavior of a shock where
particle injection and acceleration are coupled
to turbulence generation and flow modification.  Hopefully, a
clearer view of particle injection by self-generated turbulence in a
strongly magnetized subshock will become available when
relevant full particle PIC or hybrid
simulations are performed.

With the general trends observed here in mind,
we now show how \NL\ effects modify the effect dissipation has
on injection and MFA.


\subsubsection{Fully Nonlinear Model}

\label{nlresults}

In this section we demonstrate the results of the fully nonlinear
models, in which the flow structure, compression ratio, magnetic
turbulence, and particle distribution are all determined \SCly, so
that the fluxes of mass, momentum and energy are conserved across the
shock.

We use two sets of parameters, one with the far upstream gas
temperature $T_0=10^4$~K and the far upstream particle density $n_0 =
0.3$~\pcc, typical of the cold interstellar medium (ISM), and one with
$T_0=10^6$ K and $n_0=0.003$ \pcc, typical of the hot ISM.
In both cases we assumed the shock speed $u_0 = 5000$ \kmps, and the
initial magnetic field $B_0 = 3$ \muG\ (giving an equipartition of
magnetic and thermal energy far upstream, $n_0 k_B T_0 \approx
B_0^2/(8\pi)$). The corresponding sonic and \Alf\ Mach numbers are $M_s
\approx M_A \approx 400$ in both cases).  The size of the shocks
was limited by a FEB located at $\xfeb = -10^5 \rgzero\approx -3 \cdot
10^{-4}$ pc.
For both cases, we ran seven
simulations with different values of the dissipation rate $\heatpar$,
namely $\heatpar \in \{ 0; 0.1; 0.25; 0.5; 0.75; 0.9;
1.0\}$. Also, for the \hotism\ case we ran a
simulation neglecting the streaming instability effects, i.e., keeping
the magnetic field constant throughout the shock and assuming that the
precursor plasma is heated only by adiabatic compression (this model
will be referred to as the `no MFA case').

\begin{table}
\caption{Summary of Non-linear Simulation in a Cold ISM \label{sumnl_cold}}
\begin{center}
\begin{small}
\begin{tabular}{ lrrrrrrr }
\hline
$\heatpar$         & 0.00  & 0.10  & 0.25  & 0.50  & 0.75  & 0.95  &1.00 \\
\hline
$\rtot$            & 16.0  & 16.2  & 14.5  & 14.6  & 14.0  & 13.2  & 13.0  \\
$\rsub$            & 2.95  & 2.83  & 2.75  & 2.59  & 2.50  & 2.50  & 2.51  \\
$\Befftwo$, \muG   & 345   & 323   & 284   & 232   & 158   & 71    & 21    \\
$\Btrend$, \muG    & 345   & 327   & 299   & 245   & 174   & 80    & 21    \\
$\left<T(x<0)\right>$, $10^4$ K
                   & 1.06  & 4.3   & 9.0   & 17    & 26    & 37    & 56    \\
$T_1$, $10^4$ K    & 3.3   & 68    & 160   & 330   & 490   & 610   & 650   \\
$T_2$, $10^4$ K    & 1400  & 1500  & 1600  & 1600  & 1800  & 2000  & 2200  \\
$\Msone$           & 44    & 9.5   & 6.3   & 4.2   & 3.5   & 3.3   & 3.2   \\
$\fcr$, \%         & 1.0   & 1.2   & 1.6   & 2.1   & 2.6   & 3.1   & 3.2   \\
$\pmax / m_pc$     & 500   & 450   & 400   & 350   & 250   & 150   & 80    \\
$\left<\gamma(x<0)\right>$
                   & 1.33  & 1.33  & 1.33  & 1.33  & 1.34  & 1.34  & 1.34  \\
$\gamma_2$         & 1.38  & 1.38  & 1.38  & 1.39  & 1.39  & 1.40  & 1.41  \\
$\xtr  / \rgzero$  & -0.005&-0.001 & -0.001& -0.001&-0.002 & -0.004& -0.02 \\
$\xFP  / \rgzero$  & -0.04 &-0.06  & -0.06 & -0.09 & -0.23 &-0.57  & -2.1  \\
\hline
\end{tabular}
\end{small}
\end{center}
\footnotesize{See text and \cite{VBE2008} for notation}
\end{table}

\begin{table}
\caption{Summary of Non-linear Simulation in a Hot ISM \label{sumnl_hot}}
\begin{center}
\begin{small}
\begin{tabular}{ lrrrrrrrr }
\hline
$\heatpar$         & 0.00  & 0.10  & 0.25  & 0.50  & 0.75  & 0.95  & 1.00  & No MFA\\
\hline
$\rtot$            & 8.1   & 8.2   & 8.3   & 8.0   & 7.8   & 7.4   & 7.3   & 13     \\
$\rsub$            & 2.92  & 2.75  & 2.55  & 2.44  & 2.22  & 2.15  & 2.12  & 2.75   \\
$\Befftwo$, \muG   & 62    & 60    & 55    & 44    & 32    & 17    & 14    & 21     \\
$\Btrend$, \muG    & 62    & 59    & 54    & 45    & 33    & 19    & 13    & -      \\
$\left<T(x<0)\right>$, $10^6$ K
                   & 1.04  & 1.3   & 1.7   & 2.3   & 3.1   & 3.9   & 4.2   & 1.1    \\
$T_1$, $10^6$ K    & 2.0   & 6.0   & 13    & 23    & 34    & 42    & 43    & 2.7    \\
$T_2$, $10^6$ K    & 53    & 49    & 47    & 55    & 62    & 72    & 75    & 22     \\
$\Msone$           & 10.9  & 5.8   & 3.7   & 2.6   & 2.1   & 1.9   & 1.9   & 4.7    \\
$\fcr$, \%         & 1.2   & 1.6   & 2.5   & 4.0   & 6.4   & 6.9   & 6.4   & 2.4    \\
$\pmax / m_pc$     & 150   & 120   & 110   & 100   & 90    & 70    & 60    & 80     \\
$\left<\gamma(x<0)\right>$
                   & 1.34  & 1.34  & 1.34  & 1.34  & 1.34  & 1.34  & 1.35  & 1.34   \\
$\gamma_2$         & 1.43  & 1.43  & 1.43  & 1.44  & 1.45  & 1.45  & 1.45  & 1.41   \\
$\xtr  / \rgzero$  & -0.04 &-0.02  & -0.02 & -0.02 &-0.03  & -0.07 & -0.05 & -0.02  \\
$\xFP  / \rgzero$  & -0.1  &-0.1   & -0.2  & -0.2  & -0.4  & -0.9  & -1.4  & -0.1   \\
\hline
\end{tabular}
\end{small}
\end{center}
\footnotesize{See text and \cite{VBE2008} for notation}
\end{table}

Tables~\ref{sumnl_cold} and \ref{sumnl_hot} summarize
some of the results of these models.
The effect of the turbulence dissipation into the thermal plasma is
evident in the values of the pre-subshock temperature $T_1$, the
downstream temperature $T_2$, and the volume-averaged precursor
temperature $\left< T(x<0) \right>$ (the averaging takes place between
$x=\xfeb$ and $x=0$).
The value of $T_1$ depends drastically on the
level of the turbulence dissipation $\heatpar$, increasing from
$\heatpar=0$ to $\heatpar=0.5$ by a factor of 100 in the
\coldismNoT\ case, and by a factor of 11 in the
\hotismNoT\ case.
The values of the temperature as high as $T_1$ are achieved upstream
only near the subshock; the volume-averaged upstream temperature,
$\left< T(x<0) \right>$, is significantly lower. 
The downstream temperature, $T_2$, varies less with changing $\heatpar$,
because it is largely determined by the compression at the subshock,
which is controlled by many factors.
It is worth mentioning the case without
MFA reported in Table~\ref{sumnl_hot}.  Besides having a much larger
compression factor than the shocks with MFA ($\rtot=13$ as opposed to
$\rtot\lesssim 8$), it has a much smaller downstream temperature
($T_2=2.2\cdot 10^7$~K as opposed to $T_2 \gtrsim 5.3\cdot 10^7$~K)
These effects of dissipation on the precursor temperature may be
observable.

In Figure~\ref{fignlsum} we show results for $\fcr$, $\Msone$,
$\Befftwo$, and $\Btrend$ which can be compared to the results for
unmodified shocks shown in Figure~\ref{fig_unmod_inj}. For the modified
shocks, the fraction of the thermal particles crossing the shock
backwards for the first time, $\fcr$, clearly increases by a large
factor with $\heatpar$, which can be explained by the connection
between $T_1$ and the injection rate.
One could expect that the amplified
effective magnetic field $\Befftwo$ would behave similarly to the
$\rtot=3.5$ case in Section~\ref{injectionvsdissipation}, i.e. that
$\Befftwo$ would not decrease or even would increase for larger
$\heatpar$.  Instead, $\Befftwo$ behaves approximately according to the
trend~(\ref{btwotrend}), as the values of $\Btrend$ from
Tables~\ref{sumnl_cold} and \ref{sumnl_hot} show and the bottom panel of
Fig.~\ref{fignlsum} illustrates. The important point is that, even
though precursor heating causes the {\it injection efficiency} to
increase substantially, the {\it efficiency of particle acceleration}
(i.e., the fraction of energy in CRs)
and magnetic turbulence generation is hardly changed.
We base this assertion on the fact that $\Befftwo$ remains
close to $\Btrend$, which was derived under the assumption that changing
$\heatpar$ preserves the total energy generated by the instability, but
re-distributes it between the turbulence and the thermal particles. 

Considering how much the injection rate $\fcr$ increases with
$\heatpar$, and how much the upstream temperature
of the thermal plasma, $T_1$, is affected by the heating,
it is somewhat surprising that the
trend of the amplified effective field $\Befftwo$ is unaffected. The
mechanism by which the shock adjusts to the changing heating and
injection in order to preserve the MFA efficiency can be understood by
looking at the trend of the total compression ratio $\rtot$ and the
subshock compression ratio $\rsub$ in Tables~\ref{sumnl_cold} and
\ref{sumnl_hot}: they both decrease significantly for higher
$\heatpar$.
The decrease in $\rsub$ is easy to understand:
with the turbulence dissipation operating in the precursor
$\Msone$ goes down, which lowers $\rsub$. Additionally,
decreasing $\Pwone$ helps to reduce $\rsub$, and
with a boost
of the particle injection rate, the particles returning for the first
time increase in number and build up some extra pressure just upstream
of the shock, which causes
the flow to slow down in that region, thus reducing the ratio
$\rsub$.

\begin{figure}[htbp]
\centering
\vskip 0.5in
\includegraphics[height=7.0in]{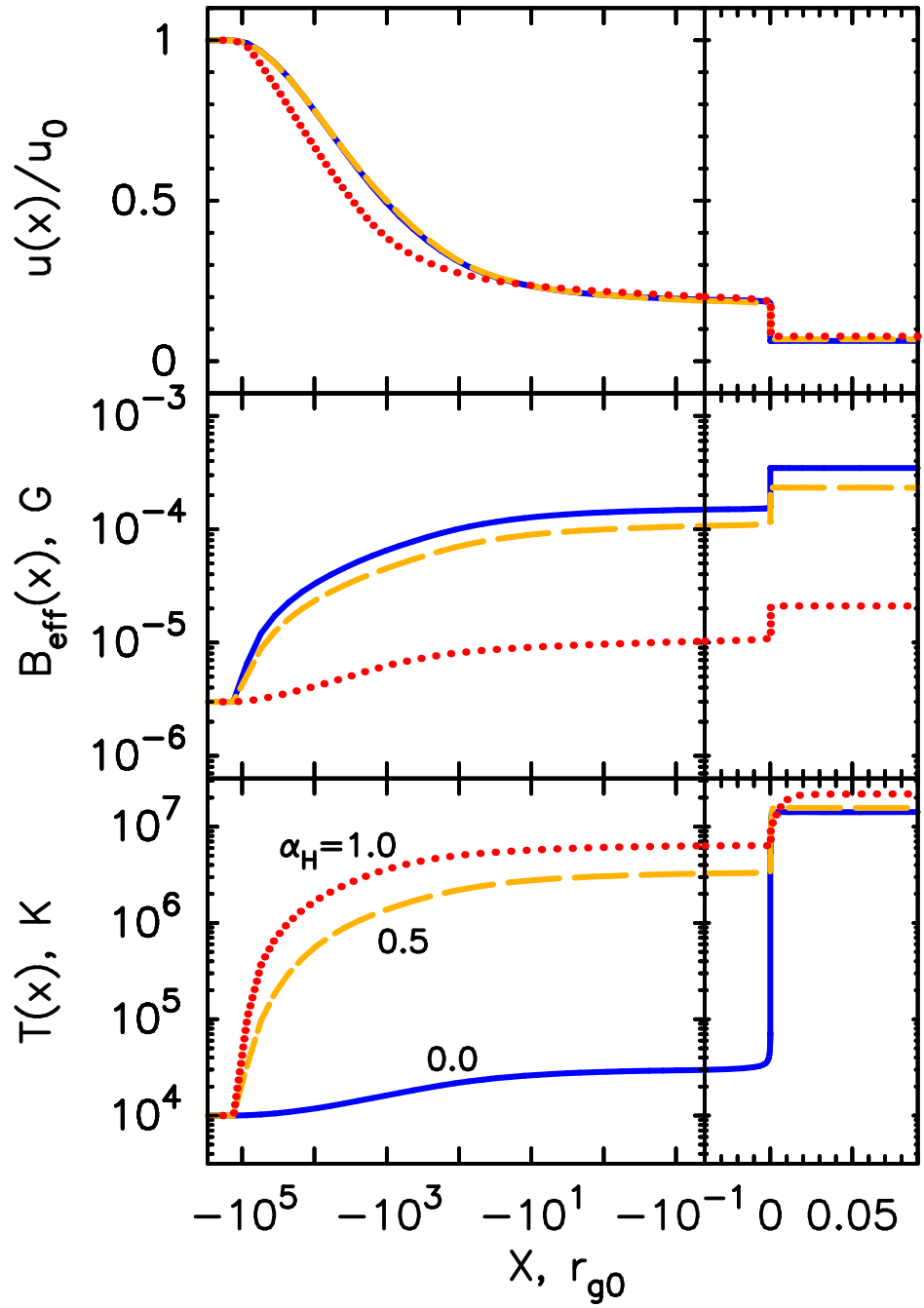}
\caption{$ $ Nonlinear shocks with dissipation, \coldismNoT \label{fig_ubt_1e4_1e4}} 
\end{figure}

\begin{figure}[htbp]
\centering
\vskip 0.5in
\includegraphics[height=7.0in]{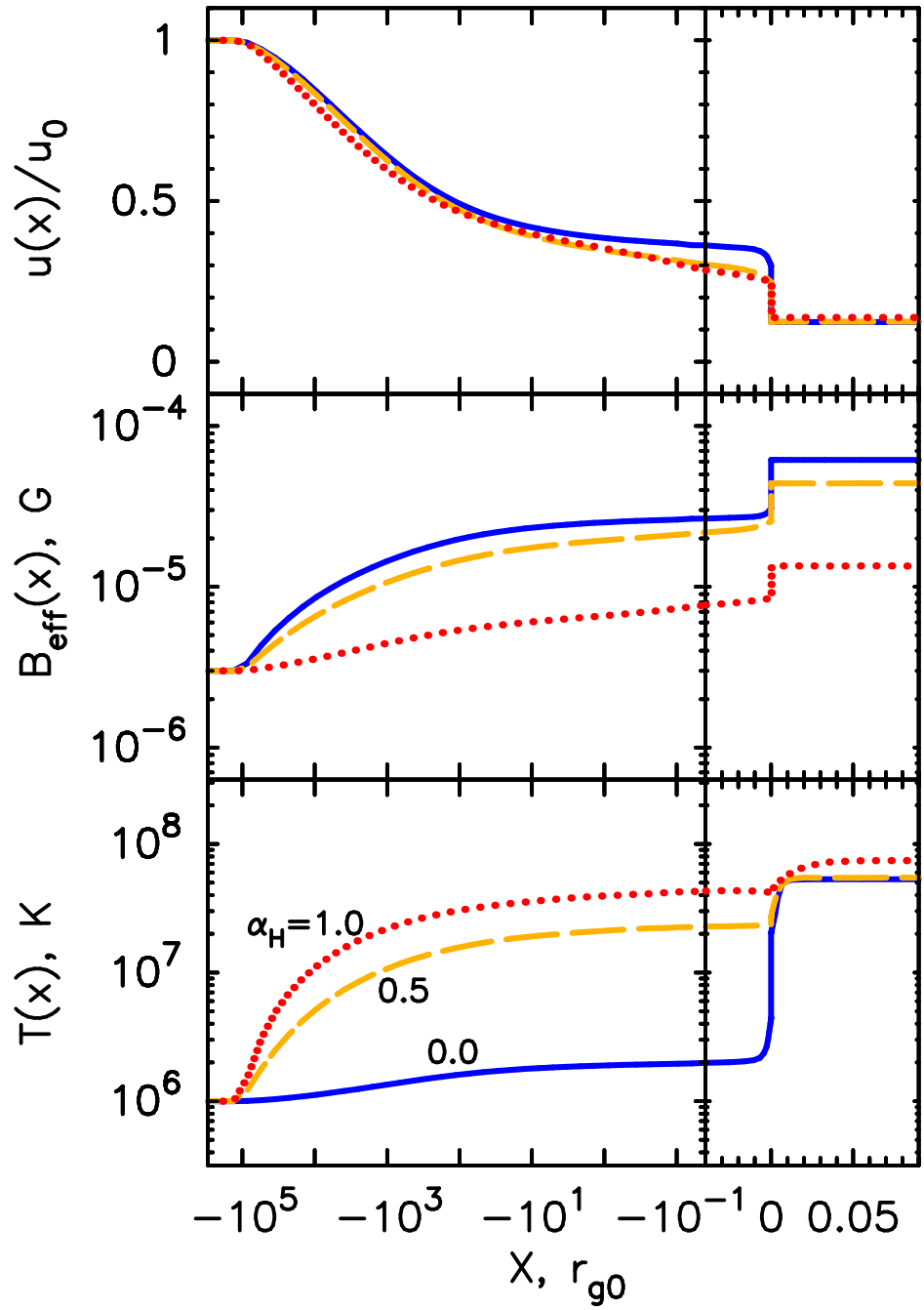}
\caption{$ $ Nonlinear shocks with dissipation, \hotismNoT \label{fig_ubt_1e4_1e6}} 
\end{figure}

\begin{figure}[htbp]
\centering
\vskip 0.5in
\includegraphics[height=7.0in]{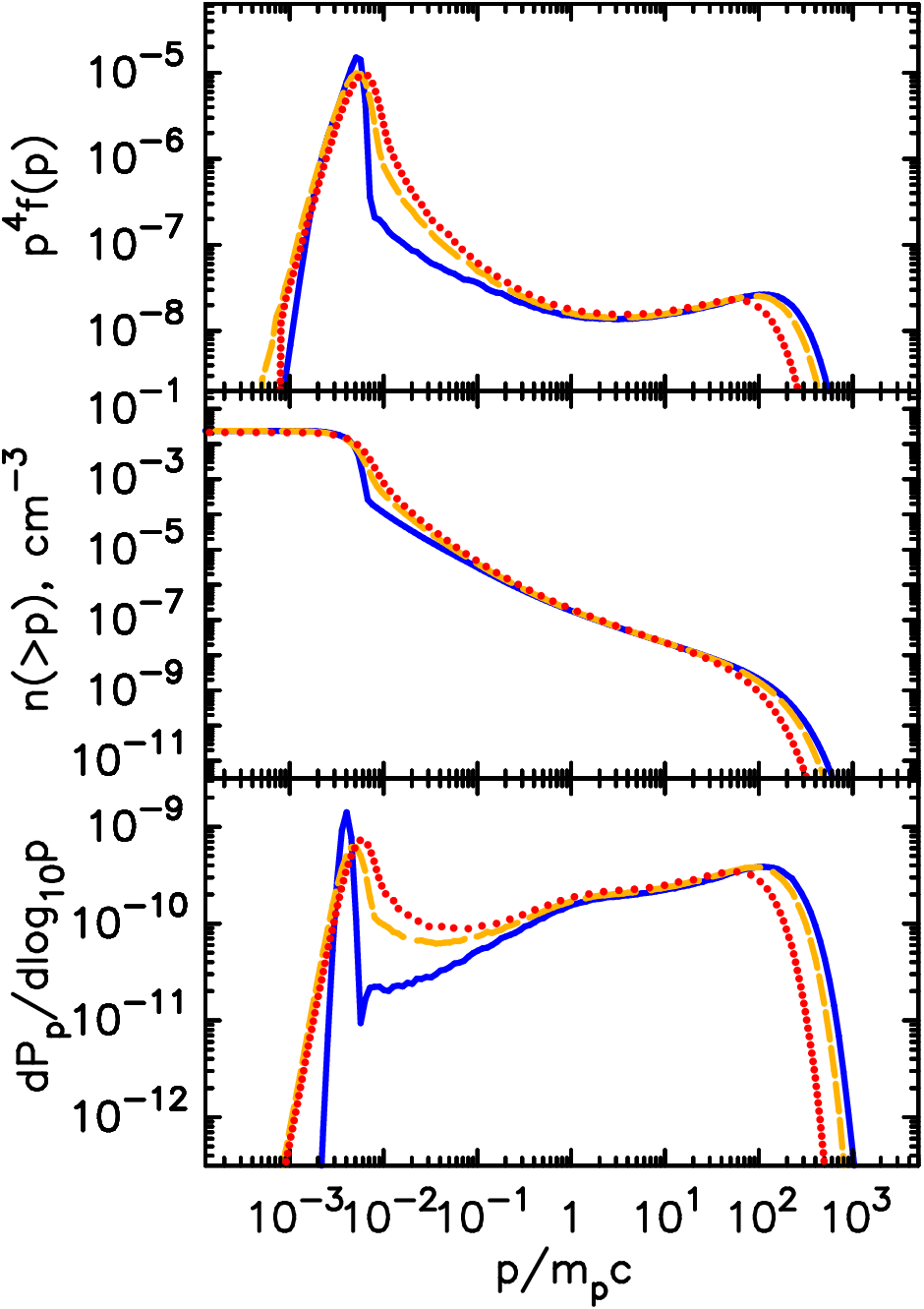}
\caption{$ $ Particle distribution with dissipation, \hotismNoT \label{fig_fnp_1e4_1e6}} 
\end{figure}

Further understanding of the shock adjustment to the changing
dissipation can be gained by studying Figures \ref{fig_ubt_1e4_1e4} -
\ref{fig_fnp_1e4_1e6}, in which we plot the spatial
structure and the momentum-dependent quantities of the shocks in the
\coldismNoT\ and the \hotismNoT\ cases for $\heatpar \in \{0; 0.5; 1\}$.

\begin{figure}[htbp]
\centering
\vskip 0.5in
\includegraphics[height=7.0in]{images/p3_f5a.eps}
\caption{$ $ Particle distribution with dissipation, \coldismNoT \label{fig_fnp_1e4_1e4}} 
\end{figure}

Figures~\ref{fig_ubt_1e4_1e4} and \ref{fig_ubt_1e4_1e6} show 
an overlap in the curves for the flow
speed $u(x)$ in the $\heatpar=0$ and $\heatpar=0.5$ models, differing only
close to the subshock, where $u(x)$ falls off more rapidly towards the subshock
in the $\heatpar=0.5$ case, resulting eventually in a lower
$\rsub$. This means that for the high energy particles, which diffuse
far upstream, the acceleration process will go on in about the same
way with and without moderate turbulence dissipation
(and the acceleration efficiency will be preserved with changing
$\heatpar$).
For lower energy particles, however, there will be observable differences in the
energy spectrum.  The $\heatpar=1.0$ case has a significantly smoother
precursor, which is not unusual, given the lower maximal energy of the
accelerated particles in this case (because of the magnetic field
remaining low).  The thermal gas temperatures $T(x)$, plotted in the
bottom panels of Figures~\ref{fig_ubt_1e4_1e4} and \ref{fig_ubt_1e4_1e6},
show that the temperature becomes high well in front of the subshock.

The low energy parts of the particle distribution functions shown in
Figures~\ref{fig_fnp_1e4_1e4} and \ref{fig_fnp_1e4_1e6} 
are significantly different for models with
and without dissipation in both the \coldismNoT\ and the \hotismNoT\
cases. The apparent widening of the thermal peak reflects the increase
in the downstream gas temperature $T_2$. The differences extend from the
thermal peak to mildly superthermal momenta $0.2\: m_p c$, indicating
an increased population of the `adolescent' particles with speeds up to
$v\approx 0.2 c \approx 12 u_0$ when the turbulence dissipation
operates. The high energy ($p>0.2 \: m_p c$) parts of the spectra for
$\heatpar=0$ and $\heatpar=0.5$ are similar (except
for a lower $\pmax$ due to a lower value of the amplified field in the
$\heatpar=0.5$ case), confirming our
assertion about the preservation of the particle acceleration
efficiency. The increased population of the low-energy particles just
above the thermal peak should influence the shock's X-ray emission.

The characteristic concave curvature of the particle spectra above
the thermal peak is clearly seen in the top panels of
Figures~\ref{fig_fnp_1e4_1e4} and \ref{fig_fnp_1e4_1e6}. 
These shocks are strongly \NL\ and, as
the pressure spectra in the bottom panels show, most of the
pressure is in the highest energy particles. For these examples, 60 to
80 percent of the downstream momentum flux is in CR particles. The
number of particles producing this pressure is small, however, and as
the plots in the middle panels show, the fraction of particles above
the thermal peak is on the order of $10^{-3}$, and the fraction of
particles above 1~GeV is around $10^{-6}$ in all cases. In addition to
the pressure (and energy) in the distributions shown, a sizable
fraction of shock ram kinetic energy flux escapes at the FEB.

\begin{figure}[hbtp]
\centering
\vskip 0.5in
\includegraphics[height=7.0in]{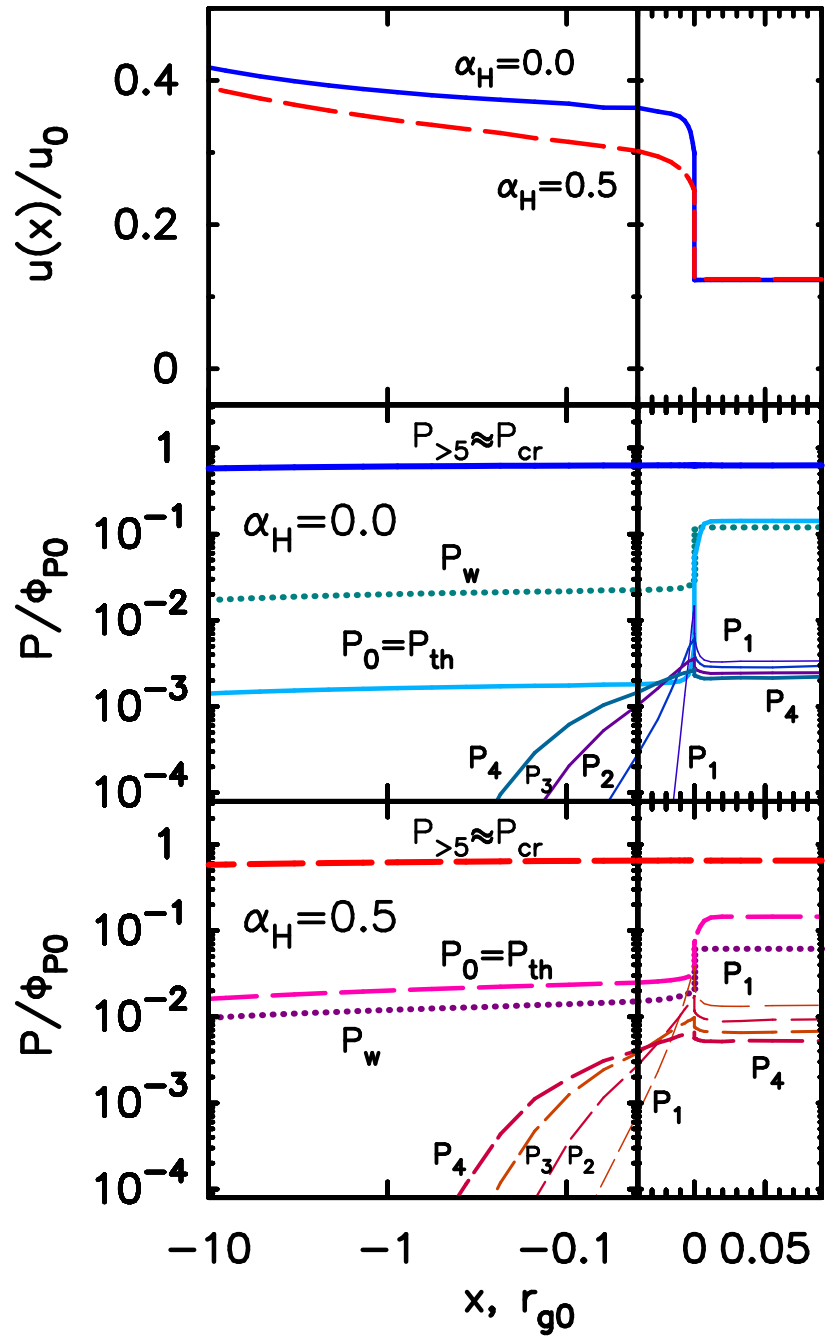}
\caption{$ $ Enlarged subshock region in the
  \hotismNoT\ case.\label{fig_partpress}}
\end{figure}

In Figure~\ref{fig_partpress} the subshock region for the \hotismNoT\
case is shown enlarged for the models with
($\heatpar=0.5$) and without dissipation ($\heatpar=0.0$).  
For $\heatpar=0$, the thermal pressure $\Pth$ remains
low upstream (middle panel), and the subshock transition is
dominated by the magnetic pressure $P_w$. For $\heatpar=0.5$
(bottom panel) the thermal pressure $\Pth$ just before the shock
becomes comparable with $P_w$, but
also the heating-boosted particle injection brings up the pressures of
the `adolescent' particles. For $\heatpar=0.5$ the
pressures produced by the first and second time returning particles
($P_1$ and $P_2$) are not small compared to $\Pth$ and $P_w$ just
upstream of the shock, which contributes to the reduction of $\rsub$
described above. However, the pressure of the `mature' particles,
$P_\mathrm{>5}$, doesn't change much, due to the
non-linear response of the shock structure to the increased injection.

To summarize,
for both the unmodified (Fig.~\ref{fig_unmod_inj}) and modified
(Fig.~\ref{fignlsum}) cases, $\Msone$ drops and
$\fcr$ grows as $\heatpar$ increases. The surprising result is that
$\Befftwo$ can increase in the unmodified shock as $\heatpar$ goes up if
$\rtot$ is large enough. This indicates that the boosted injection
efficiency (i.e., larger $\fcr$) outweighs the effects of field
damping. This doesn't happen in the modified case (top panel of
Fig.~\ref{fignlsum}) because of the \NL\ effects from the increased
injection. From Fig.~\ref{fig_partpress} we see that the boosted
injection results in a smoother subshock and this makes it harder for
low energy adolescent particles to gain energy. Once particles reach a
high enough momentum ($p\gtrsim 0.2 m_pc$; see the top panel of
Fig.~\ref{fig_fnp_1e4_1e6}) they diffuse far enough upstream
where the boost in injection has a lesser effect.

We must emphasize again that these results are very sensitive to
the physics of particle injection at the subshock. It is difficult to
predict how the \NL\ results would change if a different model of
injection was used, but we can refer the reader to the analytic
model \cite{AB2006} that uses the
threshold injection model with a different diffusion coefficient.

\subsection{Discussion}

Our two most important results are, first, that even a small rate
($\sim 10$\%) of turbulence dissipation can drastically increase the
precursor temperature, and second, that the precursor heating boosts
particle injection into DSA by a large factor. The increase in particle
injection modifies the low-energy part of the particle spectrum but, due
to \NL\ feedback effects, does not significantly change the overall
efficiency or the high energy part of the spectrum. Both the precursor
heating and modified spectral shape that occur with dissipation may have
observable consequences.

The parameterization we use here is a simple one and a more advanced
description of the turbulence damping may change our results. In our
model the energy drained from the magnetic turbulence, at all
wavelengths, is directly `pumped' into the thermal
particles. Superthermal particles only gain extra
energy due to heating because the thermal particles were more likely to
return upstream and get accelerated.  In a more advanced model of
dissipation, where energy cascades from large-scale turbulence harmonics
to the short-scale ones, the low energy CRs might gain energy directly
from the dissipation. It is conceivable that cascading effects might
increase the overall acceleration efficiency, the magnetic field
amplification, and the maximum particle energy a shock can
produce.

It is also possible that non-resonant
turbulence instabilities play an important role in magnetic
field amplification (e.g., \cite{PLM2006}).  This opens another
possibility for the turbulence dissipation to produce an increase in the
magnetic field amplification.  For instance, \cite{BT2005} proposed a
mechanism for generating long-wavelength
perturbations of magnetic fields by low energy particles. If such a
mechanism is responsible for generation of a significant fraction of the
turbulence that confines the highest energy particles, then the
increased particle injection due to the precursor heating may raise the
maximum particle energy and, possibly, the value of the amplified
magnetic field.

\newpage

\section{Bell's nonresonant instability and cascading in nonlinear model}

\label{res2009}

These results are currently in preparation for publication
by Vladimirov, Bykov and Ellison.
We will present the results of the model of nonlinear shock acceleration 
with amplification of strong stochastic magnetic fields by
Bell's nonresonant streaming instability.
We compare the assumption that the spectral
energy transfer in the generated MHD turbulence is suppressed to
the assumption that the Kolmogorov cascade determines the transfer.

The results confirm that the nonresonant instability alone
may produce a steady state shock structure with a very
strong effective magnetic field. In addition, we find that,
in the absence of cascading, the spectrum of the MHD
turbulence is not a power law, as usually assumed, but has a prominent
multiple-peak structure. The sharp peaks indicate the
presence of eddies of different distinct scales. 
Also, the precursor of the shock is no longer smooth, 
but has several layers (i.e., it is stratified), where lower
and lower energy cosmic rays are 
overtaken by the eddies and quickly accelerated.
However, if the Kolmogorov cascade is assumed, the amplification
of magnetic field is not as efficient, but the stratification
is eliminated.

We argue that the physically realistic solution 
is in between the two extreme cases that we presented here,
and discuss the consequences of both scenarios for the process
of particle acceleration by shocks and for the observable features
of emitted radiation.

\subsection{Model}

We describe the evolution of turbulence by
equation~(\ref{eq-genmfa}) with boundary
condition~(\ref{bc-genmfa}), and set $\parcompamp=\parcompwav=0$,
and $\paramplif=\parcasc=\pardiss=1$. 
For the turbulence amplification model, we
choose the Bohm nonresonant instability,
i.e., with $\nrgrowthrate$ given by
Equation~(\ref{bell_increment}).

The flux of energy along the spectrum, $\Pi(x,k)$, reflects the
cascade of turbulent structures. Cascade of MHD turbulence may 
be anisotropic \cite{GS95}, harmonics with wavenumbers transverse
to the uniform magnetic field experiencing a Kolmogorov-like
cascade, while the cascade in wavenumbers parallel to the field
is suppressed. The waves generated in the nonresonant instability 
are transverse, so the diffusion coefficient for particle 
transport parallel to the flow depends on the wavenumbers parallel to
the magnetic field. It is uncertain whether the regime in which
the instability operates will lead to a Kolmogorov cascade, or
to a suppression of the parallel cascade. We therefore consider
two extreme cases: Model A, in which the cascading is fully
suppressed, i.e., $\Pi_A=0$, and Model B,
in which the cascading is efficient and has the Kovazhny form
(e.g., \cite{VBT93}) given by equation~(\ref{pi_kolmogorov}),
i.e., $\Pi_B = \Pi_K$.
The dissipation term, $L$, is assumed to be zero for Model A,
and to have the form~(\ref{diss_ksq}) for Model B,
i.e., $L_B = L_V$.
For Model B, we also assume that the seed wave spectrum represents 
linear waves that are not subject to cascading or dissipation, and
that the transition to the turbulent regime takes place at a point
$x_0$ where the amplified wave spectrum reaches the value
$kW(x_0,k)=B_0^2/4\pi$ at some $k$. At this point,
$\Pi$ and $L$ are set from zero to the values
(\ref{pi_kolmogorov}) and (\ref{diss_ksq}). The wavenumber at which
the dissipation begins to dominate, $\kdiss$, is identified
with the inverse of a thermal proton gyroradius:
$\kdiss = eB_0/(c\sqrt{m_pk_BT})$,
where $m_p$ is the proton mass, $k_B$ is the
Boltzmann constant and $T=T(x)$ is the local gas temperature
determined from the gas heating induced by $L$,
as described in Section~\ref{turb_effects}.

Particle transport is described by the hybrid 
model of diffusion in strong turbulence, laid
out in Section~\ref{advanced_transport}.

To calculate the diffusive current $j_d(x)$, we propagate the
particles using the diffusion properties described above, and then
compute the moment of the particle distribution function
$j_d(x) = e \int v_x f(x,{\bf p}) d^3 p$
by summing over all particles crossing certain positions.

In order to determine the minimal particle gyroradius, $\rgone$, that
limits the long-wavelength generation by the instability as defined by
Equation~(\ref{bell_increment}), we
define the lowest CR momentum at the current position, $p_1$, 
as the momentum below which the CRs contribute $1\%$ of the
total CR pressure. Then $\rgone$ is defined as
$\rgone = c p_1/(e \Bls)$
with $\Bls$ calculated for the momentum $p_1$.

We use the iterative procedures~(\ref{iteration_rtot}) 
and (\ref{iteration_ux}) to
achieve a self-consistent shock structure, in which  particle
distribution, turbulence spectrum and flow structure are all
consistent with each other, and the fundamental conservation laws are fulfilled.

\newpage 
\subsection{Results}

We ran the Monte Carlo simulations of a shock with a 
speed $u_0=10^4$~\kmps\ propagating along a uniform 
magnetic field $B_0=3$~\muG\ in a plasma with a proton density 
$n_0=0.3$~cm$^{-3}$ and a temperature $T_0 = 10^4$~K. We 
assumed that the seed magnetic fluctuations have an effective value
$\Bseed=B_0$, and that the acceleration process is size-limited with
a free escape boundary located at $x=-10^7\;\rgzero$, where
$\rgzero \equiv mu_0c/eB_0 \approx 3.5\cdot 10^{10}\;\mathrm{cm}$.
Two simulations were performed, which as described in
the previous section, we will call Model~A and
Model~B. 

\begin{figure}[hbtp]
  \centering
  \vskip 0.5in
  \includegraphics[angle=-90, width=5.0in]{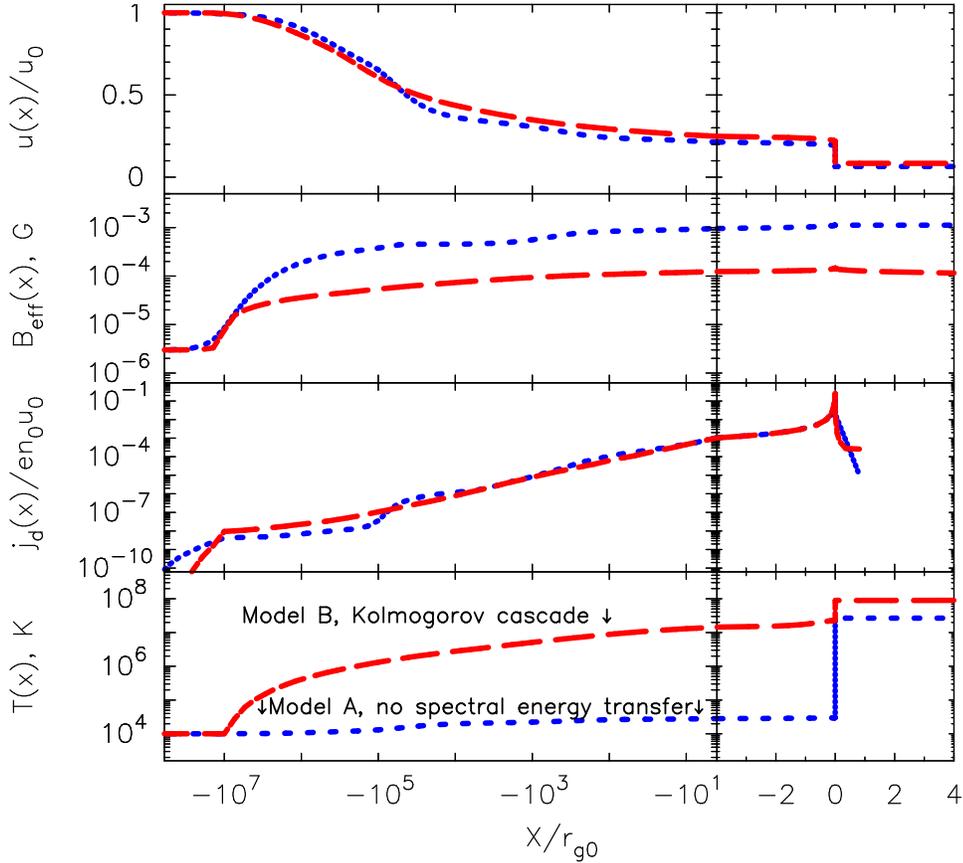}
  \caption{$ $ Shocks with turbulence generation by Bell's nonresonant
  instability
  \label{fig_ubj}}
\end{figure}

The result for model A was the steady-state structure of a 
shock modified by efficient particle acceleration and magnetic field amplification,
with a self-consistent compression ratio $\Rtot=u_0/u_2\approx 15$, 
a downstream magnetic field $\Beff(x>0)\approx 1000$~\muG, and particle acceleration
up to a maximum momentum $\pmax \approx 10^5 \; m_p c$.
Model B predicted a lower compression ratio, $\Rtot\approx 11$, lower
magnetic field $\Beff(x>0)\approx 120$~\muG, and a maximum
momentum $\pmax \approx 2\cdot 10^4 \; m_p c$.

\begin{figure}[hbtp]
\centering
\vskip 0.5in
\includegraphics[angle=-90, width=5.0in]{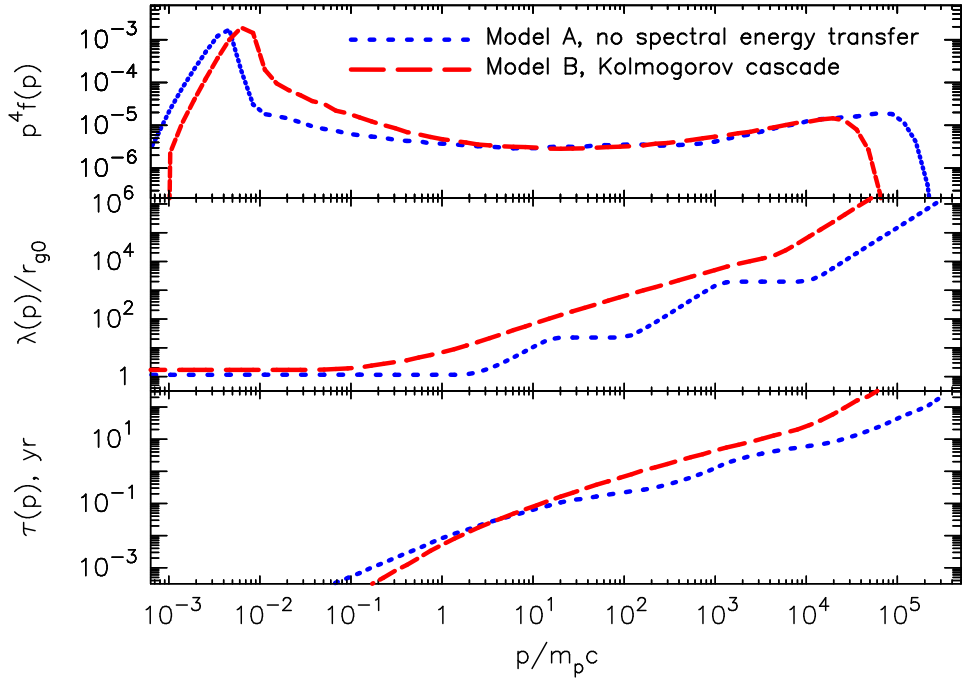}
\caption{$ $ Spectral properties of shocks shown in Figure~\ref{fig_ubj}
  \label{fig_fl}}
\end{figure}

The self-consistent structure of the shocks (the flow speeds, the 
effective magnetic field, the diffusive CR current and the thermal
plasma temperature) are shown in Figure~\ref{fig_ubj}.
Besides the above mentioned differences in the compression ratio
and the amplified magnetic fields, one may notice the difference
in the $j_d(x)$ plot. While the $j_d(x)$ curve is smooth for Model B,
it has an uneven structure for Model A, which reveals the
stratification that becomes more apparent
in the spectra of the self-generated
magnetic turbulence (see below). Another prominent difference is the significantly
increased temperature $T(x)$ in the precursor of the Model B shock,
which comes about due to the dissipation of cascading turbulence
at large $k$.

In Figure~\ref{fig_fl}, we show the particle distribution function
$f(p)$, the dependence of proton mean free path on momentum $\lambda(p)$, 
and the acceleration time to a certain momentum, $\tau(p)$.
The plots of $f(p)$ show that shocks with either model
of spectral energy transfer remain efficient particle accelerators:
the concave shape indicates the nonlinear modification of the
shock structure, also apparent in the plots of $u(x)$. The thicker
thermal peak and the higher low energy parts of the spectrum
in the model with cascading are due to the increased
turbulence dissipation, similarly to what was
observed in \cite{VBE2008}. The mean free path $\lambda(p)$ for model B
is a smooth function of $p$, with $\lambda \propto p$ for intermediate
and $\lambda \propto p^2$ for the highest energy particles, 
but for model A it has plateaus
that correspond to the trapping of particles by turbulent
vortices of different scales. A similar uneven structure
is seen in the acceleration time $\tau(p)$: it has regions of 
rapid and slow acceleration.

\begin{figure}[hbtp]
\centering
\vskip 0.5in
\includegraphics[angle=-90, width=5.0in]{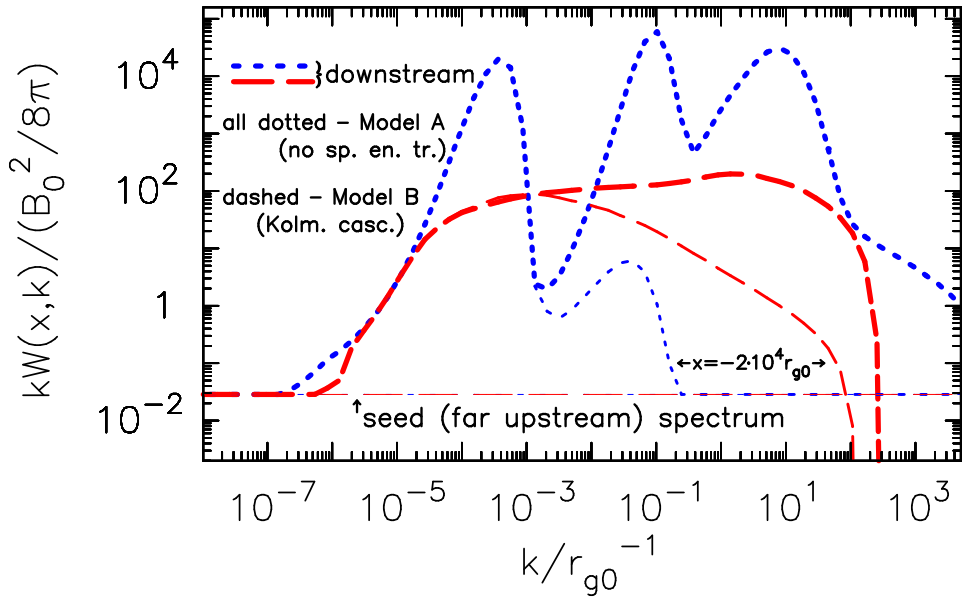}
\caption{$ $ Turbulence spectrum at  different spatial locations.
  \label{fig_wp}}
\end{figure}

The most intriguing result of this simulation is shown 
in Figure~\ref{fig_wp}. Plotted there are the turbulence
spectra, $W(x,k)$, multiplied by $k$ (so that a horizontal
line represents a Bohm spectrum $W\propto k^{-1}$).
As in the other figures, the dotted lines represent Model A,
and the dashed lines -- Model B, but here we also show the
spectra at different locations, which adds 
lines of the same styles, but different thicknesses.
The thickest lines correspond to the downstream region
$x>0$, the medium thickness lines -- a point upstream
of the subshock, $x=-2\cdot 10^{4}\;\rgzero$, and the
thinnest lines -- to the unshocked interstellar
medium (i.e., very far upstream). In both models,
the spectra of stochastic magnetic fields are
described by (\ref{eq_seed_turb}) and represented by the thin
horizontal lines. Closer to the shock, where a small
current of streaming accelerated protons is present, fluctuations
around $k=10^{-3}\;\rgzero^{-1}$ are amplified by the
nonresonant instability. In model A, the energy spectrum of these
fluctuations peaks around the value $k_c/2$ corresponding
to the maximum of $\nrgrowthrate$ from (\ref{bell_increment}),
but in model B, cascading spreads this energy
over an extended inertial range of $k$ (see the medium
thickness lines). Closer to the shock, where lower
energy particles appear, the generation of waves
at $k=10^{-3}\;\rgzero^{-1}$ shuts down, according to
the limits of applicability in (\ref{bell_increment}),
but the increased number of the streaming particles
and accordingly raised diffusive current now corresponds to a greater
$k_c$, and shorter wavelength structures get amplified,
around $k=10^{-1}\;\rgzero^{-1}$. In model A,
this results in a second peak of the turbulence
spectrum at that wavenumber, but in model B, cascading
smoothes out the spectrum. By the time the plasma reaches
the downstream region (thick lines), three distinct peaks get generated with this
mechanism in Model A, while Model obtains an
amplified turbulence spectrum close to $W\propto k^{-1}$.

The peaks occur because of the coupling of particle transport with
magnetic turbulence amplification. The first (smallest $k$) peak forms
far upstream, where only the highest energy particles are present, and
their current $j_d$ is low. 
These particles diffuse in the $\lambda\propto p^2$ regime, 
scattered by the short-scale magnetic field
fluctuations that they themselves generate. As the plasma moves toward
the subshock, advecting the turbulence with it,
lower energy particles appear. At some $x$, particles with energies low
enough to resonate with the turbulence generated farther upstream (in
the lowest $k$ peak) become dominant.  This strong resonant scattering
leads to a high gradient of $j_d$ (seen at $x \sim -10^5\;\rgzero$ in
the third panel of Fig.~\ref{fig_ubj}), and the wavenumber $k_c/2$, at
which the amplification rate $\nrgrowthrate$ has a maximum, increases
rapidly.
The increased value of $k_c/2$ leads to the emergence of the second peak
between $10^{-2}$ and $10^{-1}\,\rgzero^{-1}$, as seen in
Fig.~\ref{fig_wp}.  Similarly, the third peak is generated at distances
closer to the subshock than $\sim -2\xx{4}\,\rgzero$ and this is seen
in the thick dotted line in Fig.~\ref{fig_wp} at $k \sim
10\,\rgzero^{-1}$.

The number of peaks depends on the dynamic range, i.e., on
$\Lfeb$. A smaller $\Lfeb$ can result in two peaks, while a larger
  $\Lfeb$, and therefore a larger $\pmax$,
can yield four or more peaks in the downstream region.

The formation of the spectrum with discrete peaks occurs simultaneously
with the stratification of the shock precursor into layers (see the
plots of $j_d$), in which vortices of different scales are formed. The
peaks are a direct result of Bell's nonresonant instability, but they will
not show up unless $\lss$ and $\nrgrowthrate$ are calculated consistently,
and the simulation has a large enough dynamic range in both $k$ and $p$.

\subsection{Discussion}

Our results show that, similarly to the model with a Bohm diffusion
coefficient and a resonant streaming instability \cite{VEB2006,
  VBE2008}, the predictions of efficient particle acceleration,
shock modification and magnetic field amplification by a
large factor remain in force.
However, compared to the previous results, the precursor structure is strikingly different: 
instead of a smooth, gradual variation of all quantities in the precursor, we observe
a stratification into layers, in which vortices of distinct sizes are subsequently generated. 
The resulting turbulence spectrum has 3 sharp peaks, including
one at very short wavelengths. This stratification process is
eliminated if the rapid Kolmogorov cascade of turbulence structures is
assumed. In the latter case, the amplified turbulence spectrum
downstream becomes a power law $W\propto k^{-1}$, and the
variation of all quantities in the shock precursor reverts to being smooth.
The amplified effective magnetic field, the shock compression 
ratio and the maximum energy of accelerated particles are smaller in
the model with Kolmogorov cascade.
Cascading of MHD turbulence with respect to the
wavenumbers parallel to the mean magnetic field must be 
suppressed \cite{GS95}, and the two models (without
cascading and with the Kolmogorov cascade) should be
perceived as the extremes, between which the more
physically realistic answer lies.

If the situation without cascades and with precursor
stratification is the better approximation of physical reality,
what consequences for astrophysical observations and theory of
cosmic accelerators might it have?
The calculation performed here
derived a steady-state structure of a size-limited particle accelerator, but the information
about mean particle acceleration time, $\tacc(p)$ (bottom
panel of Figure~\ref{fig_fl}) allows a peek into the time-dependent
process. The time of acceleration to a certain momentum is, on the
average, proportional to the momentum, $\tacc \propto p$, for $p > m_p c$ is, 
but there are periods of slow acceleration, when $d\ln{\tacc}/d\ln{p}>1$,
and fast acceleration, when $d\ln{\tacc}/d\ln{p}<1$. Does it mean
that in a time-dependent calculation, one would observe quiet
periods, when the highest energy particles escape ahead of the shock
into the interstellar medium and generate large-scale turbulent vortices, 
intermittent with bursts of particle acceleration, when the lower energy particles 
are trapped by these vortices and vigorously accelerated?
The large amount of energy observed in the shortest-scale (largest $k$) peak
may influence the synchrotron radiation of electrons. Its
location was around 
$0.1 \; \rgzero \approx 3.5 \cdot 10^{9} \; \mathrm{cm}$, and it
contained roughly $1/3$ of the magnetic field energy corresponding
to the $1000$~\muG\ magnetic field. Will this rapidly varying
field affect the radio or the X-ray (e.g., \cite{BUE2008})
part of the SNR shock synchrotron spectrum?

On the other hand, if the Kolmogorov cascade is the better
representation of the spectral energy transfer in the
problem of diffusive shock acceleration, it means that
the amplified magnetic fields may not be as large as the
quasi-linear theory suggests. Also, the heating of the
shock precursor by the dissipation of turbulence must
have significant effects. Indeed, in the model
with cascades, the upstream plasma temperature $T(x)$ is
increased to values above $10^6$~K (bottom panel 
of Figure~\ref{fig_ubj}), and the accelerated particle
spectrum is elevated up to $p\approx m_p c$ with
respect to model A. These features of the solution indicate
that the X-ray emission of shocks must carry the fingerprints
of the turbulent cascade.

\newpage

\section{Fits for the nonlinear shock structure}

\label{res_scalings}

The predictions of the model: how efficient magnetic field
amplification and particle acceleration are, how much the
nonlinearly modified shocks compress and heat the medium
that they propagate in, are important for many applications
where strong shocks exist. Although nonlinear shock structure is likely
to emerge in many problems, there exists no simple description
for physicists working in other areas for to incorporating
the nonlinear effects into  calculations.

In an effort to fix this situation, I derived simple
scaling laws to replace the Hugoniot adiabat, when
nonlinearly modified shocks are considered (see also
one such scaling in \cite{Bykov2002}). I did it 
by performing the derivation of the self-consistent shock
structure for a set of input parameters
spanning a certain range.
After that, using the least squares method, I derived the
best fit coefficients for power-law scalings fitting the
obtained data.

\subsection{Model}

The nonlinear shock model used here is identical to
that described in Section~\ref{res2008}.

\subsection{Results}

I ran a total of 81 Monte Carlo simulations with different parameters and
obtained a self-consistent solution in each case. I chose
the parameter range that represents the conditions in galaxy
gluster shocks: $\heatpar=0 \dots 1$, $T_0=2\times 10^4$~K,
$B_0=0.1 \dots 1.0$~\muG, $n_0=10^{-5}\dots 10^{-4}$~\pcc,
and $u_0=1000 \dots 3000$~\kmps. For densities at the lower
end of the range, magnetic field was only varied between 
$0.1$~\muG\ and $0.5-0.6$~\muG, because \Alf\ Mach numbers 
are too low for high $B_0=1$~\muG\ for low density. The upstream
free escape boundary was chosen as $\xfeb=-10^7\,\rgzero$.

The raw data I collected are shown in the tables
\ref{clusterdata_0p0}, 
\ref{clusterdata_0p5} and 
\ref{clusterdata_1p0}. In these tables, the input parameters of the models
are listed to the left of the vertical divider:
$\heatpar$ is the dissipation rate parameter, $u_0$ is the 
shock speed, $T_0$ is the far upstream gas temperature, $n_0$ is the far
upstream plasma density, and $B_0$ is the far upstream magnetic field.
The rest of the columns are the self-consistent results of the
simulation: $\rtot$ and $\rsub$ are
the self-consistent total and subshock compression ratios, $T_1$ is the temperature
right before the subshock, 
$\Befftwo$ is the downstream amplified effective
magnetic field and $T_2$ is the downstream gas temperature.

I fitted $T_2$ and $\rtot$ from these data with power law fits in the form:
\begin{eqnarray}
\label{Tfitform}
T_2  =A u_0^{a_1}B_0^{a_2} n_0^{a_3}, \\
\label{Rfitform}
\rtot=C u_0^{c_1} B_0^{c_2} n_0^{c_3}.
\end{eqnarray}
I chose to fit the $\heatpar=0$, $\heatpar=0.5$ and $\heatpar=1$ cases separately.
The results are shown below. The temperature scales as
\begin{scriptsize}
\begin{eqnarray}
  \label{Tfit0p0}
  \left.
  \left(
  \frac{T_2}{10^6 \, \mathrm{K}} 
  \right)
  \right|_{\heatpar=0.0}
  & = & 
  0.99 
  \left( \frac{u_0}{10^3\, \mathrm{km\,s}^{-1}} \right)^{ 1.40 \pm 0.17}
  \left( \frac{B_0}{1   \, \mathrm{\mu G}     } \right)^{ 0.47 \pm 0.10}
  \left( \frac{n_0}{1   \, \mathrm{cm^{-3}}   } \right)^{-0.22 \pm 0.08}, \\
  \label{Tfit0p5}
  \left.
  \left(
  \frac{T_2}{10^6 \, \mathrm{K}} 
  \right)
  \right|_{\heatpar=0.5}
  & = & 
  1.11 
  \left( \frac{u_0}{10^3\, \mathrm{km\,s}^{-1}} \right)^{ 1.34 \pm 0.07}
  \left( \frac{B_0}{1   \, \mathrm{\mu G}     } \right)^{ 0.45 \pm 0.04}
  \left( \frac{n_0}{1   \, \mathrm{cm^{-3}}   } \right)^{-0.23 \pm 0.03},\\
  \label{Tfit1p0}
  \left.
  \left(
  \frac{T_2}{10^6 \, \mathrm{K}} 
  \right)
  \right|_{\heatpar=1.0}
  & = & 
  1.51 
  \left( \frac{u_0}{10^3\, \mathrm{km\,s}^{-1}} \right)^{ 1.40 \pm 0.06}
  \left( \frac{B_0}{1   \, \mathrm{\mu G}     } \right)^{ 0.44 \pm 0.03}
  \left( \frac{n_0}{1   \, \mathrm{cm^{-3}}   } \right)^{-0.21 \pm 0.03}.
\end{eqnarray}
\end{scriptsize}
The compression ratio scales as
\begin{scriptsize}
\begin{eqnarray}
  \label{Rfit0p0}
  \left.
  \rtot
  \right|_{\heatpar=0.0}
  & = & 
  12.9 
  \left( \frac{u_0}{10^3\, \mathrm{km\,s}^{-1}} \right)^{ 0.34 \pm 0.10}
  \left( \frac{B_0}{1   \, \mathrm{\mu G}     } \right)^{-0.25 \pm 0.06}
  \left( \frac{n_0}{1   \, \mathrm{cm^{-3}}   } \right)^{ 0.12 \pm 0.05}, \\
  \label{Rfit0p5}
  \left.
  \rtot
  \right|_{\heatpar=0.5}
  & = & 
  12.0 
  \left( \frac{u_0}{10^3\, \mathrm{km\,s}^{-1}} \right)^{ 0.37 \pm 0.06}
  \left( \frac{B_0}{1   \, \mathrm{\mu G}     } \right)^{-0.25 \pm 0.03}
  \left( \frac{n_0}{1   \, \mathrm{cm^{-3}}   } \right)^{ 0.13 \pm 0.03},\\
  \label{Rfit1p0}
  \left.
  \rtot
  \right|_{\heatpar=1.0}
  & = & 
  10.7 
  \left( \frac{u_0}{10^3\, \mathrm{km\,s}^{-1}} \right)^{ 0.35 \pm 0.04}
  \left( \frac{B_0}{1   \, \mathrm{\mu G}     } \right)^{-0.25 \pm 0.02}
  \left( \frac{n_0}{1   \, \mathrm{cm^{-3}}   } \right)^{ 0.12 \pm 0.02}.
\end{eqnarray}
\end{scriptsize}
The deviations of the parameters shown above come from the
standard least squares method and are $2\sigma$ (95\% confidence).
To estimate the fit quality, I also calculated the mean square relative error
and the maximum error of the fits in each case. For the temperature fits 
the mean square deviations of the fits 
(\ref{Tfit0p0}), (\ref{Tfit0p5}) and (\ref{Tfit1p0}) from the data 
were 18\%, 7\% and 6\%, respectively,
and the maximum errors were 41\%, 25\% and 17\%, respectively.
For the compression ratio fits 
the mean square deviations of the fits 
(\ref{Rfit0p0}), (\ref{Rfit0p5}) and (\ref{Rfit1p0}) from the data 
were 12\%, 7\% and 5\%, respectively,
and the maximum errors were 42\%, 26\% and 18\%, respectively.

\subsection{Discussion}

I calculated the self-consistent structure of nonlinear
shocks that power particle acceleration and magnetic field
amplification. The parameter range I spanned makes these
calculations applicable to cosmological shocks
\cite{BDD2008}, except for free escape boundary location, $\xfeb$.
Galaxy cluster formation shocks may have much larger
spatial scale than defined by $\xfeb$, but running the
Monte Carlo simulation with a much greater $\xfeb$ is
too time consuming. However, I argue and cans show with simulation
results that as soon as $\xfeb$ is large enough
to ensure that the fraction of ultra-relativistic
particles in the shock precursor is significant,
increasing $\xfeb$ further does not affect the
self-consistent compression ratio and the downstream
temperature too much (see also \cite{VBE2008}).

By fitting the results of a number of simulations,
I derived simple scaling laws for the downstream temperature
and the shock compression ratio, expressed by the equations
(\ref{Tfit0p0}) -- (\ref{Rfit1p0}).

These predictions are, of course, very different from the
hydrodynamic shock solution. For instance, the 
Hugoniot adiabat (\ref{hugoniot_u}) and
(\ref{hugoniot_T}) provides the following scaling
for $M_s \gg 1$:
\begin{eqnarray}
\left( \frac{T_2}{10^6\,\mathrm{K}}\right) &=&
 1.2 \cdot 10^2 \left( 
   \frac{u_0}{10^3 \, \mathrm{km\,s}^{-1}}
     \right)^2, \\
\rtot &=& 4.
\end{eqnarray}
In the fits that I found, the temperature is orders of magnitude
lower, and the compression ratio several times
greater, than in the hydrodynamic shock solution.

This method can be extended to different parameter ranges,
and our model can make similar predictions of 
macroscopic parameter scalings for shocks in other systems.
For example, the emission spectra of radiative shocks (from
the infrared to the X-ray ranges) may be influenced by
particle acceleration and magnetic field amplification.

I am grateful to A.~M.~Bykov for the idea of this
direction of research.

\begin{table}
\begin{center}
\caption{$ $ Self-consistent shock parameters for $\heatpar=0.0$}
\label{clusterdata_0p0}
\begin{scriptsizetabular}{ccccc|cccccccccc}
\hline
$\heatpar$ & $u_0$, \kmps & $T_0$, K & $n_0$, \pcc & $B_0$, G &
  $\rtot$ & $\rsub$ & $T_1$, K & $\Befftwo$, G & $T_2$, G \\
\hline  
0.0 & 1.0E+03 & 2.0E+04 & 1.0E-04 & 1.0E-07 & 7.19 & 2.87 & 3.63E+04 &  2.25E-06 & 2.70E+06 \\
0.0 & 2.0E+03 & 2.0E+04 & 1.0E-04 & 1.0E-07 & 8.50 & 2.91 & 4.06E+04 &  3.84E-06 & 7.94E+06 \\
0.0 & 3.0E+03 & 2.0E+04 & 1.0E-04 & 1.0E-07 & 8.96 & 2.94 & 4.25E+04 &  5.13E-06 & 1.64E+07 \\
0.0 & 1.0E+03 & 2.0E+04 & 1.0E-04 & 3.0E-07 & 6.61 & 2.86 & 3.50E+04 &  2.76E-06 & 3.16E+06 \\
0.0 & 2.0E+03 & 2.0E+04 & 1.0E-04 & 3.0E-07 & 7.13 & 2.89 & 3.71E+04 &  4.70E-06 & 1.10E+07 \\
0.0 & 3.0E+03 & 2.0E+04 & 1.0E-04 & 3.0E-07 & 8.12 & 2.92 & 4.09E+04 &  6.82E-06 & 1.94E+07 \\
0.0 & 1.0E+03 & 2.0E+04 & 1.0E-04 & 1.0E-06 & 3.92 & 2.81 & 2.54E+04 &  3.30E-06 & 8.29E+06 \\
0.0 & 2.0E+03 & 2.0E+04 & 1.0E-04 & 1.0E-06 & 5.69 & 2.89 & 3.28E+04 &  5.74E-06 & 1.70E+07 \\
0.0 & 3.0E+03 & 2.0E+04 & 1.0E-04 & 1.0E-06 & 6.39 & 2.91 & 3.69E+04 &  8.20E-06 & 3.07E+07 \\
0.0 & 1.0E+03 & 2.0E+04 & 3.0E-05 & 1.0E-07 & 6.95 & 2.85 & 3.60E+04 &  1.31E-06 & 2.83E+06 \\
0.0 & 2.0E+03 & 2.0E+04 & 3.0E-05 & 1.0E-07 & 8.05 & 2.92 & 4.00E+04 &  2.57E-06 & 8.90E+06 \\
0.0 & 3.0E+03 & 2.0E+04 & 3.0E-05 & 1.0E-07 & 9.01 & 2.91 & 4.37E+04 &  3.42E-06 & 1.57E+07 \\
0.0 & 1.0E+03 & 2.0E+04 & 3.0E-05 & 3.0E-07 & 5.68 & 2.83 & 3.22E+04 &  1.65E-06 & 4.13E+06 \\
0.0 & 2.0E+03 & 2.0E+04 & 3.0E-05 & 3.0E-07 & 6.81 & 2.90 & 3.66E+04 &  2.98E-06 & 1.20E+07 \\
0.0 & 3.0E+03 & 2.0E+04 & 3.0E-05 & 3.0E-07 & 7.20 & 2.92 & 3.90E+04 &  4.14E-06 & 2.43E+07 \\
0.0 & 1.0E+03 & 2.0E+04 & 3.0E-05 & 6.0E-07 & 3.62 & 2.80 & 2.42E+04 &  1.85E-06 & 9.53E+06 \\
0.0 & 2.0E+03 & 2.0E+04 & 3.0E-05 & 6.0E-07 & 5.60 & 2.88 & 3.28E+04 &  3.19E-06 & 1.73E+07 \\
0.0 & 3.0E+03 & 2.0E+04 & 3.0E-05 & 6.0E-07 & 6.13 & 2.92 & 3.67E+04 &  4.52E-06 & 3.32E+07 \\
0.0 & 1.0E+03 & 2.0E+04 & 1.0E-05 & 1.0E-07 & 6.57 & 2.83 & 3.56E+04 &  8.83E-07 & 3.11E+06 \\
0.0 & 2.0E+03 & 2.0E+04 & 1.0E-05 & 1.0E-07 & 7.52 & 2.90 & 3.86E+04 &  1.60E-06 & 9.99E+06 \\
0.0 & 3.0E+03 & 2.0E+04 & 1.0E-05 & 1.0E-07 & 8.00 & 2.91 & 4.09E+04 &  2.24E-06 & 1.99E+07 \\
0.0 & 1.0E+03 & 2.0E+04 & 1.0E-05 & 3.0E-07 & 4.02 & 2.82 & 2.59E+04 &  1.02E-06 & 7.89E+06 \\
0.0 & 2.0E+03 & 2.0E+04 & 1.0E-05 & 3.0E-07 & 5.96 & 2.88 & 3.41E+04 &  1.83E-06 & 1.55E+07 \\
0.0 & 3.0E+03 & 2.0E+04 & 1.0E-05 & 3.0E-07 & 6.41 & 2.91 & 3.77E+04 &  2.54E-06 & 3.03E+07 \\
0.0 & 1.0E+03 & 2.0E+04 & 1.0E-05 & 5.0E-07 & 2.63 & 2.60 & 2.01E+04 &  1.20E-06 & 1.47E+07 \\
0.0 & 2.0E+03 & 2.0E+04 & 1.0E-05 & 5.0E-07 & 4.51 & 2.88 & 2.95E+04 &  1.95E-06 & 2.62E+07 \\
0.0 & 3.0E+03 & 2.0E+04 & 1.0E-05 & 5.0E-07 & 5.49 & 2.92 & 3.51E+04 &  2.75E-06 & 4.09E+07 \\
\hline
\end{scriptsizetabular}
\end{center}
\end{table}

\begin{table}
\begin{center}
\caption{$ $ Self-consistent shock parameters for $\heatpar=0.5$}
\label{clusterdata_0p5}
\begin{scriptsizetabular}{ccccc|cccccccccc}
\hline
$\heatpar$ & $u_0$, \kmps & $T_0$, K & $n_0$, \pcc & $B_0$, G &
  $\rtot$ & $\rsub$ & $T_1$, K & $\Befftwo$, G & $T_2$, G \\
\hline  
0.5 & 1.0E+03 & 2.0E+04 & 1.0E-04 & 1.0E-07 & 6.68 & 2.36 & 1.28E+06 &  1.54E-06 & 3.30E+06 \\
0.5 & 2.0E+03 & 2.0E+04 & 1.0E-04 & 1.0E-07 & 7.95 & 2.55 & 1.95E+06 &  2.36E-06 & 8.68E+06 \\
0.5 & 3.0E+03 & 2.0E+04 & 1.0E-04 & 1.0E-07 & 8.74 & 2.62 & 2.87E+06 &  3.11E-06 & 1.62E+07 \\
0.5 & 1.0E+03 & 2.0E+04 & 1.0E-04 & 3.0E-07 & 5.39 & 2.24 & 2.18E+06 &  1.68E-06 & 4.91E+06 \\
0.5 & 2.0E+03 & 2.0E+04 & 1.0E-04 & 3.0E-07 & 6.73 & 2.34 & 4.71E+06 &  3.07E-06 & 1.24E+07 \\
0.5 & 3.0E+03 & 2.0E+04 & 1.0E-04 & 3.0E-07 & 7.44 & 2.42 & 7.49E+06 &  4.43E-06 & 2.27E+07 \\
0.5 & 1.0E+03 & 2.0E+04 & 1.0E-04 & 1.0E-06 & 3.65 & 2.40 & 2.40E+06 &  2.99E-06 & 8.58E+06 \\
0.5 & 2.0E+03 & 2.0E+04 & 1.0E-04 & 1.0E-06 & 4.96 & 2.26 & 9.29E+06 &  4.19E-06 & 2.18E+07 \\
0.5 & 3.0E+03 & 2.0E+04 & 1.0E-04 & 1.0E-06 & 5.72 & 2.26 & 1.67E+07 &  5.66E-06 & 3.82E+07 \\
0.5 & 1.0E+03 & 2.0E+04 & 3.0E-05 & 1.0E-07 & 5.98 & 2.29 & 1.68E+06 &  8.36E-07 & 3.97E+06 \\
0.5 & 2.0E+03 & 2.0E+04 & 3.0E-05 & 1.0E-07 & 7.45 & 2.40 & 3.53E+06 &  1.59E-06 & 1.02E+07 \\
0.5 & 3.0E+03 & 2.0E+04 & 3.0E-05 & 1.0E-07 & 8.32 & 2.48 & 4.85E+06 &  2.10E-06 & 1.77E+07 \\
0.5 & 1.0E+03 & 2.0E+04 & 3.0E-05 & 3.0E-07 & 4.66 & 2.21 & 2.74E+06 &  1.16E-06 & 6.06E+06 \\
0.5 & 2.0E+03 & 2.0E+04 & 3.0E-05 & 3.0E-07 & 5.94 & 2.25 & 7.20E+06 &  1.92E-06 & 1.59E+07 \\
0.5 & 3.0E+03 & 2.0E+04 & 3.0E-05 & 3.0E-07 & 6.63 & 2.29 & 1.21E+07 &  2.77E-06 & 2.89E+07 \\
0.5 & 1.0E+03 & 2.0E+04 & 3.0E-05 & 6.0E-07 & 3.44 & 2.46 & 2.04E+06 &  1.72E-06 & 9.52E+06 \\
0.5 & 2.0E+03 & 2.0E+04 & 3.0E-05 & 6.0E-07 & 4.82 & 2.25 & 9.95E+06 &  2.43E-06 & 2.28E+07 \\
0.5 & 3.0E+03 & 2.0E+04 & 3.0E-05 & 6.0E-07 & 5.58 & 2.30 & 1.68E+07 &  3.14E-06 & 3.98E+07 \\
0.5 & 1.0E+03 & 2.0E+04 & 1.0E-05 & 1.0E-07 & 5.36 & 2.20 & 2.30E+06 &  5.46E-07 & 5.05E+06 \\
0.5 & 2.0E+03 & 2.0E+04 & 1.0E-05 & 1.0E-07 & 6.70 & 2.29 & 5.58E+06 &  1.04E-06 & 1.29E+07 \\
0.5 & 3.0E+03 & 2.0E+04 & 1.0E-05 & 1.0E-07 & 7.52 & 2.36 & 7.68E+06 &  1.42E-06 & 2.16E+07 \\
0.5 & 1.0E+03 & 2.0E+04 & 1.0E-05 & 3.0E-07 & 3.78 & 2.38 & 2.47E+06 &  9.24E-07 & 8.21E+06 \\
0.5 & 2.0E+03 & 2.0E+04 & 1.0E-05 & 3.0E-07 & 4.96 & 2.27 & 9.51E+06 &  1.30E-06 & 2.19E+07 \\
0.5 & 3.0E+03 & 2.0E+04 & 1.0E-05 & 3.0E-07 & 5.77 & 2.26 & 1.64E+07 &  1.77E-06 & 3.81E+07 \\
0.5 & 1.0E+03 & 2.0E+04 & 1.0E-05 & 5.0E-07 & 2.62 & 2.58 & 1.41E+05 &  1.20E-06 & 1.46E+07 \\
0.5 & 2.0E+03 & 2.0E+04 & 1.0E-05 & 5.0E-07 & 4.17 & 2.37 & 9.86E+06 &  1.68E-06 & 2.84E+07 \\
0.5 & 3.0E+03 & 2.0E+04 & 1.0E-05 & 5.0E-07 & 4.93 & 2.30 & 2.04E+07 &  2.09E-06 & 4.86E+07 \\
\hline
\end{scriptsizetabular}
\end{center}
\end{table}

\begin{table}
\begin{center}
\caption{$ $ Self-consistent shock parameters for $\heatpar=1.0$}
\label{clusterdata_1p0}
\begin{scriptsizetabular}{ccccc|cccccccccc}
\hline
$\heatpar$ & $u_0$, \kmps & $T_0$, K & $n_0$, \pcc & $B_0$, G &
  $\rtot$ & $\rsub$ & $T_1$, K & $\Befftwo$, G & $T_2$, G \\
\hline  
1.0 & 1.0E+03 & 2.0E+04 & 1.0E-04 & 1.0E-07 & 5.86 & 2.12 & 2.63E+06 &  3.82E-07 & 4.72E+06 \\
1.0 & 2.0E+03 & 2.0E+04 & 1.0E-04 & 1.0E-07 & 7.52 & 2.29 & 4.66E+06 &  4.57E-07 & 1.04E+07 \\
1.0 & 3.0E+03 & 2.0E+04 & 1.0E-04 & 1.0E-07 & 8.35 & 2.44 & 5.96E+06 &  4.93E-07 & 1.84E+07 \\
1.0 & 1.0E+03 & 2.0E+04 & 1.0E-04 & 3.0E-07 & 5.08 & 2.07 & 3.46E+06 &  1.03E-06 & 5.88E+06 \\
1.0 & 2.0E+03 & 2.0E+04 & 1.0E-04 & 3.0E-07 & 5.91 & 2.02 & 1.10E+07 &  1.16E-06 & 1.82E+07 \\
1.0 & 3.0E+03 & 2.0E+04 & 1.0E-04 & 3.0E-07 & 6.96 & 2.23 & 1.36E+07 &  1.30E-06 & 2.80E+07 \\
1.0 & 1.0E+03 & 2.0E+04 & 1.0E-04 & 1.0E-06 & 3.43 & 2.25 & 4.06E+06 &  2.74E-06 & 1.00E+07 \\
1.0 & 2.0E+03 & 2.0E+04 & 1.0E-04 & 1.0E-06 & 4.49 & 2.07 & 1.66E+07 &  3.18E-06 & 2.89E+07 \\
1.0 & 3.0E+03 & 2.0E+04 & 1.0E-04 & 1.0E-06 & 5.17 & 2.09 & 3.01E+07 &  3.50E-06 & 5.12E+07 \\
1.0 & 1.0E+03 & 2.0E+04 & 3.0E-05 & 1.0E-07 & 5.48 & 2.07 & 3.07E+06 &  3.63E-07 & 5.16E+06 \\
1.0 & 2.0E+03 & 2.0E+04 & 3.0E-05 & 1.0E-07 & 6.87 & 2.25 & 5.70E+06 &  4.26E-07 & 1.24E+07 \\
1.0 & 3.0E+03 & 2.0E+04 & 3.0E-05 & 1.0E-07 & 7.69 & 2.28 & 9.80E+06 &  4.67E-07 & 2.22E+07 \\
1.0 & 1.0E+03 & 2.0E+04 & 3.0E-05 & 3.0E-07 & 4.25 & 2.08 & 4.74E+06 &  9.19E-07 & 7.98E+06 \\
1.0 & 2.0E+03 & 2.0E+04 & 3.0E-05 & 3.0E-07 & 5.32 & 2.08 & 1.30E+07 &  1.07E-06 & 2.19E+07 \\
1.0 & 3.0E+03 & 2.0E+04 & 3.0E-05 & 3.0E-07 & 5.96 & 2.10 & 2.18E+07 &  1.16E-06 & 3.89E+07 \\
1.0 & 1.0E+03 & 2.0E+04 & 3.0E-05 & 6.0E-07 & 3.32 & 2.38 & 3.47E+06 &  1.63E-06 & 1.05E+07 \\
1.0 & 2.0E+03 & 2.0E+04 & 3.0E-05 & 6.0E-07 & 4.42 & 2.12 & 1.67E+07 &  1.90E-06 & 2.93E+07 \\
1.0 & 3.0E+03 & 2.0E+04 & 3.0E-05 & 6.0E-07 & 5.01 & 2.11 & 3.21E+07 &  2.06E-06 & 5.47E+07 \\
1.0 & 1.0E+03 & 2.0E+04 & 1.0E-05 & 1.0E-07 & 4.95 & 2.04 & 3.85E+06 &  3.40E-07 & 6.33E+06 \\
1.0 & 2.0E+03 & 2.0E+04 & 1.0E-05 & 1.0E-07 & 6.04 & 2.12 & 1.01E+07 &  3.92E-07 & 1.75E+07 \\
1.0 & 3.0E+03 & 2.0E+04 & 1.0E-05 & 1.0E-07 & 6.96 & 2.18 & 1.42E+07 &  4.32E-07 & 2.81E+07 \\
1.0 & 1.0E+03 & 2.0E+04 & 1.0E-05 & 3.0E-07 & 3.56 & 2.29 & 4.12E+06 &  8.44E-07 & 9.65E+06 \\
1.0 & 2.0E+03 & 2.0E+04 & 1.0E-05 & 3.0E-07 & 4.55 & 2.06 & 1.68E+07 &  9.66E-07 & 2.88E+07 \\
1.0 & 3.0E+03 & 2.0E+04 & 1.0E-05 & 3.0E-07 & 5.24 & 2.09 & 2.95E+07 &  1.06E-06 & 4.95E+07 \\
1.0 & 1.0E+03 & 2.0E+04 & 1.0E-05 & 5.0E-07 & 2.62 & 2.56 & 3.71E+05 &  1.20E-06 & 1.44E+07 \\
1.0 & 2.0E+03 & 2.0E+04 & 1.0E-05 & 5.0E-07 & 3.87 & 2.25 & 1.64E+07 &  1.47E-06 & 3.47E+07 \\
1.0 & 3.0E+03 & 2.0E+04 & 1.0E-05 & 5.0E-07 & 4.51 & 2.17 & 3.49E+07 &  1.60E-06 & 6.35E+07 \\
\hline
\end{scriptsizetabular}
\end{center}
\end{table}

\newpage

\section{Spectrum and angular distribution of escaping particles}

\label{res_angular}

Using our model, we calculated the spectra
of particles escaping from the shock at the free escape boundary.
We provide simple fits to the energy and angular 
distribution of the escaping particles.

\subsection{Model}

For Bohm diffusion, the momentum distribution $f(p)$ of the escaping
particles can be described as 
\begin{equation}
\label{escfofp}
f(p) \propto p^{-s} 
\frac{
  \exp{\left(
    -s \int\limits_{0}^{p/\pmax} \frac{dx/x}{e^{1/x}-1}
    \right)}
}
{
  \exp{\left(\pmax/p\right)} - 1
}.
\end{equation}
Here $s$ is the power-law index corresponding to the compression
ratio $r$, $s=3r/(r-1)$ and $\pmax$ is defined below. The
numerator describes the exponential turn-over of the high energy
particles, and the denominator
describes the low-energy part of the escaping particle distribution.
Equation (\ref{escfofp}) is similar 
to equations (7) and (8) of \cite{ZP2008},
but assumes a diffusion coefficient $D(p)\propto p$, 
as opposed to $D(p) \propto p^2$ assumed in \cite{ZP2008}.

To find the normalization of the escaping particle distribution
$f(p)$, one needs to use the quantity $\qesc$ self-consistently
defined by the simulation as (\ref{qescrtot}):
\begin{equation}
4\pi \int\limits_0^{\infty} p^2 dp \int\limits_{-1}^0 d\mu
f(p) g(\mu) c p = -\qesc \rho_0 u_0^2,
\end{equation}
where $\qesc$ is the fraction of energy flux carried away by escaping
particles, and $g(\mu)$ is their angular distribution. The latter
function is defined so that $\int_{-1}^{+1}g(\mu)d\mu=1$,
and $\mu = p_x/p$.

The quantity $\pmax$ (the maximum particle momentum) can
be estimated from the test-particle theory of particle
acceleration as 
$\pmax \approx 3 u_0 e B_0 |\xfeb|/c^2$ (i.e., the momentum
at which the Bohm diffusion length equals $|\xfeb|$).
The function $g(\mu)$ is the distribution of particles
incident on a fully absorbing boundary in a flow moving
at a speed $u$. It can be estimated
using the Monte Carlo simulation, as shown
below.

\newpage

\subsection{Results}

\begin{figure}[hbtp] 
\centering
\vskip 0.5in
\includegraphics[width=5.0in, trim = 0.0in 7.8in 0.0in 0.29in, clip=true]{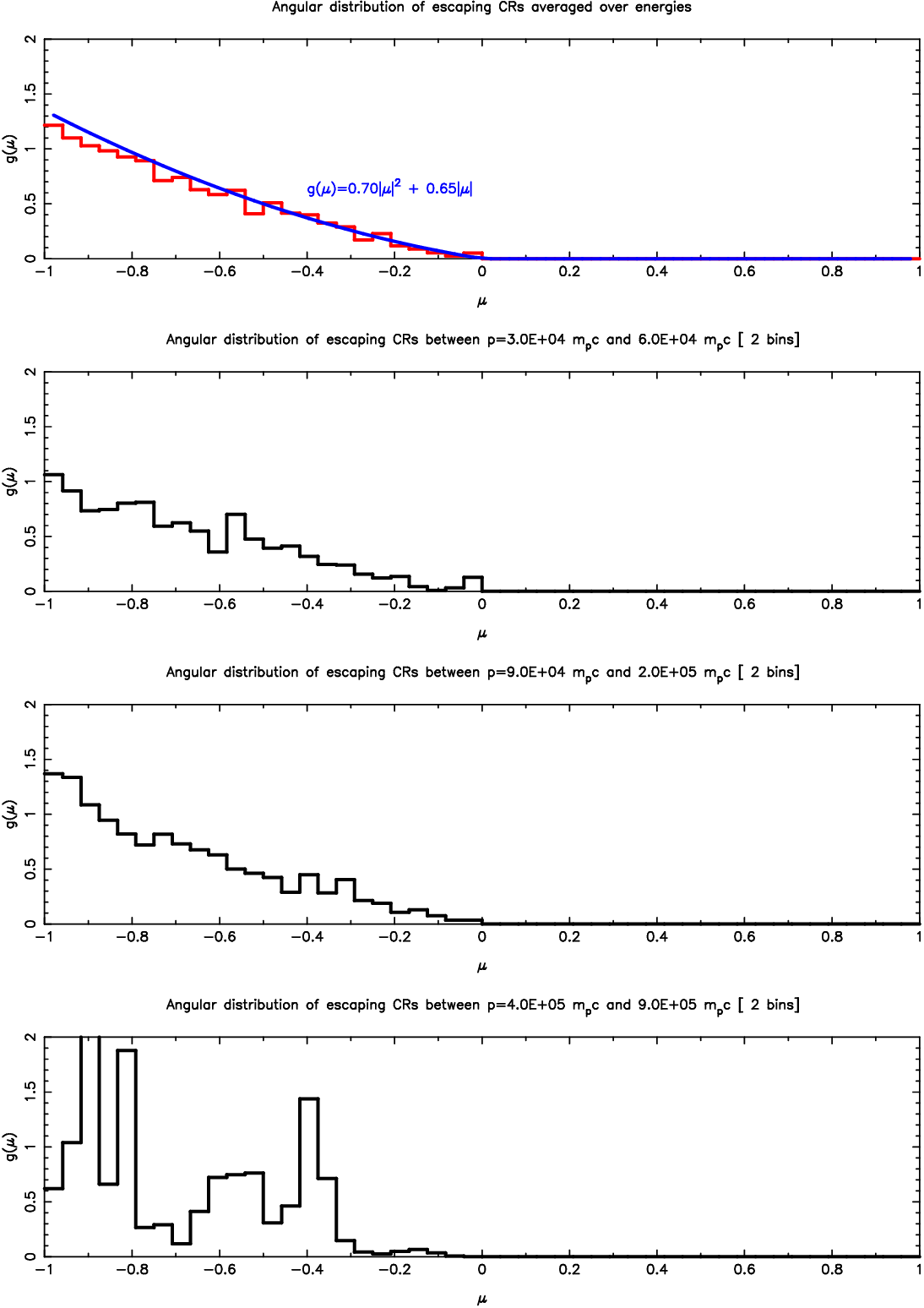}
\vskip -0.4in
\caption{$ $ Angular distribution of escaping particles\label{escmu}}
%
%
\vskip 0.5in
\includegraphics[width=5.0in, trim = 0.0in 0.0in 0.0in 8.1in, clip=true]{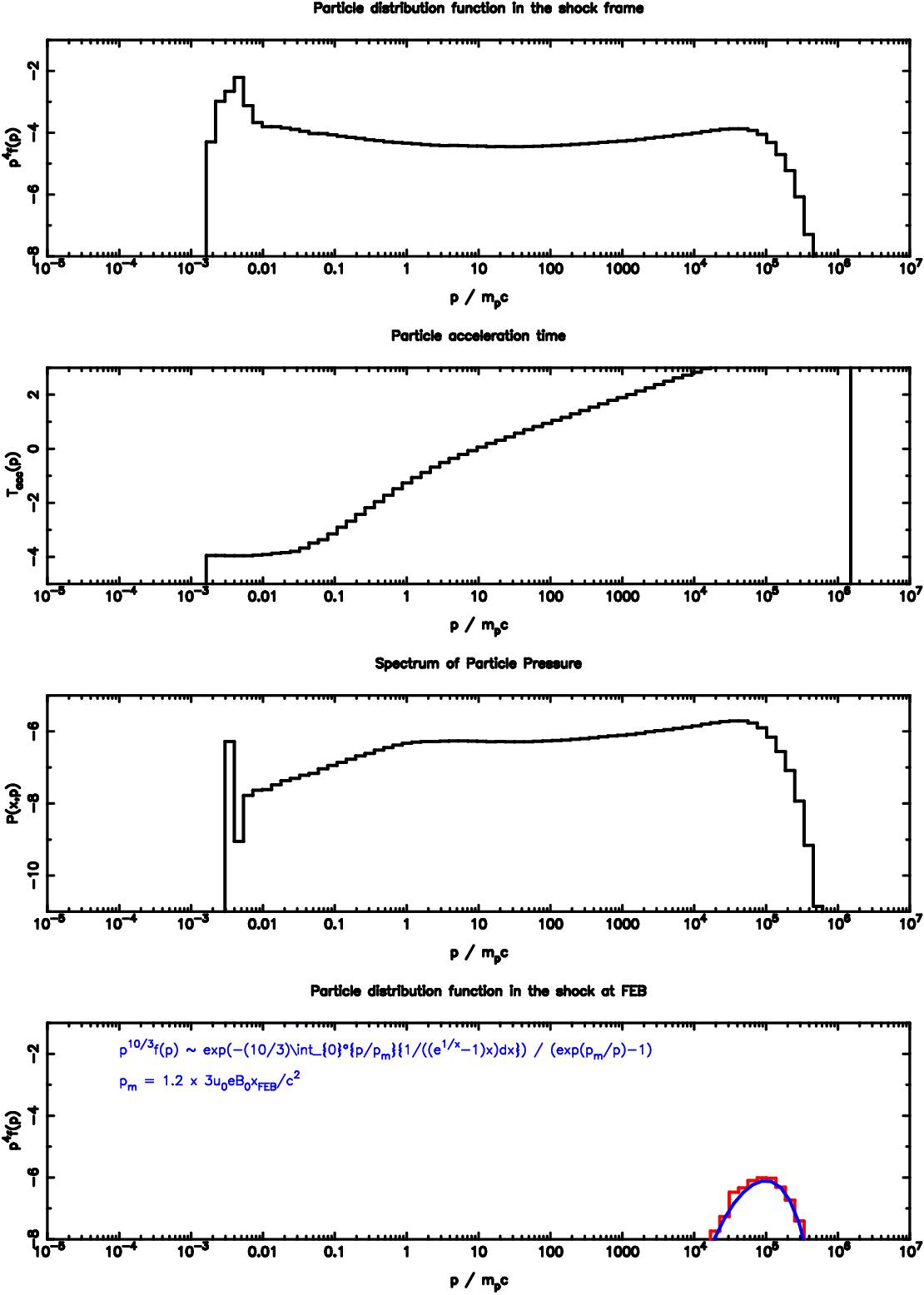}
\vskip -0.3in
\caption{$ $ Momentum distribution of escaping particles\label{escmom}}
\end{figure}

I ran a simulation of a nonlinearly modified shock with a 
speed $u_0=5000\;$\kmps, compression
ratio $r\approx 10$, no magnetic field amplification,
and Bohm model of diffusion. Then I plotted and fitted the
particle distribution (angular and momentum-space) 
determined by the simulation at the free escape
boundary located at $\xfeb$, as
as shown in Figures~\ref{escmu} and \ref{escmom}.
The histograms shown in these figures are the results of the Monte
Carlo simulation, and the smooth lines are the fits, equations for
which are provided in the figures.

It turns out that the angular distribution of the particles
can successfully be fitted with the equation shown in
Figure~\ref{escmu}:
\begin{equation}
g(\mu) = \left\{ 
      \begin{array}{l}
         0.70 |\mu|^2 + 0.65 |\mu|, \; \mathrm{if} \; \mu<0, \\
         0, \; \mathrm{if} \; \mu>0.
      \end{array}
      \right.
\end{equation}
This represents the case when all the particles are moving to
the left, i.e., away from the shock (because $g(\mu>0)=0$).
The momentum distribution of escaping particles has the
shape described by equation~(\ref{escfofp}) with
\begin{equation}
\pmax = 1.2 \frac{3 u_0 e B_0}{c^2}|\xfeb|.
\end{equation}
The factor $1.2$ is a minor correction to the above
mentioned analytic estimate.

I would like to thank T.~Kamae and S.-H.~(Herman)~Lee
for the discussions that have led to the results presented here.

\subsection{Discussion}

I used our model to quantitatively describe the
spectrum and the anisotropic angular distribution
of particles leaving the shock at the upstream
free escape boundary. The calculations accounted for
the nonlinear modification of the shock by
efficient particle acceleration.

While the particular results derived here 
have limited applicability, because they
apply to just a single set of parameters
of a plane shock, I provided them to indicate a possible
direction of research applying the model presented
in this dissertation.

One application  may be the description
of the interaction between the CRs escaping from a
shock and the interstellar medium. For instance,
the streaming of these particles, carrying a large
fraction of the shock's energy flux, may amplify
magnetic field fluctuations in the interstellar medium.

Another interesting astrophysical application of these results is
the recently discovered diffuse gamma ray sources identified
as molecular clouds illuminated by cosmic ray particles
produced in a nearby supernova remnant shock wave (e.g.,
\cite{IC443_MAGIC}). A major uncertainty for the interpretation
of these observations is whether the particle accelerating
shocks are traversing through the cloud, or located far away
from it (see, e.g., \cite{GAB2007}). In the first case, the angular distribution of 
accelerated protons is far from isotropic, while in the second
case the CRs may have had time to isotropize in the interstellar
medium before they reach the cloud. The gamma ray emission
of these protons via the decay of pi-mesons produced in
collisions with the cloud gas protons will be different
in these two cases, because the relativistic process produces
gamma rays with strong dependence of energy spectrum on the
angle of emission.

\chapter{Conclusions}

I have developed a 
model of nonlinear shock acceleration
that self-consistently
includes the amplification of
stochastic magnetic fields in the shock precursor by
the accelerated particles produced in the first order
Fermi process. The model is based on the Monte Carlo simulation
of particle transport developed by Ellison and colleagues,
and my contribution to the model was the incorporation
of the analytic description of magnetic turbulence
amplification and evolution, and the implementation 
of particle transport consistent with the generated
magnetic turbulence.

In this dissertation, I provided the details of the
model to a degree that, I believe, make it reproducible.
I presented the tests of the various parts of the
simulation, which confirm that the results of the
computer code I built agree with the known analytic results.
This dissertation also contains an outline
of our three refereed publications, in which we presented 
the applications of our model. It also features some
results that have not yet been published.

The applicability of the Monte Carlo model
to shock acceleration in space was tested
well before my work in this project by
Ellison, Baring and others
\cite{Ellison85, BOEF97}, who used the most direct data available -- 
spacecraft observations of the Earth's
bow shock and of interplanetary shocks. The physical correctness
of the magnetic field amplification that I implemented
is yet to be tested. Nevertheless, the predictions
of the model are able to explain the observations
that inspired it (i.e., the large magnetic fields
and increased shock compression ratios in SNRs).

The most important limitations of the model
are the uncertainty of the extrapolations of
linear models of magnetic 
turbulence evolution into the nonlinear
regime, and the statistical description of particle transport
in stochastic magnetic fields,
which is subject to various conjectures.
However, the strength of the presented model
compared to the simplified analytic treatments of particle acceleration
and magnetic field amplification in shocks is
the self-consistency. Our model allows one
to determine the shock structure, the accelerated
particle spectrum and the turbulence generation
all consistently with each other, through
an iterative procedure. Even compared
to the advanced analytic nonlinear models of particle
acceleration (e.g., \cite{AB2006}), our
simulation stands out because the Monte Carlo technique can handle
anisotropic particle distributions,
which is essential for a more precise description of plasma physics;
for instance, the injection of particles into the
acceleration process is predicted self-consistently in
our model, yet it requires an additional parameter in others.
Also, the inclusion of various factors that determine
the plasma physics (e.g., turbulent cascades)
is straightforward in our approach, and may be complicated
in the analytic calculations.

The applications of the model developed here are numerous
and exist wherever strong non-relativistic
collisionless shocks are present.
This includes, to some degree, shocks in interplanetary space,
supernova remnants, shocks in galaxy clusters, etc.
Our results help answer 
questions regarding the sources of galactic cosmic rays
up to the `knee' of the CR spectrum,  and
they may be used in the modeling of supernova remnants,
galaxy clusters and other objects.

The dissertation contains the results of the model
in Chapter~\ref{ch-applications}. Sections~\ref{res2006}
presents our first refereed publication \cite{VEB2006}
featuring the model. In this work we studied the self-consistent
structure of particle accelerating shocks in the presence
of the resonant CR streaming instability 
(see Section~\ref{res_desc}). We confirmed that
the efficient particle acceleration and strong magnetic 
field amplifications can exist in collisionless shocks
in a wide range of the possible rates of nonlinear 
development of the streaming instability. In Section~\ref{res2007}, 
I present our article \cite{EV2008} that discusses
the impact of magnetic field amplification on the maximum energy
of the accelerated particles. We showed that the amplified
magnetic field does increase the maximum achievable particle
energy, but by a smaller factor the increase of the field.
Section~\ref{res2008} contains the results of our investigation
of the effect of turbulence dissipation in the shock precursor.
These results, presented in \cite{VBE2008}, show that
the conversion of turbulent energy into heat increases
the pre-shock temperature, which affects particle
injection into the acceleration process. In addition to
presenting the published articles, I included some work
in progress in this dissertation. 
In Section~\ref{res2009} I demonstrate the results of
the simulation of the nonlinear shock structure with
the nonresonant Bell's instability (see Section~\ref{bells_nonres_desc})
and the hybrid model of particle diffusion (see
Section~\ref{subsec_hybr_diff}). We find that, if turbulent
cascade is suppressed, the self-consistent steady state
shock structure has a stratified precursor, and the
spectrum of turbulence has an unusual multiple-peak structure.
Section~\ref{res_scalings} contains an outline of
the calculations that can be done with the simulation
in order to obtain simple power-law fits to the results
of nonlinear DSA. Such fits can be used in the models
of supernova remnants, galaxy cluster shocks, or other
objects where strong particle-accelerating shocks are present.
Finally, in Section~\ref{res_angular} I describe how
the escaping particle distribution can be fitted with simple
functions and where these fits may be used.

Considering the rapid growth of observational 
X-ray and gamma ray facilities
that reveal the `high energy Universe',
such as Chandra, Fermi, H.E.S.S., etc., I believe that
the development of this model is very timely and beneficial for the
research in different areas of astrophysics.

\bibliographystyle{plain}
\bibliography{etd}


\newpage
~

\vspace{0.2in}

\appendix
\addappheadtotoc

\begin{center}
{\Large
{\bf APPENDICES}
}
\end{center}

\newpage

\addcontentsline{toc}{subsection}{A. Numerical integrator
  for model with isotropization}
\headsep 0.8in
\section*{Appendix A \\ Numerical integrator for model with isotropization }

\setcounter{equation}{0}  
\renewcommand{\theequation}{A-\arabic{equation}}

In our works \cite{VEB2006}, \cite{EV2008} and \cite{VBE2008}, we
were including the generation by the streaming instability
of waves traveling in both direction, and used the version
of the wave amplification equation that accounts for the interaction
between these waves. In fact, this effect is only important
for the weaker shocks that we did not consider, and in the
more recent version we neglected the waves traveling downstream.
In order to make a record of the previous work,
I provide here the numerical integrator of the previously used model.

The equations we will now consider are (\ref{uminuskp}) and
(\ref{upluskp}). They include the resonant streaming instability
(generating and damping waves traveling in both direction),
the effect of wave amplitude increase with plasma compression,
and the nonlinear interactions between the waves traveling
in different directions.

These equations, by introducing quantities
\begin{eqnarray}
\xi(x,k_{j}) & = & \int\limits_{\Delta k_j}(U_-(x,k) + U_+(x,k))\: dk, \\ 
\eta(x,k_{j})& = & \int\limits_{\Delta k_j}(U_-(x,k) - U_+(x,k))\: dk,
\end{eqnarray}
are transformed into
\begin{eqnarray}
\label{xieq2007_copy}
u\xi' - v_w\eta' + \frac32u'\xi - v_w'\eta - 
        \frac{\eta}{\xi}v_w\frac{dP}{dx}&=&0, \\
\label{etaeq2007_copy}
u\eta' - v_w\xi' + \frac32u'\eta - v_w'\xi -
        v_w\frac{dP}{dx} + \frac{2}{\tau_r}\eta &=&0.
\end{eqnarray}
Here $dP/dx$ is the gradient of CR pressure produced by particles
resonant with waves in the bin $\Delta k_j$ (see resonance condition below).
Then the latter are re-written as
\begin{equation}
\label{compact_copy}
A \vec{y}' + B\vec{y}+\vec{c}=0,
\end{equation}
where
\begin{equation}
\begin{array}{c}
\vec{y}=\left( \begin{array}{cc} \xi \\ 
                                 \eta \end{array} \right), \quad
\\
A=\left( \begin{array}{cc} u & -v_w \\ 
                        -v_w & u \end{array} \right),\quad
B=\left( \begin{array}{cc} 
        \displaystyle\frac32u' & -v_w' \\ 
                         -v_w' & \displaystyle\frac32u'+ \frac{2}{\tau_r} \end{array} \right),\quad
\vec{c}=\left(\begin{array}{cc} \displaystyle-\frac{\eta}{\xi}v_w\frac{dP}{dx} \\
                                 \displaystyle-v_w\frac{dP}{dx} \end{array}
\right).
\end{array}
\end{equation}

If equation~(\ref{compact_copy}) was linear (i.e.,
the matrices $A$, $B$ and $\vec{c}$ did not depend on $\vec{y}$), 
then solving it would be straightforward. I will skip
the details and leave it to the reader to verify
that the solution of equation~(\ref{compact_copy})
with the initial condition 
\begin{equation}
\label{xinit}
\vec{y}(0)=\vec{y}_0,
\end{equation}
assuming that $A$ is reversible (otherwise, the system
is not consistent) is
\begin{equation}
\label{simpsolform}
\vec{y}_0(x) = \exp \left( -A^{-1}Bx \right) \vec{y}_0 
               - \left[ \int\limits_0^x \exp \left( - A^{-1}Bs \right) ds \right]
               A^{-1} \vec{c}.
\end{equation}
However, $\vec{c}$ explicitly depends on $\vec{y}$, and for $\fAlf>0$,
the magnetic field determining $v_w$ depends on the integral of
$U_{\pm}$ with respect to $k$ (see Section~\ref{res2006}), 
making the matrices $A$ and $B$
depend on $\vec{y}$ in a non-trivial way, and therefore
a numerical solution is required.

To integrate (\ref{compact_copy}), let us start off by assuming that 
$\vec{y}(x)=\vec{y}_0$ for any $x$.
Then let the integrator perform a `{\it level-1}' iterative procedure, the purpose of which is to deal with the fact that
in (\ref{compact_copy}) the matrices $A$ and $B$ depend on the unknown functions
$\xi(x,k)$, $\eta(x,k)$ in all $x$-space and $k$-space through $v_w(x)$ depending on $U_{\pm}(x,k)$. 
Here is what the `{\it level-1}' iterative procedure involves.
Given a $k$-bin, integrate
the equations (\ref{compact_copy}) for that bin from far upstream to downstream. 
The values of $U_-(x,k)$ and $U_+(x,k)$ used to form matrices $A$ and $B$ are the ones obtained from
the previous iteration. After all $k$-bins have been integrated, the iteration is over. Then the just obtained 
values of $\vec{y}(x,k)=(\xi(x,k), \; \eta(x,k))^T$ are 
used to run the next iteration. Iterating
on `{\it level-1}' ends when the current iteration gives results that are close enough to the results
of the previous iteration.

Integrating from one grid plane (at $x_0$) to the next one (at $x_1$),
the routine is not likely to encounter a very strong variation of $v_w$
determining the matrices $A$ and $B$ (because the latter depend
on an integral of $\vec{y}$ with respect to $k$), but
the quantities $\eta$ and $\xi$ that enter $\vec{c}$ may vary
by a large factor, and care must be take with using (\ref{simpsolform}).
I employ another iterative procedure
('{\it level-2}') to tend to the dependence of the vector $\vec{c}$,
on $\xi(x,k)$, $\eta(x,k)$ in (\ref{compact_copy}).
This iterative procedure is described below. First, divide the step from $x_0$ to $x_1$ 
into $N_{sub}$ equal substeps between the following points:
\begin{equation}
  x_{i_{sub}} = x_0 + (x_1 - x_0) \frac{i_{sub}}{N_{sub}}, \quad i_{sub} = 0\: .. \: N_{sub}.
\end{equation}
Then use (\ref{simpsolform}) to obtain $\xi(x_{i_{sub}},k)$ and 
$\eta(x_{i_{sub}},k)$ from $\xi(x_{i_{sub}-1},k)$ and $\eta(x_{i_{sub}-1},k)$.
Here $i_{sub}$ is the number of the substep. When $x_1$ is reached,
remember the values $\xi(x_1, k)$, $\eta(x_1, k)$
and increase $N_{sub}$ twice and do another iteration. Eventually,
stop iterating on `{\it level-2}' after
$N_{sub}$ becomes large enough so that the resulting pair $\xi(x_1, k)$, $\eta(x_1, k)$
obtained at the current iteration is close enough to that from the previous iteration.

What values should the integrator use at the `{\it level-2}' iteration in the vector $\vec{c}$ to
obtain $\vec{y}(x_{i_{sub}},k)$ from $\vec{y}(x_{i_{sub}-1},k)$?
The easiest way would be to form $\vec{c}$ from $\xi(x_{i_{sub}-1},k)$ and $\eta(x_{i_{sub}-1},k)$.
But expecting this explicit method to have little stability, as typical of such methods, I decided
to use an implicit method and to form $\vec{c}$ from $\xi(x_{i_{sub}},k)$ and $\eta(x_{i_{sub}},k)$.
Of course, the code does not know the values at $x_{i_{sub}}$ when it integrates from $x_{i_{sub}-1}$
to $x_{i_{sub}}$, which is what the explicit methods are all about. So I use a `{\it level-3}' iterative
procedure for that matter. First, assume that values of $\xi$ and $\eta$ at the end of the substep
are the same as at the beginning, and obtain the preliminary values at $x_{i_{sub}}$. These
values are then used to repeat the substep as many times as it takes to get $\vec{y}(x_{i_{sub}})$
at the current iteration close enough to the one in the previous iteration.

Throughout the solution, for I assume the following:
\begin{itemize}
  \item{Quantities $u(x)$, $v_w(x)$, $\Pcr(x, p)$, $\tau_r(x)$, which are
        defined at x-grid planes, are interpolated
        linearly in between the planes;}
  \item{Spatial derivatives of the above quantities, $u'(x)$, $v'_w(x)$, $P'_{cr}(x,p)$, 
        are uniform between the grid planes. 
        Their values correspond to the slopes of the linear interpolation
        of the above quantities;}
  \item{Wave speed $v_G$ according to Equation~(\ref{VG});}
  \item{The resonant wavenumber $\kres$ is
        related to the momentum $\pres$
        as $\kres\frac{c \pres}{e B_0}=1$. }
\end{itemize}

\newpage

\addcontentsline{toc}{subsection}{B. Diffusive flow incident on a moving absorbing boundary}
\headsep 0.8in
\section*{Appendix B \\ Diffusive flow incident on a moving absorbing boundary}

\appendix
\setcounter{equation}{0}  
\renewcommand{\theequation}{B-\arabic{equation}}

It was mentioned in Section~\ref{subsec_particleintro}
that particles must be introduced into the simulation as if they are
crossing the position, at 
which they are placed, for the first time in their
histories. The angular distribution of these particles is thus
equal to the angular distribution of particles diffusively
moving with respect to a flowing background medium,
and incident on a fully absorbing boundary (the flow
is directed into the boundary in our one-dimensional case).
If the speed of the flow, $u$, is greater
than the speed of the particles, with respect to the flow, $v$, then,
assuming isotropic distribution of particles in the reference frame
tied to the flow, their angular distribution may be written
as (\ref{fastpushslowdistr}). 
This is simple, because for $v<u$ all particles
cross every position in the flow just once, because
they cannot move upstream in the stationary reference frame.
However, for $v>u$, this situation becomes more complicated,
because particles are able to move forward as well as backward
(against the flow) and the flux of such particles on a fully
absorbing boundary is more difficult to estimate.

To solve this problem, I ran the Monte Carlo simulation,
injecting the particles far upstream and propagating them 
downstream till they cross the fully absorbing boundary at $x=0$
for the first time. I recorded the angular distribution of these
particles and fitted them with a simple scaling.
I chose to search for the distribution in the power law form
\begin{equation}
  F(v_\mathrm{sf,\:x}) =
  \left\{ \begin{array} {l l} 
    C v_\mathrm{sf,\:x}^{\alpha}, & \; \mathrm{if} \; v_\mathrm{min} < v_\mathrm{sf,\:x} < v_\mathrm{max},\\
    0, & \;\mathrm{otherwise}.
  \end{array} \right.
\end{equation}
where $v_\mathrm{sf,\:x}$ is the $x$-component of the incident
particle velocity measured in the shock frame (i.e., in the frame
in which the absorbing boundary is at rest),
$v_\mathrm{min}=0$, $v_\mathrm{max}=u+v$, and $C$ is found from condition
\begin{equation}
  \int\limits_{v_\mathrm{min}}^{v_\mathrm{max}} F(v_x) \; dv_x = 
  \int\limits_{v_\mathrm{min}}^{v_\mathrm{max}} C v_\mathrm{sf,\:x}^{\alpha} \; dv_x = 1
\end{equation}
as
\begin{equation}
  C = \frac{\alpha + 1}{v_\mathrm{max}^{\alpha + 1} - v_\mathrm{min}^{\alpha + 1}},
\end{equation}
making
\begin{equation}
  \label{singleplang}
F(v_\mathrm{sf,\:x}) = 
  \left\{ \begin{array} {l l} 
    \displaystyle\frac{\alpha + 1}{v_\mathrm{max}^{\alpha + 1} - v_\mathrm{min}^{\alpha + 1}} v_\mathrm{sf,\:x}^{\alpha}, 
    & \; \mathrm{if} \; v_\mathrm{min} < v_\mathrm{sf,\:x} < v_\mathrm{max},\\
    0, & \; \mathrm{otherwise}.
  \end{array} \right.
\end{equation}
In order to derive $\alpha$, one needs to minimize the following function of $\alpha$ to find
the least squares fit:
\begin{equation}
  \Delta(\alpha) = 
    \sum\limits_{i = 1}^{N} 
      \left[ \frac{\alpha + 1}{v_\mathrm{max}^{\alpha + 1} - v_\mathrm{min}^{\alpha + 1}} v_i^{\alpha} - f_i \right]^2,
\end{equation}
where the index $i$ runs over all numerical bins of the speed
$v_\mathrm{sf,\:x}$, the values $v_i$ are the centers of these bins,
and $f_i$ is the properly normalized fraction of incident particles 
that had the $x$-component of velocity in the $i$-th bin upon their
incidence. I used a bracketing method to find the minimum, 
searching for it in the region $\alpha \in [0.5, 2.0]$.

In order to collect the data, I  ran 30 simulations, introducing
particles that were mono-energetic in the plasma frame, with a 
speed $v$, into a flow with speed $u$. The angular distribution that
I used for these particles did not matter, because they
were given enough time to scatter in the flow an isotropize
before they reached the absorbing boundary. 
I covered the range $v/u \in [1 \dots 15]$. 
For each such run, I fitted the angular distribution of
particles first entering the shock with a single parameter power law and derived an $\alpha$ for
this run. Then I plotted the resulting
power law index $\alpha$ versus the ratio $v/u$ of a run. 
The resulting curve $\alpha(v/u)$ can be
described by the following simple equation:
\begin{equation}
  \label{fittedalpha}
  \alpha\left(\frac{v}{u}\right) = 
  1.5 - 0.5\cdot\left(\frac{v}{u}\right)^{-1.15}.
\end{equation}

The angular distribution function~(\ref{singleplang})
with $\alpha$ given by (\ref{fittedalpha}) is simulated in 
the code in order to introduce particles with a plasma
frame speed $v$ greater than the local flow speed $u$.
Note that for $v \gg u$, the power law index approaches $\alpha \to 1.5$, and for
$v \to u$ the power law index approaches $\alpha \to 1.0$, and it stays $\alpha = 1$
for $v < u$, where (\ref{fastpushslowdistr}) becomes applicable. 
The last statement was demonstrated separately, in other simulations, and is obvious: 
for small $v$ there are no backward-moving particles in the shock frame, 
so every particle crossing a plane crosses it for the first and the
last time.

\end{document}